\documentclass[a4paper,12pt,twoside]{report}
\usepackage[left=2.5cm,right=2cm,top=2cm,bottom=2cm]{geometry}

\usepackage[dvipsnames]{xcolor} 

\usepackage[normalem]{ulem}
\useunder{\uline}{\ul}{}
\usepackage{longtable}

\usepackage{tcolorbox}
\tcbuselibrary{skins}

\newtcolorbox{conversationbox}{
  enhanced,
  colback=white,
  colframe=blue!75!black,
  size=small,
  sharp corners,
  boxrule=1pt,
  fonttitle=\bfseries,
  title=Using vignettes from \cite{de2016crowd} to prompt GPT-4
}

\usepackage[utf8]{inputenc}
\usepackage{url}
\usepackage{hyperref}
\usepackage{amsfonts}
\usepackage{amssymb}
\usepackage{graphicx}
\usepackage{float} 
\usepackage{subcaption}
\usepackage{listings}
\usepackage{color}
\usepackage{fancyhdr} 
\usepackage{bibentry} 
\usepackage{enumitem}
\usepackage{setspace}               

\definecolor{mylightgrey}{HTML}{DCDCDC}
\definecolor{mydarkgrey}{HTML}{808080}

\usepackage{xcolor}
\usepackage{dirtytalk}
\usepackage{csquotes}
\usepackage{makecell}
\usepackage{svg}
\usepackage{booktabs}

\usepackage{makecell}

\nobibliography* 

\newtcolorbox{boxA}{
    boxrule = 1pt,
    colframe = black 
}

\definecolor{mygreen}{rgb}{0,0.6,0}
\definecolor{mygray}{rgb}{0.5,0.5,0.5}
\definecolor{mymauve}{rgb}{0.58,0,0.82}
\definecolor{minor}{HTML}{000000}
\definecolor{major}{HTML}{000000}

\definecolor{thesis}{HTML}{000000}

\newcommand{\reporttitle}{Incorporating Different Verbal Cues to Improve  Text-Based Computer-Delivered Health Messaging}
\newcommand{\reportauthor}{Samuel Rhys COX}
\newcommand{\supervisor}{Associate Professor\\ Wei Tsang OOI
\\\textit{Co-supervisor:}\\
Assistant Professor\\ Brian Y. LIM
\\\textit{Examiners:}\\
Associate Professor Zhao Shengdong\\
Associate Professor Bimlesh Wadhwa}

\begin{document}

\begin{titlepage}

\newcommand{\HRule}{\rule{\linewidth}{0.5mm}} 


\includegraphics[width = 4cm]{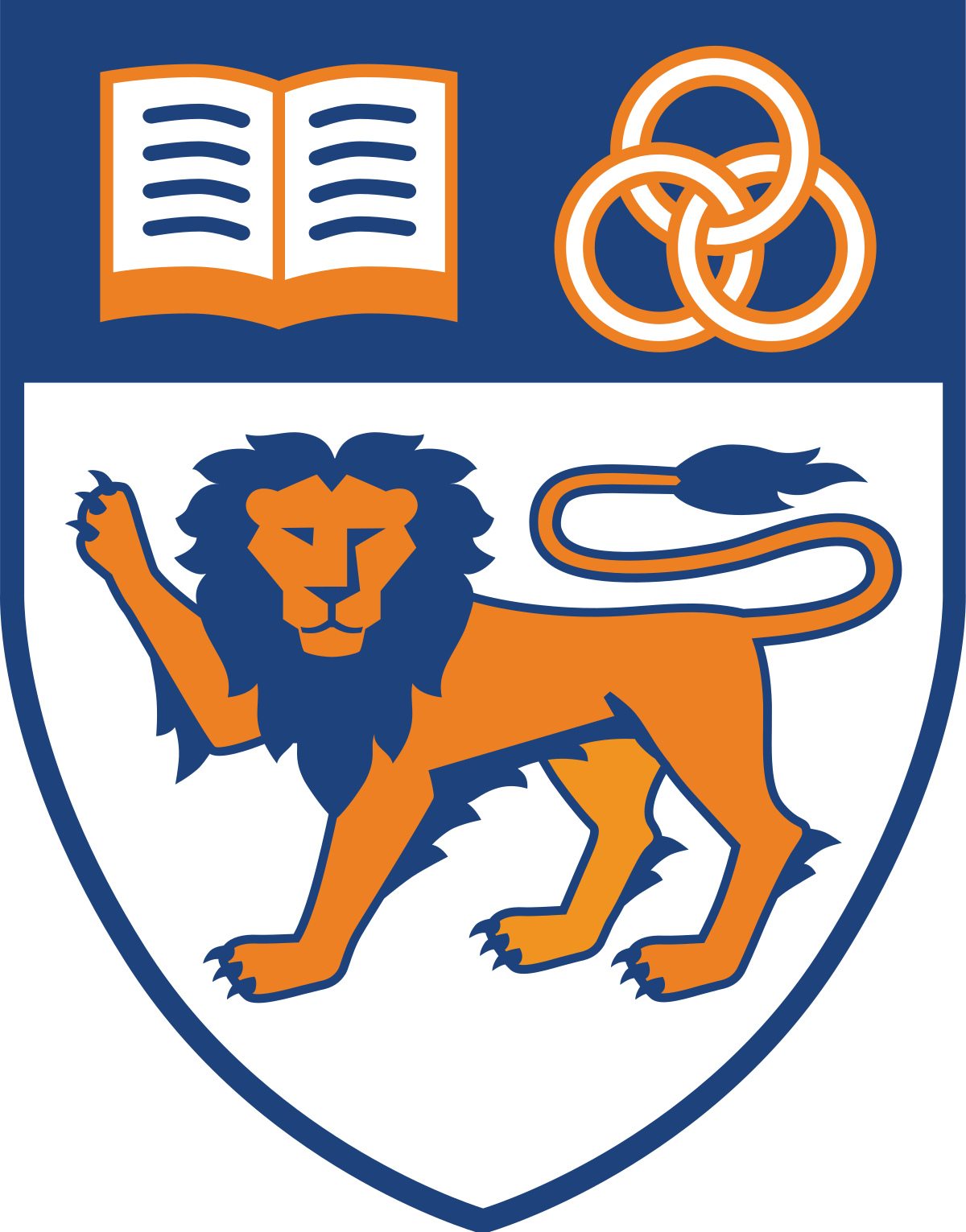}\\[0.5cm] 

\center 

A THESIS SUBMITTED 
FOR THE DEGREE OF DOCTOR OF PHILOSOPHY
\textsc{\large Department of Computer Science}\\
\textsc{\large National University of Singapore}\\[0.5cm] 


\HRule \\[0.4cm]
{ \huge \bfseries \reporttitle}\\ 
\HRule \\[1.5cm]
 

\begin{minipage}{0.4\textwidth}
\begin{flushleft} \large
\emph{Author:}\\
\reportauthor 
\end{flushleft}
\end{minipage}
~
\begin{minipage}{0.4\textwidth}
\begin{flushright} \large
\emph{Supervisor:} \\
\supervisor 
\end{flushright}
\end{minipage}\\[4cm]

\vfill 


\makeatletter
November, 2023 
\makeatother

\end{titlepage}

\begin{abstract}


The ubiquity of smartphones has led to an increase in on demand healthcare being supplied. 
For example, people can share their illness-related experiences with others similar to themselves, and healthcare experts can offer advice for better treatment and care for remediable, terminal and mental illnesses. 
As well as this human-to-human communication, there has been an increased use of human-to-computer digital health messaging, such as chatbots. 
These can prove advantageous as they offer synchronous and anonymous feedback without the need for a human conversational partner. 
However, there are many subtleties involved in human conversation that a computer agent may not properly exhibit. 
For example, there are various conversational styles, etiquettes, politeness strategies or empathic responses that need to be chosen appropriately for the conversation. 
Encouragingly, computers are social actors (CASA) posits that people apply the same social norms to computers as they would do to people. 
On from this, previous studies have focused on applying conversational strategies to computer agents to make them embody more favourable human characteristics.
However, if a computer agent fails in this regard it can lead to negative reactions from users.
Therefore, in this dissertation we describe a series of studies we carried out to lead to more effective human-to-computer digital health messaging. 


In our first study, we use the crowd to write diverse motivational messages to encourage people to exercise.
Yet, it is not a simple case of asking the crowd to write messages, as people may write similar ideas leading to duplicated and redundant messages.
To combat this, previous work has shown people example ideas to direct them to less commonly thought of ideas. However, these studies often relied on manually labelling and selecting example ideas.
Therefore, we developed \textit{Directed Diversity}, a system which uses word embeddings to automatically and scalably select collectively diverse examples to show to message writers.
In a series of user studies we asked people to write messages either using examples selected by \textit{Directed Diversity}, random examples, or no examples.
We evaluated the diversity of messages using a selection of diversity metrics from different domains, human evaluations, and expert evaluations, and found that our method led to more diverse messages compared to the baseline conditions.
Additionally, subjective evaluations of the messages found that directed diversity generated more informative and equally motivating messages compared to the baseline conditions.

Our second study investigates the effect of a health chatbot’s conversational style on the quality of response given by a user. 
The tone and language used by a chatbot can influence user perceptions and the quality of information sharing. This is especially important in the health domain where detailed information is needed for diagnosis. 
In two user studies on Amazon Mechanical Turk, we investigated the effect of a chatbot's conversational style (formal vs casual) on user perceptions, likelihood to disclose, and quality of user utterances. In general, results indicate the benefits of using a formal conversational style within the health domain. 
In the first user study, 187 users were asked for their likelihood to disclose either their income level, credit score, or medical history to a chatbot to assist finding life insurance. When the chatbot asked for medical history, users perceived a formal conversation style as more competent and appropriate. 
In the second user study, 156 users discussed their dental flossing behaviour with a chatbot which used either a formal or casual conversational style. We found that users who do not floss every day wrote more specific and actionable user utterances when the chatbot used a formal conversational style.

In our final study, we investigate the format used by a chatbot when referencing a user's utterance from a previous chatting session.
Our motivation for this study is driven in part by the Personalisation-Privacy Paradox: a tension between providing personalised services to users to improve their experience, and providing less personalised services so as not to seem invasive to users.
To explore this trade-off, we investigated how a chatbot references previous conversations with a user and its effects on a user’s privacy concerns and perceptions.
In a three-week longitudinal between-subjects study, 169 participants talked to a chatbot that either, (1-None): did not explicitly reference previous user utterances, (2-Verbatim): referenced previous utterances verbatim, or (3-Paraphrase): used paraphrases to reference previous utterances.
Participants found Verbatim and Paraphrase chatbots  more intelligent and engaging. However, the Verbatim chatbot also raised privacy concerns with participants.
Upon study completion, we conducted semi-structured interviews with 15 participants to gain insights such as to why people prefer certain conditions or had privacy concerns.
These findings can help designers choose an appropriate form of referencing previous user utterances and inform in the design of longitudinal dialogue scripting.

In summary, we have researched how to create more effective digital health interventions starting from generating health messages, to choosing an appropriate formality of messaging, 
and finally to formatting messages which reference a user's previous utterances.

\end{abstract}

\renewcommand{\abstractname}{Acknowledgements}
\begin{abstract}
I would like to offer my thanks to the following for their continued guidance and support:

\begin{itemize}
\item Associate Professor Wei Tsang Ooi
\item Assistant Professor Brian Y. Lim
\item My parents, Grace, Ashraf, Bob, Therisa and Anna.
\end{itemize}

\end{abstract}

\pagenumbering{roman}
\tableofcontents
\newpage

\listoffigures

\newpage


\pagenumbering{arabic}

\chapter{Introduction}
\label{ch:introduction}


This thesis investigates the use of \textcolor{thesis}{verbal cues} in computer driven text-based messaging to assist users in developing healthy habits, and describes several of our studies that answer related research questions.
In this introduction chapter, Section \ref{ch1:motivation} describes the motivation behind our work, and the context behind each of our studies.
Section \ref{sec:RQs} lays out our research questions, and explains the methods used to answer them.
Finally, Section \ref{ch1:outline} provides an outline of the chapters in this report.

\section{Motivation}
\label{ch1:motivation}

As part of a healthy lifestyle and to avoid chronic diseases, it is recommended that we adhere to various health guidelines \cite{rollo2020whole,fiuza2013exercise,herforth2019global,healthhub3,ohayon2017national}.
These could include getting sufficient exercise \cite{rollo2020whole,fiuza2013exercise}, eating a balanced and healthy diet \cite{herforth2019global,healthhub3},
\textcolor{thesis}{flossing our teeth \cite{schuz2006adherence,graves1989comparative,corby2008treatment}}
or getting enough sleep \cite{ohayon2017national}.
To help us adhere to these recommendations, we can turn to social support: the provision of assistance or comfort, such as sharing advice or emotional support \cite{slaughterSocialSupport1990,brown1986social}.
Social support can help us perform more healthy behaviours \cite{brown1986social,cohen2007social}, in part by improving our motivation \cite{schaefer1981health} and self-efficacy \cite{duncan1993social}.



Traditionally, this social support would come from human connections such as health experts, friends, or family \cite{griffith1985social}, and could be provided in person or via messages \cite{irwin1994support}.
However, this support pool is finite and may not be readily available to everyone due to lack of social connections, financial constraints, or time constraints \cite{fiorillo2020reasons}.
Fortunately, a solution to this gap in human support can come from the use of health messaging technologies \cite{adams2014staccato,ta2020user,benavente2023replacing}, 
which can provide immediacy in their responses to users \cite{pereira2019using}.



Health messaging technologies are already publicly available \cite{wasil2022there,tudor2020conversational,chew2022use}, and could message users via nudging interventions (e.g., push-notifications to remind someone to exercise \cite{rabbi2015mybehavior,gouveia2015we,hardeman2019systematic}), or via longer two-way conversations (e.g., chatbots to discuss someone's cancer treatment \cite{chaix2019chatbots,bibault2019chatbot}).
Additionally, the technology that makes this possible is becoming accessible to more people, and (as of 2020) nearly 50\% of the world population owns a smartphone \cite{odea_2021}.
This gives the option for low-cost and highly accessible support for people with their health needs \cite{pereira2019using,kowatsch2017design}, as the technology is scalable provided that infrastructure (reliable Internet or phone service) exists.
Furthermore, health messaging technologies have seen some acceptance among both end-users \cite{nadarzynski2019acceptability,griffiths2020acceptability,hardeman2019systematic} (such as chatbots for anonymous symptom sharing \cite{nadarzynski2019acceptability}) and medical professionals \cite{palanica2019physicians} (such as for information sharing and appointment scheduling). 
However, both physicians and consumers have raised lack of empathy and inability to understand human emotions as potential barriers to engagement with health messaging technology \cite{nadarzynski2019acceptability,palanica2019physicians}.

These concerns lead us to a potential impasse. Namely, the question of ``\textit{How} should technology communicate with us?''.
We want technology which people perceive as understanding their emotions and that responds empathetically. 
However, we do not want technology that enters the uncanny valley \cite{ho2008human,wozniak2021creepy,stein2017venturing} where communication is seen as \textit{almost} human-like, but something is not quite right.
Equally we do not want technology that responds in a way seen as inappropriate to a given situation \cite{cox2022does,chaves2019s}.
\textcolor{thesis}{For example, users may find it less appropriate if a chatbot uses a casual conversational style when discussing their sensitive health information \cite{cox2022does}.
In this case conversational style is an example of a \textit{social cue}, a signal that saliently conveys meaning to people and leads them to apply social rules, expectations and scripts to interactions \cite{feine2019taxonomy}.
Social cues can be split into those that are \textit{verbal cues} (such as \textbf{conversational style} \cite{cox2022does} and \textbf{conversational content} \cite{cox2023referencing,cox2021diverse,recanati2001said}); \textit{auditory cues} \cite{burgoon2016nonverbal} (such as voice qualities and vocalisations); \textit{visual cues} \cite{leathers2015successful} (such as agent appearance, proxemics, and kinesics); and \textit{invisible cues} \cite{feine2019taxonomy} (such as chronemics and haptics).}

\textcolor{thesis}{Effects of social cues and their resulting social scripts led to the development of the interaction paradigm \textit{Computers are Social Actors} (CASA) \cite{nass1994computers,gambino2020building}.
CASA states that people mindlessly apply the same social heuristics to computers as they do to humans, by following associated social cues.}
For example, people may carry forward their own existing gender biases when interacting with an agent \cite{mcdonnell2019chatbots}, or may expect politeness conventions to be upheld under certain contexts \cite{brown_politeness_1987,chaves2021should}.
The CASA paradigm has thereby encouraged people to develop computers that either: communicate using favourable human traits (such as a tone-aware chatbot \cite{hu2018touch}), or exploit human biases (such as a chatbot that mirrors a user's race \cite{liao_racial_2020}) to improve the effectiveness of interactions.

This desire to create more effective (such as motivational or appropriate) computer-delivered communication is a primary motivation behind our studies. Specifically, we want to create computer-delivered health messaging that is perceived more positively by people, so as to better help them meet their health needs.
We focus on text-based computer-delivered verbal cues used in messaging both at the conception stage (with the generation and curation of efficacious messages), and the end-user stage with empirical studies to investigate verbal cues that make messaging more effective. 
We focus on verbal cues in particular as the content and style of conversational messages is a primary driving force behind user perceptions and behaviour in response to computer-delivered health messaging. On from this, in our studies we investigate both the conversation content (\textit{what} is said), as well as the conversational style (\textit{how} it is said).

\section{Research Questions}
\label{sec:RQs}

This dissertation aims to improve the effectiveness of verbal cues in text-based computer-delivered health messaging.
We believe that messaging can be improved: (1) in the way it is generated; (2) by adopting an appropriate conversational style for health communication; and (3) by altering the presence and format of a chatbot's references to previous user utterances.
The high-level research question of this dissertation is therefore:

\begin{displayquote}
    \textbf{How can we make text-based computer-delivered verbals cues in health messaging more effective?}
\end{displayquote}

Below we will outline our three sets of studies in more detail, alongside their high-level research questions.

\subsection{Generating Diverse and Efficacious Health Messages}

Our first study investigates how text-based health messages can be generated that are both \textit{efficacious} and \textit{diverse}.
By generating diverse text-based health messages, we collect diverse verbal cues: i.e., conversational content and conversational styles of messages.
One method to generate large amounts of health content is to use collections of people hereby referred to as ``the crowd'' \cite{de2016crowd}.
Previous work has studied how to use the crowd to generate health content such as motivational messages \cite{de2016crowd}, exercise plans \cite{agapie2016plansourcing}, and solutions to back pain \cite{hosio2018crowdsourcing}. 
Using the crowd is low cost (compared to using experts), and can generate personable and relatable content by drawing on people's personal experiences \cite{coley2013crowdsourced}.
The ability to produce large quantities of diverse and efficacious health content is particularly useful for interventional messaging or chatbots, where large numbers of messages are needed both to avoid user boredom in repeated interventions \cite{kocielnik2017send,cacioppo1979effects,strecher2005randomized}, and in order to train algorithms for health chatbot systems \cite{yaghoub2020user}.



However, the crowd may produce similar or identical ideas, that leads to redundancy and wasted human effort \cite{bayus2013crowdsourcing,bjelland2008inside}.
Previous work within crowdsourcing and creativity have shown example ideas to the crowd to help them generate diverse ideas \cite{siangliulue2016ideahound,huang2017bluesky}, but these systems often rely on manual human effort to run. Therefore, we want to use the crowd to generate \textit{diverse} and \textit{efficacious} health messages, while also using a \textit{scalable} and more \textit{automatic} system.
This leads us to our first research question:

\begin{displayquote}
    \textit{\textbf{RQ1:} How can we use the crowd to automatically and scalably generate diverse verbal cues in health messaging?}    
\end{displayquote}

To help answer this we developed \textit{Directed Diversity}  \cite{cox2021diverse}, a method that harnesses embedding distances to scalably and automatically allocate diverse example ideas to crowd-workers.
By giving diverse examples to workers, we generate a \textit{collectively diverse} corpus of messages.
In our user studies we showed users short phrases related to physical exercise (e.g., ``hard workout may feel'') to inspire workers to write motivational messages to encourage exercise.
To automate our process, we used existing online health content to generate the examples shown to workers, and developed an algorithm to select semantically diverse examples to show to workers.
To measure the efficacy of our crowd-generated messages, we asked people to hypothetically answer how they themselves would find the message if it was a notification on their phone.
We also used human evaluations and objective measures to evaluate the diversity of messages.
It was found that our method helped the crowd generate diverse content when compared to both randomly selected examples and no examples, while still generating efficacious motivational messages. This study is described in detail in Chapter \ref{ch:DirectedDiversity}.


\subsection{Choosing an Appropriate Conversational Style for a Health Chatbot}

In our second study \cite{cox2022does}, we investigated the effect of a health chatbot's conversational style on a user's quality of information disclosure.
In order to help people improve their health, chatbots need to be given enough information. However, people may be unwilling to disclose their information if a chatbot is perceived negatively \cite{palanica2019physicians}.
Previous work has investigated the effect of conversational style on quality of multiple-choice survey responses, both with a standard survey and with a survey chatbot \cite{kim_comparing_2019,celino_submitting_2020,wambsgnass_conversational_2020}.
However, we are interested in the effect of conversational style on the quality of disclosure to a health chatbot, in terms of likelihood to disclose and quality of user utterances.
This led us to our second research question:

\begin{displayquote}
    \textit{\textbf{RQ2:} How does the conversational style (i.e., language formality) of a health chatbot affect the quality of a user's self-disclosure?}    
\end{displayquote}

On from this, we investigated the effect of a health chatbot's conversational style on the quality of responses from users \cite{cox2022does}.
We conducted two user studies to investigate the impact of formal and casual conversational styles on human perceptions (such as likelihood to disclose, and perceptions of the chatbot) and the quality of responses given to the chatbot. 
In our studies users perceived a formal conversational style to be more appropriate and competent, and a formal conversational style was found to produce higher quality user utterances. This study is described in detail in Chapter \ref{ch:FormalOrCasual}.


\newpage
\subsection{Choosing an effective format for a health chatbot to reference a user's previous utterances}


Health interviewers have been found to be effective both by adopting higher levels of social presence \cite{tsaihuman,xiao2020tell} (such as paraphrasing a user's utterances) or by adopting lower levels of social presence \cite{schuetzler2018influence,chen2021you,ng2020simulating} (such as by providing high-level references to user utterances).
These two conflicting schools of thought lead us to investigate the effect of varying levels of a health chatbot's social presence 
by changing the format used when referencing a user's previous utterances.
This leads us to our third research question:

\begin{displayquote}
    \textit{\textbf{RQ3:} How does the format used when a health chatbot references a user's utterances from previous chatting sessions impact a user's perception of a chatbot and why?}    
\end{displayquote}


As a result of this, in our third study \cite{cox2023referencing}, we conducted a between-subjects longitudinal study where we compared three different formats that a chatbot could use when referencing user utterances from previous chatting sessions.
We found that a format that referenced users' past utterances explicitly made the chatbot seem more intelligent and engaging, but also raised privacy concerns. We then conducted semi-structured interviews to further investigate \textit{why} people viewed the referencing formats differently.
This study is described in Chapter \ref{ch:Personalisation}.


\section{Thesis Outline}
\label{ch1:outline}

In this thesis, I will outline background literature, as well as the three sets of studies that we have completed. 
The structure of this report is outlined below:

\subsubsection{Chapter \ref{ch:background}: Background Research and Related Work}

This chapter investigates the topics related to the project and outlines any relevant research. 
In our related work chapter, we contextualise our three studies by providing a background overview of health communication technology.
As we are interested in text-based health messaging, we first start by giving an overview of health message generation and curation.
We then describe previous work that has attempted to improve health messaging technology, and more specifically are interested in text-based communication such as nudges or conversations to encourage healthy behaviours.




\subsubsection{Chapter \ref{ch:DirectedDiversity}: Generating Diverse and Efficacious Health Messages}

Our first studies in Chapter \ref{ch:DirectedDiversity} draw on the crowd to generate diverse and efficacious health messages, that could be used as interventions or to train chatbots. 
For this, we developed and evaluated a method to scalably and automatically generate more diverse and efficacious health messages by using embedding distances to send crowd-workers collectively diverse phrases to inspire their ideation \cite{cox2021diverse}.
This study is published as:

\bibentry{cox2021diverse}.

\subsubsection{Chapter \ref{ch:FormalOrCasual}: The Effects of a Chatbot's Language Formality on a Users' Information Sharing}

Our studies in Chapter \ref{ch:FormalOrCasual} draw a connection between health chatbots and their users. 
More specifically, we investigated the effect of a health chatbot's conversational style, and found that a formal conversational style may lead to higher quality user responses \cite{cox2022does}. 
This study is published as:

\bibentry{cox2022does}.

\subsubsection{Chapter \ref{ch:Personalisation}: The Effect of how a Health Chatbot Formats and References a User's Previous Utterances}

Finally, our study in Chapter \ref{ch:Personalisation} investigated the effect of varying how a chatbot refers to user utterances from a previous chatting session, and we found that, while people think more positively of chatbots that explicitly reference their previous utterances, they also find them more privacy violating.
This study is published as:

\bibentry{cox2023referencing}.

\subsubsection{Chapter \ref{ch:conclusion}: Conclusion and Future Work}

In Chapter \ref{ch:conclusion}, we offer concluding remarks, we reflect on our findings and implications, and offer perspective about future work.

\vspace{10mm}


\newpage
\pagestyle{fancy}
\fancyhf{}
\fancyhead[EL]{\nouppercase\leftmark}
\fancyhead[OR]{\nouppercase\rightmark}
\cfoot{\thepage}

\chapter{Background Research and Related Work}
\label{ch:background}


In this chapter, we will outline the scope of literature in relation to the research that we have undertaken. 
We will provide an overview of literature from the message generation process, to the content of messages themselves, and to the appropriate timing for messaging to take place.

\section{How should messaging be generated?}
\label{sec:HowMessagesGenerated}

Before any health messaging can take place, we need either a bank of existing health messages or an appropriate method to create health messages.
Messages could be in the form of one time interventions to encourage a healthy behaviour (such as SMS messages or app notifications \cite{coley2013crowdsourced,rabbi2015mybehavior,li_micro-randomized_2020}); or messages that emulate a conversation with another human (such as chatbot utterances \cite{park_designing_2019,liao_racial_2020}).
These messages can be generated by humans \cite{coley2013crowdsourced,hosio2018crowdsourcing,kristan2015using,agapie2016plansourcing,agapie2018crowdsourcing}
or by machines through natural language generation (NLG) \cite{lin2019caire,lin2020caire,rashkin2018towards,lee2023benefits,bubeck2023sparks}.

\subsection{Machine-generated messages}
\label{sec:machine-generated}
Machine-generated messages have been used to generate empathic responses such as for customer service \cite{hu2018touch} and general chitchat \cite{lin2020caire,lin2019caire}.
While NLG does not rely on people to write messages directly, it requires a large body of training data \cite{rashkin2018towards}, human feedback (such as subjective human ratings for reinforcement learning from human feedback used in commercial LLMs \cite{zhao2023survey,christiano2017deep,ziegler2019fine}), and the reliability of messages (such as appropriateness, grammaticality, tone and accuracy) may be difficult to ensure \cite{smith2020can,lee2023benefits}.

However, with the rapid rise and advancement of large language models (LLMs) \cite{zhao2023survey}, chatbots have become increasingly capable in responding to user queries and can be successfully applied to a variety of creative tasks and problem solving \cite{bubeck2023sparks,taylor2022galactica,cox2023gaming}, such as providing medical advice \cite{yunxiang2023chatdoctor,chen2023utility}. 
Yet (at the time of writing), while these LLMs (such as GPT-4 \cite{bubeck2023sparks}, LaMDA \cite{thoppilan2022lamda}, and more \cite{zhao2023survey}) are able to adapt to a variety of user tasks, ``hallucinated'' responses may be generated whereby the LLM generates often plausible sounding but factually incorrect statements \cite{bubeck2023sparks}.
Cases of AI hallucination within healthcare are already documented \cite{bubeck2023sparks,lee2023benefits,taylor2022galactica,chen2023utility,jeblick2022chatgpt}, such as LLMs misunderstanding medical vocabulary or providing advice that does not follow medical guidelines \cite{chen2023utility,jeblick2022chatgpt}.
For example, Lee et al. described the use of GPT-4 to act as a chatbot for medicine, whereby the chatbot claimed to have and a master's degree and family members suffering from diabetes \cite{lee2023benefits} (see also Figure 1.8 p. 12  \cite{bubeck2023sparks} for two examples of GPT-4 hallucination in the healthcare domain). 

These potential inaccuracies put the user at risk if incorrect health advice is given, and lower the acceptability of AI-driven health chatbots \cite{nadarzynski2019acceptability} or health interventions \cite{sekhon2017acceptability}. 
Nadarzynski et al. conducted in person interviews and an online survey to assess the acceptability of AI-driven health chatbots \cite{nadarzynski2019acceptability}. Interviews found that participants may be hesitant to adopt healthcare chatbots as there was a perceived risk due to miscommunication between the chatbot and its users, with some participants describing their concerns that users may not be able to describe their health issues to a chatbot in a way which the chatbot can understand. Additionally, some interviewees perceived a risk of harm to users if inadequate or inaccurate information was provided by a chatbot, and expressed concern that AI-driven chatbots may not understand the emotions of users.
Yet, as limitations (such as ``hallucinations'') are inevitably overcome the adoption of LLMs to healthcare is a promising and exciting development, and (with ongoing positive mainstream news) acceptance of healthcare chatbots may become more commonplace.

\subsection{Human-generated messages}
Human-generated messages can either be generated by experts (such as health experts or a research team), or by a collection of non-experts which will we refer to as ``the crowd'' \cite{de2017experts}. 
Expert-generated health messages have been used in studies, such as for interventions \cite{li_micro-randomized_2020,rabbi2015mybehavior} or retrieval-based chatbot scripts \cite{park_designing_2019}.
This content has the reliability of an expert, and may be grounded in health theory. 
However, it is expensive to generate using experts and not easily scalable.

\section{Crowd-generated messages: Can the crowd perform as well as experts?}
\label{sec:crowd-generated}
In contrast, the crowd can be used to generate large quantities of content more scalably (although caution needs to be made to ensure the quality of content generated \cite{kittur2013future}).
The crowd could consist of a collection of peers who suffer from similar health issues \cite{coley2013crowdsourced,hosio2018crowdsourcing,kristan2015using,agapie2016plansourcing}, or a collection of individuals who are paid to complete a task online (such as over the platforms Amazon Mechanical Turk or Prolific) \cite{agapie2018crowdsourcing,de2016crowd}.
Past studies have used the crowd to generate generate utterances for chatbots \cite{yu2016chatbot,jonell2018fantom,xiao2020tell}, and motivational messages to encourage healthy behaviours \cite{de2016crowd,coley2013crowdsourced,kocielnik2017send,kristan2015using,de2019you}.

To compare these two methods of human generation, past studies have compared the efficacy and diversity of crowd and expert-generated health content \cite{coley2013crowdsourced,kristan2015using,de2017experts,agapie2018crowdsourcing,agapie2016plansourcing}.
Coley et al. asked both experts and smokers (the crowd) to write smoking cessation messages \cite{coley2013crowdsourced}, and found that crowd-generated messages were more diverse than those written by experts. They hypothesised that this is due to the crowd incorporating more diverse past experiences into their writing.
Kristan et al. asked both experts and young adults (the crowd) to write messages to reduce alcohol consumption \cite{kristan2015using}. 
Young adults then rated messages as more motivating if they were peer-generated by the crowd, leading Kristan et al. to posit that messages written by peers more ``authentic'' and less ``manipulative'' than those from experts.
De Vries et al. used both experts and the crowd to generate messages to motivate physical activity, and found that (for certain users) motivational messages written by the crowd are as motivating as those written by experts \cite{de2017experts}.

\section{Downsides of the crowd, and how to overcome them}
While these studies may suggest that the crowd should be used in lieu of experts, importance needs to be placed on designing crowdsourcing systems to avoid low quality input through in-built quality control \cite{kittur2013future}. 
Past studies have used fault-tolerant sub-tasks \cite{bernstein2010soylent,kittur2011crowdforge,noronha2011platemate} and applied agreement filters and peer-review \cite{bernstein2010soylent,kittur2011crowdforge,von2004labeling,dow2012shepherding} to counteract this. 


Giving feedback to the crowd can be used to improve the quality of generated content \cite{dow2012shepherding}. 
Mamykina et al. suggest showing expert and peer generated feedback, or showing answers from other users to improve crowdsourcing \cite{mamykina2016learning}. 
Agapie et al. found that exercise plans crowdsourced using scaffolding were in some cases as (or more) effective as those produced by experts \cite{agapie2018crowdsourcing}. 
Using multiple design iterations within crowdsourcing has been investigated in a number of studies \cite{dow2010parallel,little2010exploring}. ``Serial'' iteration can be used to repeatedly refine a piece of content. However this can lead to reaching a local rather than global optimal solution \cite{buxton2010sketching}. Alternatively, refining multiple alternatives in ``parallel'' iterations has been shown to increase content diversity \cite{dow2010parallel}, and generated more authentic and diverse feedback from individuals \cite{tohidi2006getting}. As well as this, previous work has investigated the effectiveness of providing live feedback to crowdworkers \cite{dow2012shepherding,lee2018exploring}. Lee et al. generated design guidelines to assist in the running of crowdsourcing studies featuring live-feedback to collaborate \cite{lee2018exploring}.


Additionally, the crowd can produce similar or identical ideas, which leads to redundancy and wasted human effort \cite{bayus2013crowdsourcing,siangliulue2016ideahound,bjelland2008inside,klein2015high,riedl2010rating}.
This causes additional issues with regards to health messaging.
Since behaviour change is typically a long-term process, messaging needs to be repeated over a long period of time \cite{consolvo2014designing}. However, seeing the same or similar messages repeatedly can lead to boredom, annoyance, content blindness or even avoidance, resulting in poor adherence and dropouts \cite{cacioppo1979effects,strecher2005randomized}. Hence, researchers have called for the need to diversify messaging \cite{kocielnik2017send,pechmann1988advertising} and have offered practical methods to create diversity, such as Kocielnik and Hsieh's \cite{kocielnik2017send} use of strategies based on cognitive space.
Kocielnik and Hsieh \cite{kocielnik2017send} used writing prompts to produce diverse messages to encourage physical exercise based on target-diverse (cognitively close to target action) and self-diverse (cognitively close to message recipient) messages. They found cognitively diverse messages, maintained the same quality and meaning, but did not encounter the same negative effects of repeated exposure.
Additionally within creativity literature methods have been proposed which give crowd-workers examples to help generate diverse content \cite{siangliulue2016ideahound,huang2017bluesky,girotto2019crowdmuse}.
For example, Siangliulue et al. asked participants to manually categorise ideas as they generated them, which allowed their system to show participants example ideas that were different to their previously generated ideas \cite{siangliulue2016ideahound}.
BlueSky used the crowd to construct taxonomies to constrain sub-topics for ideation \cite{huang2017bluesky}.

However, these methods to diversify content generated by the crowd are not fully-automated and rely on manual effort to label idea categories, or are restricted to a discrete number of idea categories.
Therefore we developed \textit{Directed Diversity}  \cite{cox2021diverse}, a method which scalably and automatically allocates diverse prompts to crowd-workers. 
Our method extracted a corpus of example ideas, and used embedding distances to select diverse examples for prompting ideation. 
Our study is described in detail in Chapter \ref{ch:DirectedDiversity}.




\section{How should messages be formatted and stylised?}

Now that we've described how messaging can be generated, we will discuss the content of this messaging.
From Section \ref{sec:crowd-generated} we have heard how the written style of messaging may vary between experts and the crowd \cite{coley2013crowdsourced,de2017experts}, with crowd-generated content perhaps using less technical and more everyday language \cite{coley2013crowdsourced,kristan2015using,de2017experts}. 
Additionally, we have discussed how previous work aimed to generate health messaging that adopted a semantically diverse range of content \cite{kocielnik2017send}.
However, it may not always be clear what style or message content would be appropriate for a given situation.
The range of message content could be vast such as messaging to encourage a particular behaviour (e.g., reminders \cite{stawarz2014don}, action plans \cite{agapie2016plansourcing}, implementation intentions \cite{milne2002combining}, recommendations \cite{rabbi2015mybehavior}, or acts of deception \cite{cox2023elder}), reflective messaging \cite{kocielnik2018reflection}, and motivational messages \cite{de2016crowd,milne2002combining}, while conversational styles can also vary greatly by changing the formality \cite{heylighen1999formality,cox2022does}, readability, verbosity \cite{aicher2023towards,ding2018increasing}, and sentiment \cite{wang2021towards} of language.

A useful paradigm to call upon is the computers as social actors (CASA) paradigm \cite{nass1994computers}.
CASA posits that people are likely to apply similar social norms to computer interactions, as they do for human interactions.
This has led to a desire to create computer messaging that exhibits favourable human-like qualities such as empathy \cite{lin_caire_2020}, listening \cite{xiao2020tell}, differing conversational styles and tone \cite{liu2018should}, and adopting favourable politeness strategies \cite{brown_politeness_1987, jenkins2007analysis} or behaviour change techniques \cite{michie2011refined,ulmer2022keep}.


Interventional messaging (via smartphone notifications and chatbots) has been used to increase physical activity \cite{gouveia2015we,kocielnik2017send,rabbi2015mybehavior}. 
Gouveia et al. developed Habito \cite{gouveia2015we} to send people persuasive (e.g., ``\textit{try walking longer while on the phone}'') and informative (e.g., ``\textit{you've burned 200 calories by walking today}'') messages. They found that participants receiving persuasive messages were more likely to start walking and to walk further.
Additionally, there have been HCI studies to encourage reflection on past physical activity to produce insights or encourage change \cite{tang2017harnessing,kocielnik2018reflection,li2011understanding,lee2015personalization}. 
Lee et al. found that personalised prompts can be delivered to individuals to motivate them to exercise more \cite{lee2015personalization}. Similarly to this, Kocielnik et al. developed \textit{Reflection Companion} a conversational agent to assist people in reflecting on their physical activity \cite{kocielnik2018reflection}. They found that delivering reflection prompts through a conversational agent increased motivation, empowerment, and adoption of new behaviours during a 2-week field study.


Recently, chatbots have been used in the healthcare domain. Chatbots offer scalability, while having the potential for the flexibility and personability advantages of interviews with health experts \cite{pereira2019using}.
Commercially available chatbots include \textit{Vik}, which disseminates information for breast cancer support and can give medication reminders \cite{chaix2019chatbots,bibault2019chatbot}. 
Bibault et al. found that messages from \textit{Vik} were as informative and helpful as those from physicians \cite{bibault2019chatbot}.
Tschanz et al. propose a chat-like smartphone app (eMMA) to manage patient medication through a chatbot \cite{tschanz2018using} focused on patient interactions and medication management.
Chatbots have also been used for mental health support  \cite{liao_racial_2020,park_designing_2019,abd2020effectiveness,pham2022artificial}.
For example, Park et al. found a chatbot using motivational interviewing techniques can be helpful to encourage reflection and stress reduction \cite{park_designing_2019}.
On from this, physicians agree that chatbots can help in most automatic simple tasks in healthcare scenarios such as disseminating information or arranging appointments \cite{palanica2019physicians}.
Additionally, the increased anonymity of chatbots can lead to greater disclosure and openness among users \cite{pereira2019using}.

Previous work has also investigated the conversational style of chatbots \cite{cox2022does,celino_submitting_2020,xiao2020tell,kim_comparing_2019}.
A chatbot using a casual conversational style was found to improve the quality of course evaluations when compared to a standard survey \cite{wambsgnass_conversational_2020}.
Kim et al. surveyed adolescents about their Internet usage (via multiple-choice questions), and compared a chatbot using casual language against a chatbot using language taken from a standardised survey\cite{kim_comparing_2019}.
People found a casual conversational style more engaging, and the responses were found to be equally efficacious between both conditions.
Similarly, Celino et al. found people preferred a casual survey and it showed same level of reliability and increased level of response quality (reduced satisficing) compared to traditional multiple-choice surveys \cite{celino_submitting_2020}.
Kocielnik et al. \cite{kocielnik2019harborbot} compared a chatbot against a standard survey for health screening and found that users perceived a chatbot as more engaging.
Xiao et al. found that an AI-driven chatbot produced better quality open text responses and high levels of participat engagement compared to a standard survey \cite{xiao2020tell}.

These studies investigated the use of a chatbot as a data collection tool. This was either by comparing multiple-choice responses given to a chatbot using a casual conversational style against a standard survey \cite{kim_comparing_2019,celino_submitting_2020,wambsgnass_conversational_2020}, or by comparing the quality of open-text responses given to a chatbot compared to those elicited by a standard survey \cite{xiao2020tell}.
Yet the baseline conditions using standarised surveys are often written in concise language that may not mirror more conversational formal language.
Therefore, in Chapter \ref{ch:FormalOrCasual} we compare the use of casual and formal conversational styles of a health chatbot, and assess its effect on the both the quality of a user's open-text responses, and on a user's perceptions \cite{cox2022does}.
Furthermore, to investigate how to improve user perceptions further, in Chapter \ref{ch:Personalisation} we describe an experiment where we vary the presence and format of references to previous user utterances and measure the impact on a user's privacy concerns and perceptions (such as intelligence and engagement) of the chatbot \cite{cox2023referencing}.


\section{How should messaging be selected?}

Next, we will discuss how messages can be selected.
Messages can be selected manually by human users (such as a research team or health experts) or algorithmically by a computer.
However, while messages can be chosen manually based on their expected effectiveness, this is not scalable, is labour intensive, and could be skewed by human biases.
Due to these issues of scalability, we will therefore focus on literature describing algorithmic message selection techniques.

\subsection{Selecting messages algorithmically}
\label{sec:selecting}
Messages can be chosen algorithmically either: (1) for ongoing turn-taking conversations between a human and a chatbot or (2) for one time interventions to nudge a user.
Messages selected for turn-taking conversations are typically based on a dataset of messages, by selecting the message that is most likely to follow on from the previous messages in a conversation. 
This could then lead to a retrieval based method (where an existing message is used verbatim) or generative methods as described in Section \ref{sec:HowMessagesGenerated}.

Messages selected for one-time interventions can be chosen based on past rewarding behaviour. 
For example, by looking at previous outcomes, we could choose messages which have led to the biggest increase in a desired behaviour, or messages which have been most likely to result in a desired behaviour.
Within healthcare, studies have demonstrated the usefulness of using reinforcement learning techniques to select messaging. 
Rabbi et al. developed MyBehavior \cite{rabbi2015mybehavior} to use multi-armed bandits to exploit messages intended to maximise calorie loss and physical activity.
Similarly, Yom-Tov et al. developed a system that used a bandit-style algorithm to send messages to patients who are following a personalised plan for physical exercise \cite{yom2017encouraging}. 



\section{When should messages be sent to users?}
Message selection as described in Section \ref{sec:selecting} can also be based on a user's context.
For example, messages could be selected using data such as a user's location or mood, a user's demographics, the time of day, or meta-data collected by a device such as applications used on a smartphone \cite{liao2020personalized,xu2020multi}.

Micro-randomised trials can be used to select the optimal intervention for a given user based on contextual data and prior intervention success. 
DIAMANTE \cite{xu2020multi} delivered motivational messages to increase physical activity, and used reinforcement learning to adaptively display more effectively performing messages more often.
They experimented with different message factors (such as behaviour change technique deployed and message timing) to find and select the most effective messages for different contexts.

Liao et al. \cite{liao2020personalized} selected notifications based on a user's context (such as location and time of day).
Petovic et al. \cite{pejovic2014interruptme} developed an interruptibility model to infer the most opportune time to send a notification based on a user's past behaviour.
Kunzler et al. \cite{kunzler2019exploring} investigated how different factors influence the receptivity of health interventions by considering intrinsic factors (i.e., device type, age, gender, and personality) and contextual factors (i.e., GPS location, physical activity, date/time, Battery status, Lock/Unlock events, Screen on/of events, Wi-Fi connection state and Proximity sensor).

\section{Summary of Related Work}

This preceding related work section outlined the use of computer-delivered health messaging. 
In order to contextualise our upcoming chapters, we gave an overview of health messaging, beginning first with the source of the messages (be it human- or machine-generated), before discussing the potential content type and style of messages, and finally discussing the medium of message delivery.
As discussed above, this medium of delivery can either be computer-delivered interventional messages (such as those investigated further in Chapter \ref{ch:DirectedDiversity} where we investigated message generation) or two-way conversations between a human and an agent (such as the studies in Chapters \ref{ch:FormalOrCasual} and \ref{ch:Personalisation} where we investigated the conversational style of health chatbots). 
In the following chapters, we describe empirical studies that we conducted in order to further knowledge in these areas.

\chapter{Generating Diverse and Efficacious Health Messages}
\label{ch:DirectedDiversity}


%

In Chapters \ref{ch:introduction} and \ref{ch:background}, we gave an overview of the thesis and highlighted relevant literature. In doing so, we discussed the importance of generating a diverse range of non-repetitive health messages. Diverse messages reduce negative effects such as perceived boredom and annoyance and increase positive effects such as perceived informativeness and helpfulness \cite{kocielnik2017send,ghandeharioun2019emma}. 
In this chapter, we describe a method to generate diverse verbal cues or more specifically, conversational content (in this case motivational messages to encourage physical activity). 
Our method, \textit{Directed Diversity}, uses language embedding distances to allocate a collectively diverse range of prompts to users to aid in the generation of diverse motivational messages.
We ran a series of user studies to generate and validate messages generated using our method.
Our method produced more diverse messages, whilst also maintaining or improving more abstract human concepts such as motivation, helpfulness and informativeness. Abstract concepts such as these are useful in creating a more effective human-like chatbot.
This chapter is based on one publication:

\bibentry{cox2021diverse}.

\section{Introduction}

Crowdsourcing has been used to harness the power of human creativity at scale to perform creative work such as text editing \cite{bernstein2010soylent,clark2018creative,shah2018building}, iterating designs \cite{dow2010parallel}, information synthesis \cite{luther2015crowdlines}, and motivational messaging \cite{kocielnik2017send,coley2013crowdsourced,de2016crowd}. In such tasks, empowering crowd workers to ideate effectively and creatively is key to achieving high-quality results. Different prompting techniques have been proposed to stimulate creativity and improve the diversity of ideas \cite{agapie2018crowdsourcing,dow2010parallel,kocielnik2017send,de2016crowd}, but they suffer from ideation redundancy, where multiple users express identical or similar ideas \cite{bjelland2008inside,klein2015high,riedl2010rating,siangliulue2016ideahound}. Current efforts to avoid redundancy include iterative or adaptive task workflows \cite{yu2011cooks}, constructing a taxonomy of the idea space \cite{huang2017bluesky}, and visualizing a concept map of peer ideas \cite{siangliulue2016ideahound}, but these require much manual effort and are not scalable. 

Instead, we propose an automatic prompt selection mechanism — \textbf{Directed Diversity} — to scale crowd ideation diversity. Directed Diversity composes prompts with one or more phrases to stimulate ideation. It helps to direct workers towards new ideas and away from existing ideas with the workflow: 1) \textbf{extract phrases} from text corpuses in a target domain, 2) \textbf{embed phrases} into a vector embedding, and 3) automatically \textbf{select phrases} for maximum diversity. These phrases are then shown as prompts to ideators to stimulate ideation. The phrase embedding uses the Universal Sentence Encoder (USE) \cite{cer2018universal} to position phrases within an embedding vector space. Using the embedding vectors, we calculated distances between phrases to optimally select phrases that are farthest apart from one another; this maximizes the diversity of the selected phrases. Hence, Directed Diversity guides ideators towards under-utilized phrases or away from existing or undesirable phrases.

The embedding space provides a basis to calculate quantitative, distance-based metrics to estimate diversity in selected phrases and prompts, and subsequently ideated messages. These metrics can complement empirical measurements from user studies to evaluate prompts and ideations. We curate multiple measures and evaluation techniques and propose a Diversity Prompting Evaluation Framework to evaluate perceived and subjective creativity and objective, computed creativity, and diversity of crowd ideations. We demonstrate the framework with experiments on Directed Diversity to 1) evaluate its efficacy to select diverse prompts in a simulation study, 2) measure the perceived diversity of selected prompts and effort to generate ideas in an ideation study, and 3) evaluate the creativity and diversity of generated ideas in validation studies using quantitative and qualitative analyses. The experiments were conducted with the use case of writing motivational messages to encourage physical activity \cite{agapie2018crowdsourcing,agapie2016plansourcing,kocielnik2017send,de2016crowd}, though we discuss how Directed Diversity can apply to other crowd ideation tasks. In summary, our contributions are:

\begin{enumerate}
    \item We present Directed Diversity, a corpus-driven, automatic approach that leverages embedding distances based on a language model to select diverse phrases by maximizing a diversity metric. Using these constrained prompts, crowdworkers are directed to generate more diverse ideas. This results in improved collective creativity and reduced redundancy.
    \item A Diversity Prompting Evaluation Framework to evaluate the efficacy of diversity prompting along an ideation chain. This draws constructs from creativity and diversity literature, metrics computed from a language model embedding, and is validated with statistical and qualitative analyses.
    \item We applied the evaluation framework in a series of four experiments to evaluate Directed Diversity for prompt selection, and found that it can improve ideation diversity without compromising ideation quality, but at a cost of higher user effort.
\end{enumerate}

\section{Background and Related Work}

We discuss related research on supporting crowd ideation with the cognitive basis for creative ideation, how creativity support tools help crowd ideation, and how artificial intelligence can help collective intelligence.

\subsection{Cognitive Psychology of Creative Ideation}

Different cognitive models of creativity have been proposed to explain how ideation works. Memory-based explanation models describe how people retrieve information relevant to a cue (prompt) from long-term memory and process it generate ideas \cite{adelman1995examining,dougherty1999minerva,lubart2001models,nijstad2003production,nijstad2002cognitive}. Since retrieval is dependent on prompts, they need to be sufficiently diverse to stimulate diverse ideation \cite{nijstad2002cognitive}, otherwise people may fixate on a few narrow ideas \cite{jansson1991design}. Ideation-based models \cite{del2015option} explain how individuals can generate many ideas through complex thinking processes, including analogical reasoning \cite{green2012neural,au2006fuzzy,meheus2000analogical}, problem constraining \cite{smith2003constraining}, and vertical or lateral thinking \cite{goel2010neural}. We focus on prompting to promote memory-based retrieval than these other reasoning processes. Besides cue-based retrieval and thinking strategies, other factors influence ideation creativity, such as personal traits, motivation to perform the task, and domain-relevant skills that can affect individual creativity \cite{taggar2002individual}. We provide technological support to improve the creative mental process, rather than to select creative personalities, recruit domain experts, or improve task motivation. Next, we discuss how different cognitive factors have been leveraged at scale to support creative ideation with the crowd.

\subsection{Creativity Support Tools for Crowd Ideation}

Creativity Support Tools have been widely studied in HCI to enable crowdworkers to ideate more effectively and at scale \cite{frich2019mapping,frich2018twenty}. Showing workers ideas from their peers has been very popular \cite{girotto2017effect,siangliulue2015toward,siangliulue2016ideahound}, but can have limited benefit to creativity if peer ideas are too distant from the ideators' own ideas \cite{chan2017semantically}. Other approaches include employing contextual framing to prompt ideators to imagine playing a role for the task \cite{oppenlaender2019design} or using avatars for virtual interactions while brainstorming \cite{marinussen2019being}. While these methods focus on augmenting individual creativity, they do not coordinate the crowd, so multiple new ideations may be redundant. More recent approaches apply provide more explicit guidance to workers. IdeaHound \cite{siangliulue2016ideahound} visualizes an idea map to encourage workers to focus on gaps between peer ideas, but does not inform what ideas or topics will fill the gaps. BlueSky \cite{huang2017bluesky} and de Vries et al. \cite{de2016crowd} use crowd or expert annotators to construct taxonomies to constrain the sub-topics for ideation, but these taxonomies require significant manual effort to construct and are difficult to scale. Chan et al. \cite{chan2018best} employed latent Dirichlet allocation (LDA) to automatically identify topics, but this still requires much manual curation which does not scale to many topics. With Directed Diversity, we automatically extract a phrase corpus and embed the phrases as vectors, and select diverse phrases for focused prompting. We employ pre-trained language model to provide crowd ideation support, thus we next discuss how artificial intelligence can support collective intelligence.

\subsection{Supporting Collective Intelligence with Artificial Intelligence}

Collective Intelligence is defined as groups of individuals (the collective) working together exhibiting characteristics such as learning, judgement and problem solving (intelligence) \cite{malone2009harnessing}. Crowdsourcing is a form of collective intelligence exhibited when crowdworkers work towards a task mediated by the crowdsourcing platform. However, managing crowdwork to ensure data quality and maximize efficiency is difficult because of the nature and volume of the tasks, and varying abilities and skills of workers \cite{weld2015artificial}. HCI research has contributed much towards this with interfaces to improve crowdworker efficiency, designing incentives for workers, and workflows to validate work quality \cite{bigham2015human,he2019collective,michelucci2016power,suran2020frameworks,weld2015artificial}. Furthermore, recent developments in artificial intelligence (AI) provides opportunities to complement human intelligence to improve the quality and efficiency of crowd work \cite{weld2015artificial,kittur2013future}, optimize task allocation \cite{dai2013pomdp,fan2014cityspectrum}, adhere to budget constraints \cite{karger2014budget}, and dynamically control quality \cite{bragg2016optimal}. With Directed Diversity, we used AI to optimize ideation diversity by shepherding the crowd towards more desired and diverse ideation with diverse prompt selection.

\section{Technical Approach}

\begin{figure}[h]
    \centering
    \includegraphics[width=1\textwidth]{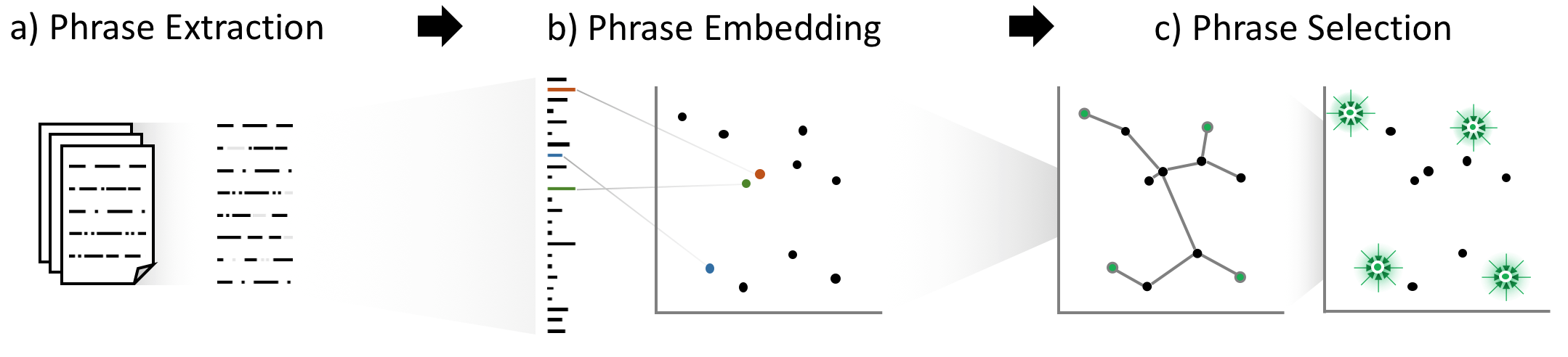}
    \caption{Pipeline of the overall technical approach to extract, embed, and select phrases to generate diverse prompts. a) Phrase extraction by collecting phrases from online articles and discussion forums (shown as pages), filtering phrases to select a clean subset (shown as the black dash for each phrase); b) Phrase embedding using the Universal Sentence Encoder \cite{cer2018universal} to compute the embedding vector of each phrase (shown as scatter plot); c) Phrase Selection by constructing the minimal spanning tree to select optimally spaced phrases (see Figure \ref{fig:Directed-Fig2} for more details).}
    \label{fig:Directed-Fig1}
\end{figure}

We aim to improve the collective diversity of crowdsourced ideas by presenting crowdworker ideators, with carefully selected prompts that direct them towards newer ideas and away from existing ones. The prompts presented to the ideators consist of one or more phrases that represent ideas that are distinct and different from prior ideas. Prompts can have one or more phrases. As a running example throughout the technical discussion and experiments, we apply our approach to the application of motivational messages for healthy physical activity, where it is important to collect diverse motivational messages \cite{kocielnik2017send,de2016crowd}. Figure \ref{fig:Directed-Fig1} shows the 3-step overall approach to extract, embed, and select phrases. We next describe each of these steps in detail.

\subsection{Phrase Extraction}
\label{sec:ch3-phrase-extraction}

\subsubsection{Phrase Selection}
We want to show people phrases relevant to exercise and motivation to inspire them to write new motivational messages. To automatically generate a large and diverse set of relevant phrases, we followed a data-driven approach by extracting phrases from a large corpus of documents about exercising, weight loss and healthy living. We created our corpus using two main sources: informative and credibly sourced health news articles, and diverse online health communities (OHCs). 

First, we collected articles from online news websites focusing on health-related topics. Such articles are written in well-formed, informative, as well as neutral and impersonal language. Using relevant keywords as search terms, we identified the most popular health news websites. We manually inspected their content and selected 3 websites based on their large number of articles and health-related categories, and their policies of specifying sources of health information: \url{www.health.harvard.edu}, \url{www.medicinenet.com}, \url{www.webmd.com}. Note that we limited ourselves to categories such as exercise, nutrition and healthy habits, and omitted categories such as pregnancy or drugs and supplements. 

In contrast to neutral and impersonal news articles, we collected user-generated posts submitted to OHCs on the popular content aggregation and discussion website Reddit (\url{www.reddit.com}). In general, user posts feature a more emotional tone, stronger sentiments and personal anecdotes. The content on Reddit is organized into ``subreddits'' that allow for organizing content and discussion around a general topic. For our corpus, we identified subreddits related to physical activity or motivation based on their popularity in terms of the number of submitted posts. We focused on subreddits where the majority of posts are user-written content in contrast to images, videos, or links to external content. Given these criteria we collected the posts from 20 relevant subreddits (90daysgoal, advancedfitness, advancedrunning, bodyweightfitness, c25k, crossfit, fitness, gainit, getmotivated, ketogains, kettlebells, leangains, loseit, motivation, powerlifting, running, selfimprovement, swimming, weightroom, xxfitness). Previous work has shown that OHCs with similar health focuses can differ in their behaviour change goals \cite{chancellor2018norms}, and by using multiple subreddits we aim to draw on more collective wisdom and diversity. 

Overall, our final corpus consists of 3,235 articles and 32,721 user posts.

\subsubsection{Phrase Cleaning}
To extract all relevant phrases from this corpus, we first split each article and post into sentences and perform Part-of-Speech (POS) tagging using an off-the-shelf text processing tool. We use the grammatical sequence of POS tags to identify and extract phrases as sequence of words forming a syntactic constituent in a sentence — in contrast to simple n-grams that might contain sequences of words of overlapping syntactic constituents. From each sentence, e.g., “Regular exercising helps to improve people’s health at any age.”, we extract verb phrases (e.g., “helps to improve”), noun phrases (e.g., “regular exercising”, “people’s health”, “age”) and prepositional phrases (e.g., “at any age”). To further expand our set of phrases, we also combine adjoining verb and noun phrases to generate noun-verb phrases (e.g. “regular exercising helps to improve”) and verb-noun phrases (e.g., “helps to improve people’s health”).
Despite limiting ourselves to health-related articles and user-generated content, not all sentences and let alone all phrases in themselves are related to this topic. For the final set of phrases to use as potential prompts for workers, we apply the following filter criteria to all phrases. 

\begin{enumerate}
    \item We remove all phrases that contain uncommon words not found a dictionary. This is particularly useful for user posts that often contain typos, slang or other stylistic devices (e.g., emoticons or expressive lengthenings).
    \item We remove all phrases with less than 3 or more than 5 words, since short phrases may not sufficiently inspire and longer phrases may lead to fixation.
    \item We remove all phrases with no word related to exercising, nutrition or health. We use WordNet lexical database \cite{miller1995wordnet} to check for each word in a phrase, if the word or any of its hypernyms is within a set of manually selected concepts. For example, the word treadmill has the hypernym exercise device, which makes it a relevant word for our phrase filter. 
    \item We remove all sub-phrases and overlapped phrases (e.g., exclude ``federal exercise recommendations'' and keep ``federal exercise guidelines rose''). 
\end{enumerate}
The application of all filter steps, resulted in 3,666 phrases that may serve as prompts for our user study.

\subsection{Phrase Embedding}

The corpus of extracted phrases provides a large set of potential phrases for prompting, but we seek to select phrases that are least similar to one another. For each phrase, we obtain a multi-dimensional vector representation, called an embedding, so that the phrase is a data point in an idea space. Similar work by Siangliulue et al. \cite{siangliulue2015toward} obtained embeddings of $N=52$ ideas by training a Crowd Kernel model \cite{tamuz2011adaptively} from 2,818 triplet annotations is not scalable to our corpus of $N=$ 3,666 phrases, since that would need $N(N-1)(N-2)/3=16.4$ million triplets. Instead, similar to Chan et al.’s \cite{chan2017semantically} use of GloVE \cite{pennington2014glove}, we use pre-trained language models based on deep learning to encode each word or sentence as a vector representation. Specifically, we use the more recent Universal Sentence Encoder (USE) \cite{cer2018universal} to obtain embeddings for phrases in our corpus, compute their pairwise distances, and selected a maximally diverse subset of phrases. Our approach is generalizable to other language embedding techniques \cite{young2018recent}.

\begin{figure}[h]
    \centering
    \includegraphics[width=1\textwidth]{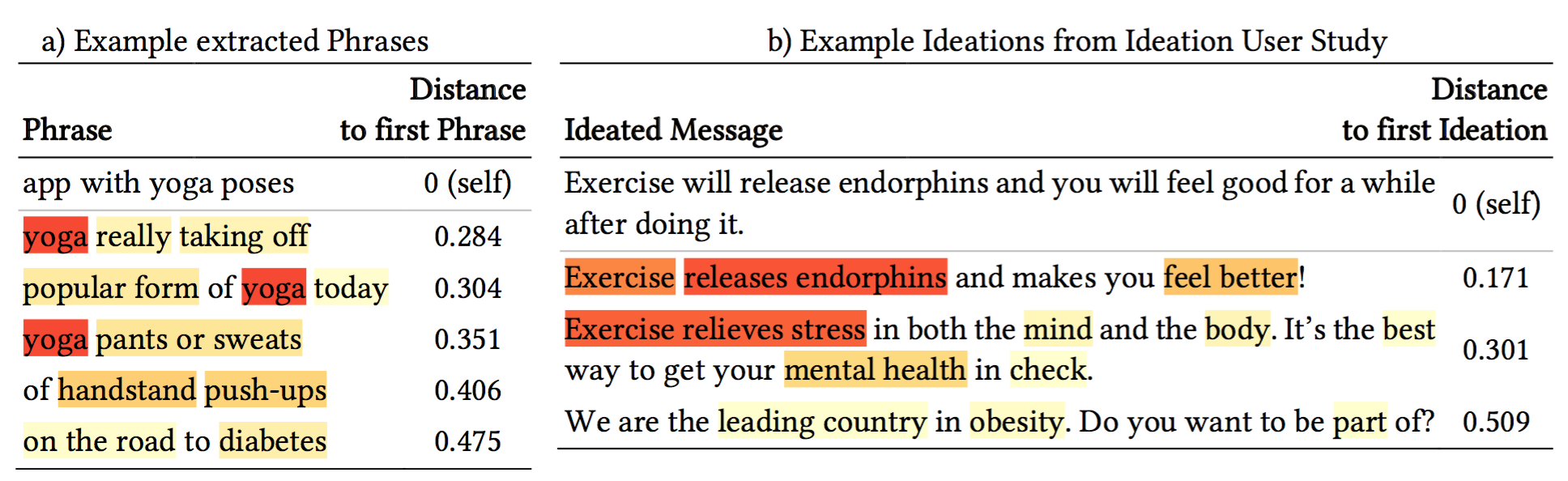}
    \caption{Demonstration of pairwise embedding angular distances between an example text items (first data row) and neighboring text items. Text items with semantically similar words have smaller distances. For interpretability, we highlighted words to indicate darker color with higher cosine similarity to the first phrase.}
    \label{tab:Directed-Tab1}
\end{figure}

To obtain the phrase embedding presentation, we use a pre-trained USE model\footnote{Pre-trained Universal Sentence Encoder model (\url{https://tfhub.dev/google/universal-sentence-encoder/4}) , which was trained using both unsupervised learning on Wikipedia, web news, web question-answer pages, and discussion forums, and supervised learning on Stanford Natural Language Inference (SNLI) corpus.} to obtain embedding vectors for each phrase. With USE, all embeddings are 512-dimensional vectors are located on the unit hypersphere, i.e., all vectors are unit length, and only their angles are different. Hence, the dissimilarity between two phrase embeddings $x_i$ and $x_j$ is calculated as the angular distance $arccos(x_i,x_j )$, which is between 0 and $\pi$. For our phrase corpus, the pairwise distance between phrases ranged from Min = 0.06 to Max = 0.58, Median=0.4, inter-quartile range 0.39 to 0.46, SD=0.043; see Appendix Figure \ref{fig:Directed10}. We use the same USE model to compute embeddings and distances for ideated messages. For a dataset of 500 motivational messages ideated in a pilot study with no prompting, the pairwise distance between ideations ranged from Min=0.169 to Max=0.549, Median=0.405, inter-quartile range 0.376 to 0.432, SD=0.043; see Appendix Figure \ref{fig:Directed11}. Table \ref{tab:Directed-Tab1} shows example phrases and messages and their corresponding pairwise dissimilarity distances. With the embedding vectors and pairwise distances for all phrases, the next step selects diverse phrases with which to prompt ideators.

\subsection{Phrase Selection}
\label{sec:ch3-phrase-selection}

\begin{figure}[h]
    \centering
    \includegraphics[width=1\textwidth]{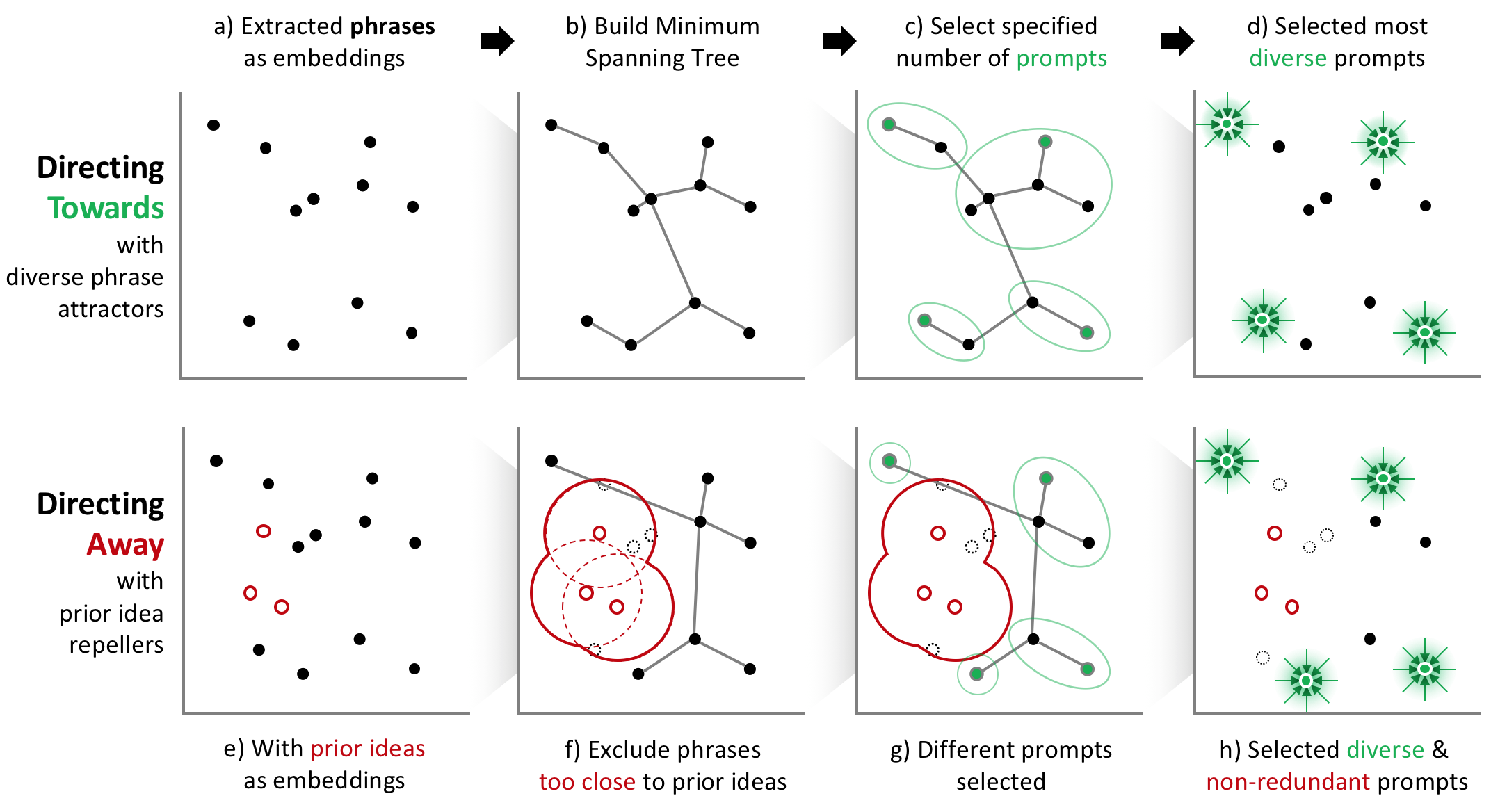}
    \caption{Procedure to direct ideation towards diverse phrases (top) and away from prior or redundant ideas
(bottom). To attract ideation with diverse prompts: a) start with embeddings of corpus-extracted phrases; b) construct minimum spanning tree (MST); c) traverse tree to select distant prompts from clusters (most distant points as green dots, in clustered phrases as green ellipses); d) selected prompts are the most diverse. To repel ideation from prior ideas, e) compute embeddings of prior ideas (red hollow dots); f) compute prompt-ideation pairwise distances of all prompts from each prior ideation, exclude phrases (dotted black circles) with pairwise distance less than a user- defined threshold (red bubble), and construct the MST with remaining phrases; g) traverse MST to select a user- defined number of prompts; h) selected prompts are diverse, yet avoids prior ideas.}
    \label{fig:Directed-Fig2}
\end{figure}

Given the embeddings of the curated phrases, we want to select the subset of phrases with maximum diversity. Mathematically, this is the dispersion problem or diversity maximization problem of “arranging a set of points as far away from one another as possible”. Among several diversity formulations \cite{chandra2001approximation}, we choose the Remote-MST diversity formulation \cite{halldorsson1995finding} (also called Remote-tree \cite{chandra2001approximation} or functional diversity \cite{petchey2002functional}) that defines diversity as the sum of edge weights of a minimum spanning tree (MST) over a set of vertices. It is robust against non-uniformly distributed data points (e.g., with multiple clusters, see Table \ref{tab:Directed-Tab4}). We construct the minimum spanning tree by performing agglomerative hierarchical clustering on the data points with single linkage \cite{sibson1973slink}. Next, we describe how we select phrases as prompts to direct ideators towards diverse phrases, or away from prior ideas. Figure \ref{fig:Directed-Fig2} illustrates the technical approach.

\subsubsection{Directing towards Diverse Phrases}

For phrase selection, we aim to select a fixed number of points n from the corpus with maximum diversity. This is equivalent to finding a maximal edge-weighted clique in a fully connected weighted graph, which is known to be NP-hard \cite{hosseinian2017maximum}. Hence, we propose a scalable greedy approach that uses the dendrogram representation of the MST resulting from the hierarchical clustering. Starting from the root, we set the number of clusters to the desired number of phrases $n$. For each cluster $C_r$, we select the phrase that is most distant from other points, with largest minimum pairwise distance from all points from outside the cluster, i.e.,

\[
x_r = argmax_{i \in C_r} (min_{j \notin C_r} d (x_i, x_j)
\]

where $x_r$ is the diverse phrase selected in cluster $C_r$, $x_i$ is a point in cluster $C_r$ and $x_j$ is a point in the corpus not in $C_r$, and d is the pairwise distance between $x_i$ and $x_j$. This method has $O(n^2)$ time complexity and runs in less than one second on a desktop PC for 3.6k phrases; it is generalizable and can be substituted for other approximate algorithms to select most diverse points \cite{chandra2001approximation,indyk2014composable}. Figure \ref{fig:Directed-Fig2} (top row) illustrates the phrase selection method to direct towards areas without ideations:

\begin{displayquote}
\textbf{a)} Start with all phrases in a corpus represented as USE embedding points.
\end{displayquote}
\begin{displayquote}
\textbf{b)} Construct a dendrogram (MST) from all points, using single-linkage hierarchical clustering.
\end{displayquote}
\begin{displayquote}
\textbf{c)} Set number of clusters equal to desired number of diverse phrases. For each cluster, find the most distant phrase.
\end{displayquote}
\begin{displayquote}
\textbf{d)} Selected phrases are the approximately most diverse from the corpus, for the desired number of phrases.
\end{displayquote}

\subsubsection{Directing Away from Prior Ideas}

Other than directing ideators towards new ideas with diverse prompts, it is important to help them to avoid prior ideas written by peers. We further propose a method to remove corpus phrases that are close to prior ideas so that ideators do not get prompted to write ideas similar to prior ones. The method, illustrated in Figure \ref{fig:Directed-Fig2} (bottom row), is similar as before, but with some changes:

\begin{displayquote}
\textbf{e)} Add the embedding points of prior ideas to the corpus.
\end{displayquote}
\begin{displayquote}
\textbf{f)} Calculate phrase-ideation distance $d(x_i^P,x_j^I )$ for each phrase $x_i^P$ and ideation $x_i^I$ and exclude phrases too close to the ideas, i.e., $d<\delta$	, where $\delta	$ is an application-dependent threshold, $\delta	=0.29$ in our case.
\end{displayquote}
\begin{displayquote}
\textbf{g)} Same as step (c), but different clusters, since fewer points are clustered.
\end{displayquote}
\begin{displayquote}
\textbf{h)} Same as step (d), but different prompts would be selected, even if the number of phrases is the same.
\end{displayquote}

\subsection{Directing with Prompts of Grouped Phrases}
\label{sec:ch3-directing-grouped}

Instead of prompting with only one phrase, prompting with multiple related terms can help ideators to better understand the concept being prompted and generate higher quality ideas \cite{chan2015impact,nijstad2006group,sio2015fixation}. We extend the phrase selection method to group multiple phrases in a single prompt using the following greedy algorithm. After step (a), we 1) sorted phrases by descending order of minimum pairwise distance for each phrase to produce a list of seed candidates, 2) for each seed phrase, perform a nearest neighbors search to retrieve a specified prompt size (number of phrases $g$ in a prompt) and remove the selected neighbors from the seed list, 3) repeat seed neighbor selection until $n$ seed phrases have been processed. We grouped the phrases into a prompt and calculate its embedding point $x_i^{Pr}$ as the angular average of all phrases $x_k^P$ in the prompt, i.e., $x_i^{Pr}= \Sigma_{(k=1)}^g x_k^P /Z$, where $Z=|| \Sigma_{(k=1)}^g x_k^P ||_2$ is the magnitude of the vector sum and $x_i^{Pr}$ is also a unit vector. We then perform steps (b) to (d) with the prompts $x_i^{Pr}$ instead of individual phrases. Note that the corpus of prompts will be smaller than the corpus of phrases. This approach has disjoint prompts that do not share phrases, but there can be alternative approaches to group phrases \footnote{An alternative approach is, after step (c), to simply group nearest neighbours. However, this will cause the prompt embeddings to be shifted after the diversity is maximized, so it may reduce the diversity of the selected prompts.}.

\section{Diversity Prompting Evaluation Framework}

\begin{figure}[h]
    \centering
    \includegraphics[width=1\textwidth]{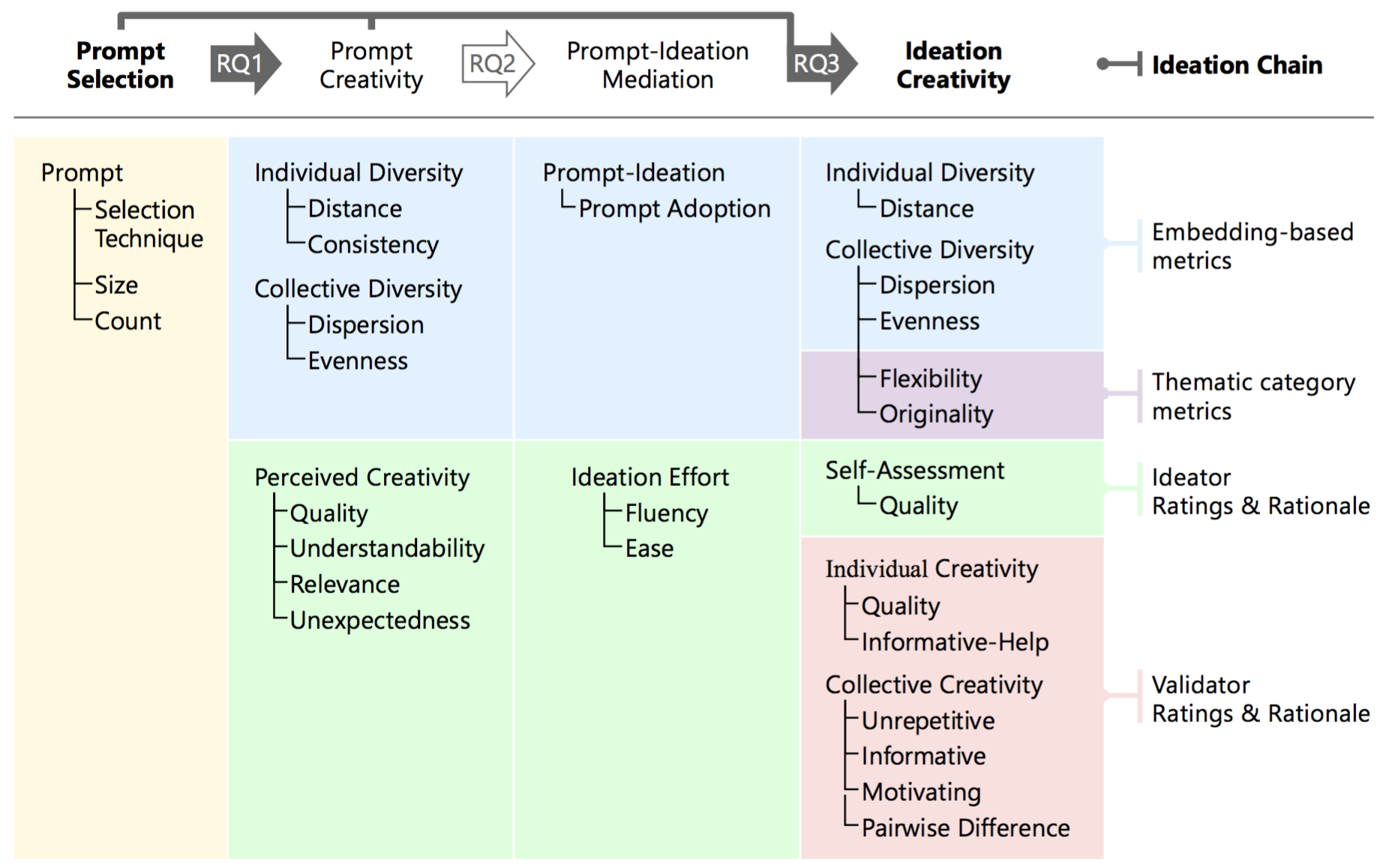}
    \caption{Diversity prompting evaluation framework to evaluate prompting to support diverse ideation along the ideation chain. We pose research questions (RQ3.1-3.3) between each step to validate the ideation diversification process. For each step, we manipulate or measure various experiment constructs to track how well ideators are prompted to generate creative ideas. Except for prompt selection, each construct refers to a statistical factor determined from factor analyses of multiple dependent variables. Constructs are grouped in colored blocks indicating different data collection methods as follows: (\textbf{\textcolor{NavyBlue}{Blue - Computed embedding-based metric}}, \textbf{\textcolor{ForestGreen}{Green - ratings from ideators}}, \textbf{\textcolor{Red}{Red - ratings from validators}}, \textbf{\textcolor{Purple}{Purple - thematic coding of ideations}}).}
    \label{fig:Directed-Fig3}
\end{figure}

To evaluate the effectiveness of the Directed Diversity prompt selection technique to improve the collective creativity of generated ideas, we define an ideation chain as a four step process (Figure \ref{fig:Directed-Fig3} top): 1) setting the prompt selection technique will influence 2) the creativity of selected prompts (prompt creativity), 3) the ideation process of the ideators (prompt-ideation mediation), and 4) the creativity of their ideation (ideation creativity). We propose a Diversity Prompting Evaluation Framework, shown in Figure \ref{fig:Directed-Fig3}, to measure and track how creative and diverse information propagates along this ideation chain to evaluate how and whether a creativity prompting technique improves various measures of creativity and diversity in outcome ideas. Note that our proposed framework is descriptive to curate many useful metrics, but not prescriptive to recommend best metrics.

\subsection{Research Questions and Experiments}

Prompt stimuli act along the ideation chain to increase ideation diversity, but it is unclear how well they work and at which point along the chain they may fail. We raise three research questions between each step in the ideation chain, which we answer in four experiments (Section \ref{sec:ch3-ideation-study}) with various measures and factors.

\textbf{RQ3.1} \textit{How do the prompt techniques influence the perceived diversity of prompts? (RQ3.1a) How do they affect diversity in prompts? (RQ3.1b) How well can users perceive differences in creativity and diversity in these prompts?} These questions relate to the prompt selection technique effectiveness and serve as a manipulation check. We answer them in a Characterisation Simulation Study (Section \ref{sec:ch3-simulation-study}) with objective diversity measures, and an Ideation User Study (Section \ref{sec:ch3-ideation-study}) with subjective measures perceived prompt diversity measures.

\textbf{RQ3.2} \textit{How does diversity in prompts affect the ideation process for ideators? (RQ3.2a) Do differences in diversity affect ideation effort? (RQ3.2b) How well do ideators adopt and apply the content of the prompts? (RQ3.2c) How does prompt creativity affect diversity in ideations?} We answer these questions as a mediation analysis in the Ideation User Study (Section \ref{sec:ch3-ideation-study}) with objective measures of task time and similarity between ideations and stimulus prompts, thematically coded creativity metrics, and perceived ease of ideation.

\textbf{RQ3.3} \textit{How do prompt selection techniques affect diversity in ideations?} Having validated the manipulation checks, we evaluate the effectiveness of prompt selection techniques in questions in the Ideation User Study (Section \ref{sec:ch3-ideation-study}) with subjective measures self-assessed creativity and thematically coded creativity metrics, and two Validation User Studies (Section \ref{sec:ch3-validation-study}) with subjective measures of perceived creativity.

\subsubsection{Independent variables of Prompt Specifications}

We manipulated \textbf{\textcolor{YellowOrange}{prompt selection}} technique, \textbf{\textcolor{YellowOrange}{prompt count}}, and \textbf{\textcolor{YellowOrange}{prompt size}} as independent variables; these are detailed in Appendix Table \ref{tab:Directed-Tab6}. We chose Random prompt selection as a key baseline where selection is non-trivial and data-driven based on our corpus, but not intelligently selected for diversity.

\subsection{Diversity and Creativity Measures of Prompting and Ideation}
\label{sec:ch3-measures}

We measured diversity and creativity for selected prompts and generated ideas with embedding-based and human rated metrics. We color code variable names based on data collection method as in Figure \ref{fig:Directed-Fig3}.

\subsubsection{Embedding-based Diversity Metrics for Prompts and Ideations}

\begin{figure}[h]
    \centering
    \includegraphics[width=1\textwidth]{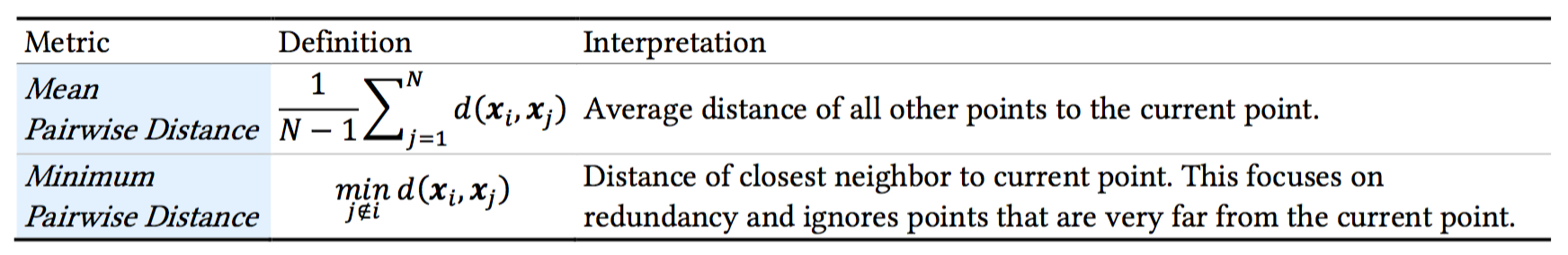}
    \caption{Metrics of distances between two points in a multi-dimensional vector space. Each metric can be calculated for an individual text item. These metrics can apply to the embedding of phrases or ideations.}
    \label{tab:Directed-Tab2}
\end{figure}

\begin{figure}[h]
    \centering
    \includegraphics[width=1\textwidth]{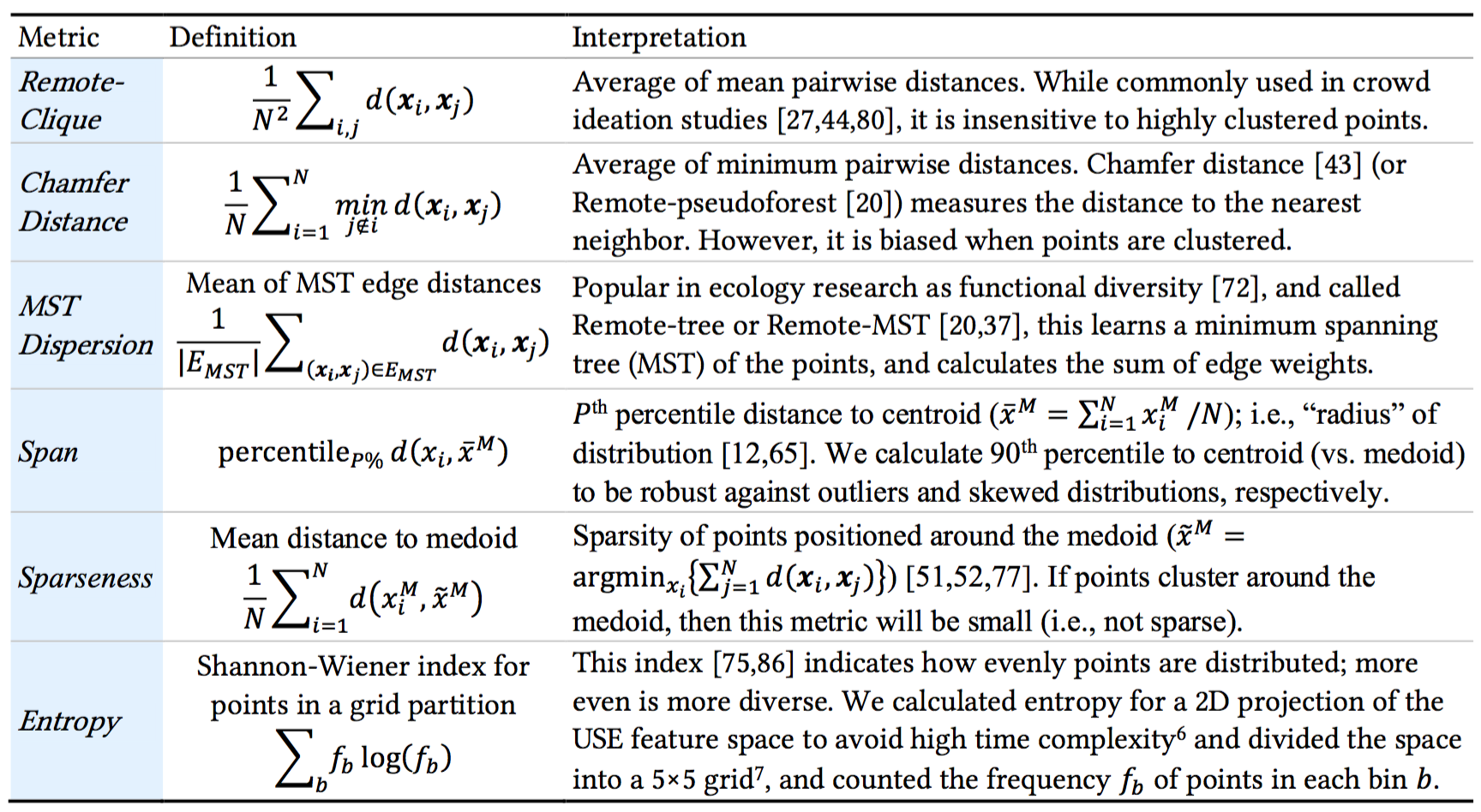}
    \caption{Metrics of diversity of phrases or ideation embeddings in a vector space. These capture more characteristics of diversity than average distances in Table 2. Each metric can only be calculated collectively for multiple items.}
    \label{tab:Directed-Tab3}
\end{figure}

\begin{figure}[h]
    \centering
    \includegraphics[width=1\textwidth]{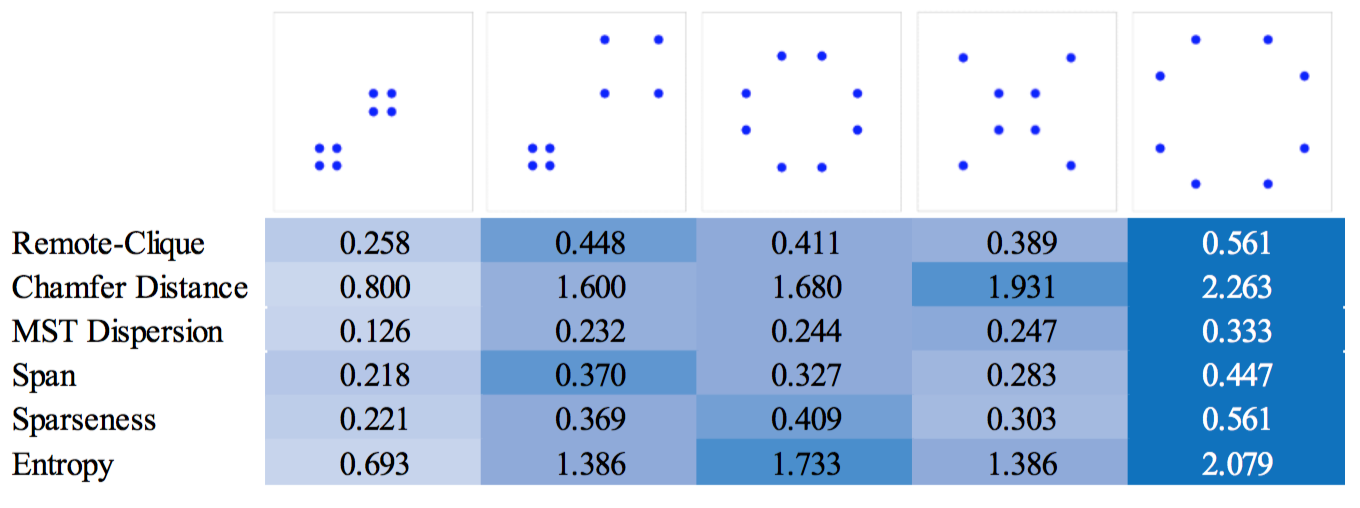}
    \caption{Comparison of diversity metrics for canonical examples of distributed points in a 2D space. Points farther apart mean higher diversity. Here, we calculate Euclidean instead of angular distance, but intuitions are similar.}
    \label{tab:Directed-Tab4}
\end{figure}

Although crowd creativity research has focused on the mean pairwise distance as a metric for idea diversity, our literature review has revealed many definitions and metrics. Here, we describe computational metrics calculated from the embedding-based distances. Inspired by Stirling’s general framework diversity framework \cite{stirling2007general}, we collect definitions from crowd ideation \cite{chan2016comparing,dow2010parallel,huang2017bluesky,siangliulue2015toward,siangliulue2016ideahound}, ecology \cite{diaz2001vive,petchey2002extinction,villeger2008new}, recommender systems \cite{fleder2007recommender,kaminskas2016diversity,ziegler2005improving,vargas2014coverage}, and theoretical computer science \cite{chandra2001approximation,halldorsson1995finding}. These cover many aspects of diversity to characterize the \textbf{\textcolor{NavyBlue}{mean distance}} and minimum \textbf{\textcolor{NavyBlue}{Chamfer distance}} between points, \textbf{\textcolor{NavyBlue}{MST-based dispersion}}, \textbf{\textcolor{NavyBlue}{sparseness}} of points around the median, \textbf{\textcolor{NavyBlue}{span}} from the centroid, and \textbf{\textcolor{NavyBlue}{entropy}} to indicate the evenness of points in the embedding vector space. Table \ref{tab:Directed-Tab2} and Table \ref{tab:Directed-Tab3} describe distance metrics for individual and collective text items, respectively. These metrics describe nuances of diversity, which we illustrate with example distributions in Table \ref{tab:Directed-Tab4}. Other measures of diversity and divergence \cite{chandra2001approximation} can be included in the framework, which we defer to future work. Next, we describe human-subjects ratings to validate these embedding-based metrics with measures that do not depend on the embeddings to avoid circular dependency.

\subsubsection{Creativity Measures for Ideations}

Along with the computed diversity metrics, we evaluate with qualitative characteristics of creativity. From creativity literature, we draw from Torrance’s \cite{torrance2018guiding} description of several measures for creativity, including quality, flexibility and originality. Quality measures whether an ideation is “usable, practical, or appropriate” \cite{mumford1996process}. We asked ideators to self-assess on a 5-point Likert scale their message’s \textbf{\textcolor{ForestGreen}{effectiveness}} (towards motivation) and \textbf{\textcolor{ForestGreen}{creativity}}. We ask validator crowdworkers to rate each individual ideation on a 7-point Likert scale whether it is \textbf{\textcolor{Red}{effective}} (motivating \cite{de2016crowd}), \textbf{\textcolor{Red}{helpful}}\footnote{Note that a message could be helpful but written in a way that participants do not find motivating.}  \cite{kocielnik2017send,strecher2005randomized}, and \textbf{\textcolor{Red}{informative}} \cite{kocielnik2017send} towards encouraging physical activity; rank collections of ideations on effectiveness, informativeness and \textbf{\textcolor{Red}{unrepetitiveness}}.; and rate the \textbf{\textcolor{Red}{pairwise difference}} between ideation pairs from each collection. Note that Directed Diversity was not designed to improve quality, since these metrics were not explicitly modeled. \textbf{\textcolor{Purple}{Flexibility}} \cite{sowden2015improvisation} measures how many unique ideas were generated, and \textbf{\textcolor{Purple}{originality}} \cite{zenasni2009perception} measures how infrequently each idea occurs. These require expert annotation to identify distinct categories. We conducted a thematic analysis on the messages using open coding \cite{glaser2017discovery} to derive categories and affinity diagramming \cite{byer1998contextual} to consolidate categories to themes (see Table \ref{tab:directed-thematic}). We calculate the flexibility and originality measures based on the coded categories (fine-grained) and themes (coarser) described in Appendix Table \ref{tab:Directed-Tab7}.

\subsubsection{Creativity Measures for Prompts}

As a manipulation check, it is important to verify that prompts that are computed as more diverse, are perceived by ideators as more creative. Since perceived creativity encompasses more qualitative effects, computed diversity may not be correlated with creativity. Thus, we measure the creativity and usefulness of prompts by asking about prompt \textbf{\textcolor{ForestGreen}{understandability}}, \textbf{\textcolor{ForestGreen}{relevance to domain topic}} (physical activity), \textbf{\textcolor{ForestGreen}{relevance to task}} \footnote{Note that a prompt could be relevant to the domain, but not motivating.}  (motivation), \textbf{\textcolor{ForestGreen}{helpfulness}} to inspire ideation, and \textbf{\textcolor{ForestGreen}{unexpectedness}} \cite{mumford1996process} along 7-point Likert scales.

\subsubsection{Mediating Variables for Prompt-Ideation Process}

Even if more diverse prompts can facilitate more creative ideation, it is important to understand whether this requires more effort and time, how the consistency of phrases within prompts affect ideation, and how well ideators adopt words and concepts from the phrases into their ideations. We measure effort as \textbf{\textcolor{ForestGreen}{ease of ideation}} with a 7-point Likert scale survey question. For individual creativity, fluency \cite{fontenot1993effects} is defined as the number of ideas an individual writes within a fixed time. Chan et al. had also measured fluency for an 8-minute crowd ideation task \cite{chan2017semantically}. In contrast, we asked ideators to only write one idea per prompt without time constraint, so we measure the inverse relation of ideation task time to generate one ideation \cite{barbot2018dynamics}. Specifically, since task time is skewed, we use –Log(ideation time) to represent \textbf{\textcolor{ForestGreen}{fluency}}. For prompts with more than one phrase, the similarity between phrases can affect their perceived consistency. Therefore, we measure the \textbf{\textcolor{NavyBlue}{intra-prompt mean phrase}} and \textbf{\textcolor{NavyBlue}{prompt average phrase Chamfer distances}} (Appendix Table \ref{tab:Directed-Tab8}) to indicate the similarity between intra-prompt phrases. We measure the adoption of the prompt ideas by calculating the proportion of words from phrases in the ideations as \textbf{\textcolor{NavyBlue}{prompt recall}} and \textbf{\textcolor{NavyBlue}{prompt precision}}, and computing the \textbf{\textcolor{NavyBlue}{prompt-ideation distance}} between the embeddings of the prompt and ideation (Appendix Table \ref{tab:Directed-Tab9}).

\subsection{Factor analyses to draw constructs from experiment variables}
\label{sec:ch3-factor-analysis}

With the numerous variables from our experiments, we observed some may be correlated since they measure similar notions or participants may confound questions to have similar meanings. We employed an iterative design-analytical process to organize and consolidate variables into factors with the following steps.

\begin{figure}[h]
    \centering
    \includegraphics[width=1\textwidth]{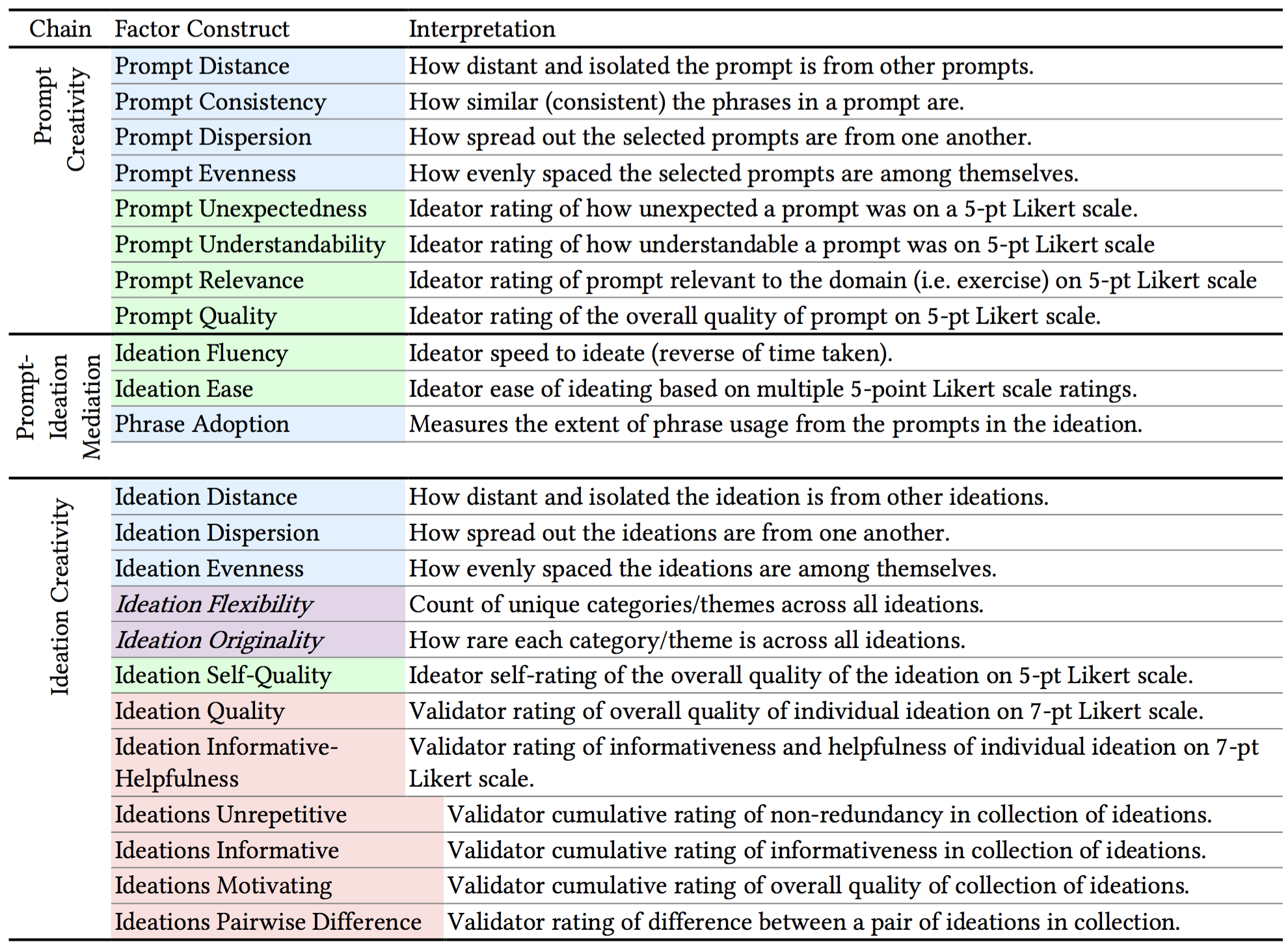}
    \caption{Constructs from factor analyses of variables along ideation chain. Factor loadings in Appendix Tables \ref{tab:Directed-Tab10} to \ref{tab:Directed-Tab17}.}
    \label{tab:Directed-Tab5}
\end{figure}

\begin{itemize}
    \item Identify metrics of creativity and diversity from a literature review from various research domains, such as ecology, creativity, crowdsourcing, theoretical computer science, recommender systems (Section \ref{sec:ch3-measures}). Ideate additional measures and questions to capture user behavior and opinions when generating and validating ideas. We refine and reduce measures based on survey pilots and usability testing.
    \item Collect measurements of each metric with different methods: 1) Compute \textbf{embedding-based metrics} from prompts shown and messages written. This was computed individually for each text item (e.g., mean pairwise distance) and collectively for all text items in each prompt technique (e.g., Remote-MST diversity). 2) Measure \textbf{perception ratings and behavioral measures} regarding reading prompts and ideating messages and rating messages. We asked text rationale to help with interpretations. 3) Measure \textbf{subjective thematic measures} to qualitatively assess the collective creativity with thematic analysis and idea counting. 
    \item Perform factor analysis on quantitative data to organize correlated variables into fewer factors. Variables are first grouped by data collection method \footnote{E.g., individual text item metrics, collective text items metrics, ratings of text item from ideators, ratings of text item from validators, ratings of collection of text items from validators.} and analyzed together. To determine the number of factors, we examined scree plots and verified grouped variables as consistent with constructs from literature. The final number of factors are statistically significant by the Bartlett Test of Sphericity (all $p<.0001$). See Appendix Tables \ref{tab:Directed-Tab10} to \ref{tab:Directed-Tab17} for the results of the factor analysis, including factor loadings and statistical significance. Table \ref{tab:Directed-Tab5} summarizes the learned factors from 42 variables that we developed.
    \item Perform statistical hypothesis testing using these learned factors to answer our research questions.
\end{itemize}

\section{Evaluation: applying framework to study Directed Diversity}

We have described a general descriptive framework for evaluating diversity prompting. We applied it to evaluate our proposed Directed Diversity prompt selection technique against baseline approaches (no prompting, random prompt selection) in a series of experiments (characterisation, ideation, individual validation, collective validation), for the use case of crowd ideating motivational messages for physical activity. Here, we describe the procedures for each experiment and their results.

\subsection{Characterisation Simulation Study}
\label{sec:ch3-simulation-study}

The first study uses computational methods to rapidly and scalably evaluate prompt selection techniques. This helps us to fine tune prompt parameters to maximize their potential impact in later human experiments.

\subsubsection{Experiment Treatments and Method}

We varied three independent variables (prompt selection, prompt count, prompt size) to measure the impact on 7 dependent variables of distance and diversity metrics. We varied Prompt Selection technique (None, Random, or Directed) to investigate how much Directed Diversity improves prompt diversity with respect to baseline techniques. For None prompt selection, we simulated ideation with 500 ideas collected from a pilot study where crowd ideators wrote messages without prompts. We simulated Random selection by randomly selecting phrases from the phrase corpus (Section \ref{sec:ch3-phrase-extraction}) and Directed selection with our technical approach (Sections \ref{sec:ch3-phrase-extraction} to \ref{sec:ch3-directing-grouped}). If we assume that prompt embeddings are an unbiased estimator for ideation embeddings, then this gives an approximation of ideation diversity due to prompting. We conducted experiments for directing towards diverse prompts and for directing away from the 500 pilot prior ideations. We varied the number of prompts (Prompt Count, $n=50$,150,…,950) to simulate how diversity increases with the number of ideation tasks performed. This investigates how diversity increases as the budget for crowd tasks increases. To investigate how well Directed selection avoids prior ideations, we varied the number of repeller prior ideations (Repeller Prior Ideations Count, $n_R=50$,100,150,200). We varied the number of phrases in prompts (Prompt Size, $g=1$ to 5) to simulate ideating on one or more phrases in each prompt. We computed the prompt embedding as the average of all phrases in the prompt. For Random selection, we randomly chose phrases to group together for each prompt. This random neighbour selection will lead to variation in prompt consistency, but does not bias the prompt embedding on average. For Directed selection, phrases in each prompt were chosen as described in Section \ref{sec:ch3-phrase-selection}.

\subsubsection{Results on Manipulation Efficacy Analysis (RQ3.1a)}

\begin{figure}[h]
    \centering
    \includegraphics[width=1\textwidth]{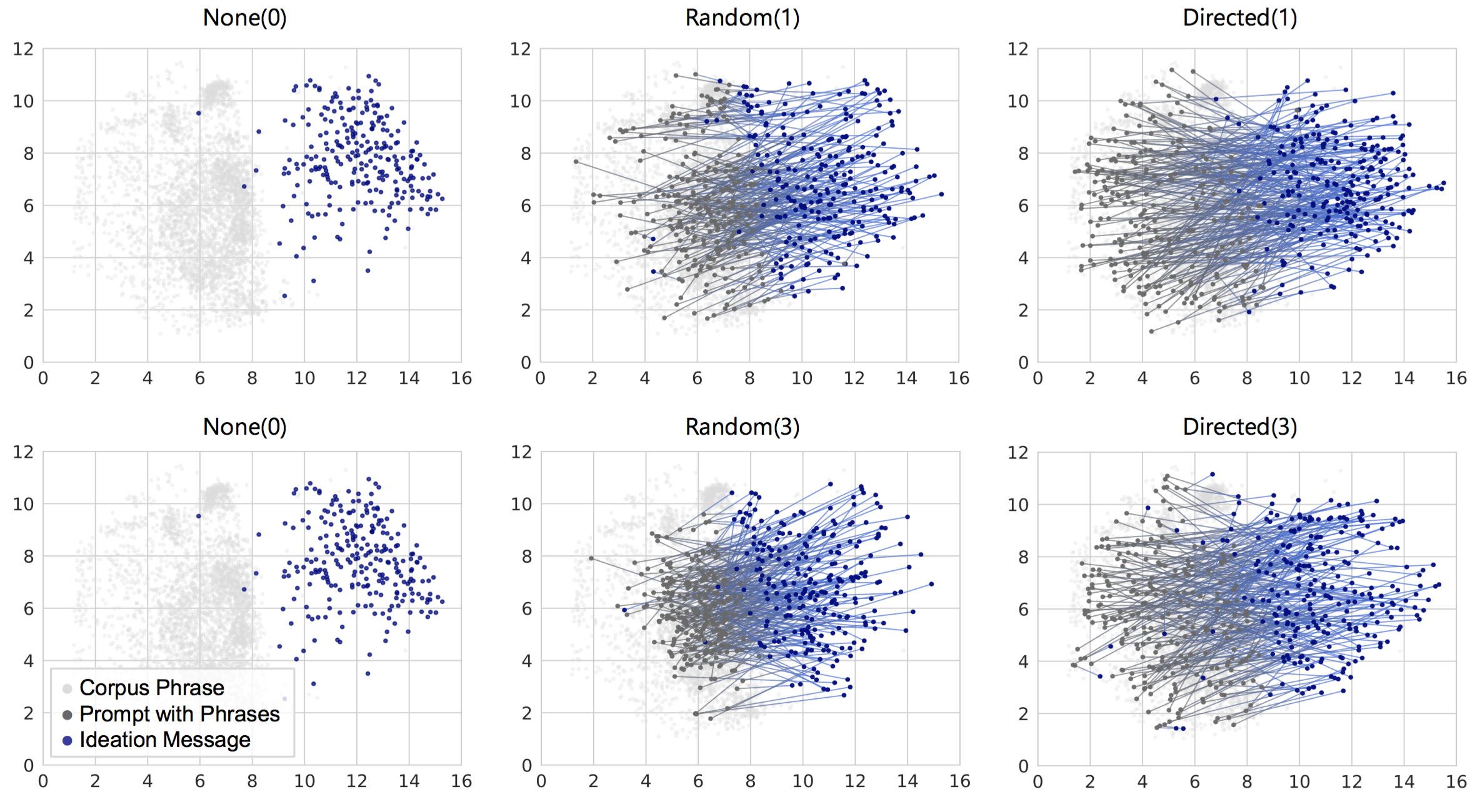}
    \caption{2D UMAP projection showing how diversely selected prompts and resulting ideation messages are distributed based on Directed or Random prompt selection technique and prompt size (number in brackets). Each point represents the embedding of a text item. \textbf{\textcolor{mylightgrey}{Light grey}} points represent all phrases in the extracted corpus, \textbf{\textcolor{mydarkgrey}{dark grey}} points represent selected phrases from the simulation study (Section \ref{sec:ch3-simulation-study}) and \textbf{\textcolor{Blue}{blue}} dots represent the ideated messages written by crowdworkers in the ideator user study (Section \ref{sec:ch3-ideation-study}). Gradient lines connect ideation messages to their stimulus prompts.}
    \label{fig:Directed-Fig4}
\end{figure}

We visualized (Figure \ref{fig:Directed-Fig4}) the phrase embeddings to help to interpret how the selected prompts are distributed, whether they are well spread out, clustered, etc. We used Uniform Manifold Approximation and Projection (UMAP) \cite{mcinnes2018umap} to reduce the 512 dimensions of USE to a 2D projection. Hyperparameters were selected such that the 2D points in UMAP had pairwise distances correlated with that of the 512-dimension USE embeddings. 

We can see that Directed prompt selection led to prompts that were more spread out, and less redundant from prior ideation. This is more pronounced for higher prompt size ($g=3$). Random(3) had lower diversity than None with tighter clustering of prompts (grey points in middle-bottom graph) than of messages (blue points in left graph). This was because Random(3) prompts averaged their embeddings from multiple phrases, such that this variance of means of points is smaller than the variance of points\footnote{This is analogous to standard error is to standard deviation.}. We further conducted a characterisation study with 50 simulations for each prompt configuration to confirm that Directed Diversity improves diversity and reduces redundancy from prior ideations for various embedding-based metrics (see Appendix \ref{Appendix:ch3-5} - Figure \ref{fig:Directed12}).

\subsection{Ideation User Study}
\label{sec:ch3-ideation-study}

The Ideation User Study serves as a manipulation check that higher prompt diversity can be perceived by ideators, and as an initial evaluation of ideation diversity based on computed and thematically coded metrics.

\subsubsection{Experiment Treatment and Procedure}

\begin{figure}[h]
    \centering
    \includegraphics[width=1\textwidth]{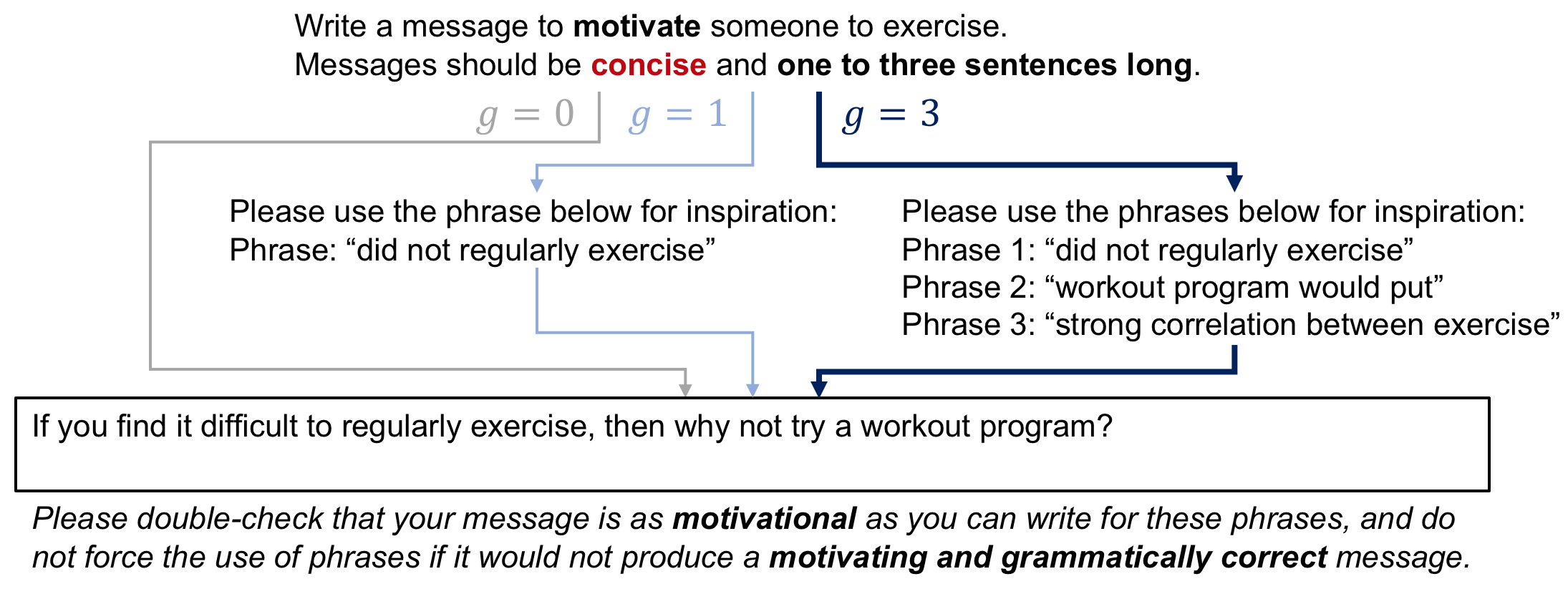}
    \caption{Example prompts shown to participants in different conditions: None (left, $g = 0$, Directed(1) (center, $g = 1$), and Directed(3) (right, $g = 3$). Phrase texts would be different for Random(1) and Random(3) selection techniques.}
    \label{fig:Directed-Fig5}
\end{figure}

We conducted a between-subjects experiment with two independent variables prompt selection technique (None, Random, Directed) and prompt size ($g=1$ and $3$), and kept constant prompt count $n=250$. The None condition (no prompt) allows us to measure if the quality of ideations become worse due to the undue influence of phrases in prompts. The Random condition provides a strong baseline since it also leverages the extracted phrases in the first step of Directed Diversity. A prompt size of $g>1$ can provide more contexts to help ideators understand the ideas in the phrases, but may also lead to more confusion if the phrases are not consistent (too dissimilar). Figure \ref{fig:Directed-Fig5} shows example prompts that ideator participants see in different conditions. The experiment apparatus and survey questions were implemented in Qualtrics (see Appendix Figures \ref{fig:Directed13} to \ref{fig:Directed19} for instructions and question interface).

\subsubsection{Experiment Task and Procedure}
\label{sec:ch3-procedure}

Participants were tasked to write motivational messages and answer questions with the following procedure:
\begin{enumerate}
    \item Read the introduction to describe the experiment objective and consent to the study.
    \item Complete a 4-item word associativity test \cite{chandler2019online} to screen for English language skills.
    \item Write 5 messages to motivate for physical activity for a fitness mobile app. For each message, one at a time,
    \begin{enumerate}[label=\alph*.]
        \item On the first page, depending on condition, see no prompt or a prompt with one or three phrases selected randomly or by Directed Diversity (see Figure 5), then write a motivational message in one to three sentences. This page is timed to measure ideation task time.
        \item Rate on a 5-point Likert scale the experience of ideating the current message: ease of ideation, self-assessed success in writing motivationally, and success in writing creatively; perception of the prompt on: understandability, relevance to domain topic (physical activity), relevance to task (motivation), helpfulness for inspiration, and unexpectedness (described in Section \ref{sec:ch3-measures}).
        \item Reflect and describe in free text on their rationale, thought process, phrase word usage, and ideation effort. We analyze these quotes to verify our understanding of the collected quantitative data.
    \end{enumerate}
    \item Answer demographics questions, and end the survey by receiving a completion code.
\end{enumerate}

\subsubsection{Experiment Data Collection and Statistical Analyses}

We recruited participants from Amazon Mechanical Turk with high qualification ($\geq$5000 completed HITs with $>$97\% approval rate). Of 282 workers who attempted the survey, 250 passed the test to complete the survey (88.7\% pass rate). They were 45.2\% female, between 21 and 70 years old (M$=38.6$); 76.4\% of participants have used fitness apps. Participants were compensated after screening and were randomly assigned to one prompt selection technique. Participants in the None condition were compensated with USD\$1.80, while others with USD\$2.50 due to more time needed to answer the additional survey questions about prompts. Participants completed the survey in median time 15.4 minutes and were compensated $>$USD\$8/hour. We collected 5 messages per participant, 50 participants per condition, 250 ideations per condition, and 1,250 total ideations.

For all response variables, we fit linear mixed effects models described in Appendix Tables \ref{tab:Directed-Tab20} to \ref{tab:Directed-Tab23}. To allow a 2-factor analysis, we divided responses in the None(0) condition (no prompt, 0 phrases) randomly and evenly to None(1) and None(3). Results are shown in Figure \ref{fig:Directed-Fig6}. We performed post-hoc contrast tests for specific differences identified. Due to the large number of comparisons in our analysis, we consider differences with $p<.001$ as significant and $p<.005$ as marginally significant. Most significant results reported are $p<.0001$. This is stricter than a Bonferroni correction for 50 comparisons (significance level = .05/50). We next describe the statistically significant results for prompt mediation check (RQ3.1b), mediation analysis (RQ3.2a, 3.2b), and ideation evaluation (RQ3.3). We include participant quotes from their rationale text response where available and relevant.

\subsubsection{Results of Manipulation Check on Creativity and Mediation on Ideation Effort (RQ3.1b, 3.2a)}

\begin{figure}[h]
    \centering
    \includegraphics[width=1\textwidth]{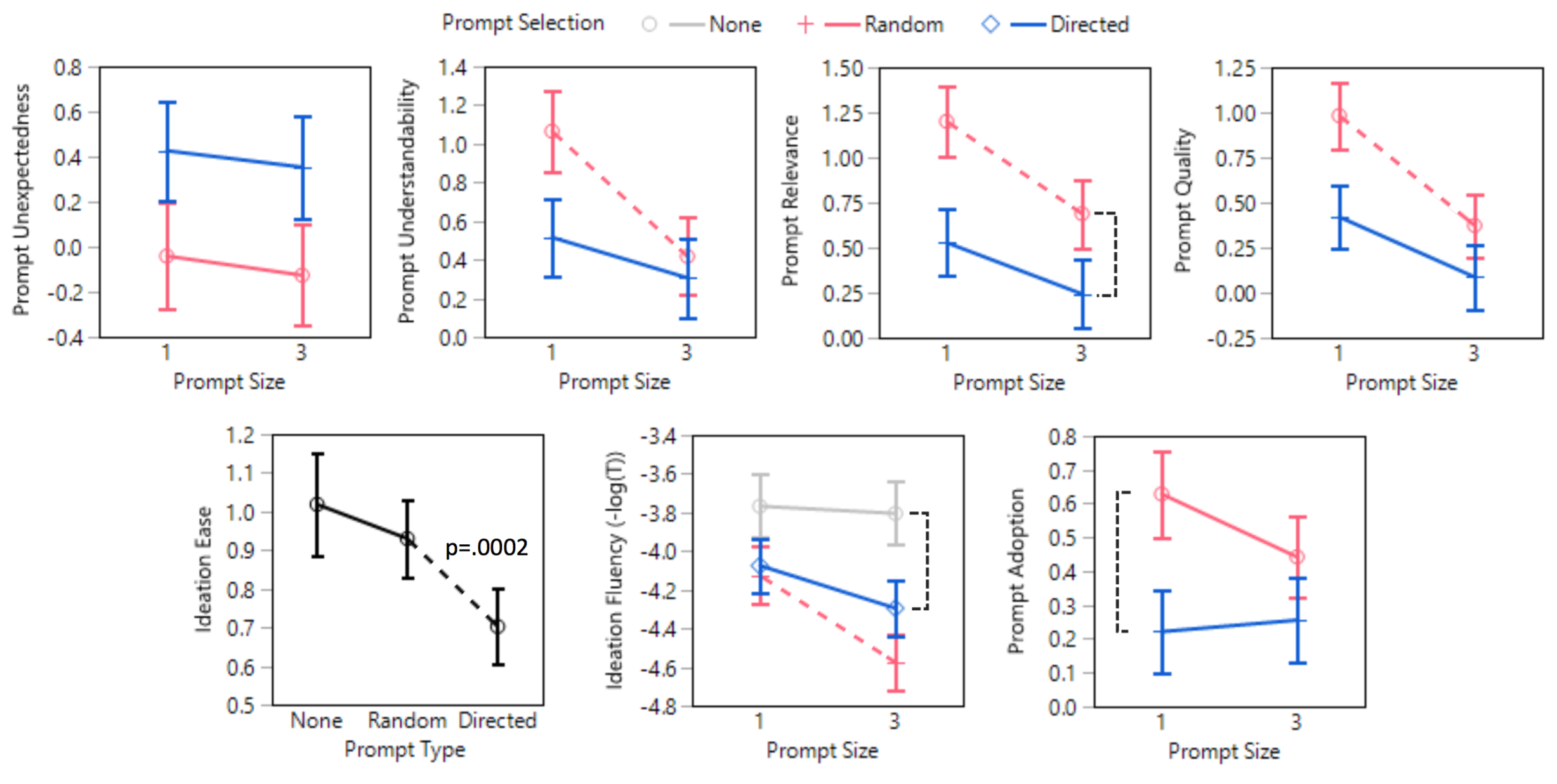}
    \caption{Results of ideators’ perceived prompt creativity (Top) and ideation effort (Bottom) for different Prompt Selection technique and Prompt Size. All factors values on a 5-point Likert scale (–2 = ``Strongly Disagree'' to 2 = ``Strongly Agree''). Dotted lines indicate extremely significant $p<.0001$ comparisons, otherwise very significant with p-value stated; solid lines indicate no significance at $p>.01$. Error bars indicate 90\% confidence interval.}
    \label{fig:Directed-Fig6}
\end{figure}

We discuss findings on how ideators perceived creativity factors in prompts and how prompt configurations affected their ideation effort. Figure \ref{fig:Directed-Fig6} (Top) shows that compared to Random, Directed Diversity selected prompts that were more unexpected (good for diversity); but were slightly more difficult to understand (by half unit on 5-point Likert scale), very slightly less relevant (1/4 unit), and of slightly lower quality (1/2 unit). However, the relevance of the selected diverse prompts was not explicitly controlled. P173 in Directed(1) felt that the phrase “\underline{first set of challenges is}” was “\textit{straightforward and gave me the idea of what to write. It was very easy}”; whereas P157 in felt that the phrase “\underline{review findings should be}” “\textit{didn't really have anything I could think to tie towards a motivational message. I tried to think of it as looking back to see progress in terms of reviewing your journey.}” Random prompts with more phrases were harder to understand, perhaps, because they were randomly grouped and are less semantically similar. P128 in Random(3) found that “\textit{these} [phrases] \textit{were hard to combine since they deal with different aspects of exercise. Also the weight lifting seems to be not the best thing for addressing obesity, so that was hard to work in.}”

We found that ideation effort was mediated by prompt factors. Figure \ref{fig:Directed-Fig6} (Bottom) shows that Directed prompts were least easy to use for ideation, and less adopted than Random selected prompts. This is consistent with Directed prompts being less understandable than Random. Ideating with 1-phrase prompts increased ideation time from 44.1s by 21.6s (48.9\%) compared to None, and viewing 3 phrase increased time further by 11.9s. In summary, Directed Diversity may improve diversity by selecting unexpected prompts, but at some cost of ideator effort and confusion. This cost compromises prompt adoption and suggests that directing diversity may not work. Yet, as we will show later, Directed Diversity does improve ideation creativity. We analyzed the confound of understandability further in Appendix Section \ref{appendix:understandability}. Next, we investigate if prompts characteristics mediate more ideation creativity.

\subsubsection{Results of Mediation Analysis of Diversity Propagation from Prompt to Ideation (RQ3.2c)}

We found that prompt configuration and perceived prompt creativity mediated the individual diversity of ideated messages (RQ3.2b). Appendix Table \ref{tab:Directed-Tab21}a (in) shows that Ideation Mean (or Min) Pairwise Distance increased with Prompt Mean (or Min) Pairwise Distance by +0.176 (or +0.146), and marginally with Intra-Prompt Phrase Mean Distance by +0.021 (or +0.020). This means that farther Prompts stimulated farther Ideations, and higher variety of Phrases within each prompt drove slightly farther Ideations too. Hence, prompt diversity (mean pairwise distance) influenced ideation diversity, and prompt redundancy (minimum pairwise distance) influenced ideation redundancy. Appendix Table \ref{tab:Directed-Tab21}b shows that as Prompt Relevance decreased by one Likert unit (on 5-point scale), ideation mean pairwise distance decreased by 0.0034 (7.9\% of ideation pairwise distance SD of 0.043) and ideation minimum pairwise distance decreases by 0.0056 (13\% of SD). This suggests that prompting with irrelevant phrases slightly reduced diversity, since users had to have to conceive their own inspiration; e.g., P165 in Directed(1) “\textit{couldn't make sense of the given messages, so I tried my best to make something somewhat motivational and correct from them.}”. Prompt understandability and quality did not influence ideation individual diversity (p=n.s.). In summary, selecting and presenting computationally diverse and less redundant prompts increased the likelihood of crowdworkers ideating messages that are more computationally diverse and less redundant.

\subsubsection{Results on Evaluating Individual, Collective Objective, Thematic Ideation Diversity (RQ3.3)}

\begin{figure}[h]
    \centering
    \includegraphics[width=1\textwidth]{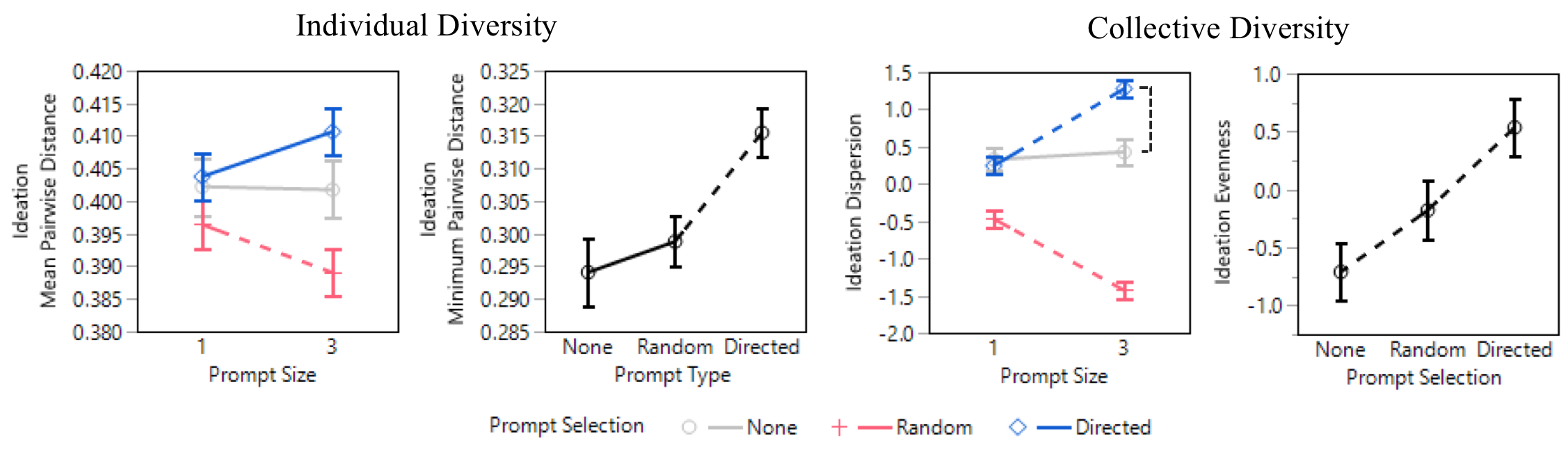}
    \caption{Results of computed individual and collective diversity from ideations for different prompt configurations. See Figure \ref{fig:Directed-Fig6} caption for how to interpret charts.}
    \label{fig:Directed-Fig7}
\end{figure}

Having shown the mediating effects of diverse prompts, we now evaluate how prompt selection techniques affect self-assessed creativity ratings, objective diversity metrics of ideations, and thematically coded diversity metrics of ideations. To carefully distinguish between the commonly used mean pairwise distance with the less used minimum pairwise distance, we performed our analyses on them separately. We calculated one measurement of each collective diversity metric in Table \ref{tab:Directed-Tab3} for all messages in each prompt selection condition, and computed uncertainty estimations from synthesized 50 bootstrap samples \footnote{For each dataset, randomly sample with replacement from the original dataset until the same dataset size is reached.} to generate 50 readings of each diversity metric. We performed factor analyses on the metrics as described in Section \ref{sec:ch3-factor-analysis}, and performed statistical analyses on these factors as described in Appendix Table \ref{tab:Directed-Tab22}. Analyses on both individual diversity and collective diversity measures had congruent results (Figure \ref{fig:Directed-Fig7}), though results for collective diversity had more significant differences ($p<.001$). For collective diversity, our factor analysis found that Ideation Dispersion was most correlated with mean pairwise distance, and Ideation Evenness with entropy and mean of Chamfer distance. Directed(3) improved Ideation Dispersion from None, while Random reduced Dispersion (even more for 3 vs. 1 phrases). Directed prompts improved Ideation Evenness more than Random with respect to None. There was no significant difference for self-assessed Ideation Quality (p=n.s., Table \ref{tab:Directed-Tab23}a in Appendix).

\subsubsection{Thematic Ideation Diversity}

The previous ideation diversity metrics were all computational. We next assess diversity with human judgement based on thematic analysis. 
By performing a thematic analysis, we can gain an understanding of the diversity of messages on a more qualitative level. Specifically, we categorised messages based on verbal cues such as behaviour change techniques from health behaviour science \cite{michie2011refined,schutzer2004barriers,orji2012towards,michie2013behavior}, or the informational content of messages (such as specific exercises, locations, times, or authorities being referenced).
To conserve manpower when evaluating ideations, we limited thematic coding and crowdworker validation to ideations from three conditions of prompts with 1 phrase, i.e., None, Random(1), and Directed(1). 

To conduct our thematic analysis of ideated messages, we used open coding of grounded theory \cite{glaser2017discovery} to derive categories. These categories were added, reduced, merged, and refined by iteratively assessing the messages. We then consolidated the categories into themes using affinity diagramming \cite{beyer1998holzblatt}. The thematic analysis was primarily performed by one co-author researcher with regular discussion with co-authors. We calculated inter-rater reliability on a random 10\% subset of messages was coded independently by another co-author to obtain a Krippendorff’s alpha with MASI distance of $\alpha = 0.82$, which indicated good agreement. Note that while thematic analyses and affinity diagramming are popular methods to interpret qualitative data, we use them here for quantifying qualitative themes and categories of messages.

Our thematic analysis resulted in 239 categories 
which we consolidated to 53 themes. 
For example categories of ``\textit{Squats}'' or ``\textit{Dance}'' are in the ``\textit{Exercise Suggestion}'' theme; and the category ``\textit{Strong immune system}'' is in the ``\textit{Health Benefits}'' theme.
See Table \ref{tab:directed-thematic} for full list of categories and themes.

\begin{longtable}[c]{p{3.5cm}p{12cm}}
\caption{Themes and categories identified with the thematic analysis of messages}
\label{tab:directed-thematic}\\
\toprule
\multicolumn{1}{c}{\textbf{Theme}} &
  \multicolumn{1}{c}{\textbf{Categories}} \\
\midrule
\endfirsthead
\multicolumn{2}{c}%
{{\bfseries Table \thetable\ continued from previous page}} \\
\midrule
\multicolumn{1}{c}{\textbf{Theme}} &
  \multicolumn{1}{c}{\textbf{Categories}} \\
\midrule
\endhead
  Ambiguous Benefits &
  Ambiguous benefits \\
  Anecdote &
  Anecdote \\
  Appeal to ``Obvious'' Knowledge &
  Appeal to ``obvious'' knowledge \\
  Appeal by Cohorts &
  Appeal to children $\mid$ Appeal to older adults $\mid$ Appeal to weight \\
  Appeal to Fear &
  Appeal to fear \\
  Appeal to Guilt &
  Appeal to guilt \\
  Appeal to Shame &
  Appeal to shame \\
  Appeal to Social Approval &
  Appeal to social approval \\
  Barrier to Ability &
  Barrier to ability \\
  Barrier to Boredom &
  Encourage exercise variety $\mid$ Prompt to try something new $\mid$ Tips to make exercise less boring/more fun  \\
  Barrier to Comfort &
  Barrier to comfort \\
  Barrier to Cost &
  Cheap exercises $\mid$ Lower healthcare/insurance costs \\
  Barrier to Effort &
  It will get easier $\mid$ Recommending less effortful routines or exercises $\mid$ Take a short break then carry on \\
  Barrier to Energy &
  Barrier to energy \\
  Barrier to Enjoyment &
  Prompt to research fun exercise $\mid$ Recommending enjoyable activity \\
  Barrier to Motivation &
  Barrier to motivation \\
  Barrier to Resources &
  At home exercises $\mid$ No equipment needed $\mid$ No gym available \\
  Barrier to Self-Efficacy &
  Don’t feel bad if confused $\mid$ Don’t need certificate or qualifications $\mid$ Improving self-confidence $\mid$ Recommending exercises within ability \\
  Barrier to Time &
  Barrier to time \\
  Call to Action &
  Call to action \\
  Call to Authority &
  Citing health experts $\mid$ Unspecified authority \\
  Collective Societal Benefits &
  Collective societal benefits \\
  Equipment &
  Bench press $\mid$ Exercise machine $\mid$ Exercise machines (unspecified $\mid$ Rubber exercise tubing/band $\mid$ Swimming gear $\mid$ Treadmill $\mid$ Vertical or horizontal press $\mid$ Work - standing desk \\
  Exercise Suggestion &
  Aerobic exercises $\mid$ Aerobics $\mid$ Anaerobic exercise $\mid$ Biking $\mid$ Body weight exercises $\mid$ Cardiovascular exercise $\mid$ Climbing $\mid$ Competitive cycling $\mid$ Dance $\mid$ Diving $\mid$ Double clean $\mid$ Exercise through chores $\mid$ Handstand $\mid$ High intensity exercise $\mid$ Hot yoga $\mid$ Internal rotation workouts $\mid$ Iron Yoga $\mid$ Jump on bed $\mid$ Jumping jacks $\mid$ Lift weights $\mid$ Lifting luggage $\mid$ Meditation $\mid$ Pull-ups $\mid$ Pushing kids on swings $\mid$ Push-ups $\mid$ Resistance exercises $\mid$ Ring Pull-ups $\mid$ Running $\mid$ Seated leg-raises $\mid$ Sit-ups $\mid$ Snatches $\mid$ Sports $\mid$ Squats $\mid$ Strength training $\mid$ Strenuous/moderate/vigorous exercise $\mid$ Stretching $\mid$ Swimming $\mid$ Tennis $\mid$ Using stairs $\mid$ Vertical and horizontal presses $\mid$ Volleyball $\mid$ Walk your dog $\mid$ Walking $\mid$ Water exercises $\mid$ Work/desk exercise $\mid$ Yoga \\
  Fear of Injury &
  Don't overexert yourself $\mid$ Recommending exercises to avoid injury $\mid$ Research good techniques to avoid injury $\mid$ Take breaks $\mid$ Tips to avoid injury for outdoor activities \\
  Food and Drink &
  Avoid steroids, pills and drugs $\mid$ Avoid unhealthy food $\mid$ Exercise supplements $\mid$ Exercise to avoid medication and drugs $\mid$ Food recommendation $\mid$ Staying hydrated $\mid$ Stress eating advice \\
  Future Life &
  Improve quality of life $\mid$ Live longer \\
  Goals &
  Journaling to track your goal $\mid$ Set Actionable goals $\mid$ Set daily exercise goal $\mid$ Set goals based on health recommendations $\mid$ Set unspecified goal $\mid$ Set weight goal $\mid$ Tips to reach goals $\mid$ Visualising meeting goals \\
  Health Advice &
  Advice for diabetics $\mid$ See a doctor if you are worried \\
  Health Benefits &
  Better mobility $\mid$ Bone health $\mid$ Cardio Health $\mid$ Fluid regulation $\mid$ Help with foot problems $\mid$ Lowers blood pressure $\mid$ More stamina $\mid$ Pain and strain relief $\mid$ Slows aging process $\mid$ Strong immune system $\mid$ Unspecified health benefit \\
  Health Risks &
  Arthritis $\mid$ Breathing difficulty $\mid$ Cancer $\mid$ Depression $\mid$ Diabetes $\mid$ Heart disease $\mid$ Hernias $\mid$ Obesity $\mid$ Unspecified health risk \\
  Improving Appearance &
  ``Look better'' $\mid$ Beach ready $\mid$ Chest and back $\mid$ Improved posture $\mid$ Look more appealing to potential partners $\mid$ Nice butt $\mid$ Six-pack abs \\
  Inspirational Phrase &
  Inspirational phrase \\
  Lack of Knowledge &
  Research exercise routines $\mid$ Research nutrition $\mid$ Study exercise form \\
  Lack of Social Support &
  Exercise with an expert $\mid$ Exercise with friends $\mid$ Family want you to be healthy $\mid$ Find places to support you $\mid$ Impress your doctor $\mid$ Interacting with others $\mid$ Join a health club $\mid$ Join an exercise class $\mid$ Meeting new people $\mid$ Playing/exercising with children $\mid$ Social competition \\
  Mental Health Benefits &
  Feeling better $\mid$ Happier $\mid$ Improved sleep $\mid$ Improves cognitive abilities $\mid$ Lower depression $\mid$ Lowers anxiety $\mid$ Relaxing $\mid$ Release endorphins $\mid$ Self-esteem $\mid$ Stress reduction $\mid$ Unspecified mental health benefits \\
  Muscle Building &
  Biceps $\mid$ Core strength $\mid$ Leg Muscles $\mid$ Physically stronger $\mid$ Shoulder muscles $\mid$ Triceps $\mid$ Unspecified Muscle building $\mid$ Upper body strength \\
  Overcoming Beliefs &
  Changing mindset \\
  Overcoming Family Obligations &
  Overcoming family obligations \\
  Overcoming Self-Consciousness &
  Overcoming self-consciousness \\
  Push to Do More &
  Prompt to increase $\mid$ Prompt to steadily increase \\
  Push to Start &
  Prompt to start exercising $\mid$ Prompt to stop sedentary lifestyle \\
  Rewards &
  Prizes from exercise competitions $\mid$ Reward with food $\mid$ Reward with new clothes $\mid$ Reward with new exercise equipment $\mid$ Unspecified reward \\
  Self-Empowerment &
  Self-empowerment \\
  Self-Forgiving &
  Self-forgiving \\
  Self-Reflection &
  Self-reflection \\
  Social Comparison &
  Avoid social comparisons $\mid$ Downwards Social comparison $\mid$ Upwards Social Comparison \\
  Specific Locations &
  Around the neighbourhood $\mid$ At desk exercises $\mid$ At home $\mid$ At school $\mid$ Beach $\mid$ Church/community centre $\mid$ Front yard $\mid$ Gym $\mid$ Outdoors $\mid$ Park $\mid$ Travelling/airport $\mid$ Walk to train station \\
  Specificity &
  Appropriate Exercises (e.g. ``Try what's best for you'') $\mid$ Developing habits $\mid$ Even small amounts of exercise $\mid$ Exercise daily $\mid$ Exercise regularly $\mid$ Follow exercise plan/routine $\mid$ Specific amount or distance to exercise $\mid$ Specific days a week to exercise $\mid$ Specific minutes to exercise \\
  Time to See Results &
  Dedicate time $\mid$ Fast results $\mid$ Promise of results $\mid$ Tips to progress faster \\
  Time to Exercise &
  Anytime $\mid$ End of the day $\mid$ Morning exercise $\mid$ Spring $\mid$ Summer/hot weather \\
  Use of Technology &
  Exergame $\mid$ Experts review your exercises from an app $\mid$ Follow videos $\mid$ Listen to music/podcast $\mid$ Reflect on progress $\mid$ Use apps for exercise tips $\mid$ Use apps for workout schedules $\mid$ Use apps to track progress $\mid$ Watch TV while exercising \\
  Weight Loss &
  Aid digestion $\mid$ Boost metabolism $\mid$ Burning calories $\mid$ Burning cellulite $\mid$ Maintaining weight $\mid$ Slimming down \\
\bottomrule
\end{longtable}

\begin{figure}[H]
    \centering
    \includegraphics[width=1\textwidth]{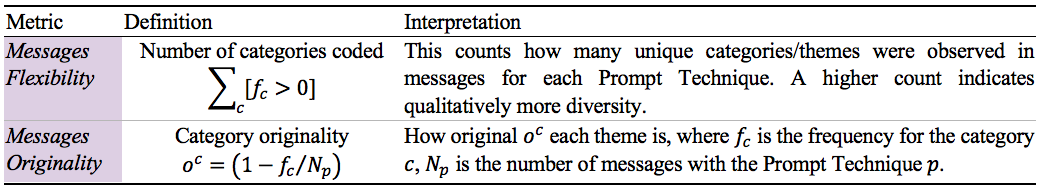}
    \caption{Metrics of creativity of ideation based on categories and themes derived from our thematic analysis of generated ideas. Metrics are shown for categories, but are the same for themes.}
    \label{tab:Directed-Tab7-Main}
\end{figure}

Next, to quantify categories and themes (and compare between conditions) we used metrics from creativity literature, namely the flexibility \cite{sowden2015improvisation} and originality \cite{zenasni2009perception} of ideations. Flexibility \cite{sowden2015improvisation} measures how many unique ideas (conceptual categories) was generated, and originality \cite{zenasni2009perception} measures how infrequently each conceptual category occurs.
Finally, we calculate the flexibility and originality measures based on the coded categories (fine-grained) and themes (coarser) described in Figure \ref{tab:Directed-Tab7-Main}.

\begin{figure}[h]
    \centering
    \includegraphics[width=1\textwidth]{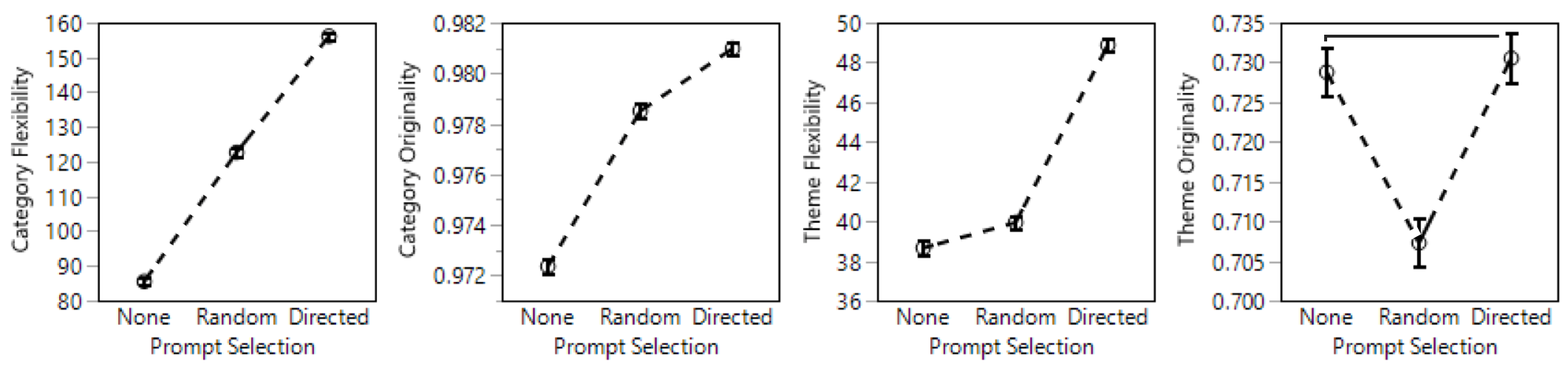}
    \caption{Results of diversity in categories and themes derived from thematic analysis of ideations.}
    \label{fig:Directed-Fig8}
\end{figure}

Figure \ref{fig:Directed-Fig8} shows results from our statistical analysis. We found that ideations generated with Directed prompts had higher Flexibility and Originality in categories and themes than with Random or None, demonstrating that Directed prompts generated more diverse verbal cues (conversational content). Ideations from Random prompts mostly had higher Flexibility and Originality compared to None, but the theme Originality was significantly lower. This could be because Random prompts primed ideators to fixate on fewer broad ideas (themes), instead of the higher number of fine-grained idea categories.

In summary, despite lower ideation ease and understandability with Directed prompts (Section \ref{sec:ch3-ideation-study}), we found objective and thematic evidence that Directed Diversity improved ideation diversity compared to Random and None. Additionally, we found that directed (through our thematic analysis) led to the generation of more diverse (conversational content) verbal cues. Next, we describe how crowdworkers rated these ideations.

\subsection{Validation User Studies}
\label{sec:ch3-validation-study}

The third and fourth studies employed third-party crowdworkers to assess the creativity of ideated messages from the Ideation User Study, to answer \textit{(RQ3.3) How do prompt selection techniques affect diversity in ideations?} This provides a less biased validation than asking ideators to self-assess. We conducted three experiments with different questioning format to strengthen the experiment design. Appendix Figures \ref{fig:Directed20} to \ref{fig:Directed24} details the questionnaires.

\subsubsection{Individual Validation: Experiment Treatment and Procedure}
\label{subsec:ch3-indivual-validation}

For the individual validation study, we conducted a within-subjects experiment with prompt selection technique (None, Random, Directed) as independent variable, and controlled prompt size ($g=1$). Each participant assessed 25 ideation messages chosen randomly from the three conditions. Participants went through the same \hyperref[sec:ch3-procedure]{procedure as in the ideation user study}, but with a different task in step 3:

\begin{itemize}
    \item Assess 25 messages regarding how well they motivate for physical activity. For each message,
    \begin{enumerate}[label=\alph*.]
        \item Read a randomly chosen message.
	    \item Rate on a 7-point Likert scale, whether the message is motivating (effective), informative, and helpful (as described in Section \ref{sec:ch3-measures}).
	    \item Reflect and write the rationale in free text on why they rated the message as effective or ineffective. This was only asked randomly two out of 25 times, to avoid fatigue.
    \end{enumerate}
\end{itemize}

As we discuss later, we found that participants confounded the three ratings questions and answered them very similarly (responses were highly correlated), thus, we designed collective validation user studies to pose different questions and distinguish between the measures.

\subsubsection{Collective Ranking Validation: Experiment Treatment and Procedure}

The collective validation study had the same \hyperref[subsec:ch3-indivual-validation]{experiment design as described above}, but with a different procedure step 3:

\begin{itemize}
    \item Complete 5 trials to rate collections of ideation messages, where for each trial:
    \begin{enumerate}[label=\alph*.]
        \item Read three groups of 5 messages 
        \item Rank message groups as most, middle or least \textit{motivating}, \textit{informative}, and \textit{repetitive}. 
    \end{enumerate}
\end{itemize}

Instead of rating messages individually, participants viewed grouped messages from each condition side-by-side and answered ranking questions. Messages in each group were selected from those ideated with the same prompt selection technique. By asking participants to assess collections rather than individual messages, we explicitly measured perceived diversity, since the user perceived the differences between all ideations in the collection; this is more direct than asking them about the “informativeness” of an ideation, since this could be confounded with “helpfulness”, “teaching something new”, “telling something different from other messages”, etc. This approach differs from the triplet similarity comparison \cite{van2012stochastic,tamuz2011adaptively} employed by Siangliulue et al. \cite{siangliulue2015toward}, and benefits from requiring fewer assessments. We asked participants to rank groups rather than rate them relatively to obtain a forced choice \cite{dougherty1999minerva}. Another method to assess diversity involves longitudinal exposure (e.g., \cite{kocielnik2017send}), but this is expensive and difficult to scale.

\subsubsection{Collective Pairwise Rating Validation: Experiment Treatment and Procedure}
\label{sec:ch3-collective}

The collective pairwise rating validation study further validates our results with an existing, commonly used measure to rate the difference between pairs of messages, both from the same prompt selection technique \cite{siangliulue2015toward,dow2010parallel}. We randomly selected 200 message-pairs from None, Random(1) and Directed(1), yielding a pool of 600 message-pairs. All steps in the procedure are \hyperref[subsec:ch3-indivual-validation]{identical as before} except for Step 3:

\begin{itemize}
    \item Rate 30 message-pairs randomly selected from the message-pair pool, where for each message-pair,
    \begin{enumerate}[label=\alph*.]
        \item Read the two messages
        \item Rate their difference on a 7-point Likert scale: 1 “Not at all different (identical)” to 7 “Very different”
    \end{enumerate}
\end{itemize}

This complements the previous study by having participants focus on two messages to compare, which is more manageable than assessing 5 messages, but is limited to a less holistic impression on multiple messages.

\subsubsection{Experiments Data Collection and Statistical Analysis}

For all validation studies, we recruited participants from Amazon Mechanical Turk with the same high qualification as the ideation study. Of 348 workers who attempted the surveys, 290 passed the screening tests to complete the surveys (83.3\% pass rate). They were 50.2\% female, between 22 and 71 years old (M=38.1); 67.5\% of participants have use fitness apps. For the individual validation study, Participants completed the survey in median time 14.7 minutes and were compensated USD\$1.50; for the collective ranking validation study, participants completed the survey in median time 12.7 minutes and were compensated US\$1.80; for the collective pairwise rating validation study, participants completed the survey in median time 8.4 minutes and were compensated USD\$1.00. In total, 740 messages were individually rated 3,375 times (M=4.56x per message), 450 message groups were ranked 1,350 times (M=3.00x per message group), and 600 message pairs were rated 2,430 times (M=4.05x per message pair). To assess inter-rater agreement, we calculated the average aggregate-judge correlations \cite{chan2017semantically} as r=.59, .62, .63 for motivation, informativeness and helpfulness for individual validation ratings, respectively; these were comparable to Chan et al.’s r=.64 for idea novelty \cite{chan2017semantically}. 

We performed the same statistical analyses as in the Ideation User Study (see Section \ref{sec:ch3-ideation-study}), report the linear mixed effects models in Appendix Table \ref{tab:Directed-Tab23}, and include participant quotes from their rationale text response where relevant. For the collective ranking validation study, we counted how often each Prompt Selection technique was ranked first or last across the 5 trials, performed factor analyses on the counts for best and worst ranks for the three metrics (motivating, informative, unrepetitive) to derive three orthogonal factors (Ideations Unrepetitive, Ideations Informative, Ideations Motivating), and performed the statistical analysis on the factors (see Table \ref{tab:Directed-Tab23}b in Appendix).

\subsubsection{Results on Evaluating Individual and Collective Ideation Creativity (RQ3.3)}

We investigated whether Directed prompts stimulate the highest ideation diversity and whether 3rd-party validations agree with our computed and thematic results. For illustration, Appendix Table \ref{tab:Directed-Tab25} shows examples of message-groups with high and low factor values. 

\begin{figure}[h]
    \centering
    \includegraphics[width=1\textwidth]{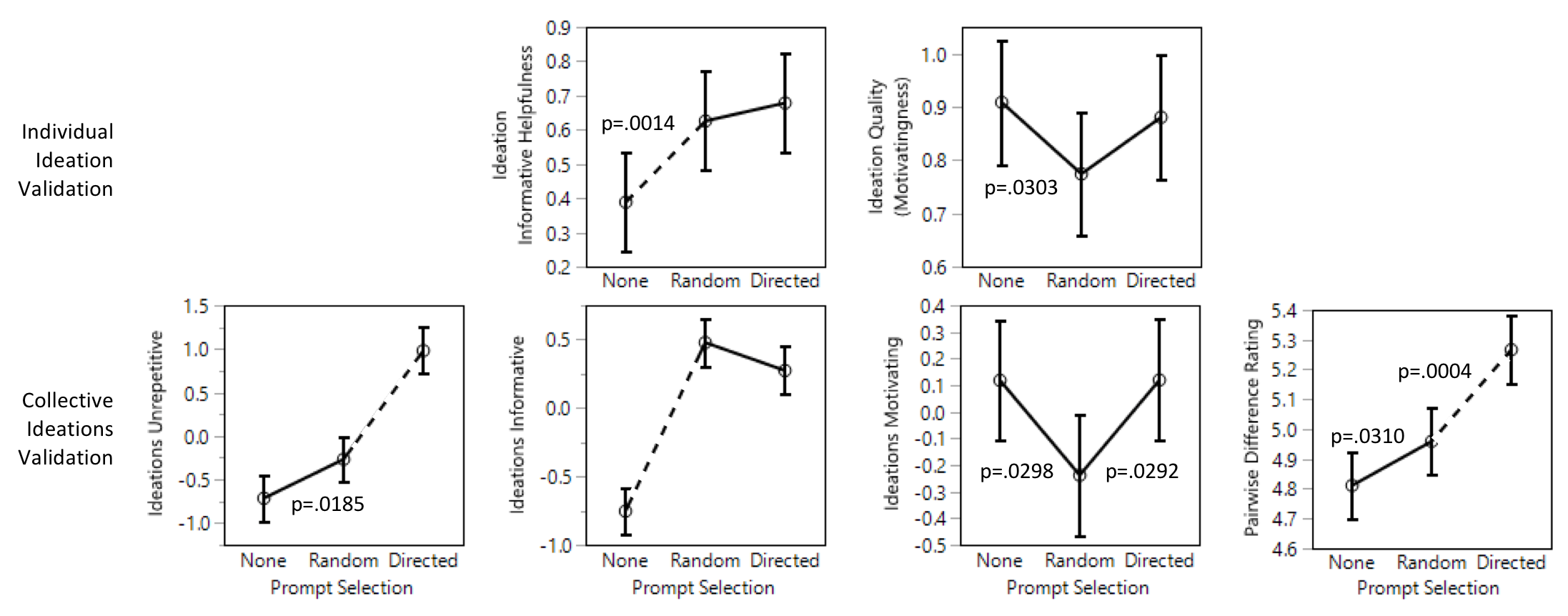}
    \caption{Results of perceived individual and collective creativity from the three validation user studies.}
    \label{fig:Directed-Fig9}
\end{figure}

Figure \ref{fig:Directed-Fig9} shows results of our statistical analysis. We found that ideations from Directed prompts were most different and least repetitive, ideations from Random were no different and as repetitive as None. Ideations generated with prompts were more informative and helpful than without prompts, but there was no difference whether the prompts were Directed or Random. For example, P4 reviewed the message “\underline{Exercise and live longer, and} \underline{prosper more!}” ideated with None, and felt that “\textit{it's basically telling you what you already know. It's a rather generic message.}”; P63 reviewed the message “\underline{Waking up early and} \underline{working out will help you get into shape, and is a great way to have more energy and better} \underline{sleep.}” from the Directed(1) prompt “into a habit of sleep” and felt “\textit{it’s effective because it gives me a goal and tells me why this is a good goal}”. There were no significant differences in ideation quality or motivation, though there was a marginal effect that Random prompts could hurt quality compared to None. Therefore, Directed Diversity helped to reduce ideation redundancy compared to Randomly selected prompts, improved informativeness, and did not compromise quality.

\subsection{Summary of Answers to Research Questions}

We summarize our findings to answer our research questions with results from multiple experiments.

\textbf{RQ3.1} \textit{How did prompt selection techniques affect diversity in prompts?} Compared to Random, Directed Diversity: 1) selected more diverse prompts, 2) with less redundancy from prior ideation, 3) that ideators perceived as more unexpectedness, but 4) of poorer quality and understandability.

\textbf{RQ3.2} \textit{How did diversity in prompts affect the ideation process for ideators?} Compared to Random, prompts selected with Directed Diversity were: 1) harder to ideate with, 2) less applied for ideation, 3) but their higher prompt diversity somewhat drove higher ideation diversity.

\textbf{RQ3.3} \textit{How did prompt selection techniques affect diversity in ideations?} Compared to None and Random, Directed Diversity: 1) improved ideation diversity and reduced redundancy, 2) increased the flexibility and originality of ideated categories, 3) without compromising ideation quality.

\section{Discussion}

We discuss the generalization of our technical approach, evaluation framework, and experiment findings.

\subsection{Need for Sensitive and Mechanistic Measures of Creativity}

We have developed an extensive evaluation framework for two key reasons: 1) to precisely detect effects on diversity, and 2) to track the mechanism of diversity prompting. We have sought to be very diverse in our evaluation of prompt technique to carefully identify any benefits or issues. We have found that some popular metrics (e.g., mean pairwise distance) were less sensitive than others (e.g., MST Dispersion / Remote-tree). Therefore, a null result in one metric (e.g., \cite{siangliulue2015toward}) may not mean that diversity was not changed (if measured by another metric). Instead of only depending on the ``black box'' experimentation of prompt treatment on ideation (e.g., \cite{chan2017semantically,huang2017bluesky,siangliulue2015toward,siangliulue2016ideahound}), investigating along the ideation chain is interpretable and helpful for us to identify potential issues or breakdowns in the diversity prompting mechanism. Had our evaluation results on ideation diversity been non-significant, this would be helpful to debug the lack of effectiveness.

Conversely, we may find that an ideation diversity effect may be due to contradictory or confounding effects. Indeed, we found that Directed Diversity improved diversity, despite poorer prompt understandability and adoption. Ideators could not directly use the selected prompts, but still managed to conceive ideas that were more diverse than not having seen prompts or seeing random ones. This suggests that they generated ideas sufficiently near the prompts. The findings also suggested that the increased effort helped to improve diverse ideation \cite{barbot2018dynamics,beaty2012ideas,wang2017neural}, but the ideator user experience should be improved. Future work is needed to improve Directed Diversity to reduce ideator effort and improve the relevance of selected prompts, such as by limiting the distance of new prompts from prior ideations, or using idea-based embeddings \cite{siangliulue2015toward,siangliulue2016ideahound} instead of language models, as discussed next.

\subsection{Generalization of Directed Diversity to other Domains}

The full process of Directed Diversity (Figure \ref{fig:Directed-Fig1}) allows us to generalize its usage to other domains, such as text creativity tasks beyond motivational messages (e.g., birthday greetings \cite{siangliulue2015toward}) by changing the document sources in the phrase extraction step. In the phrase embedding step, we used the Universal Sentence Encoder \cite{cer2018universal}, but other text embedding models (e.g., word2vec \cite{mikolov2013distributed}, GloVe \cite{pennington2014glove}, ELMo \cite{peters2018deep}, BERT \cite{devlin2018bert}) could be used that model languages slightly differently. In the third step, we selected phrases based on the Remote-tree diversity formulation using an efficient greedy algorithm that approximates the diversity maximization. Other diversity criteria and maximization algorithms could be used (see review \cite{chandra2001approximation}). Note that since USE and similar language models are domain-independent, which do not model the semantics of specific domains and semantic quality, Directed Diversity cannot guarantee improving quality. A domain-specific model trained with human-annotated labels of quality could be used to improve both diversity and quality. Furthermore, instead of representing text with language models, the idea space could be explicitly modelled to obtain embeddings from annotated semantic similarity \cite{van2012stochastic,siangliulue2015toward}. Finally, since Directed Diversity operates on a vector representation of prompting and ideations, it can also be used for ideation tasks beyond text as long as they can be represented in a feature vector by feature engineering or with deep learning approaches, such as furniture \cite{matejka2018dream}, mood boards \cite{koch2020imagesense}, and emojis \cite{zhou2017mojitalk}.

\subsection{Generalization of Evaluation Framework}

Our Evaluation Framework is a first step towards the goal of standardizing the evaluation of crowd ideation. This requires further validation and demonstration on existing methods of supporting crowd ideation. Due to the costs of engineering effort, set-up preparation, and recruitment, we defer it to future work. Just as the Directed Diversity pipeline is generalizable, we discuss how the Diversity Prompt Evaluation Framework is generalizable. We had identified many diversity metrics, but only measured some of them (see \cite{chandra2001approximation} for a review of other mathematical metrics). If applying the framework to non-text domains, the vector-based distance metrics should still be usable if the concepts can be embedded with a domain model. While we analyzed diversity in terms of mathematical metrics \cite{chandra2001approximation} and several measures for creativity \cite{torrance2018guiding}, other criteria may be important to optimize, such as serendipity for recommender systems to avoid boredom \cite{kaminskas2016diversity}.
To measure creativity, just as in prior research \cite{kocielnik2017send}, we had used several Likert scale ratings (e.g., helpfulness and informativeness) and found evidence that participants confound them. Furthermore, it may be excessive to apply all our measures, therefore the researcher is advised to use them judiciously. For example, we found that individually rating ideations tends to lead to poor statistical significance, so this data collection method should be avoided. The thematic analysis coding is also very labor intensive for the research team, but provides rich insights into the ideas generated. We had proposed using ranking and pairwise rating validations of collections of ideations as a scalable way to measure collective diversity. 

While our evaluations based on generating motivational messaging for physical activity helped to provide a realistic context, it was limited to measuring preliminary impressions of validators. The social desirability effect may have limited how accurately participants rated the effectiveness of the messages. While our focus was on evaluating diversity, future work that also seeks to improve and evaluate motivation towards behavior change should conduct longitudinal trials with stronger ecological validity \cite{kocielnik2017send}.

\section{Conclusion}

In this paper, we presented Directed Diversity to direct ideators to generate more collectively creative ideas. This is a generalizable pipeline to extract prompts, embed prompts using a language model, and select maximally diverse prompts. We further proposed a generalizable Diversity Prompting Evaluation Framework to sensitively evaluate how Directed Diversity improves ideation diversity along the ideation chain — prompt selection, prompt creativity, prompt-ideation mediation, and ideation creativity. We found that Directed Diversity improved collective ideation diversity and reduce redundancy. With the generalizable prompt selection mechanism and evaluation framework, our work provides a basis for further development and evaluations of prompt diversity mechanisms.

\chapter{The Effects of a Chatbot's Language Formality on a Users' Information Sharing}
\label{ch:FormalOrCasual}


The previous chapter described a study that used information available online to aid in the generation of diverse and efficacious health messages. 
These messages could be used for one-off interventions, or as training data (such as for a health chatbot).
However, while we generated semantically diverse messages, there can still be much variation in the conversational style of messaging.
For example, a message encouraging someone to go for a run could either be phrased in a casual tone (i.e., ``\textit{Heya! Wanna go for a quick jog?}'') or a comparatively more formal tone (i.e., ``\textit{Good morning. Would now be a good time to go for a run?}'').
To investigate the effectiveness of these varying conversational styles in more detail, our next study investigates the effect of a health chatbot's conversational style on the quality of a user's information sharing. 
This chapter is based on one publication:

\bibentry{cox2022does}.

\section{Introduction}


Chatbots can be used to replace the function of surveys and interviews as a form of data collection \cite{xiao2020tell,kim_comparing_2019}. 
In doing so, chatbots harness some of the advantages of surveys such as scalability and low-cost \cite{pereira2019using}; as well as some of the advantages of human interviews such as dialogue based interactions, and human-like characteristics.

However, user attitudes towards chatbots may be sculpted by their expectations of the most appropriate conversational style, and using the wrong conversational style for a given situation can lead to user frustration \cite{elsholz2019exploring,chaves2021should}.
%
To overcome this issue, previous work has built upon the Computers Are Social Actors (CASA) paradigm \cite{nass1994computers}, which asserts that people apply social norms from human interactions when interacting with computer agents.
This has led to the design of conversational agents that exhibit human-like characteristics that users perceive more positively.
For example, previous work has investigated the use of different conversational styles when requesting information \cite{walker1997improvising,hoegen2019end,bickmore2001relational,bickmore2000weather},
and the use of self-disclosing chatbots to encourage reciprocated user self-disclosure \cite{lee2020designing,adam_onboarding_2019,moon2000intimate}.

Despite this, the effect of a chatbot's language formality on the quality of a user's self-disclosure has not been investigated.
Previous work has already produced some conflicting results regarding the formality of a chatbot's utterances.
For example, participants criticised a HIV chatbot for being overly formal and dissimilar to how people normally talk when using chat applications \cite{van2017potential}.
In contrast, users noted that the tone of a chatbot to perform financial services was too informal and not serious enough given the task at hand \cite{duijst2017can}.
In studies more analogous to self-disclosure, it has been found that chatbots exhibiting a casual conversational style can lead to higher quality collection of multiple choice responses compared to standardised surveys \cite{wambsgnass_conversational_2020,celino_submitting_2020,kim_comparing_2019}. 
Yet, these studies did not explicitly compare different levels of language formality, and instead compared the script from a standardised (non-conversational) survey against a casual (conversational) script.
Thus, these studies motivate an interesting issue, \textcolor{major}{and inspire the \textbf{high-level research question} for our paper:}

\begin{itemize}
    \item \textbf{RQ2:} How does the conversational style (i.e., language formality) of a health chatbot affect the quality of a user's self-disclosure?
\end{itemize}

\textcolor{minor}{To evaluate this question,} we carried out two experiments on Amazon Mechanical Turk (AMT). 
Both studies investigated the impact of using either a casual or (comparatively) formal conversational style, and we used an objective measure of language formality \cite{heylighen1999formality} to design chatbot scripts.
Both studies are motivated similarly at a high-level to investigate the effect of a chatbot's language formality on self-disclosure; with \hyperref[sec:ch4-Study1]{ Study 1 }investigating \textit{likelihood} to disclose information, and \hyperref[sec:ch4-Study2]{ Study 2 } investigating the \textit{quality} of user utterances (and thereby the quality of self-disclosure to the chatbot).

\textcolor{minor}{Specifically, in Study 1 ($N = 187$) we measured user perceptions and} the likelihood of information disclosure while varying the information requested by the chatbot by its domain (health vs non-health information) and information sensitivity (sensitive vs non-sensitive information).
We used empirically validated levels of information sensitivity to differentiate our levels \cite{markos2017information}, and (as we did not want to collect user's sensitive information) we conducted a hypothetical scenario-driven experiment where participants indicated their likelihood to disclose information to the chatbot (similarly to \cite{sundar2019machine}).


\textcolor{minor}{The findings of Study 1 indicated the benefits of varying a chatbot's language formality when requesting health information.
However, as these findings were hypothetical, it motivated us to conduct} Study 2 ($N = 156$) where we analysed the quality of user utterances when people discuss health behaviours with a chatbot.

From our two user studies, we found evidence that chatbots related to the health domain could benefit by adopting a more formal conversational style. 
Specifically, Study 1 found that when a chatbot elicits sensitive health information, users perceive a formal conversational style as more competent and appropriate than a comparatively casual style.
Study 2 found that a chatbot that adopts a formal conversational style elicits higher quality user utterances compared to a comparatively casual conversational style.


\section{Related Work}








In this section we discuss how higher quality responses can be elicited from users of chatbots, and how the language used by a chatbot can affect user perceptions and engagement.



\subsection{Designing Chatbots to Act Like People}



Prior studies have created chatbots that replicate conversational styles people perceive positively.
This has in part been influenced by the CASA paradigm \cite{nass1994computers}, which states that people apply social norms from human interactions when interacting with computer agents. On from this, studies have focused on producing human-like chatbots that emulate human responses and contextual awareness, such as by creating chatbots using empathic language \cite{rashkin_towards_2019,lin_caire_2020,liu2018should,guo2021shing} or politeness strategies \cite{srinivasan2016help,danescu-niculescu-mizil_computational_2013,dippold2020turn}. 
Drawing on CASA, we can develop chatbots that incorporate recommendations for human-to-human communication, but it is uncertain whether such recommendations apply to chatbots given differences in perceived social roles and user expectations of chatbots \cite{khadpe2020conceptual}.
Conflicting findings for \cite{van2017potential,wambsgnass_conversational_2020,celino_submitting_2020,kim_comparing_2019} and against \cite{duijst2017can,jenkins2007analysis} using casual chatbot conversational styles thereby motivate us to compare the impact of a chatbot's language formality on user self-disclosure.







\subsection{Improving Self-Disclosure to Chatbots}

Now, we will discuss work that has manipulated chatbot features to increase user self-disclosure (the voluntary sharing of information from one person to another, such as feelings, opinions, or personal information \cite{pearce1973self}).
People often make decisions to disclose information based on cognitive heuristics: general rules of thumb users apply to different situations \cite{tversky1974judgment,fiske1991social,sundar2019machine,sundar2020online}. 
For example, Sundar et al. found the effect of the machine heuristic applies to disclosure to a chatbot, specifically: people were more likely to disclose to a chatbot than to a human because they believe a machine handles information more securely \cite{sundar2019machine}.




Previous studies have investigated the effects of changing a chatbot's traits such as personality \cite{zhou2019trusting}, levels of mutual disclosure \cite{lee2020designing,adam_onboarding_2019,moon2000intimate,saffarizadeh2017conversational}, and use of small talk \cite{walker1997improvising,hoegen2019end,bickmore2001relational,bickmore2000weather} on users' self-disclosure.
Zhou et al. developed chatbots for job interviews that had either a warm, cheerful personality or a serious, assertive personality, 
and found that high-stakes job interviewees were more willing to confide in a chatbot that used a serious and assertive personality \cite{zhou2019trusting}.
Adam et al. found people disclose more to a chatbot when it requests information conversationally one-by-one rather than all at once (similarly to a standard survey) \cite{adam_onboarding_2019}.
Lee et al. found people disclosed more to a mental health chatbot that disclosed information about itself \cite{lee2020designing}.
Brickmore et al. found that conversational agents engaging in social dialogue (such as small talk) lead to increased levels of rapport and self-disclosure \cite{bickmore2000weather,bickmore2001relational}. 
Xiao et al. used Gricean maxims \cite{grice1975logic} (quantity, informativeness, relevancy, manner) to measure the quality of user utterances \cite{xiao2020tell,xiao2020if}, and found people gave higher quality utterances to a chatbot using active listening techniques compared to a chatbot giving generic responses \cite{xiao2020if}.

On from this, our user studies measured the effect of a chatbot's language formality both on a user's likelihood to disclose information (similarly to Sundar et al. \cite{sundar2019machine}), and quality of user utterances (similarly to Xiao et al. \cite{xiao2020tell,xiao2020if}).




\subsection{Changing Chatbot Conversational Style}




As we want to investigate the effect of varying a chatbot's conversational style (i.e., the formality of a chatbot's utterances), we will provide an overview of related literature.

Previous work has investigated the effect of chatbots incorporating different conversational styles \cite{elsholz2019exploring,linder2020effects,wambsgnass_conversational_2020,celino_submitting_2020,kim_comparing_2019,liebrecht2021too}.
For example, Elholz et al. compared user perceptions of two theatre ticket booking chatbots: one using a modern, and one using a Shakespearean conversational style \cite{elsholz2019exploring}. They found that while the former was more user-friendly, the latter was more entertaining to users.
Kim et al. compared the effect of a chatbot using a casual conversational script (adopting conversational language, emojis and abbreviations) against a chatbot using a standardised survey script, and found that a casual conversational style leads to collection of higher quality multiple choice survey responses \cite{kim_comparing_2019}.
However, studies using casual conversational styles \cite{kim_comparing_2019,celino_submitting_2020,wambsgnass_conversational_2020} used baseline conditions of language from traditional surveys (using more terse, non-conversational language) rather than a formal but conversational style, and did not investigate the effect of asking for personally sensitive information.

We therefore conducted user studies where a chatbot elicited information using comparatively casual and formal conversational styles to investigate the effect of a chatbot's language formality on user perceptions and self-disclosure.




\section{Study 1: The Effect of Conversational Style on Disclosure}
\label{sec:ch4-Study1}

In this study, we investigate how the conversational style of a chatbot and the sensitivity of information requested by the chatbot affect user perceptions and likelihood to disclose information. 
\textcolor{major}{This gave us the research questions of:}
\begin{itemize}
    \item \textcolor{major}{\textbf{RQ2.1:} How does the domain and sensitivity of information requested by a chatbot influence a user's likelihood to disclose information?}
    \item \textcolor{major}{\textbf{RQ2.2:} How does a chatbot's conversational style (formal vs casual) influence a user's likelihood to disclose information?}
\end{itemize}


\textcolor{minor}{For this,} we conducted a between-subjects, scenario-based experiment on AMT where participants were asked to imagine they had been talking to a chatbot to help them discover life insurance products. 
Life insurance was chosen as the chatbot's use case as this allowed the chatbot to request information of varying sensitivity, and ensured the scenario was relatable given the prevalence of financial assistance chatbots.

\subsection{Experiment Conditions}

Participants were shown a screenshot of a conversation between a human and a chatbot.
Chatbot scenarios shown to participants were 2 $\times$ 3 factorial with the following factors:

\begin{itemize}
    \item \textit{Chatbot Conversational Style}: casual, formal;
    \item \textit{Domain Sensitivity (of information requested)}: Low-sensitivity non-health, High-sensitivity non-health, High-sensitivity health.
\end{itemize}

\subsubsection{Chatbot Conversational Style}

To objectively measure language formality, we use a formality measure taken from linguistics literature \cite{heylighen1999formality}, which has been used previously to differentiate between genres of texts \cite{teddiman2009contextuality,Nowson05weblogsgenres}, and detect clickbait news articles \cite{biyani20168}.
This metric (known as the F-score) calculates language formality by using the part of speech of language in the following formula: 


\begin{displayquote}
F = (noun frequency + adjective freq. + preposition freq. + article freq. – pronoun freq. – verb freq. – adverb freq. – interjection freq. + 100)/2
\end{displayquote}

This formula measures the ``deep formality'' (akin to the specificity) of a text.
From this definition, a casual style uses more deictic terms whose understanding is dependant upon context (such as ``this'', ``later''), whereas a formal style uses explicit terms such as nouns, prepositions, and articles.
A higher F-score denotes more formal language, while a lower F-score is more casual. Within Heylighen's \cite{heylighen1999formality} study (for the English language) phone conversations had the lowest F-score (scoring 36), while informational writing had the highest F-score (scoring 61). Although the F-score has limitations such as not accounting for differences in punctuation, and scoring all words as $\pm1$ (while slang could be perceived as more casual and technical language more formal), it acts as a proxy of formality to differentiate our levels of conversational style.

\begin{figure*}[!htb]
    \centering
    \begin{subfigure}[b]{0.45\textwidth}
        \includegraphics[width=\textwidth]{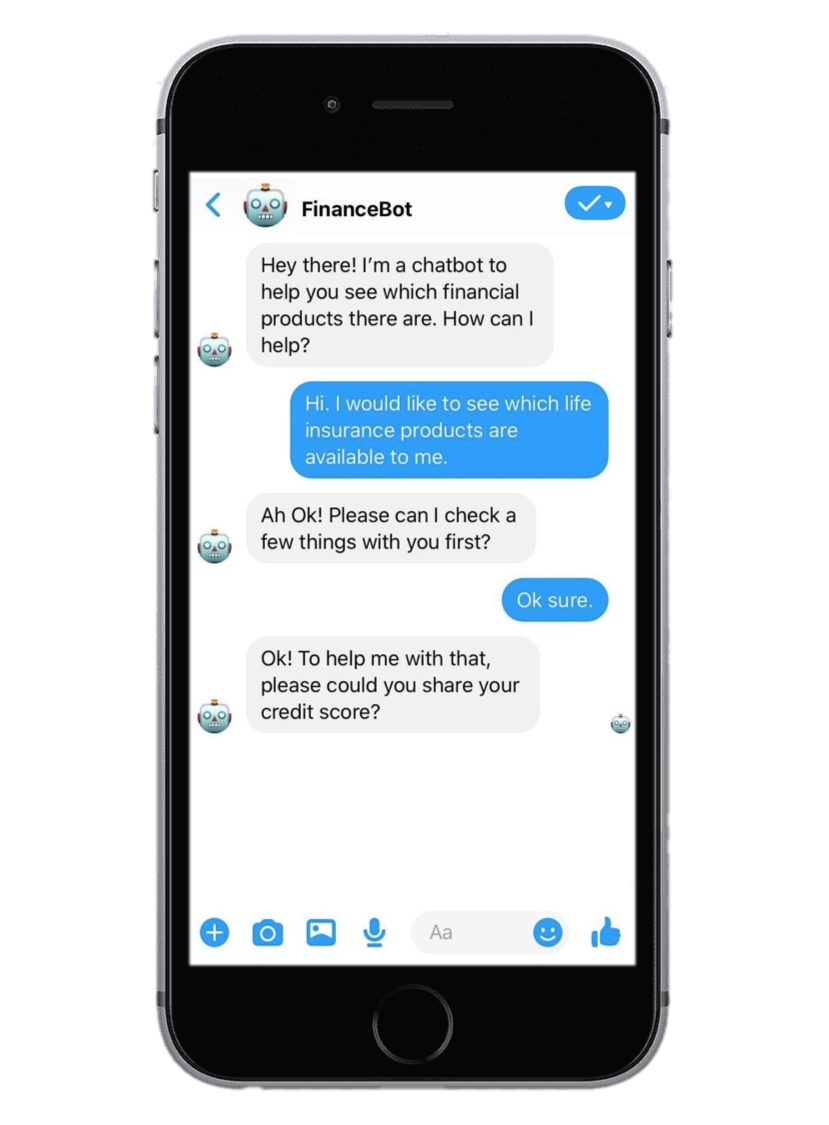}
        \caption{\textbf{Casual chat condition:}
        with F-score of 42 (similar to F-score of normal spoken English conversation \cite{heylighen1999formality})}
        \label{fig:casual-credit}
    \end{subfigure}
    \hfill 
    \begin{subfigure}[b]{0.45\textwidth}
        \includegraphics[width=\textwidth]{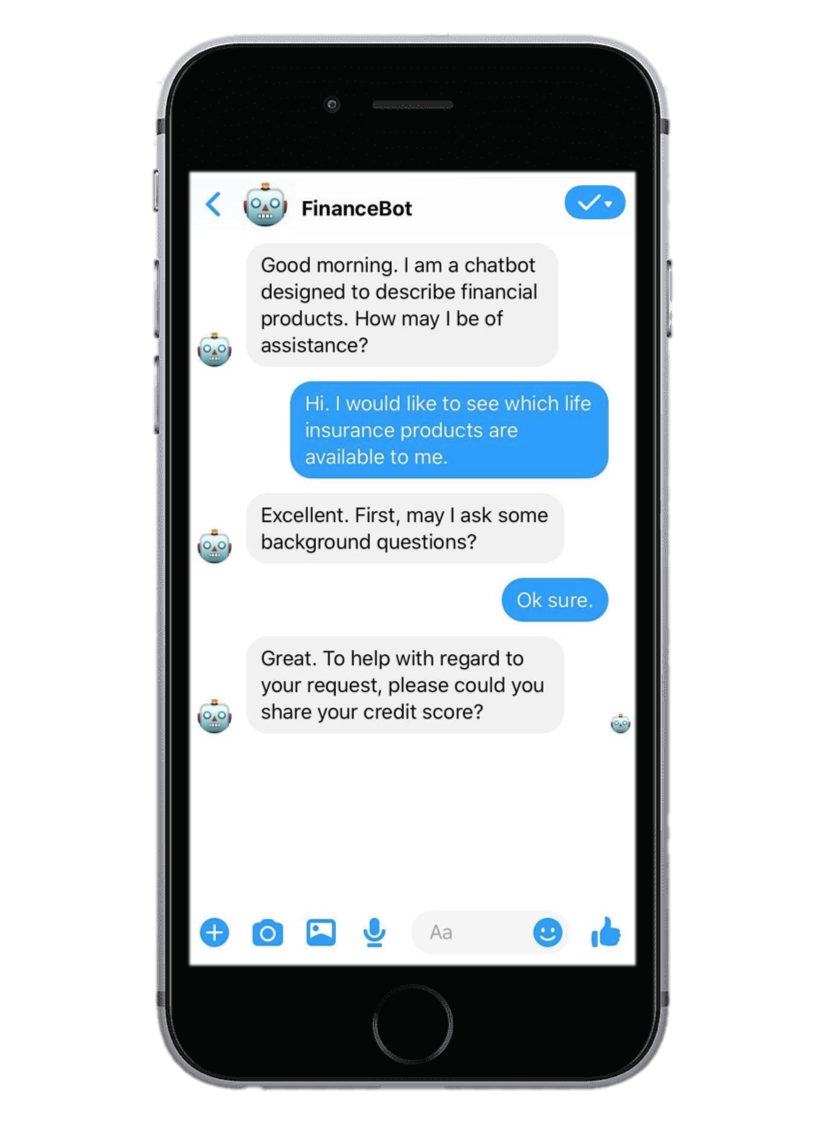}
        \caption{\textbf{Formal chat condition:} example with F-score of 50.5 (similar to F-score of prepared English speech \cite{heylighen1999formality})}
        \label{fig:formal-credit}
    \end{subfigure}
    \caption{The chatbot conversation for both casual and formal conversational styles when asking for the user's credit score.}
    \label{fig:Mock-ups}
\end{figure*}

This definition of formality should not be confused with references to casual language in other literature, which are more analogous to using politeness strategies, empathic communication, emojis, or slang \cite{kim_comparing_2019,thies2017you,duijst2017can,jenkins2007analysis}.
As we are not investigating the effectiveness of these, we scripted chatbot dialogue to be similar (in terms of meaning and politeness) between casual and formal conditions.
See Figure \ref{fig:Mock-ups} for the chat sessions shown to participants in the casual and formal conditions, along with each condition's corresponding F-score.

\subsubsection{Domain Sensitivity of Information Requested}

We compared the influence of low and high sensitivity information, and health and non-health information. 
Doing so allowed us to investigate whether people dislike a request for information due to information requested being health sensitive or information sensitive.
To choose experiment conditions, we drew from Markos et al.'s survey of US consumers' perceived sensitivity of their personal information 
\cite{markos2017information}. 
Specifically, we chose three pieces of information which affect the price of life insurance policies as our experiment conditions,
giving us conditions where the chatbot asks the user for their:

\begin{itemize}
    \item \textbf{income level} (low-sensitivity non-health \cite{markos2017information})
    \item \textbf{credit score} (high-sensitivity non-health \cite{markos2017information}).
    \item \textbf{medical history} (high-sensitivity health \cite{markos2017information}).
\end{itemize}

\subsection{Design Considerations}

To avoid racial or gender mirroring effects \cite{liao2020racial}, we used a robot icon to represent the chatbot in the mock-up scenarios (see Figure \ref{fig:Mock-ups}).
Dialogue in all conditions was the same except for the language formality of chatbot responses, and the information requested by the chatbot in the final message. 
Human responses are identical between conditions, and chatbot scripts deploy the same politeness strategies \cite{danescu-niculescu-mizil_computational_2013,brown_politeness_1987} and follow the same conversation structure.
First, the chatbot greets the user, states its purpose, and offers assistance. Next, the chatbot asks for permission to request information, before finally requesting specific information.


\subsection{Participants}
\label{sec:Study1Participants}

We recruited participants from AMT, and used selection criteria to ensure reliable results (US based, $>$98\% approval rate, $>$ 5,000 completed HITs). 
Participants were paid USD\$0.60 and the survey took a mean time of 4.22 minutes to complete.
264 participants attempted the survey, and 187 (mean age 37.4; 81 female) passed screening tasks. 
Participants were randomly assigned a condition with 95 in casual and 92 in formal.


\subsection{Procedure}

Participants completed the following procedure:
\begin{enumerate}
    \item Introduction and consent form.
    \item Pass word associativity test \cite{chandler2019online} screening for English language skills (similarly to previous crowdsourcing studies \cite{cox2021diverse, chandler2019online}).
    \item Answer demographics (age, gender, education level, prior experience with chatbots) and one instructional manipulation check (from Oppenheimer et al. \cite{oppenheimer2009instructional}), followed by more detailed instructions.
    \item Each participant sees one screenshot of a conversation between a chatbot and a user seeking information about financial products (see Figure \ref{fig:Mock-ups} for screenshots). Participants were told ``\textit{Please imagine you have been talking to the chatbot in the conversation below}'', and evaluated the chat sessions using measures described in Section \ref{sec:Study1Measures}.
    \item Post-test questions (see Section \ref{sec:Study1PostQs}).
\end{enumerate}

\subsection{Measures}
\label{sec:Study1Measures}

Participants evaluated \textcolor{minor}{the} scenario \textcolor{minor}{assigned to them} using measures on 7-point Likert scales (1 = Strongly Disagree to 7 = Strongly Agree), and were asked ``\textit{Do you personally agree or disagree that...}'' for the following:

\begin{itemize}
    \item \textbf{Likelihood to disclose}: ``I am likely to disclose my [``income level''/``credit score''/``medical history''] to the chatbot'' \cite{sundar2019machine,markos2017information}
    \item \textbf{Enjoyment}: ``I would enjoy talking to the chatbot'' \cite{celino_submitting_2020}
    \item \textbf{Warmth}: ``This chatbot was warm'', ``This chatbot was good-natured'' \cite{cuddy2008warmth} (M = 4.85, SD = 1.31, $\alpha$ = 0.865)\footnote{Here $\alpha$ denotes the Cronbach's alpha reliability score (the extent to which a measure is consistent). Values above 0.8 are considered good for empirical studies \cite{lance2006sources}.}
    \item \textbf{Competency}: ``This chatbot was competent'', ``This chatbot was capable'' \cite{cuddy2008warmth} (M = 5.26, SD = 1.16, $\alpha$ = 0.965)
    \item \textbf{Appropriateness of tone}: ``The chatbot used an appropriate tone'' \cite{ghandeharioun2019emma}
\end{itemize}

\textcolor{major}{Likelihood to disclose was used as a direct measure for RQ2.1 and RQ2.2. Other subjective measures were collected to gain additional insights, but do not explicitly measure self-disclosure}.
Similarly to previous chatbot studies \cite{khadpe2020conceptual,lopatovska2021user}, warmth and competency measures are taken from the Stereotype Content Model (SCM) \cite{cuddy2008warmth}, that posits that human stereotypes are formed along two dimensions of warmth and competence. Specifically,
warmth is the intention of someone to help you, and competence is their ability to act on these intentions.
Groups perceived as both warm and competent inspire desire to assist them.
For example, physicians are perceived as both warm and competent.


Additionally, free-text feedback was collected asking participants \textit{why} they liked or disliked chatbot conversations (``What do you like or dislike about the conversation with the chatbot?'').  See Section \ref{sec:Study1Qualitative} for qualitative findings.

\subsection{Post Interaction Survey}
\label{sec:Study1PostQs}

On experiment completion, participants answered a post-interaction survey measuring levels of privacy concerns, belief in robotic intelligence, and belief in robotic feelings (measured on 7-point Likert scales - 1 = Strongly Disagree to 7 = Strongly Agree).

Levels of privacy concerns were measured using 4 items from Dinev and Hart \cite{dinev2006extended}: ``\textit{I am concerned that the information I submit on the Internet could be misused.}'', ``\textit{I am concerned about submitting information on the Internet, because of what others might do with it.}'', ``\textit{I am concerned about submitting information on the Internet, because it could be used in a way I did not foresee.}'', and ``\textit{I am concerned that a person can find private information about me on the Internet.}'' (M = 5.01, SD = 1.60, $\alpha$ = 0.962).

To measure belief in robotic intelligence, participants responded to ``\textit{I believe robots can be intelligent}'' and to measure belief in robotic feelings participants responded to ``\textit{I believe robots can have real feelings}'' (both from Liu et al. \cite{liu2018should}).

\section{Study 1: Results and Findings}
\label{sec:Study1-Results}

\begin{figure*}[t]
    \centering
    \begin{subfigure}[b]{0.3\textwidth}
        \includegraphics[width=\textwidth]{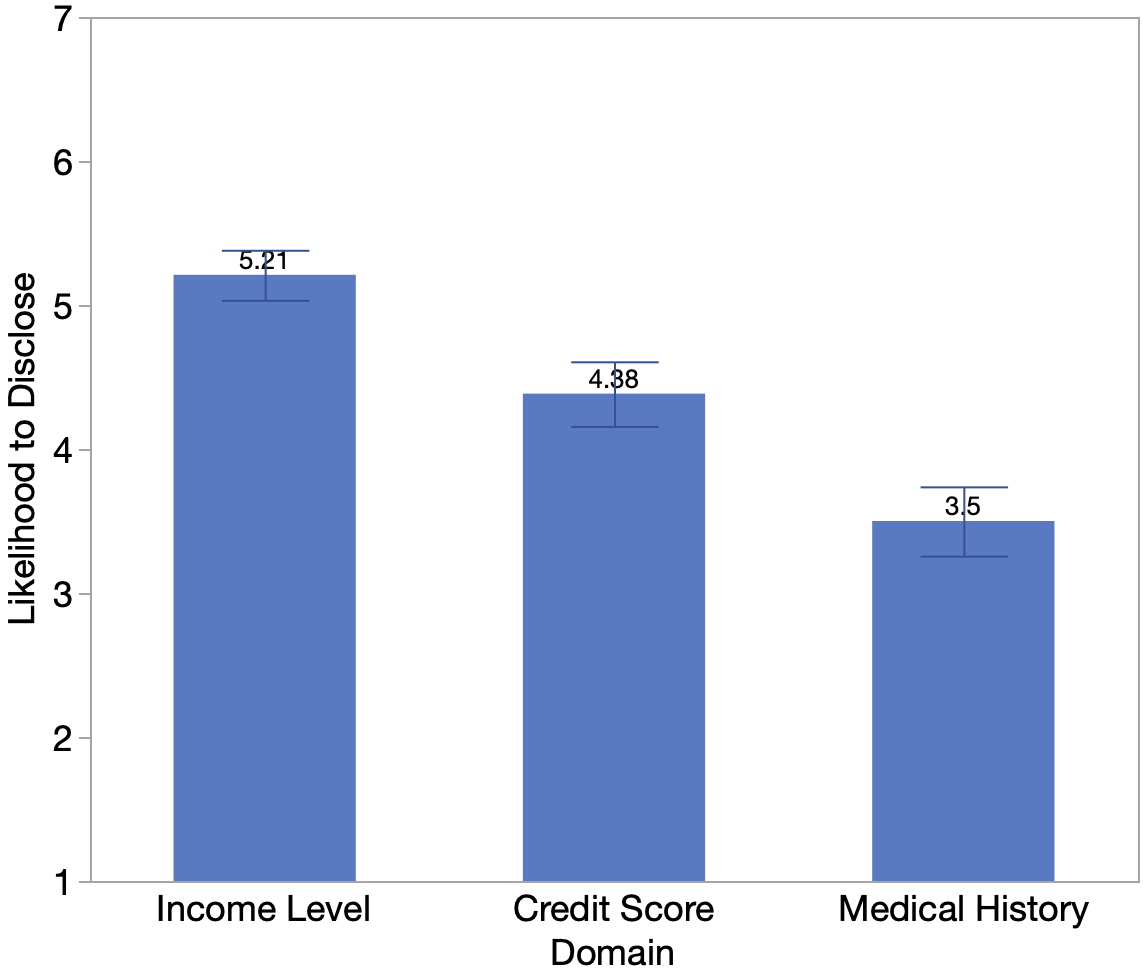}
        \caption{Participants were most likely to disclose income level and least likely to disclose medical history\\}
        \label{fig:Study1-Disclosure}
    \end{subfigure}
    \qquad 
    \begin{subfigure}[b]{0.3\textwidth}
        \includegraphics[width=\textwidth]{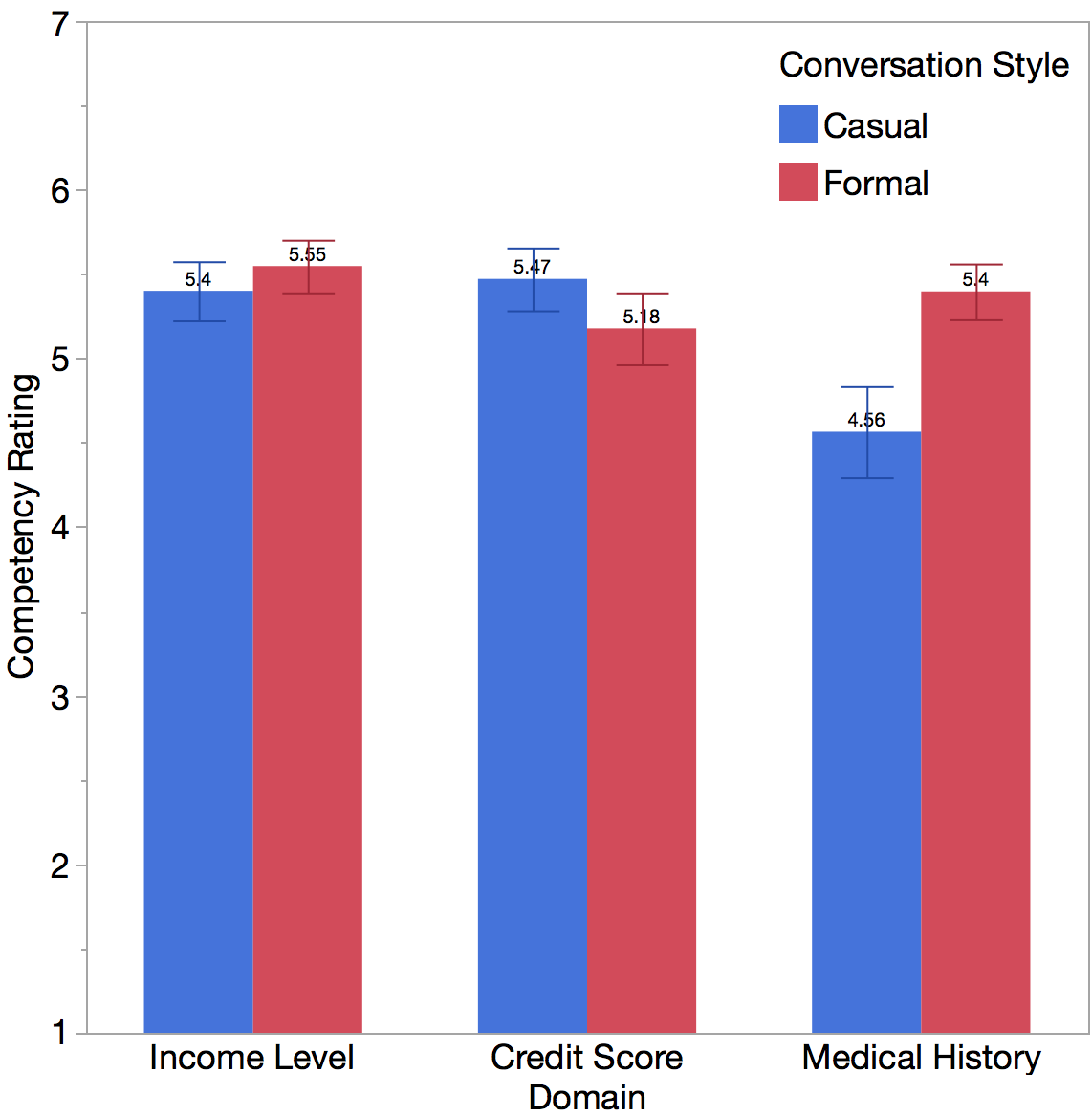}
        \caption{When asked for their medical history, participants perceived a chatbot as more \textbf{competent} if it used a formal conversational style}
        \label{fig:Study1-Competency}
    \end{subfigure}
    \qquad 
    \begin{subfigure}[b]{0.3\textwidth}
        \includegraphics[width=\textwidth]{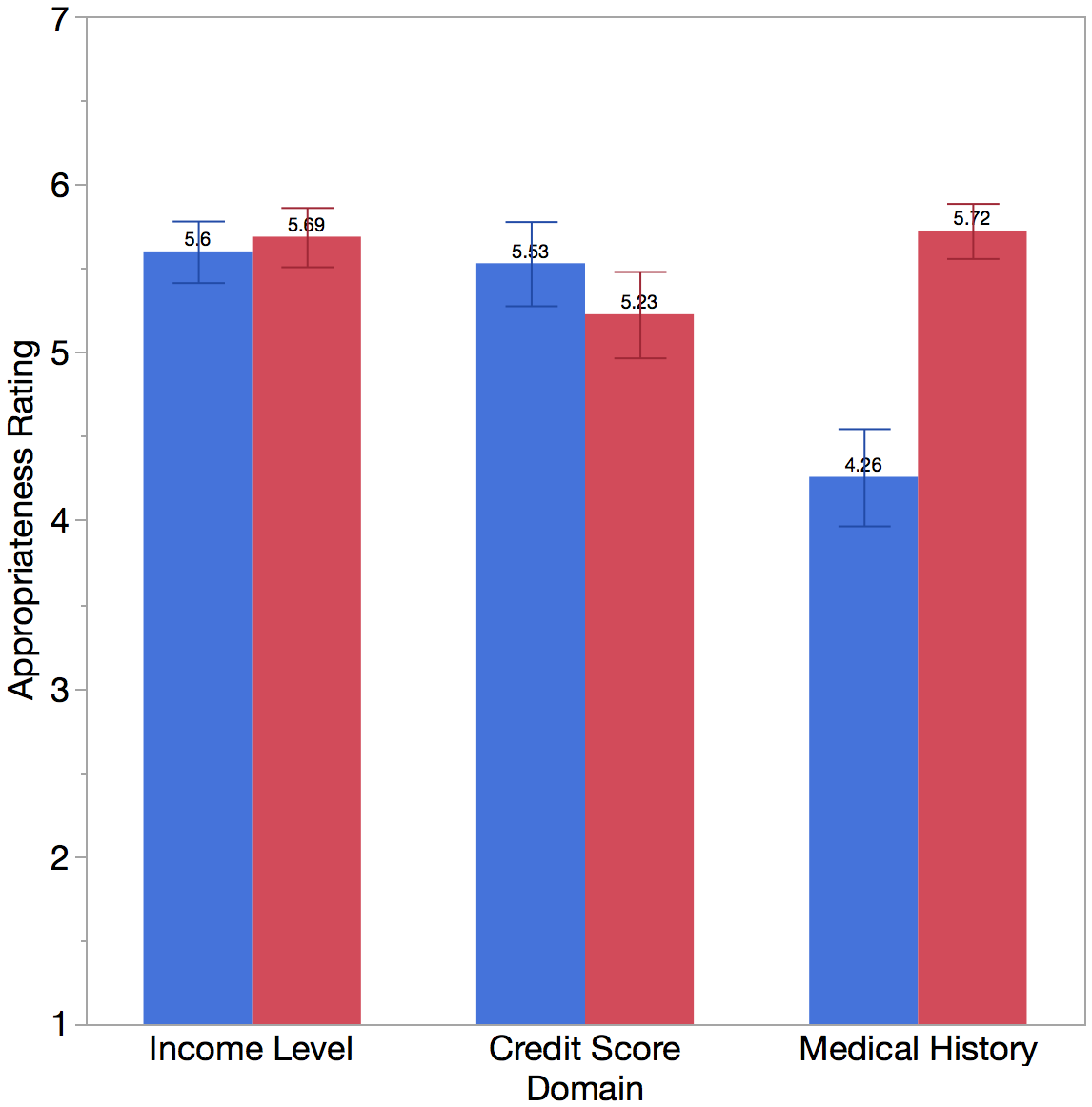}
        \caption{When asked for their medical history, participants perceived a chatbot as more \textbf{appropriate} if it used a formal conversational style}
        \label{fig:Study1-Appropriateness}
    \end{subfigure}
    \caption{Key figures for Study 1. Graphs \ref{fig:Study1-Competency} and \ref{fig:Study1-Appropriateness} show significant differences between casual and formal conversational styles when medical history was requested.}\label{fig:Study1-AllGraphs}
\end{figure*}

\begin{table*}[bt!]
  \resizebox{\textwidth}{!}{%
  \centering
  \begin{tabular}{lcccccc}
    \toprule
     & \multicolumn{2}{c}{Income Level} & \multicolumn{2}{c}{Credit Score} & \multicolumn{2}{c}{Medical History} \\
     \midrule
     & Casual & Formal & Casual & Formal & Casual & Formal \\
    \midrule
     \makecell[l]{Likelihood \\to Disclose} & 5.03 (1.58) & 5.34 (1.13) & 4.41 (1.88) & 4.35 (1.76) & 3.51 (1.95) & 3.48 (1.80) \vspace{0.6em} \\ 
     Enjoyment & 4.33 (1.40) & 4.56 (1.44) & 4.32 (1.47) & 4.13 (1.43) & 3.94 (1.70) & 4.10 (1.72) \vspace{0.6em} \\
     Warmth & 5.05 (1.23) & 4.84 (1.09) & 4.92 (1.17) & 4.60 (1.40) & 4.85 (1.37) & 4.83 (1.12) \vspace{0.6em} \\
     Competence & 5.40 (0.96) & 5.55 (0.88) & 5.47 (1.09) & 5.18 (1.19) & \textbf{4.56 (1.50)} & \textbf{5.40 (0.89) **} \vspace{6pt} \\
     \makecell[l]{Appropriateness\\ of Tone} & 5.60 (1.00) & 5.69 (1.00) & 5.53 (1.46) & 5.23 (1.43) & \textbf{4.26 (1.61)} & \textbf{5.72 (0.88) **} \\
     
  \bottomrule
\end{tabular}}
\caption{Outcome measures for Study 1 \textcolor{major}{split by both Domain and Conversational Style (values} shown as ``Mean (s.d.)''). Significant differences shown in bold ($p<.0001$).}
\label{tab:study1-quant}
\end{table*}

We fit a \textcolor{minor}{linear model} on each dependent variable with conversational style and domain as the fixed effects, and performed \textcolor{minor}{post-hoc Tukey's HSD} to identify specific differences.
See Table \ref{tab:study1-quant} results summary by domain and conversational style.
We will now discuss specific findings and their implications, before discussing qualitative findings in Section \ref{sec:Study1Qualitative}.

Similarly to previous findings within privacy literature \cite{markos2017information}, the domain of information requested influenced the likelihood of information disclosure  \textcolor{minor}{(see Figure~\ref{fig:Study1-Disclosure})}. 
Participants were most likely to disclose income level (M = 5.21), less likely to disclose credit score (M = 4.39), and least likely to disclose medical history (M = 3.50) ($p<.0001$).
While Markos et al. found participants equally likely to disclose ``medical history'' and ``credit rating'', we found participants were less likely to disclose their ``medical history''. 
This difference may be due to how concretely people can perceive the information being requested (i.e., medical history may appear more open-ended), or due to people questioning the necessity of the information being requested (similar to findings from Heckler et al. where they noted the perceived sensitivity of health information was relative to the domain from which it is being requested \cite{hecker2006ontological}).

While conversational style had no direct impact on likelihood to disclose, conversational style did influence user perceptions. There was a significant effect of (conversational style against domain) on perceived competence and appropriateness. Post-hoc tests found that when being asked for their medical history, participants found Formal more competent ($p<.0001$) and appropriate ($p<.0001$) compared to Casual (see Figures \ref{fig:Study1-Competency} and \ref{fig:Study1-Appropriateness}).
There were moderate significant effects for conversational style ($p = 0.0228$) and domain ($p = 0.0381$) on appropriateness. 
The difference for appropriateness between Casual (M = 5.14) and Formal (M = 5.54) may seem to suggest that a formal conversational style is always preferred, but the difference can be accounted for by the difference in medical history (see Figure \ref{fig:Study1-Appropriateness}).




Participants with lower levels of privacy concerns found the chatbot more enjoyable ($p < .0001$), warm ($p = 0.0045$), competent ($p = 0.0034$) and appropriate ($p = 0.0008$) and were more likely to disclose to the chatbot ($p < .0001$) than those with high levels of privacy concerns.
Participants who believe in robotic intelligence found the chatbot more enjoyable ($p < .0001$), warm ($p < .0001$), competent ($p<.0001$), and appropriate ($p = 0.0058$) than participants who do not believe in robotic intelligence.
Similarly, participants who believe in robotic feelings found the chatbot more enjoyable ($p = 0.0004$) and were more likely to disclose to the chatbot ($p = 0.0033$) than participants who do not believe in robotic feelings.

Overall, these findings suggest that the likelihood to disclose is affected by the perceived sensitivity of the information being requested rather than the conversational style of the agent requesting the information.
However, the finding that people perceived a formal conversational style as more competent and appropriate when sensitive health information was requested suggests that there may be more imperceptible effects taking place.

\subsection{Qualitative findings from user feedback}
\label{sec:Study1Qualitative}

To investigate further why people perceive conversational styles differently we coded user feedback.
All participant feedback was open-coded by a member of the research team blind to the experiment condition of the feedback.
Due to varying levels of detail provided in the feedback, each response could be coded into one, multiple, or no categories.
This gave an initial set of 29 categories, which were discussed and consolidated into 8 categories.
\textcolor{major}{A second researcher from outside the project then blind labelled the feedback using the 8 consolidated categories. After this, 187 of the labels were identical between the two labellers, while 29 labels conflicted (i.e., an alternative or additional category was labelled for a given piece of feedback).
These conflicting labels were discussed and finalised between the two labellers, giving us the final category count found in Table \ref{tab:study1-quote-frequencies}.}

\begin{table}
  \centering
  \caption{Frequency count of categories within user feedback. Casual was more often described as human-like or unprofessional, while Formal as polite or professional.}
  \label{tab:study1-quote-frequencies}
  \begin{tabular}{lcccc}
    \toprule
    Category & Total & Casual & Formal \\
    \midrule
    Human-like & 41 & 25 & 16\\
    Uncanny & 19 & 9 & 10\\
    Friendly & 56 & 31 & 25\\
    Unfriendly & 6 & 1 & 5\\
    Professional & 41 & 9 & 32\\
    Unprofessional & 12 & 12 & 0\\
    Polite & 26 & 5 & 21\\
    Helpful & 15 & 7 & 8\\
  \bottomrule
\end{tabular}
\end{table}

From this coding, we found that Casual was more often described as human-like:

\begin{displayquote}
    \textit{I like how the chatbot uses informal syntax to make it feel more human-like.} -P125-Casual(Income Level)
\end{displayquote}

\noindent
Several Casual participants enjoyed the friendly nature of conversation, but expressed concern for level of professionalism:

\begin{displayquote}
    \say{\textit{I like that the chat bot seems friendly and personable but I would not want that to go too far and become unprofessional.}} -P222-Casual(Income Level) 
\end{displayquote}

\noindent
This perceived unprofessionalism was reinforced by several participants who said the casual tone was not domain appropriate, or would affect their likelihood to disclose:

\begin{displayquote}
    \say{\textit{I dislike that the language style of the chatbot is not professional and that makes me cautious of the type of information that I have to provide.}} -P82-Casual(Credit Score) 

    \noindent
    \say{\textit{It's too jovial and colloquial for such a topic.  Friendly is fine if I'm buying clothes but not when discussing finances.}} -P2-Casual(Income Level) 
    
\end{displayquote}

\noindent
In contrast, Formal was more often described as professional or polite:

\begin{displayquote}
    \say{\textit{I like how the bot was talking in a professional manner. This makes me trust it more and not think it's a terrible idea to use chat bots.}} -P96-Formal(Credit Score) 
    
    \noindent
    \say{[...] \textit{It is respectful in asking for certain information from you that is otherwise private.}} -P122-Formal(Credit Score) 
    
    
\end{displayquote}

\noindent

However, participants also noted Formal lacked empathy:

\begin{displayquote}
    \say{\textit{It’s not very reassuring or comforting}} -P152-Formal(Medical History) 

    \noindent
    \say{\textit{It seems very impersonal, which may come across as cold, but that also means it doesn't have any particular bias}} -P10-Formal(Medical History) 
\end{displayquote}


Interestingly, Formal(\textit{Medical History}) feedback suggests while participants may find tone more appropriate, they would not be willing to disclose their personal information in an online chat setting:


\begin{displayquote}
    \say{\textit{The language sounded patient and warm and the chat bot made me feel comfortable.  But, I would still feel uncomfortable sharing my medical information online in a chat room.}} -P19-Formal(Medical History) 
\end{displayquote}

\noindent
In contrast, participants described Casual(\textit{Medical History}) not meeting their expected level of appropriateness:

\begin{displayquote}
    \say{\textit{The tone of the chatbot was a little too informal for the topic it was discussing. Discussing personal finances should have a more formal tone.}} -P82-Casual(Medical History) 

\end{displayquote}

Overall, the feedback indicates the importance of a conversational style matching a user's expectations. 
A conversational style that matches user expectations can lead to greater feelings of trust, while a mismatch leads to negative user reactions (such as Casual being perceived as unprofessional).


\section{Study 2: The Effect of Conversational Style on Response Quality}
\label{sec:ch4-Study2}

\begin{figure*}[!htb]
    \centering
    \includegraphics[width=1\textwidth]{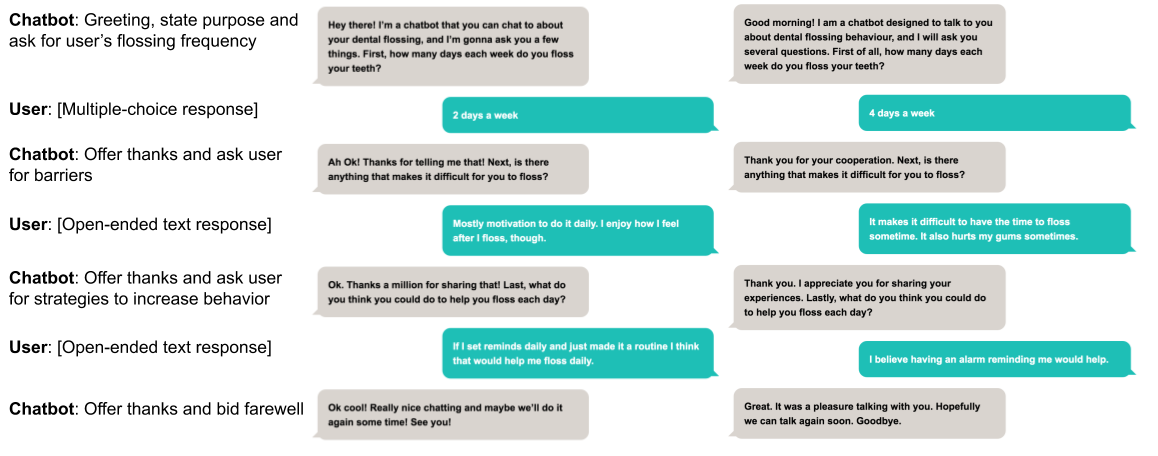}
    \caption{The Study 2 chatbot for both casual and formal conversational style conditions. The interface is as seen by participants with chatbot utterances in the grey speech bubbles and actual participant responses in green speech bubbles. 
    The F-score for the casual condition was 40.5, and the F-score for the formal condition was 47.5.}
    \label{fig:Study2-ConversationFlow}
\end{figure*}

While Study 1 results did not find a difference in likelihood to disclose between conversational styles,
we found that people perceived a formal conversational style as more appropriate and competent when being asked to disclose sensitive health information (i.e., ``medical history'').

However, as the information being requested in Study 1 was hypothetical, we next want to investigate the effect of a chatbot's language formality on the quality of user utterances.
Therefore, we expanded on the Study 1 results for health information and ran a second study where participants discussed their health-related habits (specifically, their dental flossing) with a chatbot.
We then analysed user utterances to investigate the effect of a chatbot's language formality on the quality of a user's responses. \textcolor{major}{This gave us our third research question:}
\begin{itemize}
    \item \textcolor{major}{\textbf{RQ2.3:} How does a chatbot's conversational style (formal vs casual) influence the quality of user utterances?}
\end{itemize}

We chose not to request high sensitivity information (such as medical history), or to discuss health habits which are more likely to affect a participant's health (such as medication adherence) in order to lower privacy concerns and health implications of the study.

Accordingly, we chose dental flossing as the domain of our chatbot as we wanted to investigate a use case where a participant could talk to a chatbot about a healthy behaviour to which health experts recommend regular adherence (as is the case for flossing \cite{schuz2006adherence}).
Additionally, people can have barriers to dental flossing \cite{aguirre2016identification,buglar2010role}, which was key to our chatbot's script; people are benefitted by brief interventions \cite{gillam2019brief} and daily diary keeping \cite{suresh2012exploratory} - tasks which could be facilitated by chatbots; and flossing benefits one's health and is effective at reducing disease causing bacteria \cite{corby2008treatment} and gum bleeding \cite{graves1989comparative}.



\subsection{Experiment Conditions}

We conducted a between-subjects experiment on AMT where participants talked to a chatbot using one of two conversational styles: casual or formal.

We want to control the level of formality of the responses given by the chatbot.
To achieve this level of control, machine generated responses would not be appropriate.
We therefore implemented a chatbot following rule-based logic and dialogue written by the research team. 
The chatbot was hosted within Qualtrics, and used HTML and JavaScript to emulate the look and feel of a chatbot.
While intent classification could be used to recognise user intent and give adaptive chatbot responses, we chose to use pre-scripted responses so participants would see consistent responses between conditions and to reduce the potential for negative perceptions due to user intents being misclassified. 
Therefore, questions asked by the chatbot are written to avoid dependency to one another.
Additionally, chatbot responses were given instantaneously independent of the length or complexity of the user's utterances.

Chatbot dialogue followed the same high-level structure between conditions (see Figure \ref{fig:Study2-ConversationFlow} for high-level chatbot utterances, conversational styles, and interface seen by participants).
As part of this dialogue, the chatbot asks three questions. First is a multiple choice question to ask the user for their flossing frequency.
Next, the chatbot asks two open-text questions asking the user to identify: (1) barriers to dental flossing, and (2) strategies for overcoming barriers to dental flossing.
The phrasing of this second question depends on the flossing frequency of the user. Users who do not floss each day are asked ``\textit{what do you think you could do to help you floss each day?}'' whereas those who floss every day are asked ``\textit{what do you do to help you floss each day?}''.
To ensure questions asked by the chatbot are interpreted identically across conditions, we used the same questions verbatim for each condition.
The text preceding the question asked by the chatbot was varied in conversational style to be either casual or formal.

\subsection{Participants}
We recruited participants from AMT using the selection criteria described in Section \ref{sec:Study1Participants}.
Participants were paid USD\$0.85, and the survey took a mean time of 4.02 minutes to complete.
183 participants attempted the survey, and 156 (mean age 37.4; 58 female) passed the screening tasks of whom 44 used dental floss every day.
Participants were randomly assigned a condition with 80 in casual and 76 in formal.


\subsection{Procedure}

Participants completed the following procedure:
\begin{enumerate}
    \item Introduction and consent form.
    \item The same demographics and screening questions were used as those in Study 1.
    \item Remaining participants were shown more detailed task instructions, before conversing with the chatbot about their dental flossing, and rating their experience (see Section \ref{sec:Study2-Measures}).
    \item Post-test questions (see Section \ref{sec:Study2PostQs}).
\end{enumerate}


\subsection{Subjective Measures}
\label{sec:Study2-Measures}

\textcolor{major}{Akin to Study 1, participants evaluated their chatbot interaction on 7-point Likert scales (1 = Strongly Disagree to 7 = Strongly Agree), and (while these measures are not explicitly linked to RQ2.3) they are intended to gain additional insights.}
Participants were asked ``\textit{Do you personally agree or disagree that...}'' for the following:

\begin{itemize}
    \item \textbf{Desire to continue}: ``I would want to continue using the chatbot'' \cite{khadpe2020conceptual}
    \item \textbf{Enjoyment}: ``I enjoyed talking to the chatbot'' \cite{celino_submitting_2020}
    \item \textbf{Warmth}: ``The chatbot was warm'', ``The chatbot was good-natured'' \cite{cuddy2008warmth} (M = 5.02, SD = 1.50, $\alpha$ = 0.894)
    \item \textbf{Competence}: ``The chatbot was competent'', ``The chatbot was capable'' \cite{cuddy2008warmth} (M = 4.59, SD = 1.61, $\alpha$ = 0.930)
    \item \textbf{Appropriateness of tone}: ``The tone of the chatbot was appropriate'' \cite{ghandeharioun2019emma}
\end{itemize}

We also asked participants for free-text feedback describing \textit{why} they liked or disliked the chatbot conversations (``\textit{What did you like or dislike about your conversation with the chatbot?}'').

\begin{figure}[hb]
    \centering
    \includegraphics[width=0.42\textwidth]{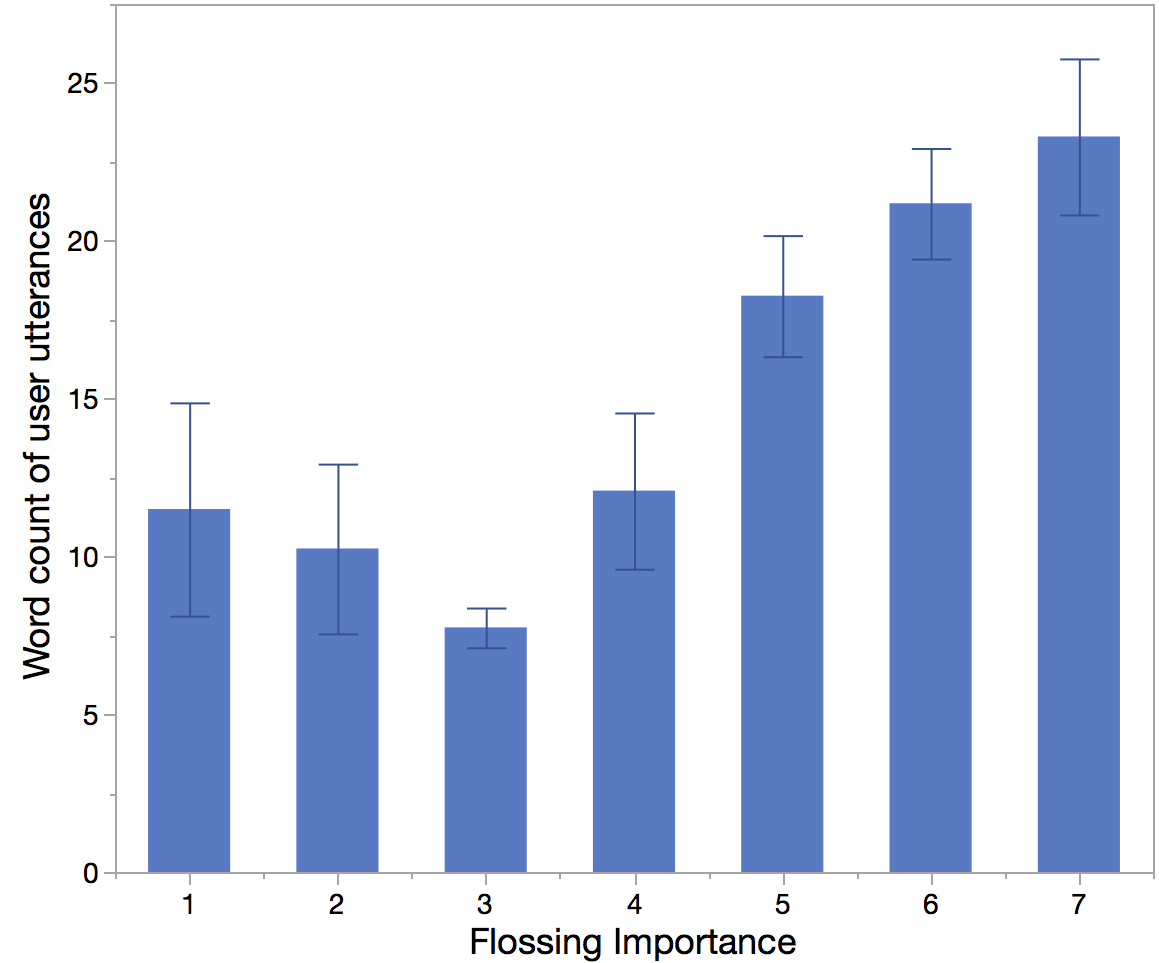}
    \caption{Users who perceived flossing as more important provided longer utterances.}
    \label{fig:Study2-Importance}
\end{figure}

\subsection{Analysing User Utterances}

We want to measure the quality of user utterances, but as utterances are brief (with each user providing two open-text responses), automated methods could not be used reliably.
Consequently, we analysed the \textit{response length} of user utterances, and manually labelled quality (in terms of \textit{specificity} and \textit{actionability} of utterances).

Response length was measured as the total word count of both user utterances. Word count has been used previously as an indicator for the quality of user utterances \cite{louw2011active,xiao2020if,rhim2022application}.

To label quality, we took inspiration from Xiao et al. and labelled utterances to give a human interpreted proxy for response quality \cite{xiao2020tell,xiao2020if}.
Xiao et al. were guided by Gricean maxims and labelled user utterances in terms of clarity, relevancy, and specificity.
Due to the clarity and relevancy of user utterances being consistently good (with no responses being gibberish or irrelevant), we chose to label only for the maxim of specificity.
From collected utterances, we want users to identify specific flossing barriers, and actionable strategies to overcome barriers.
Consequently, we labelled barrier identifying utterances in terms of \textbf{specificity} and strategy identifying utterances in terms of \textbf{actionability}.~\\


\begin{table*}[h]
  \centering
  \caption{Outcome measures for Study 2 (shown as ``Mean (s.d.)''). Significant differences shown in bold.}
  \label{tab:study2-quant}
  \begin{tabular}{llcc}
    \toprule
    &  & Casual & Formal \\
    \midrule
    Subjective Measures & Continue & 4.06 (1.64) & 4.01 (1.52) \\
    (7-point Likert) & Enjoyment & 4.45 (1.58) & 4.42 (1.62) \\
     & Warmth & \textbf{5.31 (1.31)} & \textbf{4.71 (1.47)} \\
     & Competency & 4.61 (1.64) & 4.57 (1.48) \\
     & Appropriateness & 5.70 (1.07) & 5.67 (1.19) \\
     \midrule
    \makecell[l]{Response Quality Measures} & \makecell[l]{Response length (words)} & 19.69 (12.78) & 19.82 (16.13) \vspace{2mm} \\
    & Specificity of barriers & \multicolumn{2}{c}{See Figure \ref{fig:Specificity and actionability graphs}} \\
    & Actionability of strategies & & \\
  \bottomrule
\end{tabular}
\end{table*}

\begin{figure*}
    \centering
    \begin{subfigure}[b]{0.4\textwidth}
        \includegraphics[width=\textwidth]{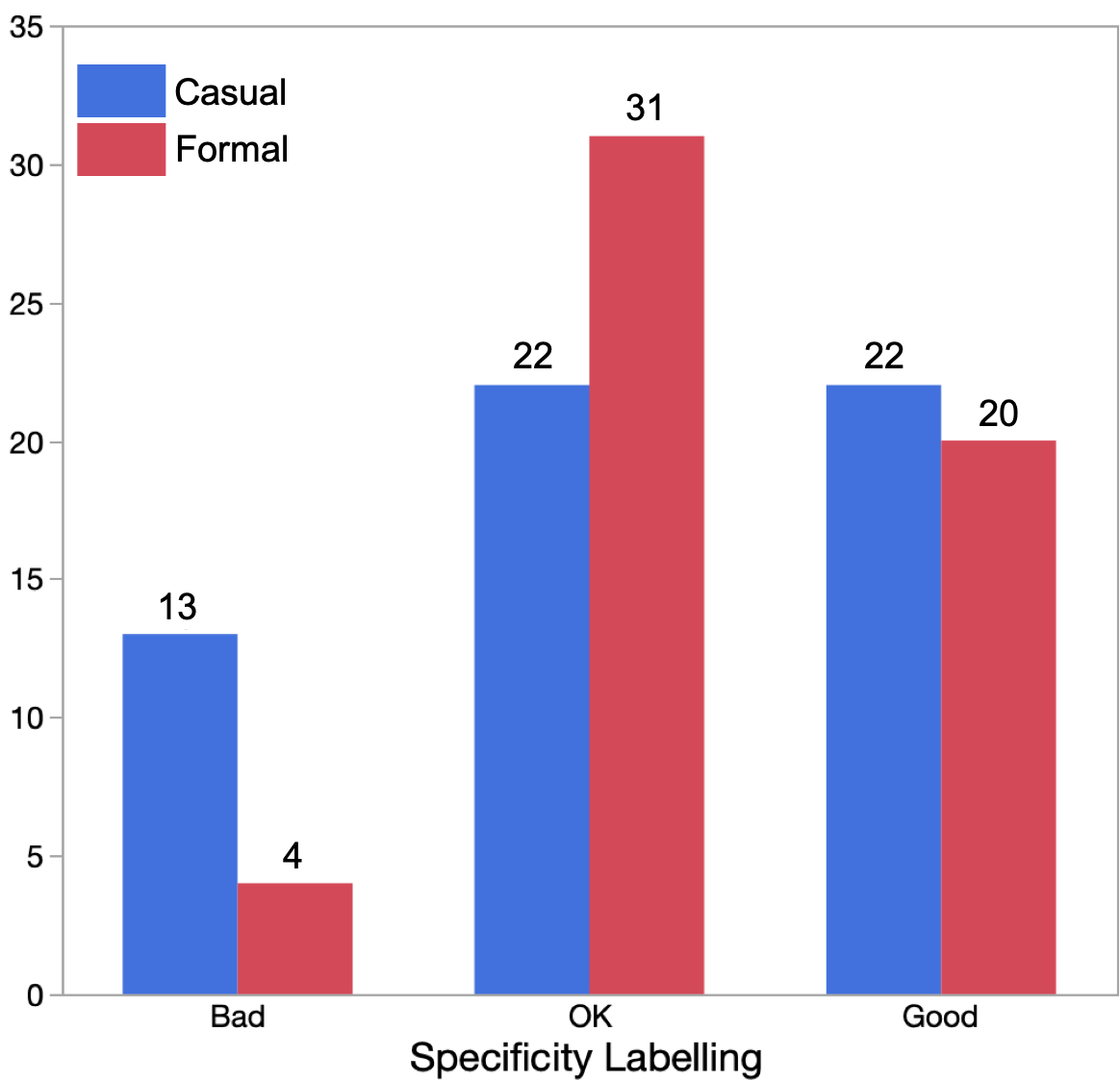}
        \caption{Formal participants provided fewer ``bad'' and more ``ok'' specificity responses than Casual participants.}
        \label{fig:Study2-Specificity}
    \end{subfigure}
    \hfill 
    \begin{subfigure}[b]{0.4\textwidth}
        \includegraphics[width=\textwidth]{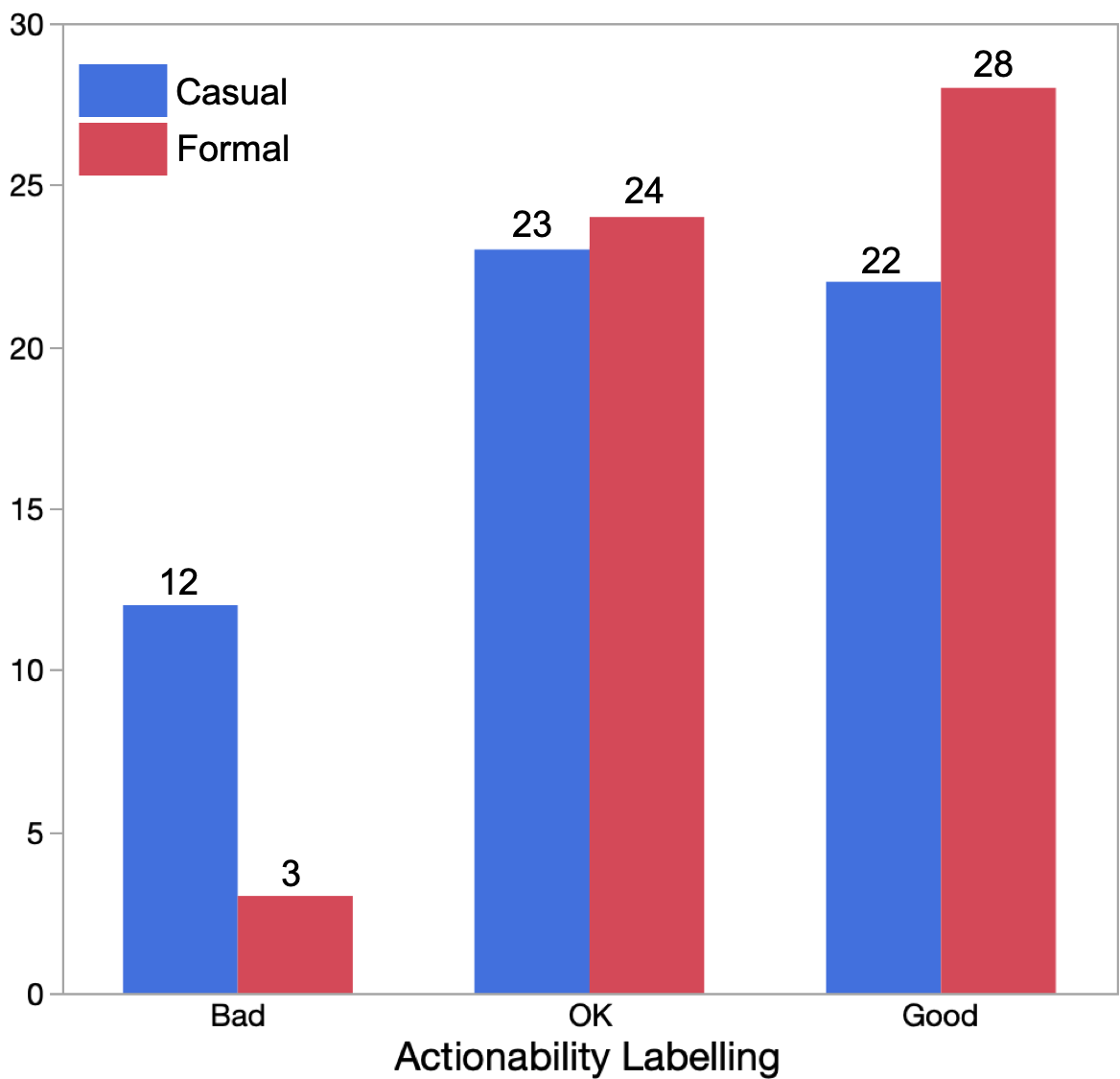}
        \caption{Formal participants provided fewer ``bad'' actionability responses than Casual participants.}
        \label{fig:Study2-Actionability}
    \end{subfigure}
    \caption{Count of specificity and actionability labelling for user utterances. Participants provide more specific and actionable utterances when conversing with a formal chatbot compared to a casual chatbot.}
    \label{fig:Specificity and actionability graphs}
\end{figure*}

\subsubsection{Labelling Specificity of Utterances} ~\\
When labelling specificity of barriers, we used the following guidelines:

\begin{displayquote}
    \say{\textit{Does it give you enough info on how to identify and then give advice on how to overcome the barrier? For example, ``I forget'' is less specific and more difficult to act on, whereas ``I forget when I'm...'' is a more specific response.}}
\end{displayquote}

\noindent
This gave us three labels:
\begin{itemize}
    \item ``\textit{\textbf{0 (bad)} - no barrier is identified in the utterance}'' such as the utterance: ``\textit{yes}''.
    \item ``\textit{\textbf{1 (Ok)} - a barrier is alluded to, but the utterance is not specific enough to allow for detailed guidance on how to overcome the barrier}'' such as the utterance: ``\textit{I forget to do it regularly}''.
    \item ``\textit{\textbf{2 (good)} - a barrier is described in a detailed enough way to provide guidance on how to overcome the barrier}'' such as the utterance: ``\textit{Simply remembering or finding the floss since I hide it from my child so it doesn't get destroyed.}''.
\end{itemize}

\hspace{2mm}
\subsubsection{Labelling Actionability of Utterances}~\\
When labelling actionability of strategies, we used the following guidelines:

\begin{displayquote}
    \say{\textit{Is the strategy described concrete enough to act upon? (Is it actionable?). A clear strategy can be followed such as ``set an alarm'' rather than something like ``just do it''.}}
\end{displayquote}
\noindent
This gave us three labels: 
\begin{itemize}
    \item ``\textit{\textbf{0 (bad)} - No strategy is identified}'' such as the utterance: ``\textit{I don't know}''.
    \item ``\textit{\textbf{1 (Ok)} - A strategy is alluded to, but may be abstract, under-explained or difficult to act upon}'' such as the utterance: ``\textit{make it easier}''.
    \item ``\textit{\textbf{2 (good)} - A strategy is identified and it is clear what is intended to be done}'' such as the utterance: ``\textit{Maybe an external reminder, like an alarm on my phone or something like that.}''
\end{itemize}


\subsection{Post Interaction Survey}
\label{sec:Study2PostQs}

On experiment completion, participants answered a post-interaction survey measuring belief in robotic intelligence and robotic feelings (using measures described in Section \ref{sec:Study1PostQs}), and perceptions of dental flossing (measured on 7-point Likert scales - 1 = Strongly Disagree to 7 = Strongly Agree).

We used 3 measures of flossing perception \cite{buglar2010role} that are influenced by the extended Health Belief Model \cite{orji2012towards}, namely: adherence: ``\textit{I try to floss my teeth every day}'', flossing importance: ``\textit{It is important to floss my teeth every day}'', and self-efficacy ``\textit{It is easy to floss my teeth every day}''.


\section{Study 2: Results and Findings}
\label{sec:Study2-Results}

We fit a \textcolor{minor}{linear model} on each of our dependent variables with conversational style as the fixed effect, and we performed \textcolor{minor}{post-hoc Tukey's HSD} to identify any specific differences. 
See Table \ref{tab:study2-quant} for a summary of the results by conversational style.
We will now discuss specific findings and support these with participant quotes.

Conversational style was only significantly different for perceived warmth (p = 0.0096), with Casual being perceived as more warm than Formal.
This contrasts with Study 1 where Formal participants perceived a chatbot as more appropriate and competent when being asked for sensitive health information.
This difference in results may be due to medical history requested in Study 1 being perceived as more sensitive than dental flossing behaviour in Study 2, and thereby requiring a more formal conversational style.
Additionally, Study 2 participants may appreciate the increased warmth of Casual when they are discussing their own personal behaviour (rather than reflecting on hypothetical relaying of information in Study 1).


Participants who believed in robotic intelligence were more likely to find the chatbot enjoyable (p = 0.0053), warm (p < .0001), competent (p < .0001) and appropriate (p = 0.0004), while those who believed in robotic feelings were more likely find the chatbot enjoyable (p = 0.0031).

Participants who perceived flossing as important had more desire to continue chatbot use (p = 0.0113), found conversations more enjoyable (p = 0.0025), found the chatbot's tone more appropriate (p = 0.0153), and wrote longer responses (p = 0.0006; see Figure \ref{fig:Study2-Importance}) than those with low perceived dental flossing importance.
These findings highlight the value of educating users on the importance of healthy behaviours, and are supported by the extended Health Belief Model \cite{orji2012towards} which describes perceived importance as a determinant of healthy behaviour.

\subsection{Response Informativeness}

Two members of the research team independently coded user utterances while blind to source condition.
The intraclass correlation (ICC) for Specificity labels was 0.902, and the ICC for Actionability labels was 0.821 (above the ``excellent'' threshold of 0.75 from Cicchetti et al. \cite{cicchetti1994guidelines}).
We excluded utterances from participants who floss each day, as they commonly described having no barriers to flossing due to their current adherence to the healthy behaviour. 
Typical utterances for fully adhering participants were: \say{\textit{No everything is smooth and easy!}}.
This excluded utterances from 44 participants (23 casual and 21 formal) leaving a total of 112 participants (57 casual and 55 formal).

For the remaining user utterances that were initially labelled identically between coders, Formal users were more likely to provide more specific ($\chi^2$(2, N = 98) = 10.153, p = .0062) and actionable utterances ($\chi^2$(2, N = 84) = 12.210, p = .0022). 
After discussing and resolving specificity and actionability labels differing between the two coders (see Figure \ref{fig:Specificity and actionability graphs}), it was still found that Formal users were more likely to provide specific ($\chi^2$(2, N = 112) = 6.612, p = .0367) and actionable utterances ($\chi^2$(2, N = 112) = 6.490, p = .0390).

Interestingly, the driving force of the difference seems to be that a formal conversational style elicits less ``bad'' utterances compared to a casual style.
These results indicate a formal style might have made participants exert more effort in providing their responses.

\section{Summary of Key Findings:}

In summary, our key findings for research questions one to three are as follows:

\begin{itemize}
    \item \textbf{RQ2.1:} Users were least likely to disclose sensitive health information, followed by sensitive non-health, and were most likely to disclose non-sensitive non-health information.
    \item \textbf{RQ2.2:} \textcolor{minor}{We did not find a difference in conversational style and a user's likelihood to disclose. However,} (when being asked for their medical history) users perceive a chatbot with a formal conversational style as more appropriate and competent.
    \item \textbf{RQ2.3:} Users who do not fully adhere to a healthy behaviour wrote higher quality utterances when prompted with a formal conversational style.
\end{itemize}

\section{Discussion}
\label{sec:Discussion}

Here we discuss the implications of both Studies 1 and 2. Interpretations for study-specific findings can be found in Sections \ref{sec:Study1-Results} and \ref{sec:Study2-Results}.
Our goal was to investigate the effect of a chatbot's \textcolor{minor}{language formality} on self-disclosure. 
Our findings provide some empirical evidence that a chatbot using a more formal conversational style could be beneficial within the health domain by improving perceptions of competency and appropriateness (\hyperref[sec:ch4-Study1]{ Study 1 }), and by eliciting higher quality user utterances (\hyperref[sec:ch4-Study2]{ Study 2 }).

\textcolor{major}{These findings could be explained due to differences from the expectations held by users. The Expectancy Violations Theory states that interactions either conform to or violate a user's expectation of how an agent should behave \cite{burgoon2016application}. 
Qualitative feedback from Study 1 shows that participants expect a chatbot to use a more formal style when requesting their health information, and similarly subjective scores show that users perceive a casual style more negatively (less appropriate and competent) when sensitive health information was being requested.
This negative violation (an undesired and unexpected act) led users to rate the casual conversational style more harshly (as opposed to rating the formal style more highly).
This finding highlights the importance of using an appropriate conversational style in relation to the sensitivity of the information being requested by the chatbot (with medical history in Study 1 being the least likely piece of information for users to disclose, and thus leading to expectations of a more formal conversational style). Designers should carefully choose an appropriate conversational style if sensitive information is being requested so as not to violate any user expectations.}

\textcolor{major}{Furthermore, greater quality elicitation in Study 2 may be due to a mirroring effect.} 
Previous work has found that people may mirror the response length of their (chatbot) interlocutor \cite{hill2015real}, and Social Exchange Theory \cite{emerson1976social} describes how people reciprocate the level of effort within a conversation.
Relevant to this, Casual users in Study 2 produced more low quality user utterances, which may be due to a similar mimicry effect, whereby users are emulating the less specific language style of the casual prompts.
In this sense, users may perceive a casual style as being lower effort (or formal as being higher effort) and mimic the style accordingly.
\textcolor{major}{Study 2 results could imply that a formal conversational style should also be used when users are discussing their health habits with a chatbot. However, as the domain of the Study 2 chatbot did not go beyond discussing healthy habits, implications can not be generalised confidently.}

\textcolor{major}{These findings could be seen as opposing previous studies that found that higher quality (multiple-choice) responses are collected by chatbots using a casual conversational style \cite{kim_comparing_2019,celino_submitting_2020}. However, the baseline conversational style of these studies were (non-conversational) standarised surveys, rather than a formal conversational style. Therefore increases in response quality could be attributed to using a conversational style rather than a terse script.
Additionally, Zhou et al. compared two chatbots that used ensembles of features to create high-level personalities.
They found that during high-stakes job interviews users are more willing to confide in a serious and assertive chatbot \cite{zhou2019trusting}. This matches our findings that under certain contexts, a more appropriate conversational style is expected (which could therefore lead to better response quality).}

\textcolor{major}{However, Study 1 findings also indicate that (when requesting less sensitive information) a formal conversational style is not expected by users, and a casual style could be adopted without negatively violating user expectations.}
This suggests chatbot conversational style could then be tailored more flexibly to the brand or user expectation of the chatbot.
For example, a chatbot for children could use a more casual conversational style to sound more straight-forward and friendly. 
\textcolor{major}{On from this, there is also evidence that some users found a formal conversational style less empathetic (anecdotally from Study 1 user quotes, Section \ref{sec:Study1Qualitative}) and warm (see Table \ref{tab:study2-quant} for Study 2 subjective ratings).
From Study 1, this could imply that (when non-sensitive information is being requested by the chatbot) a more casual conversational style could be used if designers desire a more empathetic and human-like chatbot.
Warmth measures being statistically higher for Casual in Study 2, could imply that people may appreciate a more warm chatbot when they are discussing their health behaviours (similarly to disclosing more to a family member if they are more warm due to lessened feelings of judgement \cite{howe2001siblings,dotterer2019parental}).
Adopting a more casual (and thereby warm) conversational style could then lead users to co-operate and interact with the chatbot for longer \cite{khadpe2020conceptual}.}

Building on this, it is perhaps more clear which style to use for chatbots with a narrow domain of expertise or clear user expecations. However for multi-expertise tasks (potentially involving multiple domains and levels of information sensitivity) the choice of conversational style could be more difficult.
If a different domain may necessitate a different conversational style, a single agent switching conversational style could prove jarring \cite{chaves2019s,chaves2021should}, but multi-chatbots could be used \cite{chaves2018single,jung2015voice,pinhanez2018different}, whereby the user is referred to a (supposedly) different chatbot that adopts an appropriate conversational style.

Additionally, it could be satisfying to users if chatbots adopt a similar conversational style to the user themselves.
The use of linguistic mimicry (adopting the language style of another individual) leads to feelings of empathy \cite{otterbacher2017show}; and within information-seeking conversations between people, style mismatch can lead to less satisfying conversations \cite{thomas2018style,thomas2021theories}. 
Related to this, users found a voice-based conversational agent to be more trustworthy if it matched their own conversational style \cite{hoegen2019end}; multi-lingual users preferred a chatbot that mimicked their own code-mixing language (using multiple dialects) \cite{bawa2020multilingual}; and recent work has investigated the development of style matching chatbots \cite{hoegen2019end,thomas2021theories,ritschel2019adaptive}.
On from this, the effect of style matching on a user's quality of self-disclosure to a chatbot could be investigated further.

\textcolor{major}{In addition our work highlights the importance of a user's prior belief in chatbot capabilities, with both Studies 1 and 2 finding that people perceived chatbot interactions more favourably if they believed in robotic feelings or intelligence.
This matches previous findings that people base their judgement of systems on their own pre-existing beliefs \cite{hartmann2008framing,klaaren1994role,raita2011too,wilkinson2021or}. 
While these findings might suggest that people should be primed into believing a chatbot will be intelligent, Khadpe et al. noted the reverse to be true. 
They found users perceived a chatbot more positively if they had low expectations of a chatbot's competency before interacting with it \cite{khadpe2020conceptual}.}

\section{Limitations}

We will now outline our studies’ limitations.
Study 1 elicited users' likelihood to disclose, which may differ from actual disclosure \cite{zlotowski2017can}. 
\textcolor{major}{Additionally, in Studies 1 and 2 we asked participants to provide subjective ratings such as warmth, competence, enjoyability, and desire to continue; measures that may be difficult for participants to quantify without more extended and in-depth interactions with the chatbot.}

Experiments were not field studies, rather we used crowdsourcing where participants may be less intrinsically motivated or behave differently to volunteers or field study participants.
However, crowdworkers have been found to provide accurate data \cite{rand2012promise,horton2011online,paolacci2010running}, and be representative of the US population \cite{paolacci2010running}.

Our Study 2 responses come from desktops or laptops, although one of the most popular mediums of communication is mobile phones.
External validity could be affected if people give shorter responses on mobile \cite{mavletova2013data,antoun2017effects}, or have different expectations for chatbots on mobile (e.g., a more casual style \cite{kim_comparing_2019}).

We also cannot claim generality over different modalities (such as voice), or different levels of language formality (as we did not ask questions in a wide variety of formality levels, but rather two contrasting examples within a certain domain).
\textcolor{major}{Similarly, it may be difficult to generalise results beyond the domains of Studies 1 (finance) and 2 (low-sensitivity health behaviours).}
Results could vary over a longitudinal study, such as participants reacting negatively to messaging they find more repetitive \cite{westerman2020thou,kocielnik2017send}.

\section{Conclusion}

This study investigates how the language formality of a chatbot's utterances affect the quality of self-disclosure from a user.
We reported the results of two user studies conducted over AMT.
Our findings suggest a formal conversational style may be perceived more positively when a chatbot is requesting sensitive health information, and may elicit more high quality user utterances when discussing a user's health behaviour.



\chapter{The Effect of How a Health Chatbot Formats and References a User's Previous Utterances}
\label{ch:Personalisation}

Similarly to the motivation for Chapter \ref{ch:FormalOrCasual}, we want to investigate how changing the format of \textit{how} a health chatbot stylises its messaging can impact user perceptions (such as intelligence, engagement and privacy-violations).
This is important as users who think more positively of and have lower privacy concerns of chatbots, engage with chatbots more and give potentially higher quality utterances \cite{lee2020hear,lee2020designing}.
This can in turn allow the chatbot to better understand the user and thereby provide better service (such as providing the user with health guidance).
This chapter is based on one publication:

\bibentry{cox2023referencing}.


\section{Introduction}

Advances in language models are leading to chatbot interactions that can persist across multiple sessions, and refer back to previous user utterances \cite{xu2021beyond,xu2022long,park2023generative,bae2022keep}. 
This use of long-term memory can help maintain relationships and build rapport \cite{fivush1996remembering,bluck2003autobiographical,brewer2017remember}, and can improve user experience in chatbot interactions such as in open-domain conversations \cite{xu2022long,xu2021beyond} or discussions of personal health and wellness \cite{bae2022keep,jo2023understanding,holmes2018weightmentor}.
Additionally, by giving more relevant responses \cite{schuetzler2018investigation,chaves2021should} or explicitly referencing past user utterances \cite{jo2023understanding,rourke1999assessing,lander2015building,kreijns2022social}, a chatbot could increase its social presence: the feeling that it is present in the conversation \cite{oh2018systematic,biocca2002defining,rourke1999assessing}.

While this could prove beneficial to users and improve user perceptions of the chatbot, it could also lead to feelings of privacy violations. 
This phenomenon is known as the Personalisation Privacy Paradox \cite{awadpersonalization}, where there is a tension between collecting more user data to provide personalised services, and a user's feeling of intrusiveness leading to unwillingness to share their personal information.
This trade-off could particularly be an issue when people are discussing their sensitive information \cite{gomez2023sensitive}\cite[Art.9]{GDPR}. For example, people may be less willing to disclose socially undesirable behaviours due to embarrassment \cite{tsaihuman}, and users of mHealth services have reported that concerns over use of their personal data can negatively impact service adoption and satisfaction \cite{guo2016privacy}. 

The Personalization Privacy Paradox may hold additional uncertainty when chatbot designers need to choose an appropriate referencing format given the range of styles available to them \cite{zhang2018making,el2021automatic,Cappelen1997-CAPVOQ,wilson_sperber_2000}\footnote{See \cite{wilson_sperber_2000} for a summary of the four main types of quotation from linguistics literature.}.
To explore this paradox, we investigated the level of social presence used when a chatbot references a user's utterances, and its effect on both how privacy violating, and positively (e.g., intelligent, engaging) users perceived the chatbot.
Specifically, we compared 3 referencing formats from low social presence (not explicitly referencing user utterances) to higher social presence (referencing user utterances either verbatim or via paraphrases). 
We conducted a between-subjects longitudinal study ($N = 169$) where participants talked to a chatbot about their dental flossing once a week for three weeks. Participants talked to a chatbot that either: (1-None) did not explicitly reference their previous week's utterances, (2-Verbatim) referenced their previous week's utterances verbatim (e.g.., ``\textit{Last week you said ``My teeth sometimes hurt when I floss''}''), or (3-Paraphrase) referenced their previous week's utterances using paraphrases (e.g., ``\textit{Last week you said that your teeth hurt}'').
Users found chatbots that explicitly referenced their past utterances more intelligent and engaging. 
However, explicitly referencing a user’s past utterances also lead to increased feelings of privacy violations.
To gain further insights as to \textit{why} users might have perceived chatbot referencing formats differently, we conducted semi-structured interviews ($N=15$). 
%
Finally, we discuss implications and provide recommendations for chatbot designers when scripting interactions that reference user utterances.

\begin{figure*}[h]
    \centering
    \includegraphics[width=1\textwidth]{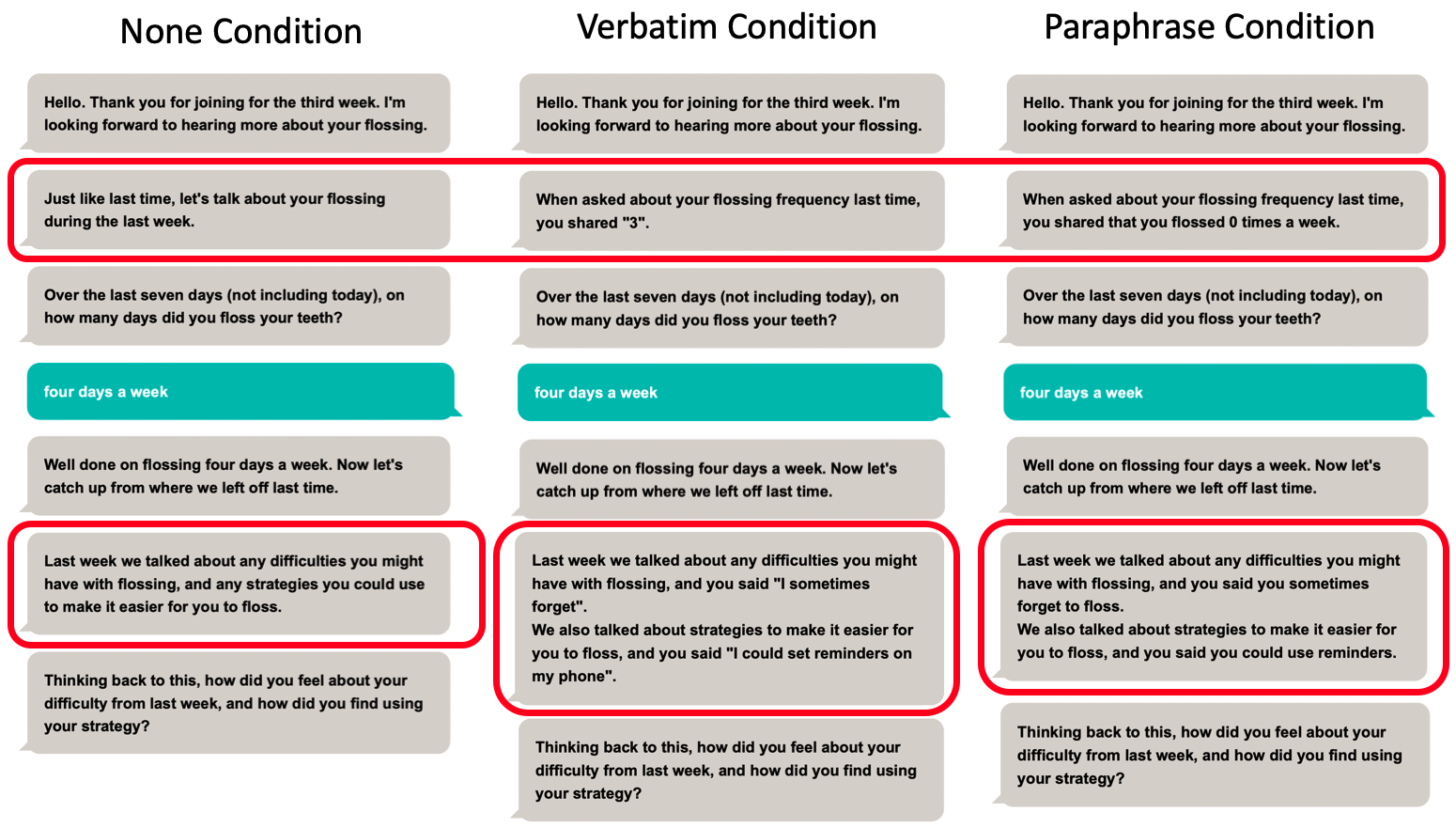}
    \caption{An extract from week 3 of the user study showing the 3 levels of chatbot referencing format.
    Grey bubbles are chatbot utterances, and teal bubbles are user utterances. 
    Differences in referencing format between conditions are circled in red.}
    \label{fig:conditions}
\end{figure*}

\section{Related Work}

We aim to investigate the effect of a chatbot referencing a user's utterances via different referencing formats.

\subsection{Increasing Social Presence and User Perceptions of Chatbots}
\label{sec:related-social}

As a chatbot uses more conversational referencing formats, it could be said to possess higher levels of social presence \cite{jo2023understanding,rourke1999assessing,lander2015building,kreijns2022social} (i.e., the feeling of being with an interlocutor \cite{oh2018systematic,biocca2002defining,rourke1999assessing}).
Previous work has investigated the user's feeling of ``being with'' a more present, engaging and human-like chatbot \cite{zheng2022ux,chaves2021should}, 
and increasing these feelings has been found to have benefits such as improved trust \cite{hagens2023trustworthy,zhou2019trusting} and desire to engage with \cite{pauw2022avatar,lee2020hear, bickmore2000weather,bickmore2001relational} chatbots.
Furthermore, chatbots have been found to benefit from various human-like qualities such as empathy \cite{lin_caire_2020,liu2018should}, listening \cite{xiao2020tell}, and differing conversational styles \cite{cox2022does,chaves2019s,yang2021effect}, personas \cite{volkel2022user,roy2021users} or politeness strategies \cite{brown_politeness_1987,ouchi2019should,miyamoto2017improving}.

Previous studies have also found benefit in interviewers that have higher levels of social presence.
For example, Xiao et al. found that people give higher quality responses to chatbots that use a battery of AI-driven techniques such as using more relevant responses to users \cite{xiao2020tell}; 
Tsai et al. found that users were more likely to disclose embarrassing behaviours related to their sexual health to a human compared to a chatbot \cite{tsaihuman}; 
and multiple studies have found that chatbots that self-disclose information lead to mutual disclosure from users and improved feelings of trust \cite{lee2020designing,adam_onboarding_2019,moon2000intimate,saffarizadeh2017conversational}.


More specifically to our study of chatbot referencing format, previous work has found benefit in chatbots that remember and reference details from previous interactions \cite{jain2018evaluating, portela2017new, chen2021you,medhi2017you,zamora2017m}.
For example, Jain et al. found chatbots that reference details from previous conversations lead to increased feelings of empathy \cite{jain2018evaluating},
and Portela and Granell-Canut reported that participants perceived a chatbot to have higher levels of affection when it remembered previous user utterances or the user's name \cite{portela2017new}.
Beyond this, we are interested in the effect on positive user perceptions caused by the format used by a chatbot when referencing a user’s previous utterances.
This gives us our first research question of:

\begin{itemize}
    \item \textbf{RQ3.1:} How does chatbot referencing format (None, Verbatim, Paraphrase) impact:
    \begin{enumerate}[label=(\alph*)]
        \item desire to continue using the chatbot?
        \item perceived chatbot engagement?
        \item perceived chatbot intelligence?
    \end{enumerate}
\end{itemize}


\subsection{Privacy Concerns Among Chatbot Users}

However, while Section \ref{sec:related-social} outlines the benefits of increased social presence, it could also lead to increased feelings of privacy concerns \cite{xu2008examining} amongst chatbot users \cite{ischen2019privacy,chen2021you}.
For example, Schuetzler et al. \cite{schuetzler2018influence} found that people were less likely to disclose to chatbots that use more relevant responses to user utterances during a small-talk session before asking (non-differentiated) health questions.
Ng et al. showed participants two hypothetical financial chatbots (one human-like and one factual) and found that, while the human-like chatbot scored higher social presence, participants were more likely to share information with the factual chatbot \cite{ng2020simulating}.
Bae et al. \cite{bae2023friendly} found that people trusted a robot-like chatbot more than a human-like chatbot when discussing positive experiences.
More analogous to our study's aim, Chen et al. investigated the perceived invasiveness of a chatbot that referenced participants' personal information (name, presence of heart disease and hand-washing frequency) \cite{chen2021you}.
While some findings indicated that people found chatbots more invasive when referencing their information, this was contrasted with a null finding once the user's perceived identity of the chatbot (human or chatbot) was taken into account.
Building on these previous findings and conflicting results gives us our second research question of:

\begin{itemize}
    \item \textbf{RQ3.2:} How does the chatbot referencing format (None, Verbatim, Paraphrase) impact the user's feelings of privacy violations?
\end{itemize}

We were interested to explore RQ3.2, as conflicting previous work indicates potential contradictory and uncertain findings. 
That is to say, by referencing user utterances in different formats, it could become more apparent to the user that the chatbot is storing or manipulating their personal information, and thereby heighten privacy concerns.
Alternatively, users could appreciate the increased levels of social presence and personalisation.
Additionally, by referencing user utterances verbatim, the chatbot could either make data storage more apparent and therefore privacy-violating to users, or it could be seen as more transparent about storing the user's data without manipulation (and by showing less advanced AI capabilities, users may perceive the chatbot more favourably by generating a metaphor of a chatbot which is less capable \cite{khadpe2020conceptual}).
Similarly, by paraphrasing user utterances the chatbot could be seen as invasive (by storing and manipulating user data), or create greater feelings of engagement with the user.
Finally, by not explicitly referencing user utterances, the chatbot could be seen as less privacy violating, but also potentially less engaging.

\begin{figure*}[h]
    \centering
    \includegraphics[width=1\textwidth]{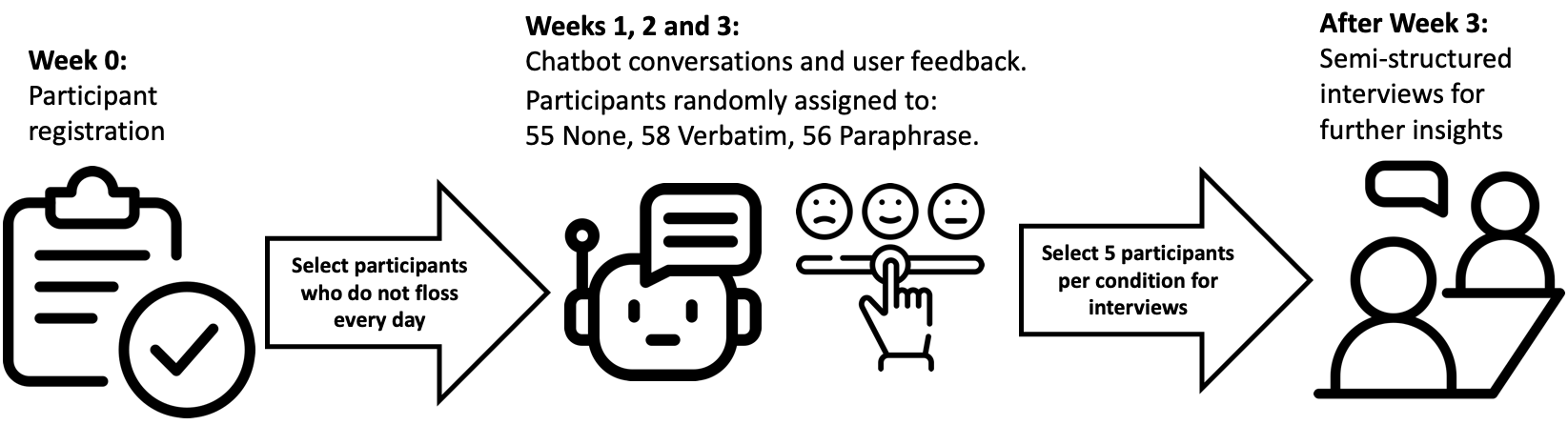}
    \caption{Experiment procedure. Participants were divided into 3 conditions, and talked to a chatbot once a week for 3 weeks about their dental flossing habits and beliefs.}
    \label{fig:individuatation-longitudinal-process}
\end{figure*}

\section{User Study}

This study investigates the effect of a health chatbot remembering (and incorporating into conversation dialogue) user utterances from a previous chatting session. 
For this, we conducted a longitudinal between-subjects experiment where participants talked to a chatbot about their dental flossing once a week for three consecutive weeks \footnote{Ethics approval received from our institutional IRB prior to study commencement.}.
Our chatbot had an independent variable of \textbf{Chatbot Referencing Format} (3 levels) which affected whether the chatbot \textit{explicitly} referenced (the previous week's) user utterances, and the format used when referencing utterances (see Figure \ref{fig:conditions} for examples of referencing format). 
The levels of \textbf{Chatbot Referencing Format} \footnote{Literature may refer to referencing formats using various terminology. In our case, verbatim is analogous to extractive summarisation \cite{zhang2018making,el2021automatic} or direct quotation \cite{wilson_sperber_2000,Cappelen1997-CAPVOQ}, and paraphrase is analogous to abstractive summarisation \cite{zhang2018making,el2021automatic} or indirect quotation \cite{wilson_sperber_2000,Cappelen1997-CAPVOQ}.} are:

\begin{itemize}
    \item \textbf{None} (control group): The chatbot did not explicitly incorporate previous user utterances into subsequent conversations, and instead referenced previous discussions at a high-level.
    \item \textbf{Verbatim}: The chatbot incorporated previous user utterances verbatim into subsequent chatbot utterances.
    \item \textbf{Paraphrase}: The chatbot incorporated paraphrased versions of previous user utterances into subsequent chatbot utterances. 
\end{itemize}

See Figure \ref{fig:conditions} for examples of the referencing format for all 3 conditions. 

\begin{figure*}[h]
    \centering
    \includegraphics[width=1\textwidth]{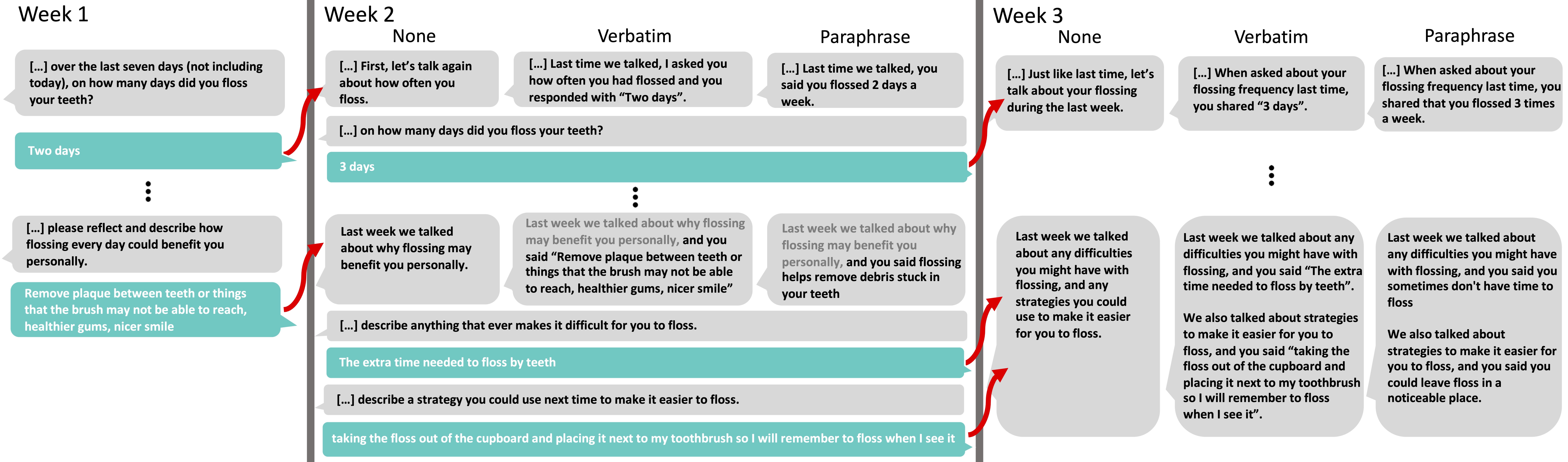}
    \caption{User utterances and their potential None, Verbatim and Paraphrase chatbot responses across all 3 weeks. Grey bubbles are chatbot utterances, and teal bubbles are user utterances. Red arrows show where a user utterance would be referenced (by Verbatim and Paraphrase) in the following week.}
    \label{fig:individuatation-longitudinal}
\end{figure*}




\subsection{Chatbot Script}
\label{sec:script}

The chatbot led a conversation with the user about their dental flossing habits and beliefs. 
We chose dental flossing as we want users to discuss something personal to themselves and (as flossing can benefit from both diary keeping \cite{suresh2012exploratory} and brief interventions \cite{gillam2019brief}) it is appropriate for short weekly personal conversations.
Additionally, dental flossing is an activity that health experts recommend daily adherence to \cite{schuz2006adherence}, and people can have barriers to dental flossing \cite{aguirre2016identification,buglar2010role}, both of which are incorporated into our chatbot's script.



The topic of conversation for each of the 3 weeks was as follows (responses elicited by the chatbot were open-ended unless specified otherwise):


\begin{itemize}
    \item \textbf{Week 1:} All participants saw the same script as the chatbot could not yet reference previous week's. Participants shared their dental flossing beliefs \cite{buglar2010role} (7-point Likert), flossing frequency, and perceived benefits of flossing.
    \item \textbf{Week 2:} The chatbot referenced flossing frequency and perceived benefits from Week 1. Participants shared their flossing frequency, barriers to flossing, and strategies to overcome barriers.
    \item \textbf{Week 3:} The chatbot referenced flossing frequency, and barriers and strategies from Week 2. Participants shared their flossing frequency, reflected on their barriers and strategies from the previous week, and shared their perceived susceptibility and perceived risks, before sharing their dental flossing beliefs \cite{buglar2010role} (7-point Likert) again.
\end{itemize}

Please see Appendix Section \ref{appendix:chatbot_script} for the full chatbot script (including highlights of user utterances that were referenced back to the user in the following week's session).


\subsection{Implementation Details}

The chatbot was hosted on Qualtrics, and used Javascript and HTML to emulate the look and feel of a chatbot.
Microsoft LUIS \footnote{\url{https://www.luis.ai/}} was used for both intent recognition (in real-time) and for selecting the most appropriate paraphrase for a given week.

Intent recognition was trained using user utterances from \cite{cox2022does} for recognising users' barriers to flossing and strategies to overcome barriers. Training data for other prompts was generated by the research team and by piloting the chatbot until a range of responses could be recognised by the chatbot. We then used data augmentation (e.g., synonym replacement) to generate additional training data.

Please see full chatbot script in Appendix Section \ref{appendix:chatbot_script} for both the chatbot utterances: with intent recognition, and varied by referencing format.

\subsubsection{Intent Recognition:}
\label{subsec:intent}
We used intent recognition (in all 3 conditions) to recognise the intent of user utterances within a week's session. An appropriate response would then be appended to start of the subsequent chatbot utterance.
For example, the chatbot could reply ``\textit{Well done on flossing five days a week}'' in response to a user's flossing frequency.




\subsubsection{Delivering Paraphrases:}
The paraphrases used in the Paraphrase condition were pre-written by the research team.
To deliver paraphrases of user utterances, we recognised the user's intent, and then assigned the user with a paraphrase (for example, see Table \ref{tab:chatbot-paraphrases} for paraphrases used in Week 2). 
For example, if the intent of a user utterance was recognised as ``\textit{Cleaner teeth}'', the chatbot would message the user: ``\textit{Last week we talked about why flossing may benefit you personally, and you said \textbf{flossing helps you have cleaner teeth}.}'' in the subsequent week (paraphrase highlighted in bold).
While this approach is limited in providing a discrete number of paraphrases and not accounting for multiple intents, it ensured that consistent and coherent paraphrases could be delivered to users.
Example script and paraphrases can be seen in Figures \ref{fig:conditions} and \ref{fig:individuatation-longitudinal}, and a full list of paraphrases and script can be found in Appendix Chapter \ref{ch:appendix-referencing_utterances}.


\begin{table}[]
\begin{tabular}{p{0.3\linewidth} p{0.5\linewidth}}
\toprule
\textbf{Recognised intent} & \textbf{Chatbot paraphrase given}                         \\ \midrule
Cleaner teeth              & flossing helps you have cleaner teeth            \\
Feel good                  & flossing helps you feel good                     \\
Helps build habits         & flossing helps you build habits                  \\
Improve appearance         & flossing helps improve your appearance           \\
Improve breath             & flossing helps improve your breath               \\
Improve gum health         & flossing improves gum health                     \\
Improve oral hygiene       & flossing helps improve oral hygiene              \\
No benefits                & flossing has no benefits                         \\
Prevent gum disease        & flossing helps prevent gum disease               \\
Prevent tooth ache         & flossing helps prevent tooth ache                \\
Prevent tooth decay        & flossing helps prevent tooth decay               \\
Remove bacteria            & flossing helps remove bacteria from your mouth   \\
Remove debris              & flossing helps remove debris stuck in your teeth \\
Remove plaque              & flossing helps remove plaque from your teeth    \\
\bottomrule
\end{tabular}%
\caption{User intents and their respective paraphrases (of perceived flossing benefits discussed in Week 1) delivered by the chatbot in Week 2. All paraphrases used can be found in Appendix Section \ref{appendix:paraphrases_used}}
\label{tab:chatbot-paraphrases}
\end{table}



\subsection{Participants}

We recruited participants using local university advertisement boards, and only selected participants who did not fully adhere to daily flossing (similarly to previous intervention studies \cite{kocielnik2017send}).
All participants completed the study in November and December 2022, and all responses were completed remotely and asynchronously.
Participants were paid S\$2 to complete the first week's session, S\$2 for the second week, and S\$3 for completing the third and final week's session. Weeks 1 and 2 took on average $\sim$3 minutes, and Week 3 on average $\sim$5 minutes.

169 participants (mean age 22.7; 64\% female) completed all 3 weeks of the study, with 7 participants completing weeks 1 and 2 only, and 4 participants completing week 1 only. We only include data from participants who completed all 3 weeks of the study (with other participants being paid for their completed time, but excluded from analysis).
This resulted in 55 None participants, 58 Verbatim, and 56 Paraphrase.



\subsection{Procedure}

Each week, participants were contacted via email and followed the procedure:
\begin{enumerate}
    \item Follow Qualtrics link to individual chatbot session.
    \item (\textit{Week 1 only}): give consent (participants informed their responses will be stored and analysed).
    \item Brief instructions recap (i.e., no right/wrong answers, responses in English).
    \item Complete weekly chatting session with chatbot.
    \item Post-test questions for respective week (see Section \ref{sec:measures}).
\end{enumerate}

Participants were invited to weeks 2 and 3 seven days after completing the previous week's session, and were given three days to complete these sessions upon receiving the email.
Responses were controlled so that only desktop or laptop devices could be used.

\subsection{Measures}
\label{sec:measures}

\subsubsection{Weekly Measures:}
At the end of each week's chatbot session, participants rated their experience on 7-point Likert scales (Strongly Disagree to Strongly Agree), and were asked ``\textit{Do you personally agree or disagree that...}'' for the following prompts:

\begin{itemize}
    \item \textbf{Interest to continue chatbot usage:} ``I would want to continue using the chatbot'' \cite{xiao2020if}.
    \item \textbf{Chatbot engagement:} ``The chatbot seemed engaged in our discussion'', ``I felt the chatbot was NOT paying attention to what I said'' \cite{shamekhi2018face}.
    \item \textbf{Chatbot intelligence:} ``The chatbot was intelligent'', ``The chatbot was competent'' \cite{cuddy2008warmth,cox2022does}.
\end{itemize}

\subsubsection{Privacy concerns, intrusiveness, and risks:}

We also wanted to compare whether chatbot referencing style had an impact on privacy-related measures.
Therefore, (at the end of week 3 only) we took measures for perceived privacy concerns, privacy intrusion and privacy risks \cite{xu2008examining}. Specifically, on a 7-point Likert scale participants were asked  if they personally agree or disagree to the following questions below.

\textbf{Privacy concerns} which refer  to concerns that inhibit users from sharing information \cite{xu2008examining}:
\begin{itemize}
    \item ``I was concerned that the chatbot was collecting too much personal information about me.''
    \item ``I was concerned about submitting my information to the chatbot.''
\end{itemize}

\textbf{Privacy intrusiveness} which refers to the unwelcome general encroachment into another’s presence or activities  \cite{xu2008examining}:
\begin{itemize}
    \item ``I feel that as a result of this interaction, information about me is out there that, if used, will invade my privacy.''
    \item ``I feel that as a result of this interaction, my privacy has been invaded.''
\end{itemize}

\textbf{Privacy risks} which refer to the uncertainty arising from the possibility of an adverse consequence \cite{xu2008examining}:
\begin{itemize}
    \item ``Personal information was inappropriately used by the chatbot.''
    \item ``Providing the chatbot with my personal information involved many unexpected problems.''
\end{itemize}

\section{User Study Results}
\label{sec:results}

\begin{figure}[h]
    \centering
    \includegraphics[width=0.85\textwidth]{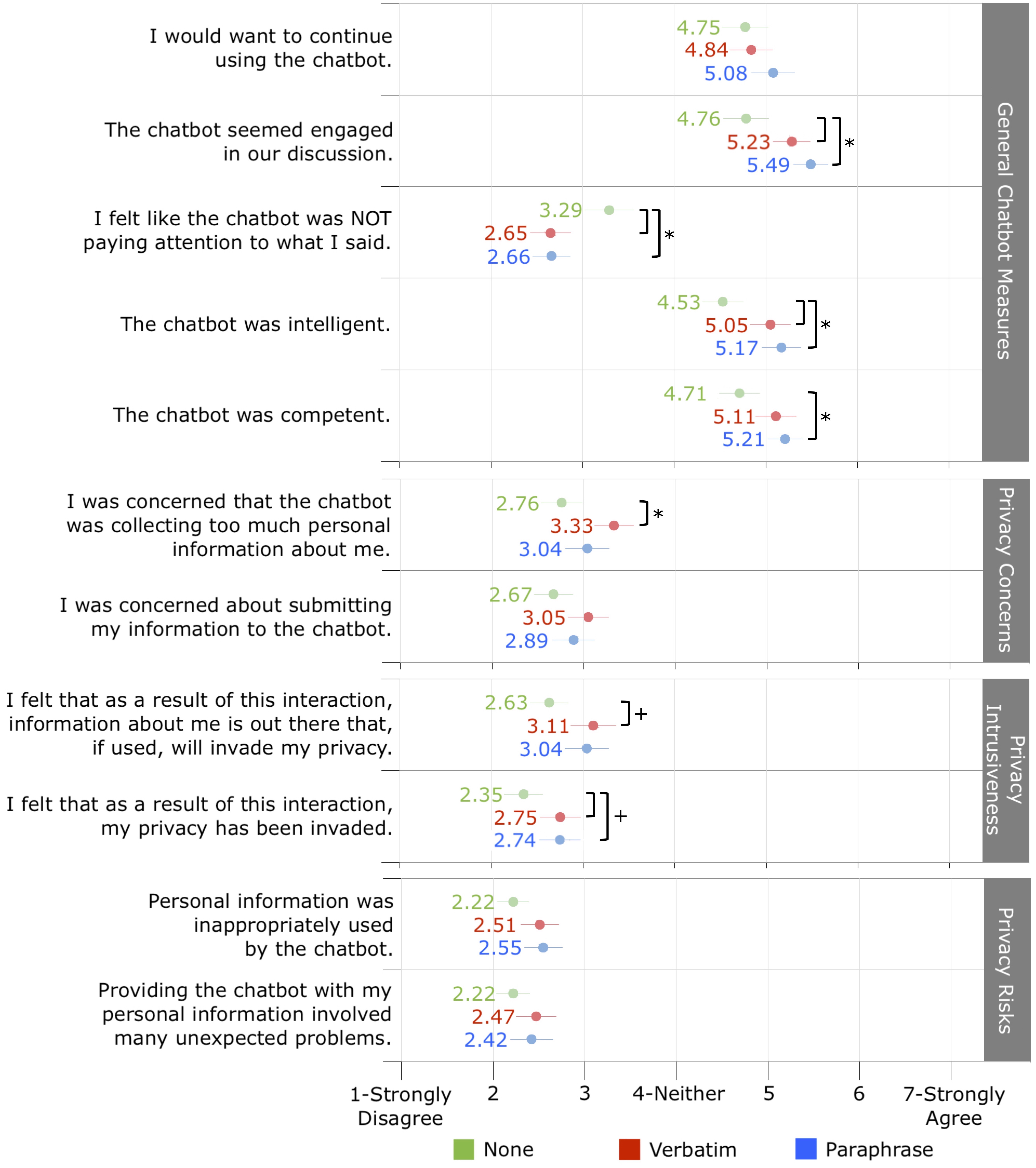}
    \caption{Outcome measures by question asked in the final week of the study. Significance $p<0.05$ indicated by \textbf{$\ast$}, and $p<0.10$ indicated by \textbf{+}.}
    \label{fig:outcomes}
\end{figure}

We fit a linear model on each dependent variable collected from the final week and Chatbot Referencing Format as the fixed effect, and performed post-hoc Student's t-tests to identify specific differences.
We excluded the Likert scale ratings of 8 participants (4 None, 1 Verbatim, 3 Paraphrase) who gave conflicting responses to ``\textit{The agent seemed engaged in our discussion}'' and ``\textit{I felt the agent was NOT paying attention to what I said}'' (e.g., both rated as Strongly Agree). This left us with 161 responses.
See Figure \ref{fig:outcomes} for summary results. 
In addition, we analysed user responses (response length before and after removing stop words), but found no difference between conditions.
We will now discuss individual findings and their significance.

\subsection{General Chatbot Perceptions}

Measures related to RQ3.1 are described below. Chatbot referencing format had no direct impact on a user's desire to continue using the chatbot, and there was no significant difference between conditions.

However, participants found the Verbatim and Paraphrase conditions to be more engaging compared to None.
Specifically, for positively perceived engagement, both Paraphrase ($p = 0.0022$) and Verbatim ($p = 0.0444$) were rated more favourably than None.
While Paraphrase scored higher than Verbatim, it was not statistically significant.
Similarly, for negatively perceived engagement, both Paraphrase ($p = 0.0125$) and Verbatim ($p = 0.0159$) were rated more favourably than None. 
These results indicate that explicitly referencing a user's previous week's utterances positively impacts the user's feelings of chatbot engagement, while not explicitly referencing negatively impacts a user's feelings of being listened to.

Participants also found the Verbatim and Paraphrase chatbots to be more intelligent. 
For perceived intelligence, both Paraphrase ($p = 0.0093$) and Verbatim ($p = 0.0301$) were rated more favourably than None.
For perceived competence, Paraphrase was rated more favourably than None ($p = 0.0327$).
These two results indicate that explicitly referencing a user's previous utterances makes a chatbot appear more intelligent and competent.

\subsection{Privacy Perceptions}
\label{subsec:ch5-privacy-concerns}

Measures related to RQ3.2 are described below.
The Verbatim chatbot was found to generate more \textbf{privacy concerns} than None for one of the measures.
Specifically, for ``\textit{I was concerned that the chatbot was collecting too much personal information about me}'' Verbatim scored higher than None ($p = 0.0227$).

For measures of \textbf{privacy intrusiveness}, there were weakly significant differences, that \textit{could} suggest participants found Verbatim or Paraphrase conditions to be more intrusive compared to None.
For the measure: ``\textit{I feel that as a result of this interaction, information about me is out there that, if used, will invade my privacy}'', Verbatim (M = 3.11) scored highest (worst) and was weakly different to None (M = 2.63) ($p = 0.0716$).
For the measure: ``\textit{I feel that as a result of this interaction, my privacy has been invaded}'', Verbatim (M = 2.75) scored highest (worst), and was weakly different to None (M = 2.35) ($p = 0.0677$).
Similarly, Paraphrase (M = 2.74) was weakly different to None ($p = 0.0866$).
For both measures of \textbf{privacy risk}, while Verbatim and Paraphrase trended above None, there were no statistically significant differences between conditions.

These results indicate that explicitly referencing a user's previous utterances may raise privacy concerns, and that this may be further exacerbated if utterances are referenced verbatim. 
That Verbatim participants were more concerned that the chatbot was collecting too much information about themselves, may also indicate that by being transparent in directly quoting a user's utterances, people were more conscious of their data being collected, and therefore had more privacy concerns.

However, it is important to note that all privacy measures averaged below ``\textit{4 - Neither Agree Nor Disagree}'' reflecting that feelings of privacy violations were still low amongst participants. This feeling may be reflected by the domain of the chatbot (dental flossing) which some participants may not find to be a very sensitive topic (discussed further in Section \ref{sec:interview_privacu_unrelated}).

\subsection{Self-declared flossing frequency and beliefs}
\label{sec:flossing-frequency}

Plots of self-declared flossing frequency and beliefs can be found in Appendix Section \ref{ch:appendix-health-outcomes}.
There was no statistically significance change in flossing frequency among participants during the study, with flossing frequencies having a slight (low-magnitude) trending increase but staying largely the same (see Figures \ref{fig:WeeklyFlossing} to \ref{fig:WeeklyFlossingDIFFbycondition}).
There was a statistically significant increase in intention to floss from week 1 to week 3 among all participants, but no difference between experiment conditions (see Figure \ref{fig:intentionDIFF}).

\section{Semi-Structured Interviews}
\label{sec:interviews}

Verbatim participants statistically significantly felt concerned that the chatbot was collected too much of their personal information compared to None. There were also trending, but weakly-significant, results to indicate that participants found the Verbatim and Paraphrase conditions as potentially more privacy-violating than None (see Section \ref{subsec:ch5-privacy-concerns}). Therefore, to gain further insights as to \textit{why} people may perceive the chatbot referencing formats differently, we conducted semi-structured interviews.


\subsection{Participants}
We recruited 5 participants per condition (N = 15; mean age 20.9; 9 female) for interviews, all of whom had completed the full 3 weeks of the study. Interviews lasted between 20 and 30 minutes, participants were reimbursed S\$5 for their time, and each interview was conducted online.

\subsection{Procedure}
At the start of interviews, participants were instructed that there are no right or wrong answers, and consent was sought to screen-record the interview.
Participants then discussed their experience taking part in the study and responded to questions pertaining to perceived effect on dental flossing, privacy violations, chatbot intelligence, chatbot warmth and the participant's perception of their assigned condition.

After these questions, the interviewer concluded by revealing and describing the 3 experiment conditions to the participant, and asking the participant to describe their feelings towards each condition, and rank their preferred referencing forms. Participants were asked to think aloud \cite{charters2003use}, and explain their reasoning towards each experiment condition.
See Appendix Section \ref{appendix:interview_guide} for the interview guide used.

\subsection{Findings}


We will now discuss the findings from our semi-structured interviews. 
We discuss affects on participant's health beliefs and habits, privacy concerns raised by participants (split between those related and not related to referencing format), chatbot intelligence, recall assistance, and chatbot naturalness. Finally we discuss the last section of the interview where participants saw all 3 conditions and explained their referencing format preferences.



\subsubsection{\textbf{Health Beliefs and Habits:}}
\label{sec:interview_health}

Supporting our findings described in Section \ref{sec:flossing-frequency}, participants described that while they generally did not change their flossing frequency during the course of the study, the discussions with the chatbot raised their consciousness of and intention to floss. Participants also described the chatting sessions increasing their belief in flossing importance, and that sessions made them aware of their behaviour (barriers and strategies to flossing) for the first time.

\subsubsection{\textbf{Privacy Concerns (Unrelated to Reference Format):}}
\label{sec:interview_privacu_unrelated}

Similarly to previous findings, the perceived sensitivity of the domain varied among participants, and affected their hesitancy in sharing information \cite{cox2022does,markos2017information}.
Some participants without privacy concerns described dental flossing as a non-sensitive domain, meaning they were not hesitant sharing their information:
\begin{displayquote}
``\textit{this topic isn't something that's very sensitive, so I wasn't particularly concerned about it.}'' - P5(Verbatim)
\end{displayquote}

In contrast, some participants were concerned to share their dental flossing behaviour as they saw it as sensitive information.
P11(Verbatim) raised this concern in addition to hesitancy sharing their information from uncertainty as to who will read their messages:

\begin{displayquote}
``\textit{dental flossing is} [...]\footnote{Due to space limitations, some transcriptions have been shortened to remove word repetitions, or thinking aloud speech before interviewees arrived at a final conclusion.} \textit{a more private, embarrassing.. umm.. thing. So I think differently sometimes, like whether telling the chatbot like how often I floss or whether I managed to achieve my goals} [...] \textit{it does feel a little scary because I'm not very sure who exactly is seeing the information.}'' - P11(Verbatim)
\end{displayquote}

On from this, participants described feeling embarrassed when sharing health behaviour that they perceived as insufficient:

\begin{displayquote}
``\textit{\textit{I was a bit embarrassed, because they asked me, uh, how many times did I floss over the week and then I was like ``0''}}'' - P4(Paraphrase)
\end{displayquote}

Furthering this, P12(Paraphrase) felt embarrassed when their flossing frequency was referenced:

\begin{displayquote}
``\textit{I was like... slightly embarrassed} $<$laughs$>$ \textit{about how I-I never floss at all.} [...] \textit{it made me health-aware about how I wasn't really flossing at all} [...] \textit{I wouldn't really say that made me feel uncomfortable. Just like a little bit embarrassed. A little bit self-aware.}'' - P12(Paraphrase)
\end{displayquote}


Similarly to previous findings on socially desirable responding \cite{schuetzler2018influence-2,schuetzler2018influence}, one participant described how they considered lying to the chatbot about their flossing behaviour:

\begin{displayquote}
``\textit{I felt guilty for like flossing my teeth like once a week. And I was like ``try to lie to them'', but then I was like: ``OK, never mind I won't lie to them''}'' - P2(None)
\end{displayquote}

The expectation of data storage and perceived sensitivity of the task also affected feelings of privacy invasion, with P1(Paraphrase) equating the task and data storage to writing a diary for themselves:

\begin{displayquote}
``\textit{it's normal for it to store information.} [...] \textit{It's equivalent as to you writing a diary, so it wasn't really something that was particularly invasive to me.}'' - P1(Paraphrase)
\end{displayquote}

\subsubsection{\textbf{Privacy Concerns (Related to Reference Format):}}

When discussing privacy concerns, several participants expressed surprise that the chatbot referenced what they said in previous weeks:

\begin{displayquote}
``\textit{at first when they repeated what I said the previous week, then I was like, ``oh shit, they record everything'' but-but it's not that big of a deal, I guess. Like it's alright, it's just dental hygiene}'' - P4(Paraphrase)
\end{displayquote}

Some of this surprise was accounted to participants' (lack of) expectations of chatbot abilities, with interviewees describing how their concerns subsided after the initial exposure to chatbot references.

However, some Verbatim participants were negatively surprised.
P11(Verbatim) found it ``unnerving'' that Verbatim remembered what they said, and found sharing flossing embarrassing:

\begin{displayquote}
``\textit{it was like a little unnerving because the chatbot remembered what I said previously.} [...] \textit{I didn't expect it to be that smart} $<$laughs$>$\textit{, so it was a little startling but, because-because we're talking about something like dental flossing so, I guess it was a little embarrassing at first.}'' - P11(Verbatim)
\end{displayquote}


Conversely, P1(Paraphrase) described how the referencing format did not raise feelings of privacy intrusion as they expected their data to be stored:

\begin{displayquote}
``\textit{I think that the chatbot just took in} [...] \textit{whatever I put in from the last time around, and} [...] \textit{data storage is} [...] \textit{a normal thing of a chatbot for me, so not really is anything that felt very intrusive}'' - P1(Paraphrase)
\end{displayquote}

P9(Verbatim) described their appreciation for utterances being unchanged, as any ``processing'' would have raised privacy concerns:

\begin{displayquote}
``\textit{I would prefer this over if they were to process my message} [...] \textit{Rather that they just feedback what I have said so my perception would be: ``}[...] \textit{they have just stored my data and then given it back to me'' as compared to them }[...] \textit{processing the background and feeding something else, which I think would have raised a bit more of a privacy issue for me.}'' - P9(Verbatim)
\end{displayquote}

When asked what made them hesitant sharing information, some None participants described how the non-explicit referencing format of None made them doubt the engagement of the chatbot and thereby be hesitant in sharing information:


\begin{displayquote}
``\textit{I think the only thing that was uncomfortable was that the chatbot} [...] \textit{didn't really seem to engage in the conversation.} [...] \textit{he made a reference to its own question. He didn't make reference to my answer.} [...] \textit{he just made the whole chat feel a bit disengaged. So like whatever answer I put down doesn't really matter to the chatbot anyway}'' - P7(None)
\end{displayquote}


This went further with some None participants describing simplifying their responses as they did not think the chatbot would understand them otherwise:

\begin{displayquote}
``\textit{probably if anything} [made me hesitant sharing information], \textit{it was maybe like how complex I structured my sentences. So I tried to keep my sentences like as simple as possible so that maybe the chatbot would be..} $<$pause$>$ \textit{easier for the chatbot to recognise the sentence structures}'' - P6(None)
\end{displayquote}

However, some None participants also described that None did not violate their privacy as it did not reference their past utterances explicitly:

\begin{displayquote}
``\textit{it was just a series of prompts that doesn't really consider any reference to my own, and so I don't really feel any breach of privacy or something}'' - P7(None)
\end{displayquote}





\subsubsection{\textbf{Perceived Intelligence and Engagement:}}

Interviewees generally found the Verbatim and Paraphrase chatbots to be more intelligent.

\begin{displayquote}
``\textit{I was like pretty pleasantly surprised that it like remembered my answers from previous weeks. Yeah, It made me think the chatbot was like a little bit more intelligent.}'' - P12(Paraphrase)
\end{displayquote}


Similarly, Verbatim and Paraphrase participants found referencing their previous utterances made the chatbot feel engaged.


However, some participants thought less of Verbatim with P5(Verbatim) stating: ``\textit{it felt like a survey}''. Others disliked Verbatim due it repeating their utterance word-for-word:

\begin{displayquote}
``\textit{it’d be good to somehow be able to paraphrase what I've said} [...] \textit{so it wouldn't feel so obvious that it's just copying and pasting what I've said previously}'' - P5(Verbatim)
\end{displayquote}


By contrast, while None participants described the intent recognition as a feature of an intelligent chatbot, they also (due to None's referencing format) questioned the intelligence of the chatbot, with some doubting the chatbot's ability to understand them.

\subsubsection{\textbf{Referencing Format and Recall:}}
\label{sec:referencing_recall}

Participants described how the referencing from both Verbatim and Paraphrase helped them remember what they wrote previously.
Verbatim was preferred by some participants as a more precise reminder of their utterance. 
For example, P15(Verbatim) equated the referencing style to a lecture recap, and valued Verbatim's consistency:

\begin{displayquote}
``\textit{Like in lectures and like videos where there's like a recap or review.} [...] \textit{I wouldn't have remembered what I said to the robot, so it kept like a certain consistency of like the interview}'' - P15(Verbatim)
\end{displayquote}




Otherwise, participants appreciated Verbatim as they:
distrust a chatbot's ability to accurately paraphrase their words (and believe paraphrasing will lose nuance); 
want to know their exact utterance so previous conversations are not repeated; 
consider Verbatim will better distinguish their own utterances from the chatbot's; 
or may desire to be held accountable to their prior utterances: 

\begin{displayquote}
``\textit{the retrieval by the chatbot to bring back exactly, especially word for word, what I said, kind of reminded me that ``ohh, I kind of agreed to this, to try this strategy'' and yeah to see one week later I actually did carry it out}'' - P9(Verbatim)
\end{displayquote}

By contrast, None participants found referencing past utterances at a high-level made it difficult to remember what they said previously:

\begin{displayquote}
``\textit{the problem in very generic statements is that} [...] \textit{I kind of like forgot what I've written, and then when they tried to resume conversation, I had no idea what I said.}'' - P2(None)
\end{displayquote}

\noindent Which led some None participants to suggest the chatbot should reference their previous utterances:

\begin{displayquote}
``\textit{it would have been better in the chatbot could like mention what I said at least}'' - P2(None)
\end{displayquote}

\begin{displayquote}
``\textit{if it's more specific, it might allow the user to remember what is their response previously and then give a better response for the following answer as well.}'' - P10(None)
\end{displayquote}

More specifically, P10(None) suggested that the chatbot could reference their previous utterances similarly to existing messaging applications:

\begin{displayquote}
``\textit{it can be more like} [...] \textit{in WhatsApp or Telegram you can reply to the message.} [...] \textit{so you can actually see that ``actually the chatbot is referring to this message that I have sent previously'', so it is clearer.}'' - P10(None)
\end{displayquote}

\subsubsection{\textbf{Naturalness of Referencing Format:}}
\label{sec:interview_natural}

Participants described Paraphrase as feeling natural and human-like. 
This went so-far as P3(Paraphrase) describing that they appreciated that the chatbot did not copy their previous utterances word-for-word, and thereby felt more engaging: 

\begin{displayquote}
``\textit{how the bot referenced it feels very natural.} [...] \textit{it didn't copy what I said verbatim. 
So like it felt as if like a friend was just like, ``Oh yeah, I remember you said something about this like last time we met'' so it felt quite natural, and} [...] \textit{I also really like the fact that they did remember} [...] \textit{because then it made me feel like ``OK, at least the bot is listening to what I say. I'm not like shouting into the abyss''}'' - P3(Paraphrase)
\end{displayquote}

Conversely, some Verbatim participants described how quoting verbatim did not feel personable:

\begin{displayquote}
``\textit{I feel like because it's.. it was quoted directly, right? I felt like there wasn't, say, a lot of personal interaction. It felt more like those things... are just coded.}'' - P5(Verbatim)
\end{displayquote}

Expanding on this P5(Verbatim) described how they would prefer it if the chatbot could paraphrased their utterances:

\begin{displayquote}
``\textit{it’d be good to somehow be able to paraphrase what I've said, or to do so without directly quoting? Yeah, so it wouldn't feel so obvious that it's just copying and pasting what I've said previously, yeah?}'' - P5(Verbatim)
\end{displayquote}

Some None participants described the condition as less natural, and suggested that explicitly referencing past utterances would make the chatbot more personable:

\begin{displayquote}
``\textit{if they reference to my difficulties directly, you feel more... personal.}'' - P7(None)
\end{displayquote}


\subsubsection{\textbf{Comparing the 3 Referencing Formats:}}

At the end of the interview, we revealed the 3 referencing formats to participants, and asked them to think-aloud and explain their preference between the formats.
This reinforced some of the previous qualitative findings, and also generated opinions from participants of their non-assigned conditions.
When ranking their preference for referencing format, all interviewees put None as their last choice, 5 interviewees put Verbatim as their first choice, and 10 interviewees put Paraphrase as their first choice.

Some user feedback, mirrored that already discussed in Section \ref{sec:interview_privacu_unrelated}, with users describing Verbatim as ``creepy'', ``scary'' and ``guilt-tripping'' them, or stating that they appreciate the fidelity to their original utterance; Paraphrase as more natural and human-like; and None as unengaging.
Interestingly, some participants who chose Verbatim as their first preference described that they see chatbots as a tool, and value their own word over that of a robot. Opposingly, those who favoured Paraphrase described seeing a chatbot as a conversational partner that they wish to be more human-like.

\section{Summary of Key Findings}

In summary, key findings for RQs one and two are:

\begin{itemize}[leftmargin=*]
    \item \textbf{RQ3.1:} Users perceived the (Verbatim and Paraphrase) chatbots that explicitly referenced their previous utterances as more engaging and intelligent compared to None.
    \item \textbf{RQ3.2:} Verbatim participants perceived the chatbot as collecting too much of their personal data compared to None participants. Potential reasoning for this can be found in qualitative findings (Section \ref{sec:interviews}).
\end{itemize}

\section{Discussion}
\label{sec:ch5-discussion}

Here we discuss the implications of our study. We aimed to investigate the impact of a chatbot's format when referencing a user's utterances from a previous chatting session. By comparing high-level non-explicit references, verbatim references, and paraphrased references, we wanted to investigate effects on both positive user perceptions and privacy-related perceptions.
Our findings provide some empirical evidence that users value Verbatim and Paraphrase as more engaging and intelligent. However, (in support of Personalisation-Privacy Paradox \cite{awadpersonalization}) there is some evidence Verbatim and Paraphrase raised privacy concerns among users.

Although we did not find measurable differences in response quality between conditions, results indicated that people receiving non-explicit or verbatim references may be hesitant in providing their personal information.
Specifically, Verbatim participants were more concerned about the quantity of personal information being collected, and our interviews found that Verbatim participants raised concerns that the referencing style was ``unnerving'' and ``creepy''. 
Some None participants were hesitant providing complex utterances (as they doubted that the chatbot could understand them).
These findings could reflect the expectations of users before interacting with the chatbot \cite{hartmann2008framing,klaaren1994role,raita2011too,wilkinson2021or}. In order to abate these concerns, (before using different referencing formats) more clear consent could be sought and explanation of privacy practices could be provided \cite{seymour2023ignorance,lau2018alexa,phinnemore2023creepy}, and the abilities of the chatbot could be more clearly advertised to avoid user disappointment \cite{khadpe2020conceptual}.

Interviewees saw chatbots along a spectrum as either more of a conversation partner, or more of a tool to be used. 
Implications from this are that those who view chatbots as conversation partners may prefer paraphrased references, while contrarily, those who view chatbots as more of a tool may prefer a chatbot that references them verbatim.
Similarly, those with more faith in their own word compared to a chatbot (or no belief in chatbot intelligence or emotions) may prefer a verbatim reference format. 
This could be taken further by investigating the role of personality in user preferences for referencing formats. For example, users who are more extroverted or agreeable may prefer a more conversational (paraphrase) format, while users who are more introverted or conscientious may prefer a more direct and factual (verbatim) format.

Our findings also indicate the contextual nature of reference format. 
For example, if the user's utterance is akin to a ``contract'' to themselves (such as a goal for a healthy behaviour), they may want to see their utterance in its entirety in order to solidify their commitment. Similarly, if there is purpose in the user revisiting and developing on previous utterances (such as for creativity tasks or goal-setting) users may prefer their words to remain unchanged so as to build on their previous interaction. Equally, certain use cases (such as in legal settings) may require chatbots to be more conservative in their use of paraphrasing, or to provide verbatim quotes alongside the chatbot's paraphrase (akin to the use of mixed quotations in linguistics literature \cite{wilson_sperber_2000,Cappelen1997-CAPVOQ}).



This implies that chatbots could, in some cases, use a mixture of paraphrased and verbatim reference formats, depending on the content of the user's utterance. 
In the case of dental flossing, the chatbot could use paraphrased responses to reference a user's previous behaviour (flossing frequency), but maintain the user's utterance when referencing the user's behaviour strategy that they devised in the previous chatting session.

Study findings also have implications for the design of chatbot interfaces. 
If chatbots are designed to reference utterances (e.g., verbatim quotes), designers need to be transparent to users, and ensure user control over their data and that user privacy is protected.
Similarly, if paraphrased references are used, the chatbot needs to ensure that the meaning of the user's original utterance is retained and that users do not feel that their utterance has been distorted.






\section{Limitations and Future Work}

The user study was conducted over 3 weeks with one chatting session per week, which was not long enough to potentially encourage health behaviour change among participants.
Furthermore, we cannot claim generality over different chatbot referencing formats \cite{zhang2018making,el2021automatic}, sensitivity and intimacy of user data in references \cite{gomez2023sensitive}, domain of conversations with the chatbot, and input modalities.

Further work could investigate the use of referencing formats across different modalities. For example, while a voice-user interface (VUI) could also reference users verbatim or via paraphrasing, verbatim references could have the added dimension of using the voice of either an agent or of the user themselves \cite{NewsApple}. The added dimension of voice playback could raise addition concerns among users.
Additionally, alternative referencing formats (such as summarisation styles \cite{zhang2018making,el2021automatic,di2014hybrid}, or use of mixed quotations \cite{wilson_sperber_2000,Cappelen1997-CAPVOQ}) could be investigated. Choice of these could depend on factors such as the length, quantity, temporal spacing and content of utterances.
For example, for longer utterances, showing the entire utterance verbatim may prove unwieldy, adding to user burden \cite{eklundh1994use,severinson2010quote}.

\section{Conclusion}
This study investigates how the format used when a chatbot references user utterances from a previous chatting session affects a user's positive perceptions (chatbot intelligence and engagement) and privacy related perceptions. 
Our findings suggest that if a chatbot references previous user utterances, both verbatim or by using paraphrases, it can lead to increased feelings of chatbot intelligence and engagement. 
Despite this, referencing user utterances can also raise privacy concerns among users. 
Our semi-structured interviews then investigated \textit{why} people have these privacy concerns. 
We discussed the implications of our findings for chatbot designers and researchers, and we provided recommendations
for the choice of referencing format.




\chapter{Conclusion and Future Work}
\label{ch:conclusion}

In conclusion, we have presented three different sets of studies, with the high-level goal of making text-based computer-delivered verbals cues in health messaging more effective.
In our Chapter \ref{ch:DirectedDiversity} studies, we generated diverse conversational content (a factor of verbal cues). Specifically, we used the crowd to produce messaging that are both efficacious (motivational and informative) and that incorporate diverse verbal cues.
In our Chapter \ref{ch:FormalOrCasual} studies, we investigated what can make messaging more effective (in terms of user perceptions and self-disclosure) by varying the conversational style (i.e., language formality) of a health chatbots.
Finally, in our Chapter \ref{ch:Personalisation} study, we investigated what can make messaging more effective (in terms of user perceptions and privacy concerns) by varying the format (both conversational style and conversational content) a health chatbot can use when referencing a user's utterances from previous chatting sessions.

Below we will describe how the three chapters highlight the importance of \emph{how} a computer communicates with its users, as well as provide discussion on additional insights not contained in a chapter's respective discussion sections, and potential future work.

\section{Generating and maintaining diverse content - LLMs or the crowd?}
\label{sec:conclusion-DDvsLLM}

Chapter \ref{ch:DirectedDiversity}, emphasised the importance of health messaging that adopts a wide range of themes (with our Directed Diversity method producing collections of \hyperref[sec:ch3-collective]{messages that were seen as less repetitive}), while still maintaining more abstract human concepts such as how motivational a message is found to be. These more abstract concepts \textit{may} require human input or feedback (such as labelled datasets and example utterances, or reinforcement learning from human feedback (RLHF) \cite{zhao2023survey}) to ensure that computer-delivered messaging systems can reply appropriately.
Additionally, LLMs require a flow of example utterances (such as purposely crowd-sourced content, or extracted online content) to keep up to date with changing cultural references, slang, humour and language style.
Prompt engineering and human judgement could also be used to produce and provide feedback on the appropriateness of responses \cite{bubeck2023sparks} (such as asking GPT-4 to write and evaluate its own message - see Figure 1.8 in \cite{bubeck2023sparks}).


In recent work, Karinshak et al. \cite{karinshak2023working} used GPT-3 to generate messages to encourage people to receive the Covid-19 vaccine. Subsequently, crowd-workers found that messages generated by GPT-3 were more persuasive than those from the CDC.
This early work indicates that health messaging generated using commercially available LLMs could prove fruitful (although it should be noted that the LLM produced messages explicitly referenced the authority of the CDC, whereas CDC messages did not explicitly reference their own authority within their messages). 
Cegin et al. \cite{cegin2023chatgpt} found that ChatGPT could generate more diverse paraphrases than the crowd.
It has also been found that LLMs function better with few-shot prompts (i.e., instructions alongside example output) \cite{brown2020language} rather than using zero-shot prompts (with no examples), indicating that crowdsourcing instructions (which often include specific language and examples) could be useful for LLM prompts.

Building on this, it could seem that (beyond some current limitations) LLMs could be used to generate health content as an alternative to the crowd.
To investigate this (inevitable) future direction further, we used instructions given to the crowd (both with our own Directed Diversity \cite{cox2021diverse,cox2023LLM} method, as well as other crowdsourcing work \cite{de2016crowd,agapie2016plansourcing}) to generate health content.

As described in Chapter \ref{ch:DirectedDiversity}, a variety of techniques have been used to help the crowd generate high-quality and diverse content. 
These instructions to the crowd, attempt to include clear and unambiguous instructions and guidelines in order to have crowd-workers produce content to a similar quality to that of a health expert \cite{de2017experts,agapie2018crowdsourcing}.
Taking advantage of these existing instructions, we will investigate the use of LLMs (i.e., ChatGPT-4) to generate health content using a number of different crowdsourcing techniques to assist in prompting.
Please see Appendix Section \ref{ch:appendix-conclusion} for prompts used with ChatGPT-4 and health content produced. 

\subsection{Directed Diversity: comparing LLM and crowd-written messages}

Firstly, our Directed Diversity method from Chapter \ref{ch:DirectedDiversity} could be used to prompt a LLM to generate a diverse collection of health messages.
On from this, we compared the collective diversity of motivational messages that were generated by our human writers (in Chapter \ref{ch:DirectedDiversity}) against those written using three different GPT-4 prompts via the ChatGPT web-UI (see Figure \ref{fig:prompt} for GPT-4 prompts).
For the \textbf{Human-Written} condition, we used 250 messages written by crowd-workers in the DD(1) condition of Chapter \ref{ch:DirectedDiversity}. 
We then used the same phrases and instructions to prompt GPT-4 to write 250 messages (in the \textbf{Phrase-GPT} condition). For comparison, we also prompted with a \textbf{Simple-GPT} condition: simply asking GPT-4 to write 250 messages; and a \textbf{Diverse-Naïve-GPT} condition: requesting GPT-4 to write 250 messages that are diverse from one another.
Some example messages produced can be seen in Table \ref{tab:conclusion-messages} within this section. 
This subsection is based on the publication \cite{cox2023LLM}:

\bibentry{cox2023LLM}

\begin{figure}
    \begin{boxA}
    {\footnotesize \textbf{Simple-GPT prompt:}\\
    You will now write 250 messages to motivate someone to exercise. Please use the following guidelines:\\
    - Write concise messages (imagine the message could be sent to people on a smartphone). Messages should generally be 1 to 3 sentences long.\\
    - Output your response with one message per line.
    \par\noindent\rule{\textwidth}{0.5pt}
    \textbf{Diverse-Naïve-GPT prompt:}\\
    You will now write 250 messages to motivate someone to exercise. Please use the following guidelines:\\
    - Write concise messages (imagine the message could be sent to people on a smartphone). Messages should generally be 1 to 3 sentences long.\\
    - Ensure the messages are diverse from one another, by limiting message repetition and similarity (such as by limiting repetition of themes or phrasing).\\
    - Output your response with one message per line.
    \par\noindent\rule{\textwidth}{0.5pt}
    \begin{spacing}{1}
    \textbf{Phrase-GPT prompt:}\\
    You will now write 250 messages to motivate someone to exercise, and you will be shown short phrases to inspire you. Please use the following guidelines:\\
    - Use the phrases for inspiration to write a message that would motivate someone to exercise.\\
    - Write concise messages (imagine the message could be sent to people on a smartphone). Messages should generally be 1 to 3 sentences long.\\
    - You can write using the entire phrases, or fragments of phrases.\\
    - Ignore phrases that you don't find relevant to writing motivational messages for exercise (but feel free to be creative if you can use any of the phrases to write your message).\\
    - Output your response in comma separated format of: Inspirational phrase, motivational message generated from phrase.\\
    Below are the 250 phrases to use:\\
    elite athletes often train\\
    pink name brand workout outfit\\
    care to use proper form\\
    $\cdots$
    \end{spacing}
    \vspace{-0.2cm}
    }
    \end{boxA}
    \vspace{-0.5cm}
    \caption{Prompt instructions given to GPT-4. Instructions for crowd-workers (similar to Phrase-GPT) found in \cite{cox2021diverse}.} 
    \label{fig:prompt}
\end{figure}


\subsubsection{Results and Discussion}

To calculate the diversity of each set of messages, we calculated the mean pairwise Euclidean distance \cite{kaminskas2016diversity,cox2021diverse} between all messages within each condition (where a higher distance reflects more diversity). From lowest to highest diversity, this gave us: 4.13 for \textbf{Simple-GPT}, 4.29 for \textbf{Naïve-Diverse-GPT}, 5.66 for \textbf{Phrase-GPT}, and 6.90 for \textbf{Human-Written}. This indicates that such a crowdsourcing pipeline could be used to increase the diversity of content generated by LLMs. While Phrase-GPT did not produce a corpus of messages as diverse as those from human-writers, this may be due to differences in message length (with Human-Written averaging 24.0 words, and Phrase-GPT 18.7 words). In addition, Simple-GPT averaged 9.2 and Naïve-Diverse-GPT 9.8 words per message (emphasising the impact of including examples when prompting LLMs).

The work involved in producing messages should also be noted.
The 250 human-written messages took on average 73 seconds each to be written, while GPT-4 took roughly 6 seconds per message.
Additionally, while some human-written messages in our Chapter \ref{ch:DirectedDiversity} studies were excluded (such as those using poor levels of English or apparent gibberish), the LLM-written messages seemingly suffered from no such issues.
Example Human-Written and Phrase-GPT messages can be seen below in Table \ref{tab:conclusion-messages} alongside their respective inspirational phrases.



\begin{table}[h!]
  \centering
  \caption{Example messages written by the crowd (in Chapter \ref{ch:DirectedDiversity}) and GPT-4 using the same phrases for inspiration.}
  \label{tab:conclusion-messages}
  \begin{tabular}{p{4cm}p{5.5cm}p{5.5cm}}
    \toprule
    Inspirational Phrase & \textbf{Human-written} message & \textbf{Phrase-GPT} message \\
    \midrule
\textbf{Example 1.} lots of fear and intimidation &
If you can't exercise outside, try listening to music while exercising   inside.  It can help relax and motivate   you at the same time. &
Overcome fear and intimidation in your exercise journey by seeking   support, educating yourself, and taking small steps towards your goals. \\
\textbf{Example 2.} homeopathic doses of exercise &
  Even small amounts of exercise can help you become healthy. You don't   need to dedicate 2 hours a day to become healthier! &
  Start with small, manageable doses of exercise to build a foundation for   a healthier, more active lifestyle. \\
\textbf{Example 3.} military press one rep &
  Don't focus on working out for 30 minutes, an hour, etc. &
  Just focus on your exercises one rep at a time!	Challenge yourself with exercises like the military press, focusing on form and gradually increasing weight for improved strength. \\
\textbf{Example 4.} rats doing regular aerobic exercise &
   Regular running and aerobic exercise is great! Never stop working out. &
   Take inspiration from studies, even those on rats, that show the benefits of regular aerobic exercise for overall health and longevity. \\
\textbf{Example 5.} vice president of program development &
   Just 30 minutes of exercise a day can change your life! &
   Seek guidance from experienced professionals like vice presidents of program development to optimize your fitness routine and approach. \\
  \bottomrule
\end{tabular}
\end{table}


These results demonstrate the potential effectiveness of using a crowdsourcing pipeline to generate more diverse messages compared to two baseline prompts. However, similar to some previous creativity tasks that require more advanced reasoning abilities \cite{stevenson2022putting,jentzsch2023chatgpt}, human-writers were more successful than GPT-4. Further investigation could alter LLM parameters such as temperature (default 1.0 on ChatGPT \cite{OpenAITemperature}).

Several additional insights are demonstrated by examples in Table \ref{tab:conclusion-messages}. Both human and LLM messages demonstrated the ability to draw metaphors from phrases (Example2). GPT-4 may have had difficulty deciding on the relevance of phrases and would generally incorporate phrases, while human writers would act more discerningly (see Example 4 and Example 5). This emphasises that LLMs follow the form rather than meaning of language \cite{bender2020climbing}, and implies that crowdsourcing pipelines could be atomised further for LLMs (e.g., including an initial step asking the LLM to judge the relevance of a phrase to physical activity). 

At times, human-writers would not incorporate phrases (perhaps if they do not have domain knowledge of more esoteric phrases) while the LLM could (Example 3).
However, risk of AI hallucination (e.g., within healthcare \cite{bubeck2023sparks,lee2023benefits,taylor2022galactica,chen2023utility,jeblick2022chatgpt} such as LLMs misunderstanding medical vocabulary or providing advice that does not follow medical guidelines \cite{chen2023utility,jeblick2022chatgpt}) should be noted, and additional measures would be needed to ensure the veracity of output.

Additionally, (while one could prompt a LLM to incorporate different conversational styles or sociocultural perspectives), attention is needed to ensure that it does not produce harmful cultural stereotypes \cite{durmus2023towards,mirowski2023co,feng2023pretraining}.
Similarly, human writers incorporate personal experiences into messages \cite{coley2013crowdsourced} (Example 1) that may not necessarily be available to a LLM (if such experiences are poorly represented).

Finally, while we used automated measures to indicate diversity of each experiment condition, further human evaluations for diversity could be conducted \cite{dow2010parallel,cox2021diverse,siangliulue2016ideahound}, in addition to human evaluations of message efficacy (e.g., motivation \cite{de2016crowd,cox2021diverse}).

\subsection{Prompt engineering with other previous crowdsourcing instructions}

Additionally, LLM prompts could include extra context and specificity in order to produce more personalised health content. For example, we could prompt LLMs more granularly by including intrinsic and extrinsic user factors. This could include specifying device received on and medium of messaging/notification (e.g., a smartphone push-notification) or user demographics and interests (e.g., a Gen-Z person who uses TikTok). 

\begin{figure*}[h!]
    \centering
    \begin{subfigure}[b]{0.46\textwidth}
        \includegraphics[width=\textwidth]{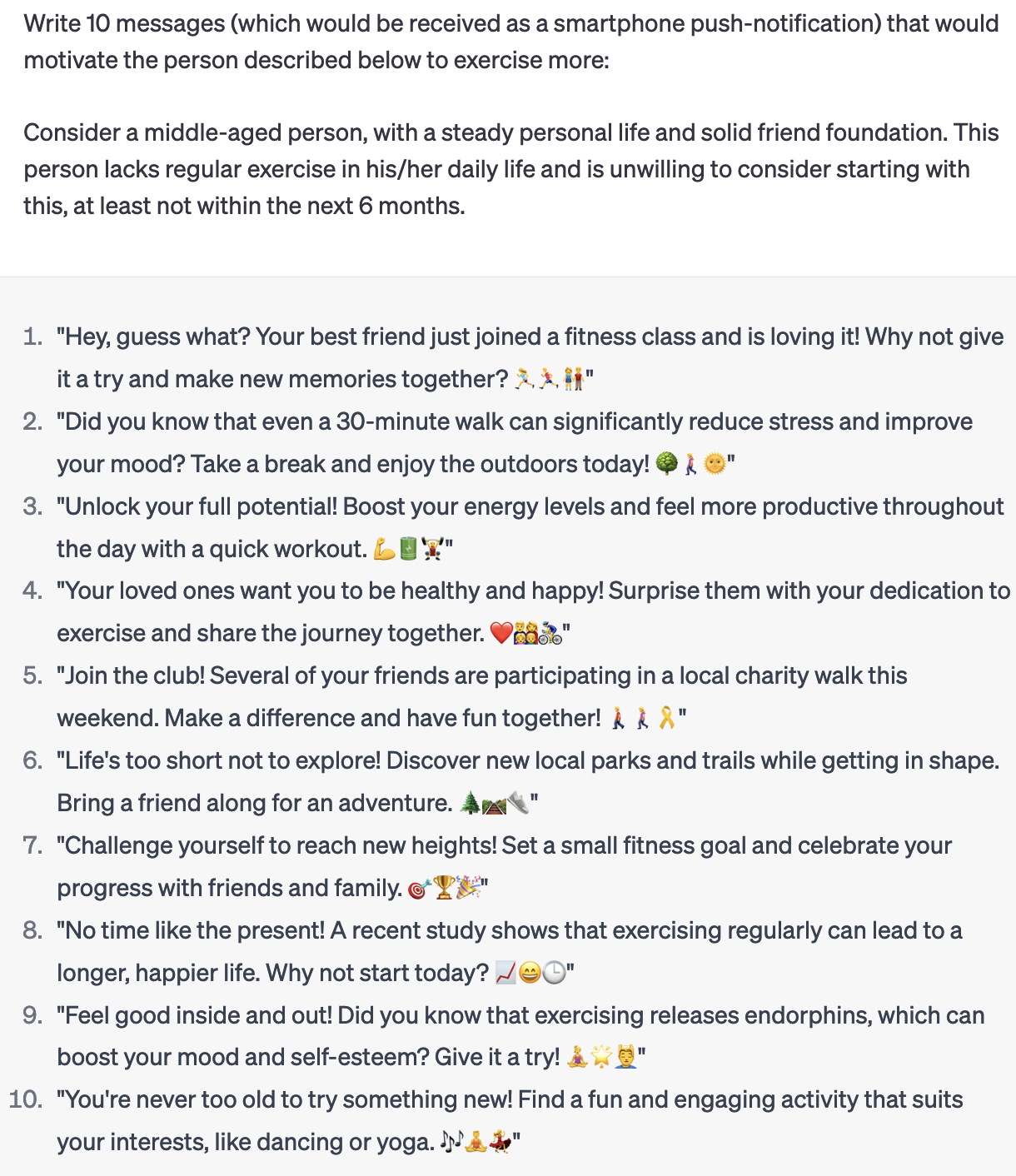}
        \caption{Using a scenario describing a precontemplation stage of change.}
        \label{fig:conclusion-precon}
    \end{subfigure}
    \hfill 
    \begin{subfigure}[b]{0.46\textwidth}
        \includegraphics[width=\textwidth]{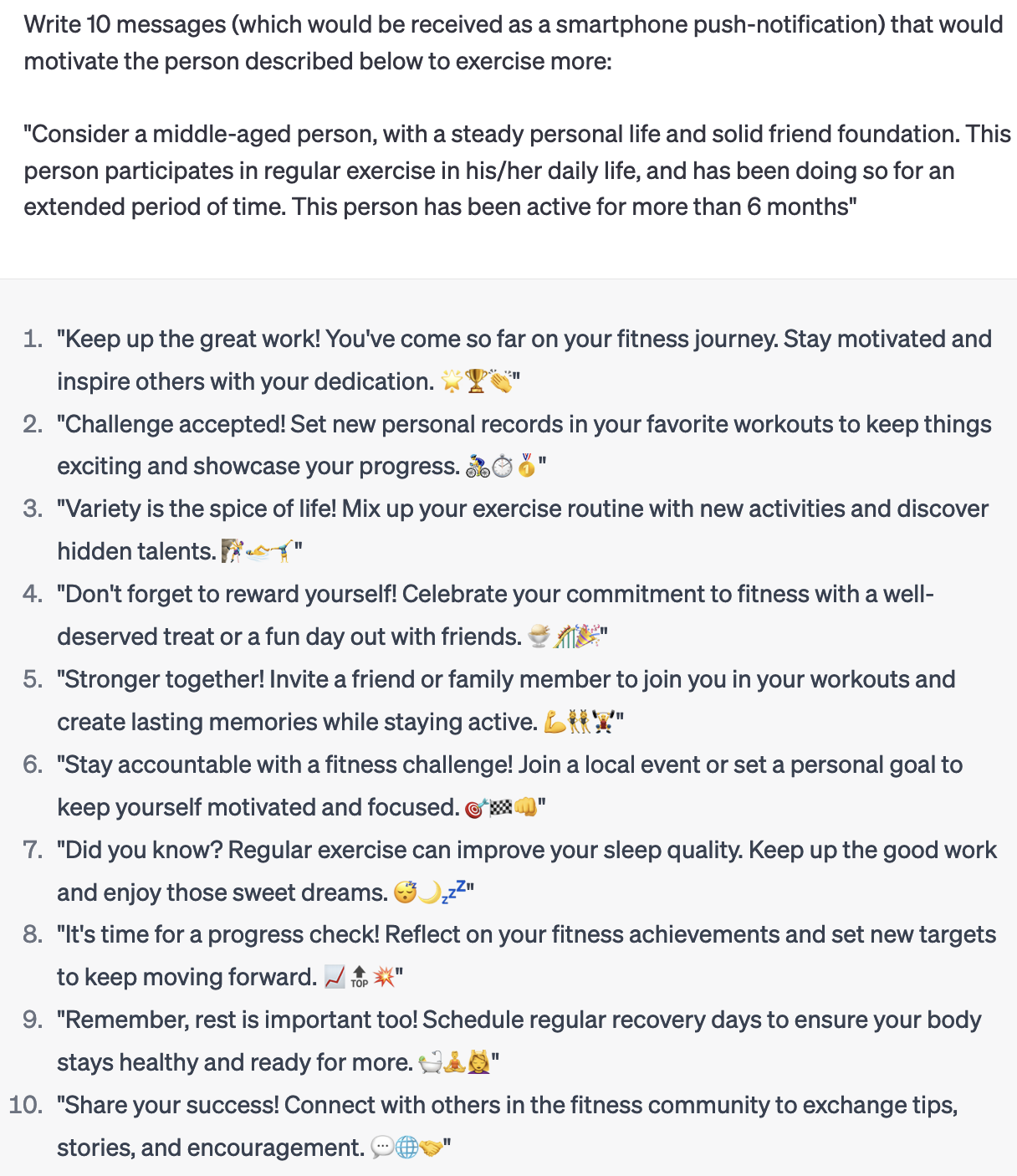}
        \caption{Using a scenario describing a maintenance stage of change.}
        \label{fig:conclusion-maintenance}
    \end{subfigure}
    \caption{Prompting GPT-4 using scenarios from De Vries et al. \cite{de2016crowd}. Scenarios describe the stage of change (see Transtheoretical Model \cite{prochaska1997transtheoretical}) of an individual}
    \label{fig:conclusion-devries}
\end{figure*}

Therefore, we investigated prompting using vignettes derived from health behaviour change theory, specifically the scenarios used by De Vries et al. when instructing the crowd \cite{de2016crowd}. 
These scenarios described the demographic features, and physical activity levels and beliefs of a hypothetical individual for which motivational messages should be written.
Please see Figure \ref{fig:conclusion-devries} for two example prompts given, and resultant messages produced by GPT-4.
These messages from GPT-4 seem to indicate that the features described in the scenario are incorporated into the messages (such as Figure \ref{fig:conclusion-precon} messages prompting the user to begin exercising, Figure \ref{fig:conclusion-maintenance} messages encouraging the user to maintain their exercise, and both sets of messages drawing on social support).

We also investigated producing exercise plans using similar instructions given to previous crowd-workers.
Agapie et al. \cite{agapie2016plansourcing} used the crowd to produce exercise plans that adhered to guidelines. Using these guidelines (that also described the goals, constraints and exercise preferences of a given user) we used GPT-4 to produce exercise plans.
These can be found in Appendix Section \ref{ch:appendix-agapie}.
Again, this health content seems (anecdotally) similar in quality to the example crowd-sourced exercise plans given by \cite{agapie2016plansourcing}.

This could indicate the potential for LLMs to produce health content that adheres to instructions, and is highly personalisable.
It should be noted however that current LLMs are prone to changes of output even when prompts are adjusted to what could be considered a small amount. For example, by specifying that the messages would be seen on a smart-phone in the Figure \ref{fig:conclusion-devries} prompts, GPT-4 included emojis at the end of all messages.

\section{The importance of message style}

Our studies in Chapters \ref{ch:FormalOrCasual} and \ref{ch:Personalisation} reflect that it is important to study how an agent should interact with someone in terms of the \textit{style of message} delivery. This is complex and depends on the intrinsic \textit{factors of the user} themselves (such as cognitive biases \cite{sundar2019machine}, gender \cite{jung2016feminizing}, age or race \cite{liao2020racial}), extrinsic user factors (such as location, date/time \cite{liao2020personalized} or physical activity), and the \textit{context and domain of the conversation} (such as domain sensitivity \cite{cox2022does}).
For example, intrinsic user factors were important in both studies, with user's preconceptions of agents affecting how users perceived difference styles of message delivery.
Beyond the findings outlined in these chapters, it is also important to consider the recent rise of LLMs. This has lead to commercially available technology that can give (mostly) accurate, appropriate, human-like, and highly customisable responses to a given user utterance.
Once issues such as hallucinations are less pervasive LLMs are expected to be applied to healthcare scenarios \cite{bubeck2023sparks,lee2023benefits,jo2023understanding}.
The next step for future usability of conversational systems such as these thereby highlights the importance of studies which investigate how users and agents interact (such as in Chapters \ref{ch:FormalOrCasual} and \ref{ch:Personalisation}). Our Chapter \ref{ch:FormalOrCasual} findings also highlight that studies are needed across domains as appropriate conversational style will vary.

Equally as these systems become more ubiquitous, user attitudes may change \cite{siau2018building} which in turn could lead to changes in user perceptions of the most appropriate interaction style of an agent. 
While our user studies were conducted in a time where mainstream use and discussion of LLMs (such as ChatGPT) was less commonplace, the increasing ubiquity of such systems could lead people to more often have faith in and accept the use of a chatbot as a conversational partner.
This could lead to certain findings (such as users that view a chatbot as more of a conversational partner  \hyperref[sec:ch5-discussion]{preferring paraphrased responses}) having increased importance. 
Finally, we investigated the modality of text-based interactions. While a similar high-level takeaway can be made (that the style and content of messages is important) individual findings may differ across different modalities (such as voice), domains (such as education), and demographic groups.

\newpage
\addcontentsline{toc}{chapter}{Bibliography}

\bibliographystyle{plain}
\bibliography{References}

\appendix

\chapter{Chapter \ref{ch:DirectedDiversity} Appendices}

Below are the appendices for \textit{Chapter \ref{ch:DirectedDiversity}: Generating Diverse and Efficacious
Health Messages}, where we developed and evaluated our Directed Diversity method.

\section{Definitions of Prompt Selection Variables}

\begin{figure}[H]
    \centering
    \includegraphics[width=1\textwidth]{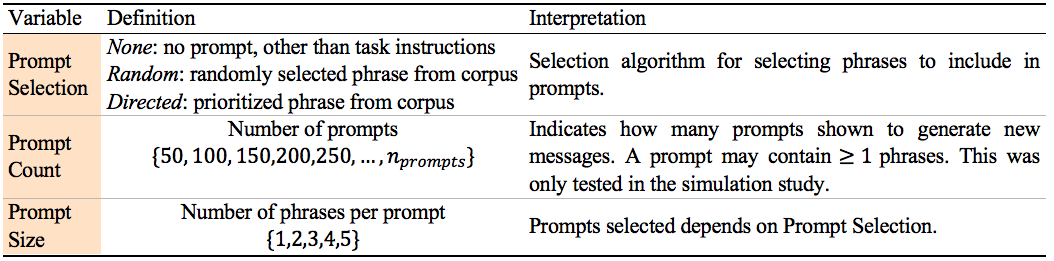}
    \caption{Independent variables used in the simulation and user studies to manipulate how prompts are shown to ideators.}
    \label{tab:Directed-Tab6}
\end{figure}

\newpage
\section{Intra-Prompt Diversity Metrics based on Embedding Distances}

\begin{figure}[H]
    \centering
    \includegraphics[width=1\textwidth]{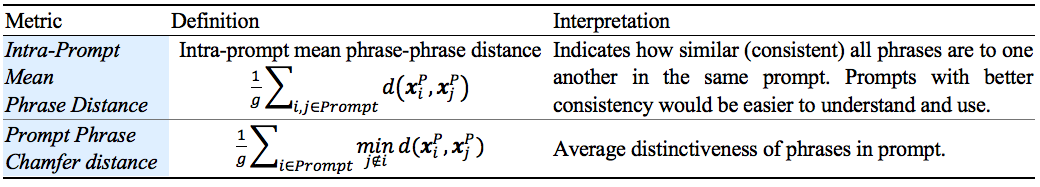}
    \caption{Metrics of prompt diversity for all phrases in a single prompt.}
    \label{tab:Directed-Tab8}
\end{figure}

\section{Definitions of Prompt Adoption Metrics}

\begin{figure}[H]
    \centering
    \includegraphics[width=1\textwidth]{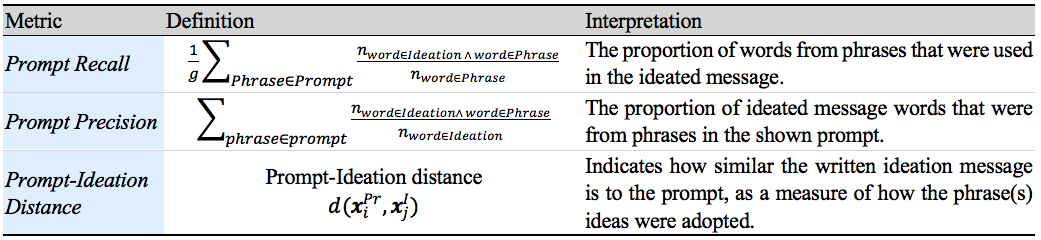}
    \caption{Metrics indicating how much of prompt text and concepts are adopted into the ideations.}
    \label{tab:Directed-Tab9}
\end{figure}

\newpage
\section{Pairwise Embedding Distances of Phrases and Messages}

These figures show the distribution of pairwise distances based on the embeddings of phrases and messages.

\begin{figure}[H]
    \centering
    \includegraphics[width=0.6\textwidth]{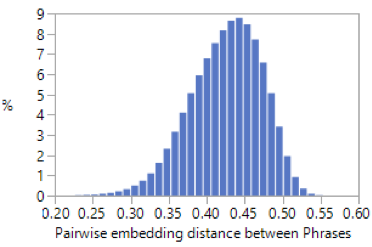}
    \caption{Distribution of pairwise distances between the extracted phrases (N=3,666). The pairwise distances ranged from Min=0.057 to Max=0.586, Median=0.430, inter-quartile range 0.394 to 0.460, SD= 0.047.}
    \label{fig:Directed10}
\end{figure}

\begin{figure}[H]
    \centering
    \includegraphics[width=0.6\textwidth]{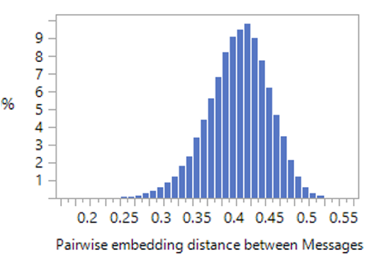}
    \caption{Distribution of pairwise distances between the messages (N=250) ideated in the pilot study with no prompting (None). The pairwise distances ranged from Min=0.169 to Max=0.549, Median=0.405, inter-quartile range 0.376 to 0.432, SD=0.043.}
    \label{fig:Directed11}
\end{figure}

\newpage
\section{Results of Characterisation Simulation Study}
\label{Appendix:ch3-5}

\begin{figure}[H]
    \centering
    \includegraphics[width=1\textwidth]{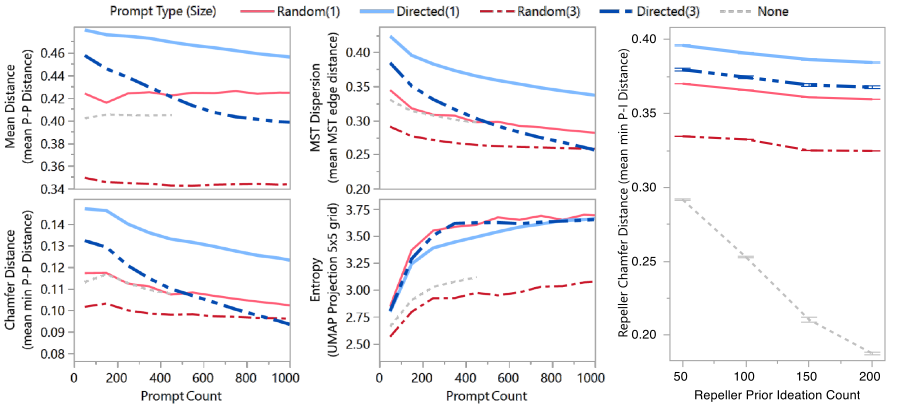}
    \caption{Influence of prompt selection technique, prompt size, and prompt count on various distance and diversity metrics. Higher values for all metrics indicate higher diversity. Span and Sparseness results are not shown, but are similar to Mean Distance. Note that we computed the mean of MST edge distances instead of sum, which is independent of number of prompts. Error bars are extremely small, and not shown for simplicity.}
    \label{fig:Directed12}
\end{figure}

We created 50 simulations for each prompts configuration to get a statistical estimate of the performance of each prompt selection technique. Figure \ref{fig:Directed12} shows the results from the simulation study. Error bars are extremely small, and not shown for simplicity. Span and Sparseness results not shown, but are similar to Mean Distance. Note that we computed the mean of MST edge distances instead of sum, which is independent of number of prompts. In general, Directed Diversity selects prompts to be more diverse for fewer prompts (smaller prompt count), but after a threshold, Random selection can provide for better diversity. This demonstrates directing is useful for small crowd budgets. Note that the actual threshold depends on corpus and application domain. We found an interaction effect where single-phrase prompts benefit most with Directed Diversity, since for low prompt count, and Directed(1-phrase) has highest diversity, followed by Directed(3), Random(3), and Random(1) with lowest diversity.

\newpage
\section{Factor Loadings from Factor Analysis in User Studies}

\begin{figure}[H]
    \centering
    \includegraphics[width=0.8\textwidth]{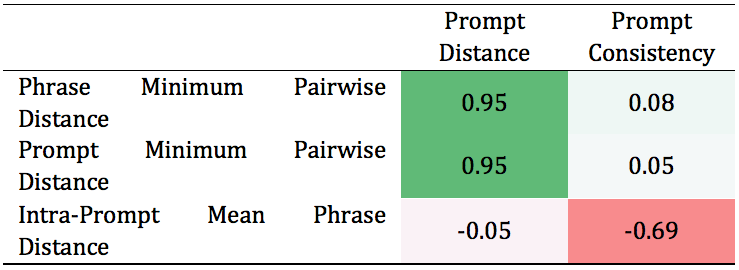}
    \caption{The rotated factor loading of factor analysis on metrics of prompt distance and consistency. Factors explained 73.6\% of the total variance. Bartlett’s Test for Sphericity to indicate common factors was significant ($\chi^2$= 5810, $p<.0001$).}
    \label{tab:Directed-Tab10}
\end{figure}

\begin{figure}[H]
    \centering
    \includegraphics[width=1\textwidth]{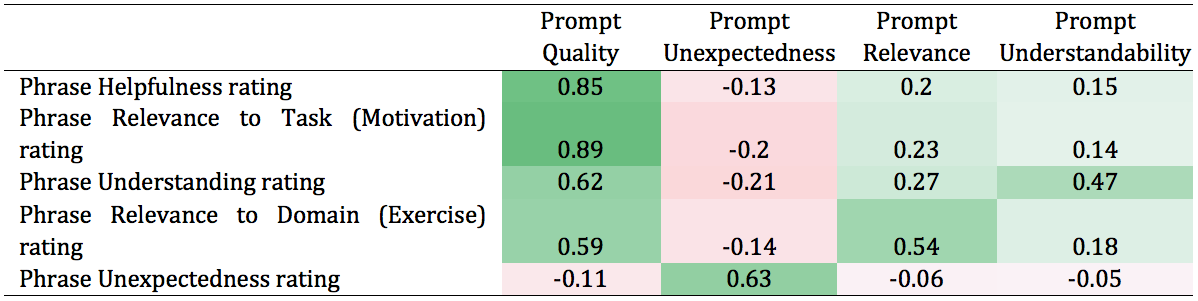}
    \caption{The rotated factor loading of factor analysis on metrics of perceived helpfulness of prompts. Factors explained 68.9\% of the total variance. Bartlett’s Test for Sphericity to indicate common factors was significant ($\chi^2$= 2575, $p<.0001$).}
    \label{tab:Directed-Tab11}
\end{figure}

\begin{figure}[H]
    \centering
    \includegraphics[width=0.7\textwidth]{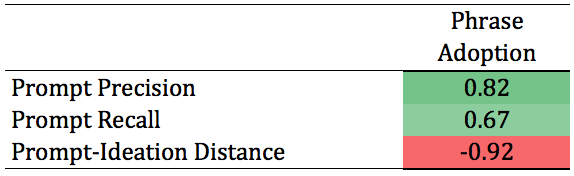}
    \caption{The rotated factor loading of factor analysis on metrics of prompt adoption. Factors explained 65.1\% of the total variance. Bartlett’s Test for Sphericity to indicate common factors was significant ($\chi^2$= 1315, $p<.0001$).}
    \label{tab:Directed-Tab12}
\end{figure}

\begin{figure}[H]
    \centering
    \includegraphics[width=1\textwidth]{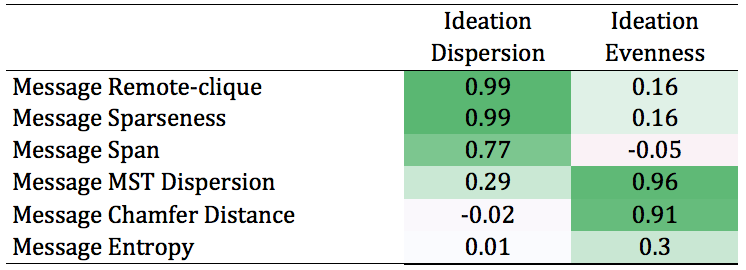}
    \caption{The rotated factor loading of factor analysis on diversity metrics of generated messages. Factors explained 75.2\% of the total variance. Bartlett’s Test for Sphericity to indicate common factors was significant ($\chi^2$= 2676, $p<.0001$).}
    \label{tab:Directed-Tab13}
\end{figure}

\begin{figure}[H]
    \centering
    \includegraphics[width=1\textwidth]{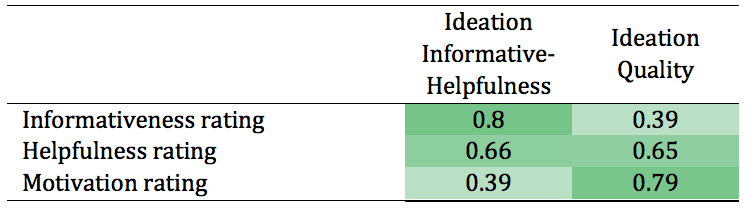}
    \caption{The rotated factor loading of factor analysis on metrics of perceived quality of the generated messages. Factors explained 80.9\% of the total variance. Bartlett’s Test for Sphericity to indicate common factors was significant ($\chi^2$= 5810, $p<.0001$).}
    \label{tab:Directed-Tab14}
\end{figure}

\begin{figure}[H]
    \centering
    \includegraphics[width=1\textwidth]{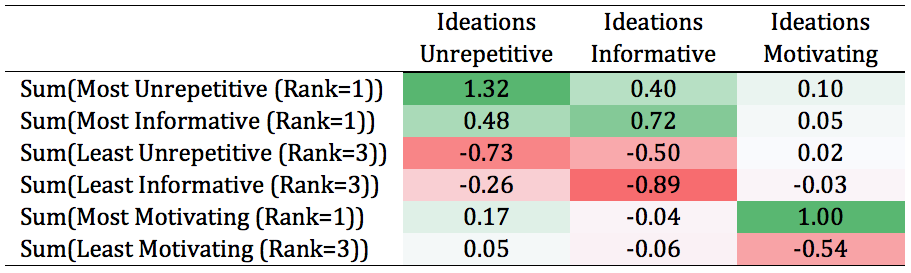}
    \caption{The rotated factor loading of factor analysis on metrics of group ranking of the generated messages. Factors explained 93.9\% of the total variance. Bartlett’s Test for Sphericity to indicate common factors was significant ($\chi^2$= 366, $p<.0001$). For usability, ``unrepetitive'' was measured with the word “repetitive” in the survey.}
    \label{tab:Directed-Tab15}
\end{figure}

\begin{figure}[H]
    \centering
    \includegraphics[width=0.7\textwidth]{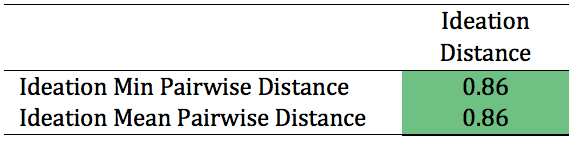}
    \caption{The rotated factor loading of factor analysis on metrics of message distinctness. Factors explained 74.8\% of the total variance. Bartlett’s Test for Sphericity to indicate common factors was significant ($\chi^2$= 1022, $p<.0001$).}
    \label{tab:Directed-Tab16}
\end{figure}

\begin{figure}[H]
    \centering
    \includegraphics[width=1\textwidth]{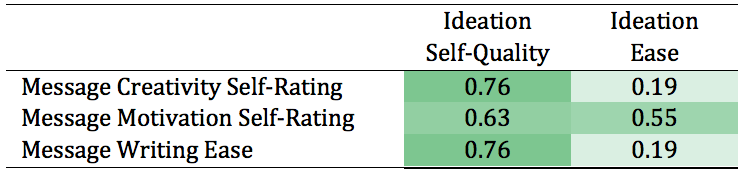}
    \caption{The rotated factor loading of factor analysis on metrics of ideation effort. Factors explained 59.0\% of the total variance. Bartlett’s Test for Sphericity to indicate common factors was significant ($\chi^2$= 1008, $p<.0001$).}
    \label{tab:Directed-Tab17}
\end{figure}

\newpage
\section{Survey Screenshots in User Studies}

\subsection{Ideation User Study}

\begin{figure}[H]
    \centering
    \includegraphics[width=1\textwidth]{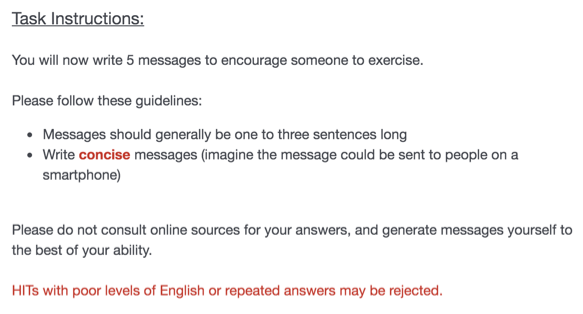}
    \caption{The instructions in the Ideation User Study for the None condition.}
    \label{fig:Directed13}
\end{figure}

\begin{figure}[H]
    \centering
    \includegraphics[width=1\textwidth]{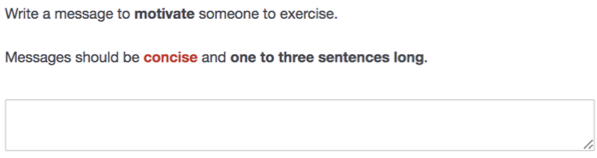}
    \caption{For the None condition, users are asked to write a message that is between one and three sentences long.}
    \label{fig:Directed14}
\end{figure}

\begin{figure}[H]
    \centering
    \includegraphics[width=1\textwidth]{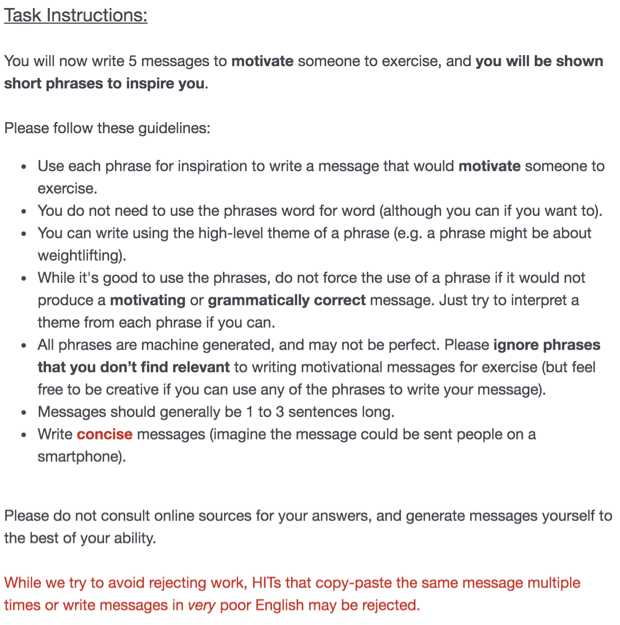}
    \caption{The instructions in the Ideation User Study for the Random(1) and Directed(1) conditions.}
    \label{fig:Directed15}
\end{figure}

\begin{figure}[H]
    \centering
    \includegraphics[width=1\textwidth]{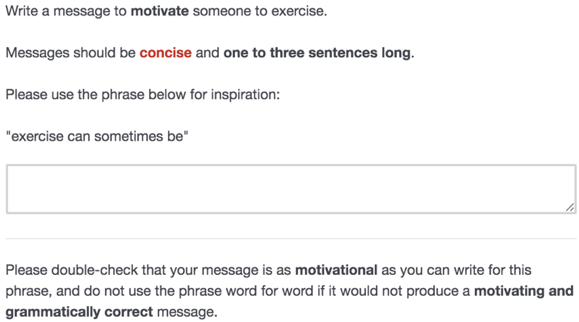}
    \caption{Random(1) and Directed(1) prompts consisted of one phrase per prompt. Different phrases were used for each trial.}
    \label{fig:Directed16}
\end{figure}

\begin{figure}[H]
    \centering
    \includegraphics[width=1\textwidth]{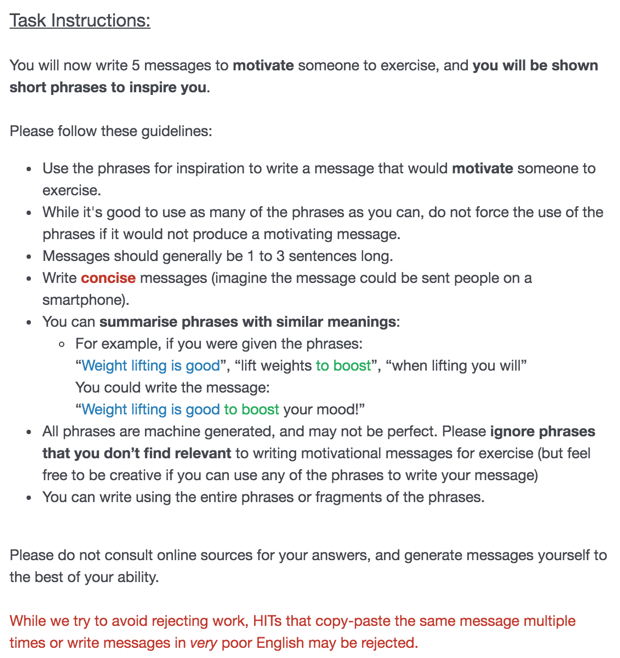}
    \caption{Random(1) and Directed(1) prompts consisted of one phrase per prompt. Different phrases were used for each trial.}
    \label{fig:Directed17}
\end{figure}

\begin{figure}[H]
    \centering
    \includegraphics[width=1\textwidth]{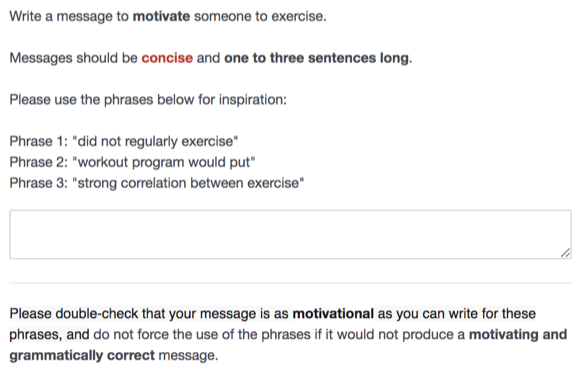}
    \caption{Random(3) and Directed(3) prompts consisted of three phrases per prompt. Different phrases were used for each trial.}
    \label{fig:Directed18}
\end{figure}

\begin{figure}[H]
    \centering
    \includegraphics[width=1\textwidth]{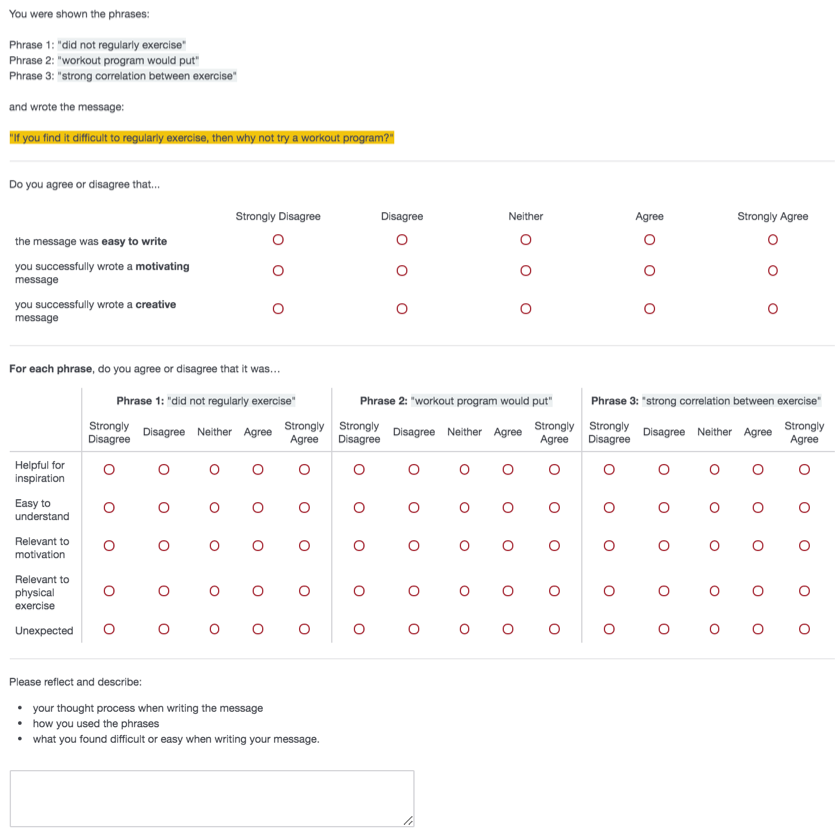}
    \caption{Ideators are asked to evaluate the message they wrote by providing Likert scale ratings for many different factors along with a short reflection about the message writing process. The screenshot above shows the evaluation screen for Directed(3).}
    \label{fig:Directed19}
\end{figure}

\newpage
\section{Validation User Studies}

\begin{figure}[H]
    \centering
    \includegraphics[width=0.8\textwidth]{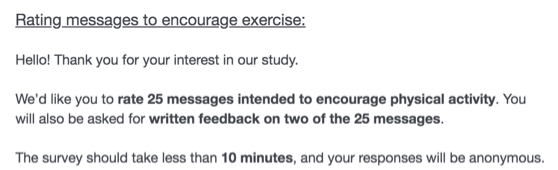}
    \caption{The instructions for individual message rating tasks.}
    \label{fig:Directed20}
\end{figure}

\begin{figure}[H]
    \centering
    \includegraphics[width=0.8\textwidth]{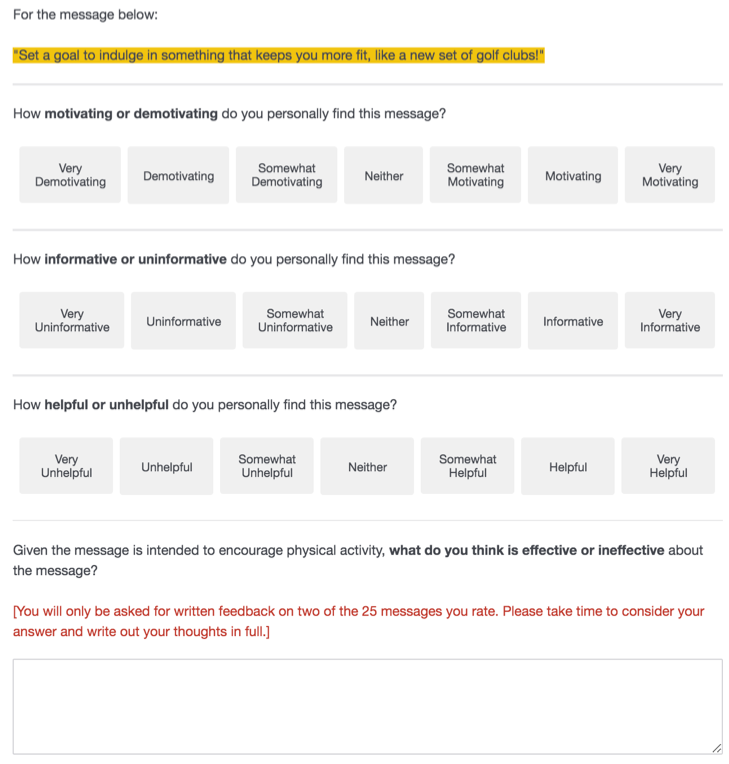}
    \caption{Validators rated a randomly selected message on a 7-point Likert scale and gave a justification.}
    \label{fig:Directed21}
\end{figure}
\newpage

\begin{figure}[H]
    \centering
    \includegraphics[width=1\textwidth]{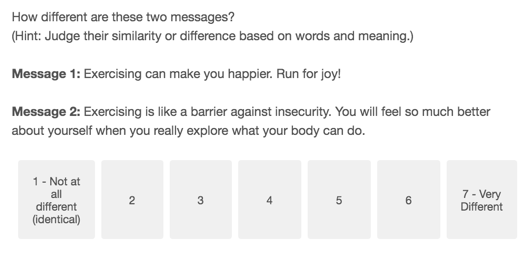}
    \caption{Validators were asked to rate the difference between two messages in a message-pair.}
    \label{fig:Directed24}
\end{figure}

\begin{figure}[H]
    \centering
    \includegraphics[width=1\textwidth]{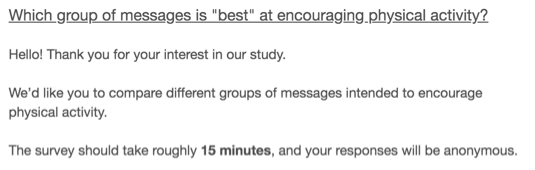}
    \caption{The instructions for group message ranking tasks.}
    \label{fig:Directed22}
\end{figure}

\newpage

\begin{figure}[H]
    \centering
    \includegraphics[width=1\textwidth]{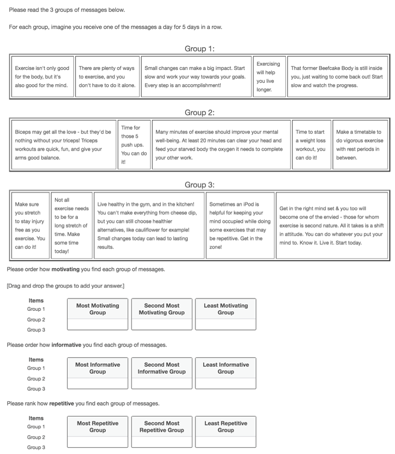}
    \caption{Validators were asked to rank groups of messages for motivation, informativeness and repetitiveness. Note that while we used the word ``repetitive'' for usability in the survey, we analysed this dependent variable as ``unrepetitive'' to be consistent with other diversity metrics.}
    \label{fig:Directed23}
\end{figure}

\newpage
\section{Examples of Prompts Shown to and Messages Written by Ideators}

\begin{figure}[H]
    \centering
    \includegraphics[width=1\textwidth]{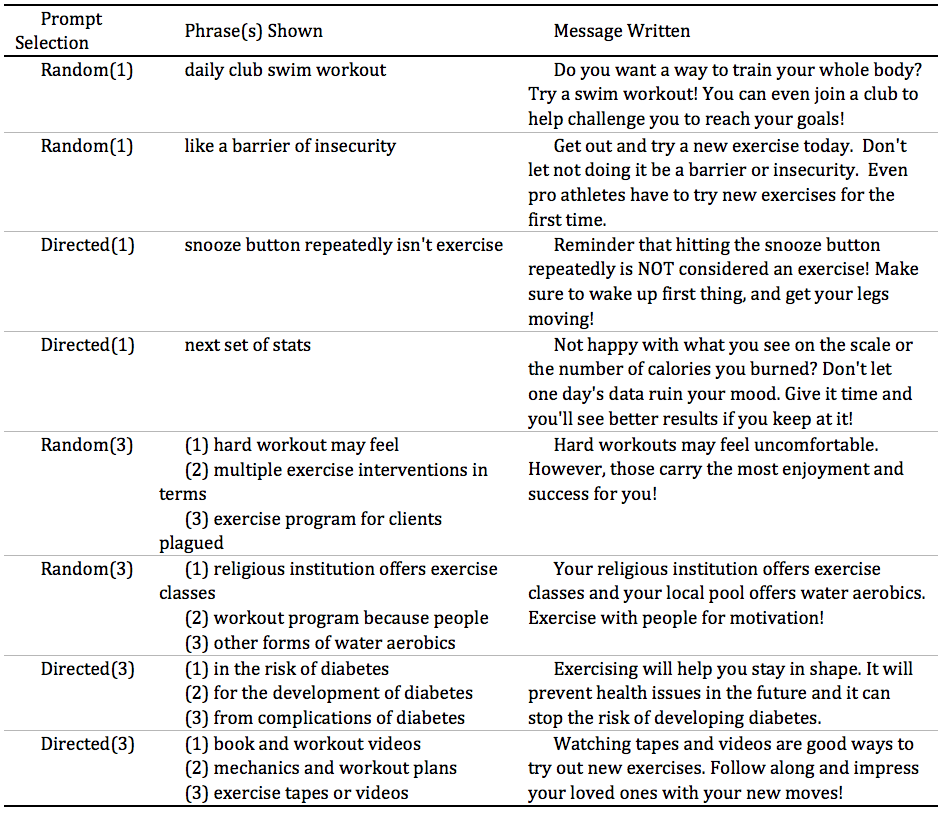}
    \caption{Messages generated in our study and the phrase prompt(s) that were shown to ideators.}
    \label{tab:Directed-Tab18}
\end{figure}

\newpage

\section{Linear Mixed Models and statistical analysis results of Prompt Creativity, Prompt-Ideation Mediation, and Ideation Diversity}

\begin{figure}[H]
    \centering
    \includegraphics[width=1\textwidth]{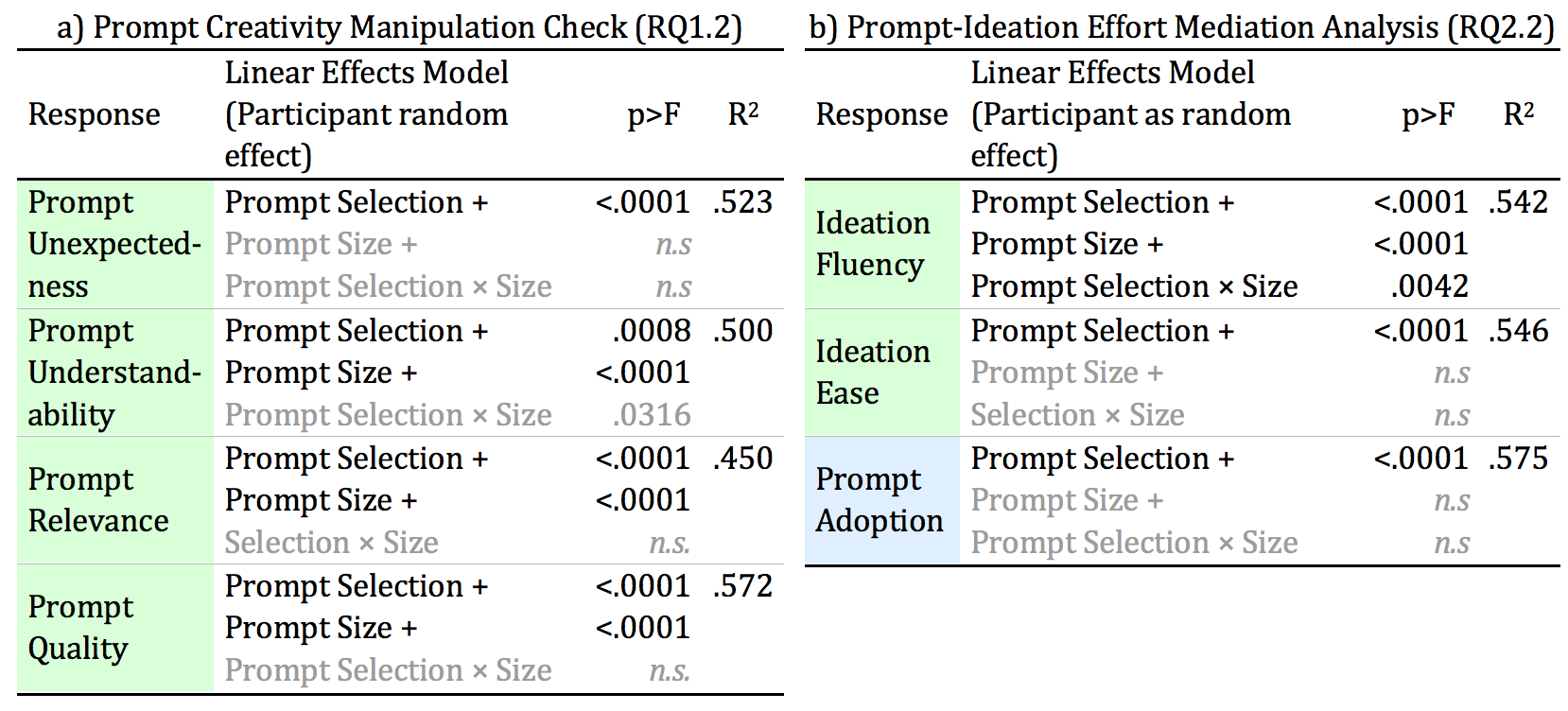}
    \caption{Statistical analysis of responses due to effects (one per row), as linear mixed eﬀects models, all with Participant as random eﬀect, Prompt Selection and Prompt Size as fixed effects, their interaction effect. a) model for manipulation check analysis of how prompt configurations affect perceived prompt creativity (RQ3.1.2); b) model for mediation analysis of how prompt configurations affect ideation effort (RQ3.2.2). \textit{n.s.} means not significant at $p>.01$. $p>F$ is the significance level of the fixed effect ANOVA. $R^2$ is the model’s coefficient of determination to indicate goodness of fit.}
    \label{tab:Directed-Tab20}
\end{figure}

\begin{figure}[H]
    \centering
    \includegraphics[width=1\textwidth]{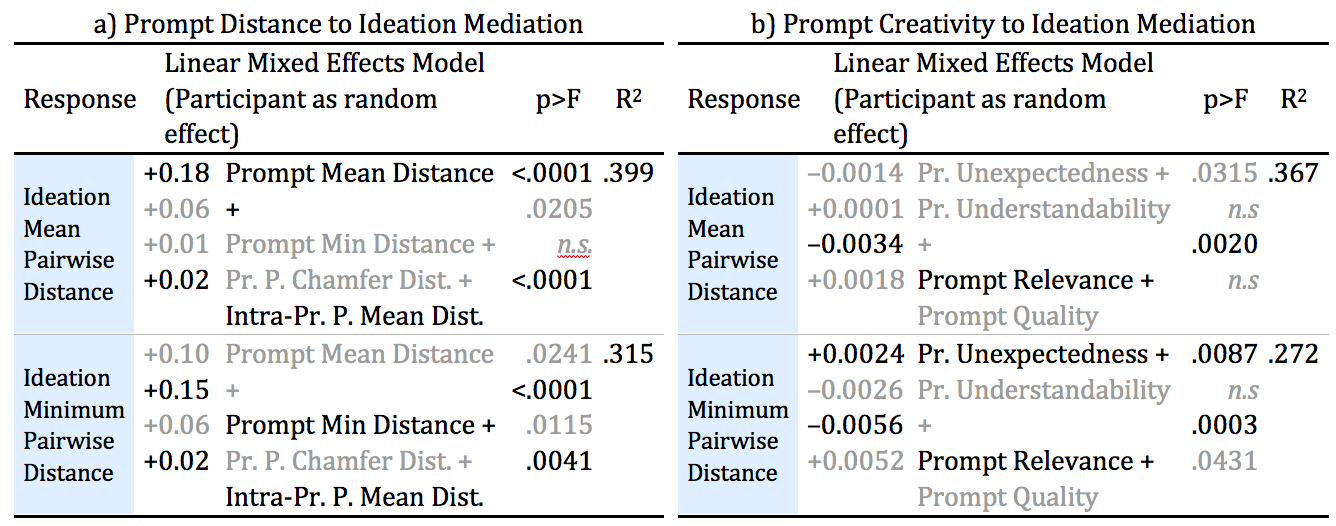}
    \caption{Statistical analysis and results of mediation effects (RQ3.2.3) of how prompt configurations (a) and perceived prompt creativity (b) affect ideation diversity. See Table \ref{tab:Directed-Tab20} caption to interpret tables. Positive and negative numbers in second column represent estimated model coefficients indicating how much each fixed effect influences the response.}
    \label{tab:Directed-Tab21}
\end{figure}

\begin{figure}[H]
    \centering
    \includegraphics[width=1\textwidth]{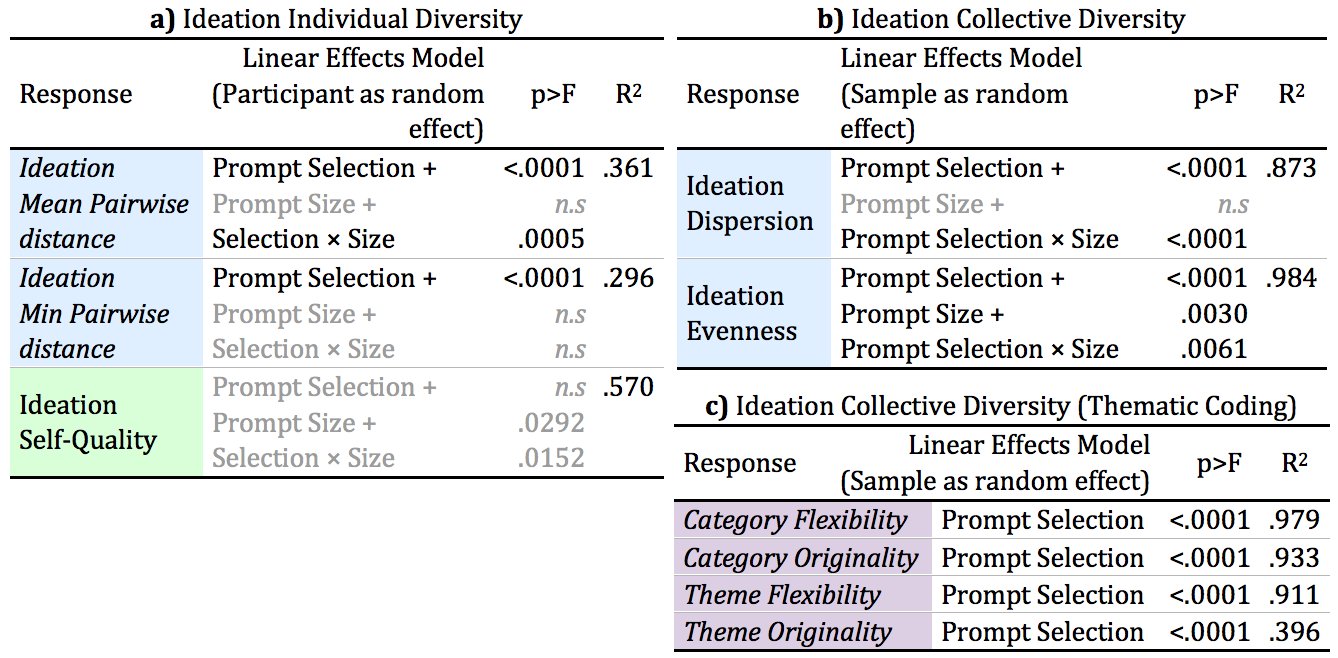}
    \caption{Statistical analysis of how prompt selection influences ideation diversity defined by different metrics (RQ3.3): a) individual diversity, b) collective diversity, and c) thematic diversity. See Table \ref{tab:Directed-Tab20} caption for how to interpret tables.}
    \label{tab:Directed-Tab22}
\end{figure}

\begin{figure}[H]
    \centering
    \includegraphics[width=1\textwidth]{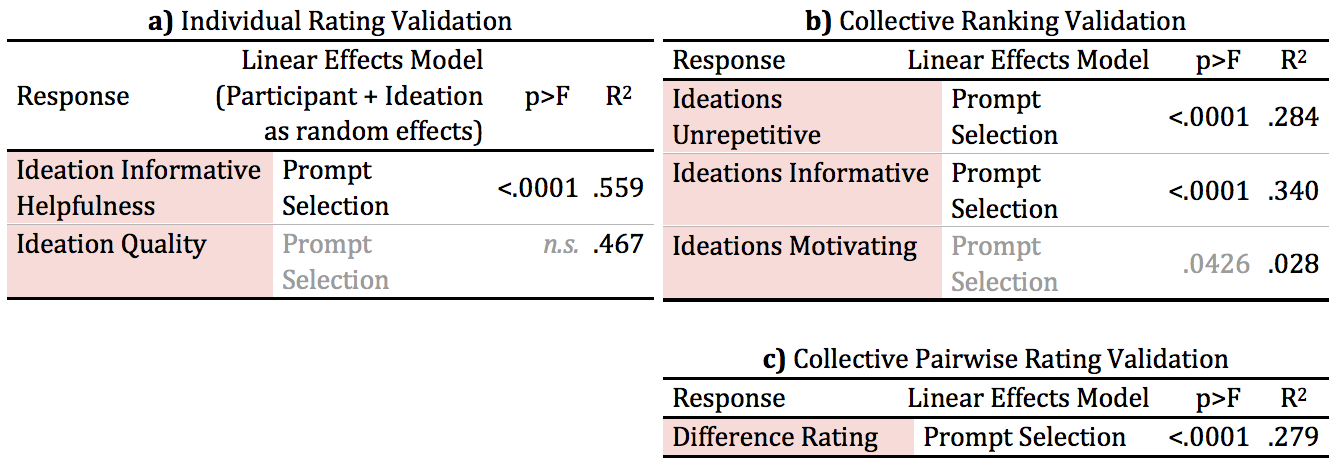}
    \caption{Statistical analysis of how prompt selection influences ideation creativity as validated by different methods (RQ3.3.1): a) individual rating, b) collective ranking, and c) collective pairwise rating. See Table \ref{tab:Directed-Tab20} caption for how to interpret tables.}
    \label{tab:Directed-Tab23}
\end{figure}

\newpage
\section{Investigating Confound of Prompt Understandability on Ideation Diversity}
\label{appendix:understandability}

Having found that prompt understanding difficulty is correlated with ideation diversity, we investigated the alternative hypothesis that the difficulty of interpreting the prompts was a key reason for improved ideation because of increased ideation determination, rather than the content diversity in phrases due to the prompt selection technique. We argue that the increase in ideation diversity due to Directed Diversity is evidenced by increased perceived diversity ratings from validators and the higher number of idea categories from the thematic analysis. This shows that Directed Diversity did stimulate more diverse ideas due to some knowledge transfer from prompt to ideations, albeit with difficulty. We identify three more sources of evidence next.

First, we qualitatively analyzed ideation rationales and found that while prompts could be rated hard to understand or irrelevant, participants still adopted some ideas. Ideators cherry-picked parts that were usable or conceived tangential ideas: e.g., P1 read ``\textit{orthopaedic surgeons and exercise specialists}'' and decided to ``\textit{cut out the bit about surgeons... I focused on the idea of specialists...}''; P2 read ``\textit{ballistic stretch uses vigorous momentum}'', commented that ``\textit{this isn't a phrase that I'm familiar with}'', yet could write about stretching: ``\textit{Stretch, breathe, and feel mindful.}''

\begin{figure}[H]
    \centering
    \includegraphics[width=1\textwidth]{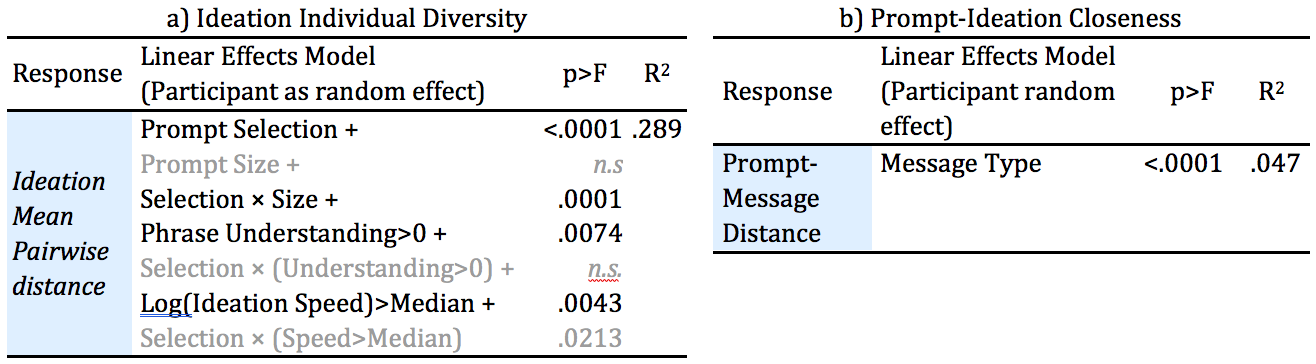}
    \caption{Statistical analysis of a) how ideators’ understanding of phrases influences ideation diversity and b) how similar Directed ideations are to their prompts compared to other None and Random Messages.}
    \label{tab:Directed-Tab24}
\end{figure}

\begin{figure}[H]
    \centering
    \includegraphics[width=1\textwidth]{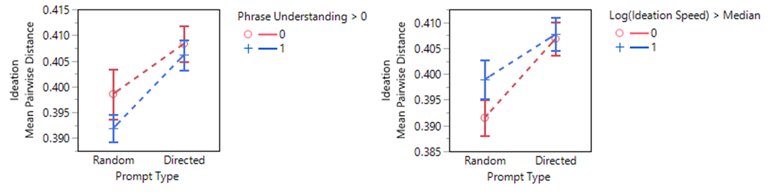}
    \caption{Results of computed individual diversity from ideations for different prompt configurations for (left) prompts that users understood ($>0$) or did not and (right) ideations that were fast or slow.}
    \label{fig:Directed25}
\end{figure}

Second, we quantitatively analyzed the Ideation Mean Pairwise Distance for prompts that participants understood (Phrase Understanding factor $>$ 0). Table \ref{tab:Directed-Tab24}a describes the statistical analysis of the linear mixed effects model. We found that although distance was slightly higher (i.e., less diverse) when ideators understood phrases less, regardless of understanding, ideations from Directed(1) prompts had higher distances than ideations from Random(1) prompts (Figure \ref{fig:Directed25}, left).  The effect due to Prompt Type was larger than due to Phrase Understanding. Furthermore, we analyzed whether the difficulty to understand may manifest as slower ideation speed due to more thinking time to ideate to lead to better diversity, but did not find a correlation between phrase understanding and ideation speed ($\rho=.046,p=n.s.$), and found the opposite effect that slower ideations led to lower distances (Table \ref{tab:Directed-Tab24}a and Figure \ref{fig:Directed25} right). These suggest that prompt selection is a primary factor.

\begin{figure}[H]
    \centering
    \includegraphics[width=0.4\textwidth]{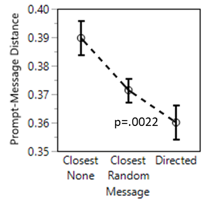}
    \caption{Results of prompt-message distance (how dissimilar a prompt is from a message) comparing different messages with respect to Directed(3) prompts.}
    \label{fig:Directed26}
\end{figure}

Third, we investigated if Directed Diversity helped to stimulate ideas closer to prompts than would be done naturally without prompts (None) or accidentally with Random prompts. We analyzed this by calculating the prompt-ideation distance between Directed prompts and their corresponding ideated message, and their closest None and Random messages. Table \ref{tab:Directed-Tab24}b describes the statistical analysis of the linear mixed effects model. Figure \ref{fig:Directed26} shows that the directed ideations were closest to the prompts, indicating the efficacy of Directed Diversity to transfer knowledge for ideation diversity.

\newpage
\section{Examples of Message-Group Ranking}

The factors of message-group ranking were derived from the sum of rankings (for each of the three condition) per validator for his five ranking trials (see the factor loadings in Table \ref{tab:Directed-Tab15}). Therefore, these factors reflect the probability of how a validator ranked the message-groups of each condition. The following table shows examples of the factors and the corresponding message-group samples.

\begin{figure}[H]
    \centering
    \includegraphics[width=1\textwidth]{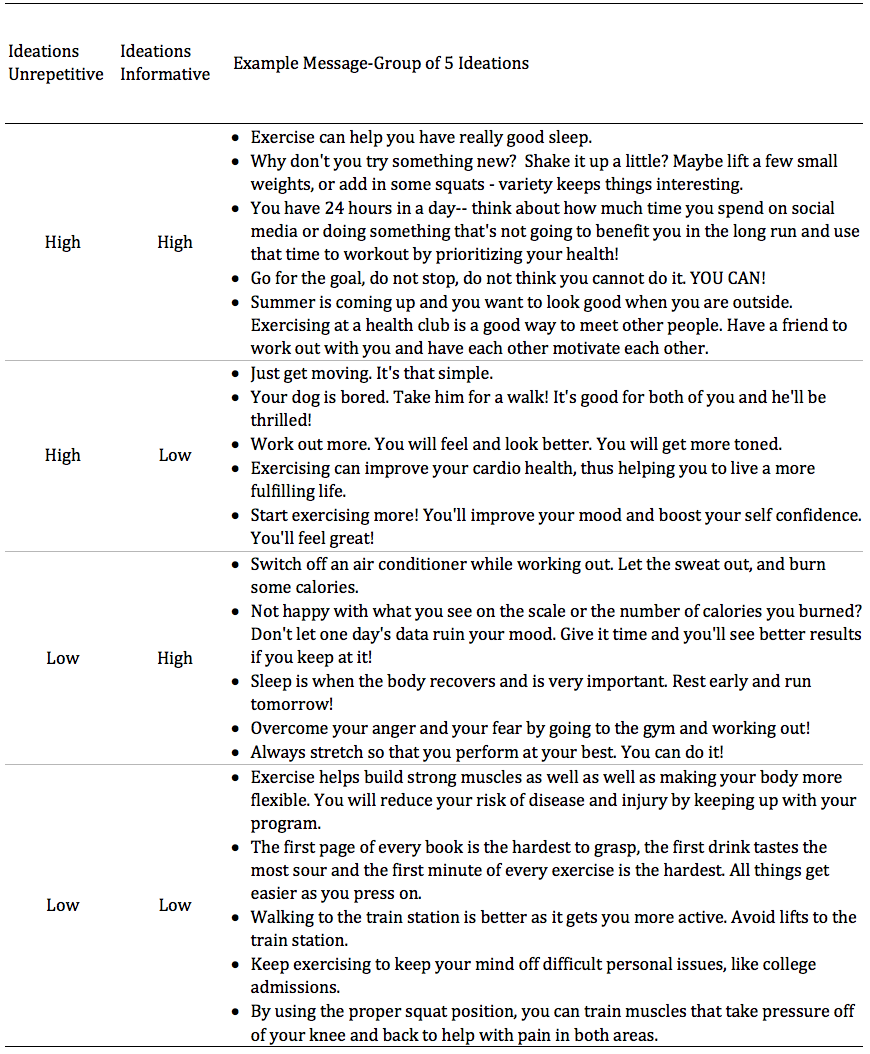}
    \caption{Examples of the factors with low $(<$ Median) and high ($\geq$ Median) scores for ``Ideations Unrepetitive'' and ``Ideations Informative''.}
    \label{tab:Directed-Tab25}
\end{figure}


\chapter{Chapter \ref{ch:Personalisation} Appendices}
\label{ch:appendix-referencing_utterances}

Below are the appendices for \textit{Chapter \ref{ch:Personalisation}: The Effect of How a Health Chatbot Formats and References a User's Previous Utterances}.

\section{Plots of weekly measures}

\begin{figure}[H]
    \centering
    \includegraphics[width=1\textwidth]{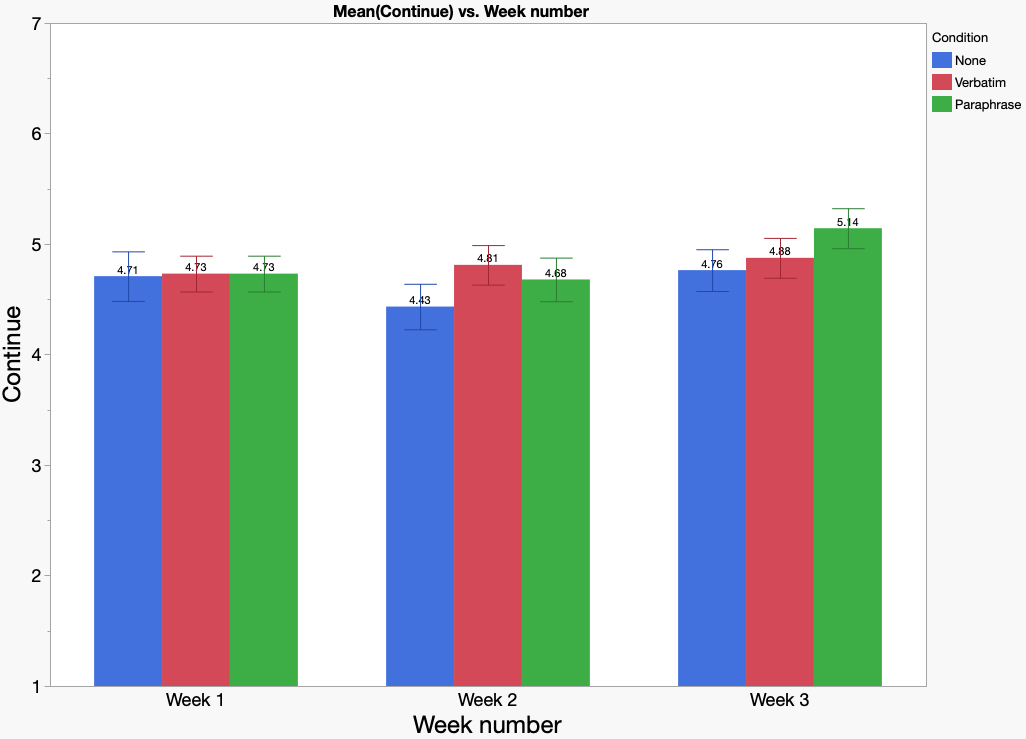}
    \caption{\textbf{Desire to continue measures} for weeks 1 to 3 by condition}
    \label{fig:WeeklyContinue}
\end{figure}

\begin{figure}[H]
    \centering
    \includegraphics[width=0.9\textwidth]{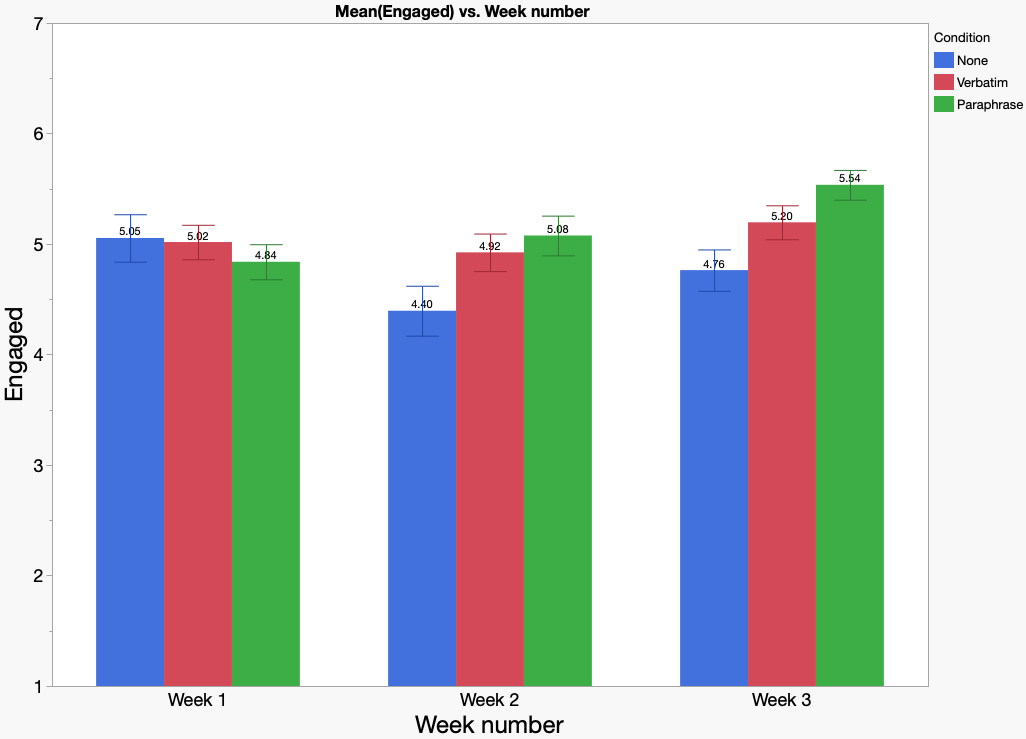}
    \caption{\textbf{Positive engagement measures} for weeks 1 to 3 by condition}
    \label{fig:WeeklyEngaged}
\end{figure}

\begin{figure}[H]
    \centering
    \includegraphics[width=0.9\textwidth]{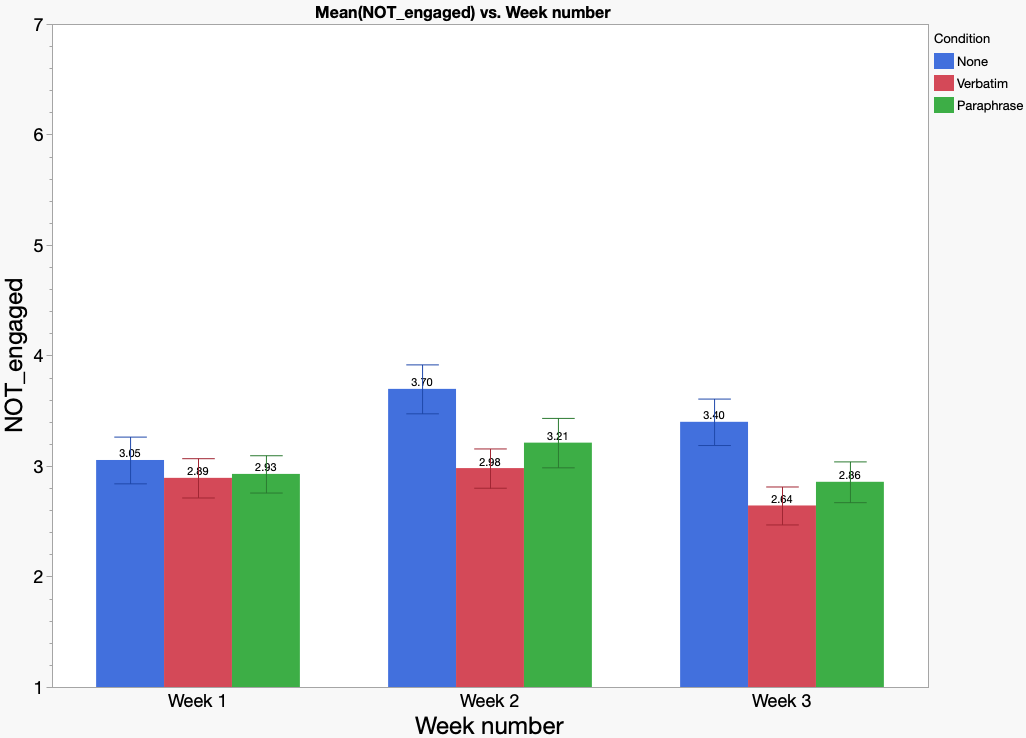}
    \caption{\textbf{Negative engagement measures} for weeks 1 to 3 by condition}
    \label{fig:WeeklyNOTEngaged}
\end{figure}

\begin{figure}[H]
    \centering
    \includegraphics[width=0.9\textwidth]{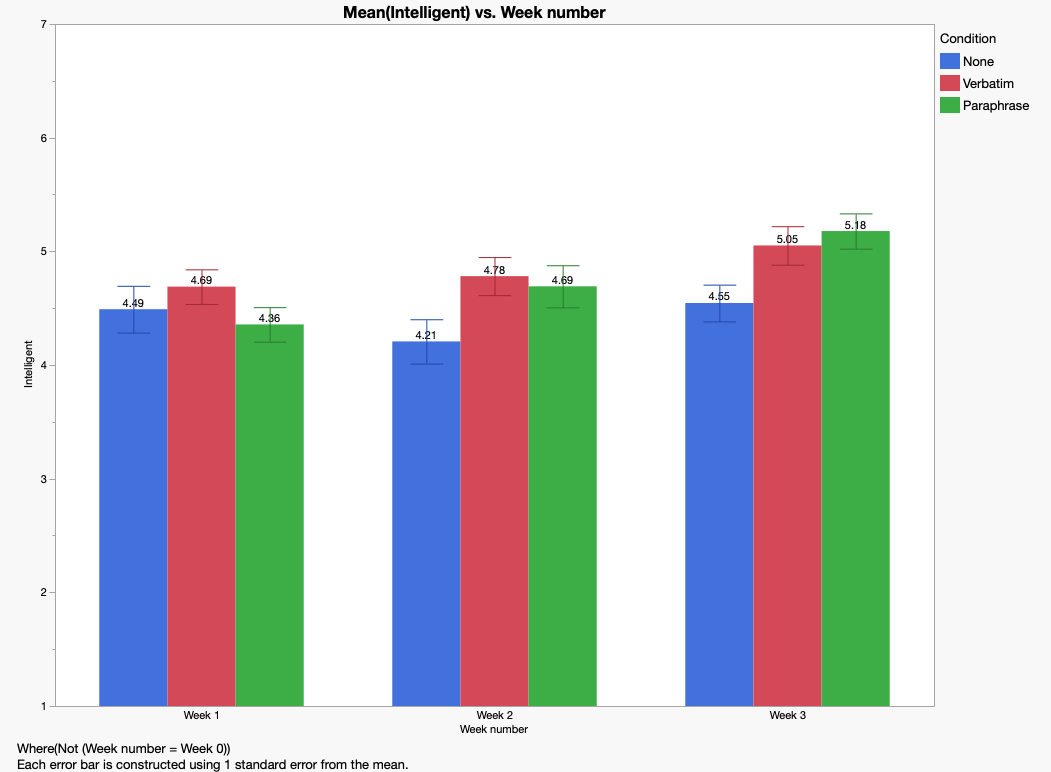}
    \caption{\textbf{Intelligence measures} for weeks 1 to 3 by condition}
    \label{fig:WeeklyIntelligence}
\end{figure}

\begin{figure}[H]
    \centering
    \includegraphics[width=0.9\textwidth]{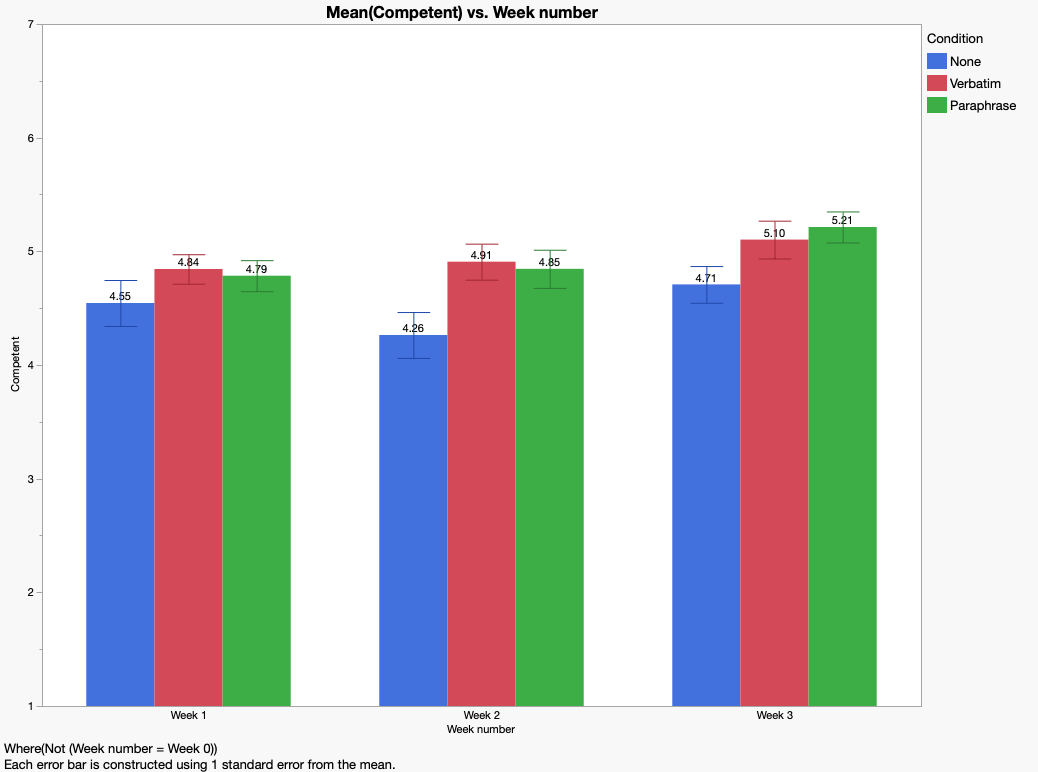}
    \caption{\textbf{Competence measures} for weeks 1 to 3 by condition}
    \label{fig:WeeklyCompetence}
\end{figure}

\begin{figure}[H]
    \centering
    \includegraphics[width=0.9\textwidth]{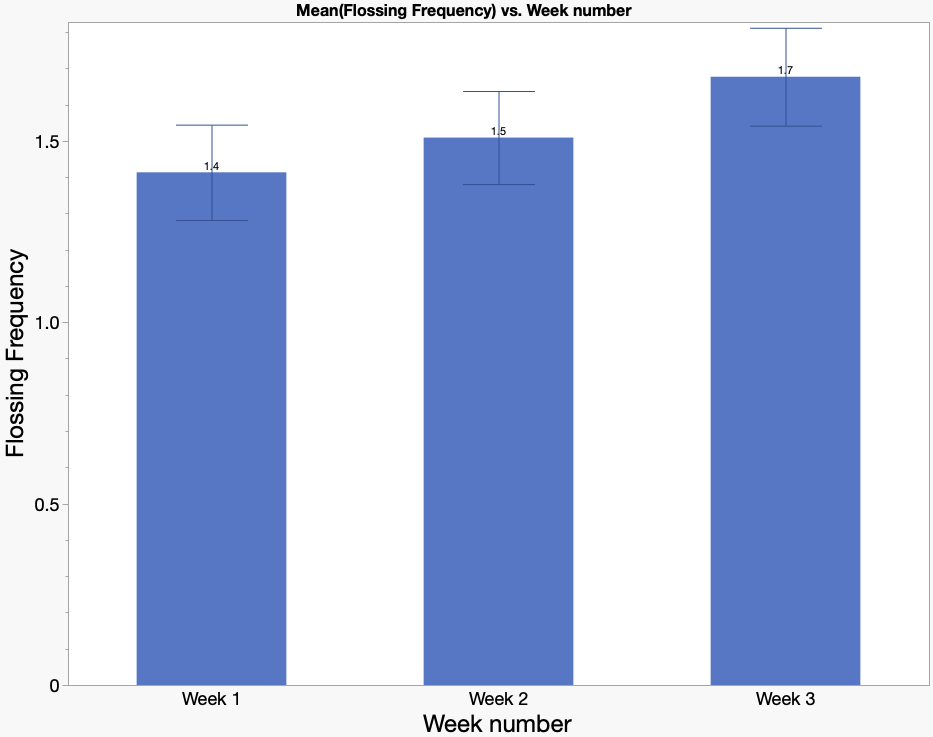}
    \caption{\textbf{Flossing Frequency} for weeks 1 to 3.}
    \label{fig:WeeklyFlossing}
\end{figure}

\begin{figure}[H]
    \centering
    \includegraphics[width=0.9\textwidth]{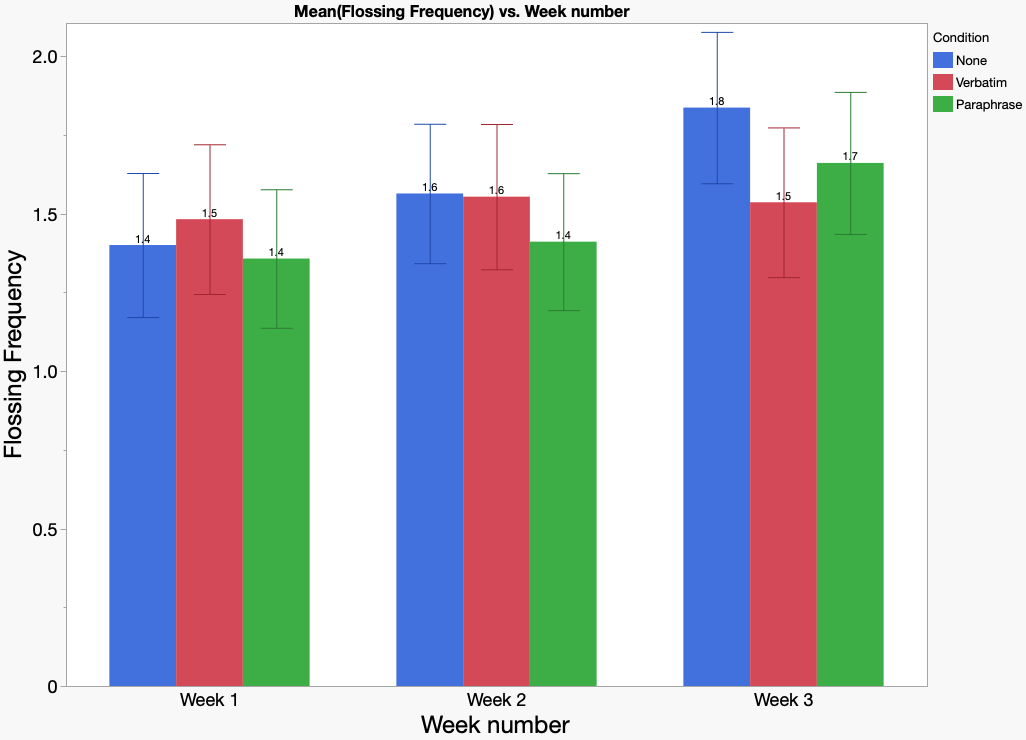}
    \caption{\textbf{Flossing Frequency} for weeks 1 to 3 by condition.}
    \label{fig:WeeklyFlossingByCondition}
\end{figure}

\newpage
\section{Health related outcomes}
\label{ch:appendix-health-outcomes}

\begin{figure}[H]
    \centering
    \includegraphics[width=0.8\textwidth]{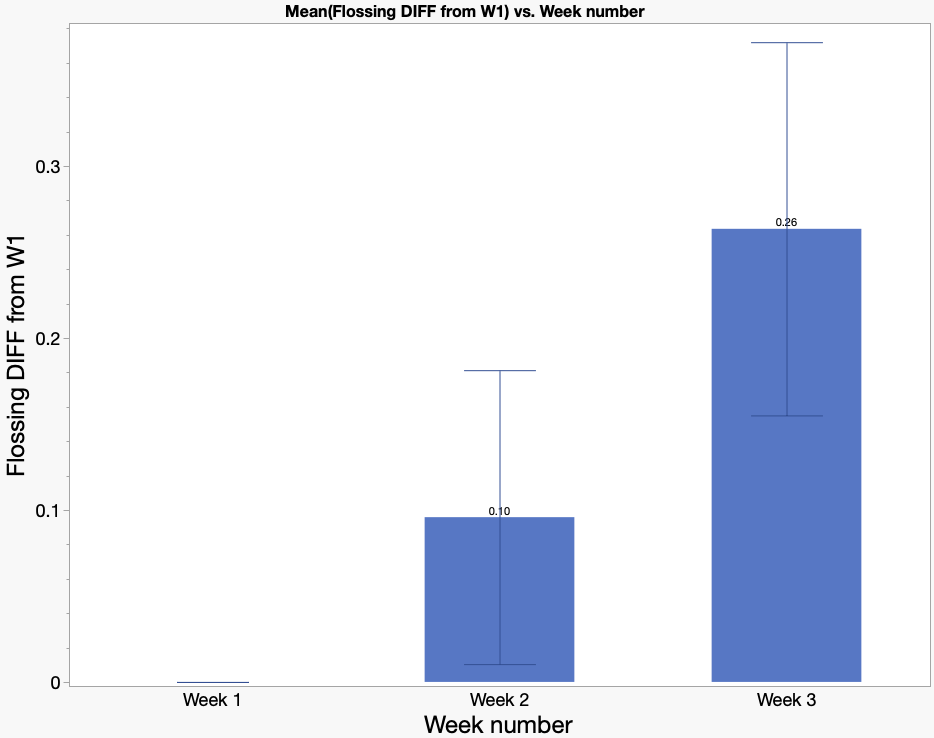}
    \caption{Weekly differences from week 1 \textbf{flossing frequency}}
    \label{fig:WeeklyFlossingDIFF}
\end{figure}

\begin{figure}[H]
    \centering
    \includegraphics[width=0.9\textwidth]{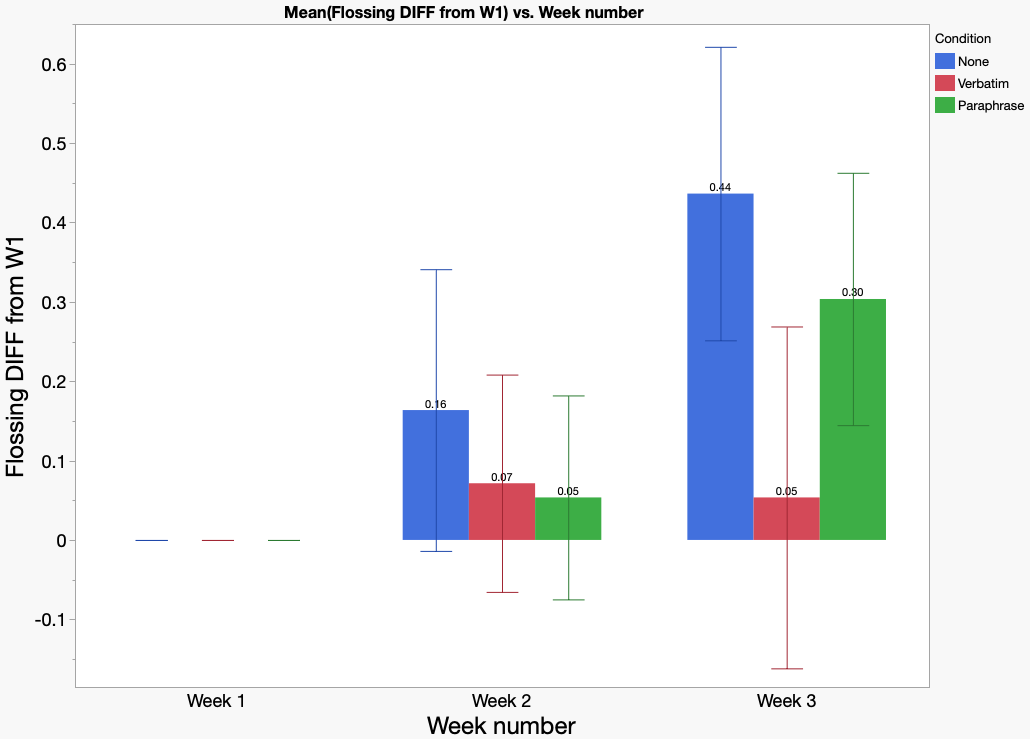}
    \caption{Weekly differences from week 1 \textbf{flossing frequency} by experiment condition}
    \label{fig:WeeklyFlossingDIFFbycondition}
\end{figure}

\begin{figure}[H]
    \centering
    \includegraphics[width=0.8\textwidth]{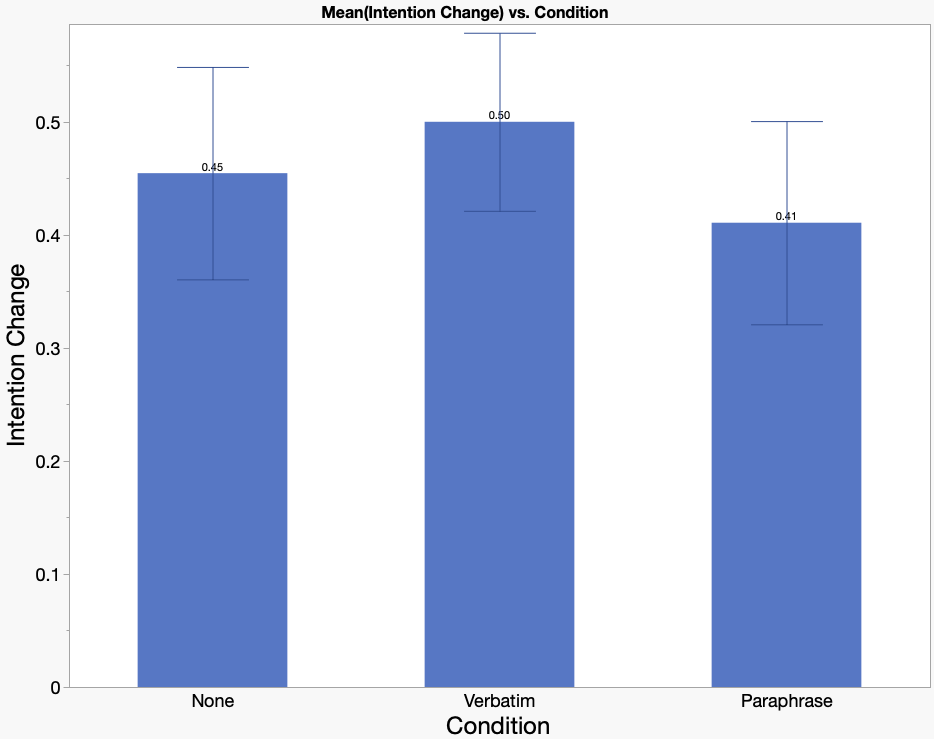}
    \caption{Change in \textbf{intention to floss}  from week 1 to week 3 by experiment condition}
    \label{fig:intentionDIFF}
\end{figure}

\begin{figure}[H]
    \centering
    \includegraphics[width=0.8\textwidth]{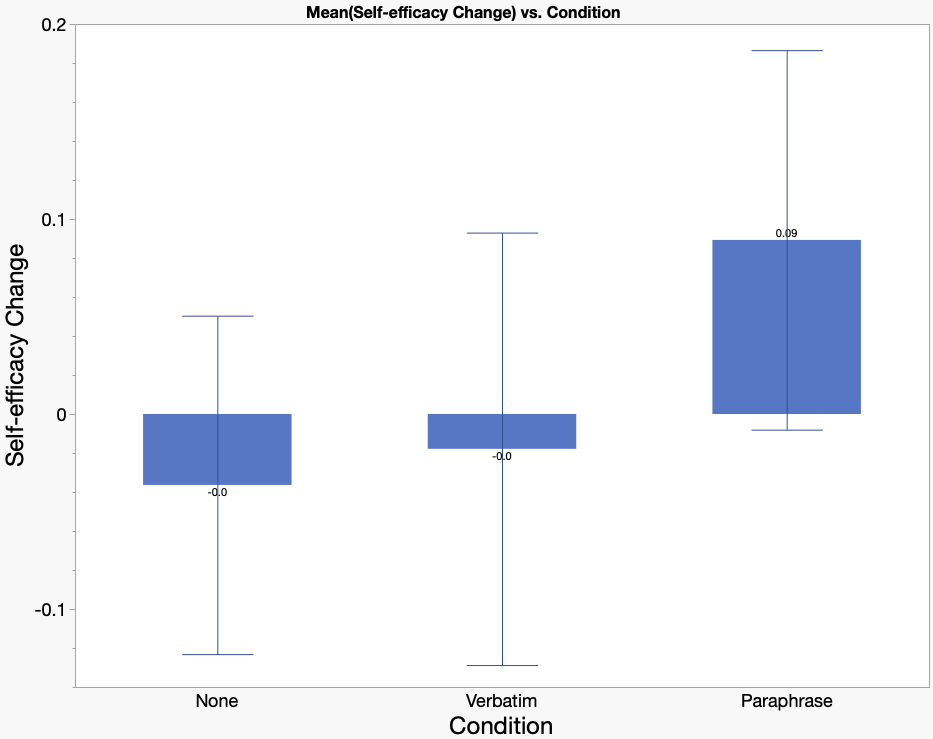}
    \caption{Change in \textbf{self-efficacy to floss} from week 1 to week 3 by experiment condition}
    \label{fig:efficacyDIFF}
\end{figure}

\begin{figure}[H]
    \centering
    \includegraphics[width=0.9\textwidth]{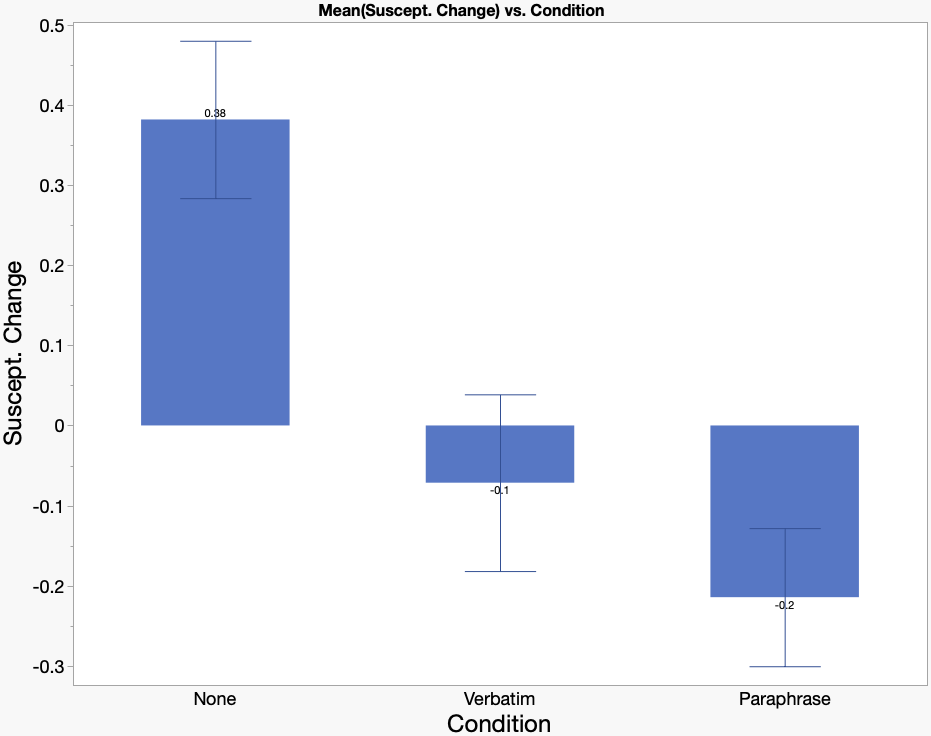}
    \caption{Change in \textbf{perceived susceptibility} from week 1 to week 3 by experiment condition}
    \label{fig:susceptDIFF}
\end{figure}

\newpage
\section{Full Chatbot Script}
\label{appendix:chatbot_script}

Below is the script of the chatbot for each of the 3 weeks of the user study. For weeks 2 and 3, text that is highlighted in yellow like this represents content which will be replaced by one of the user’s previous user utterances (either verbatim using their utterance, or using a paraphrase of their utterance).

\subsection{Week 1 Script}

\begin{figure}[H]
    \centering
    \includegraphics[width=1\textwidth]{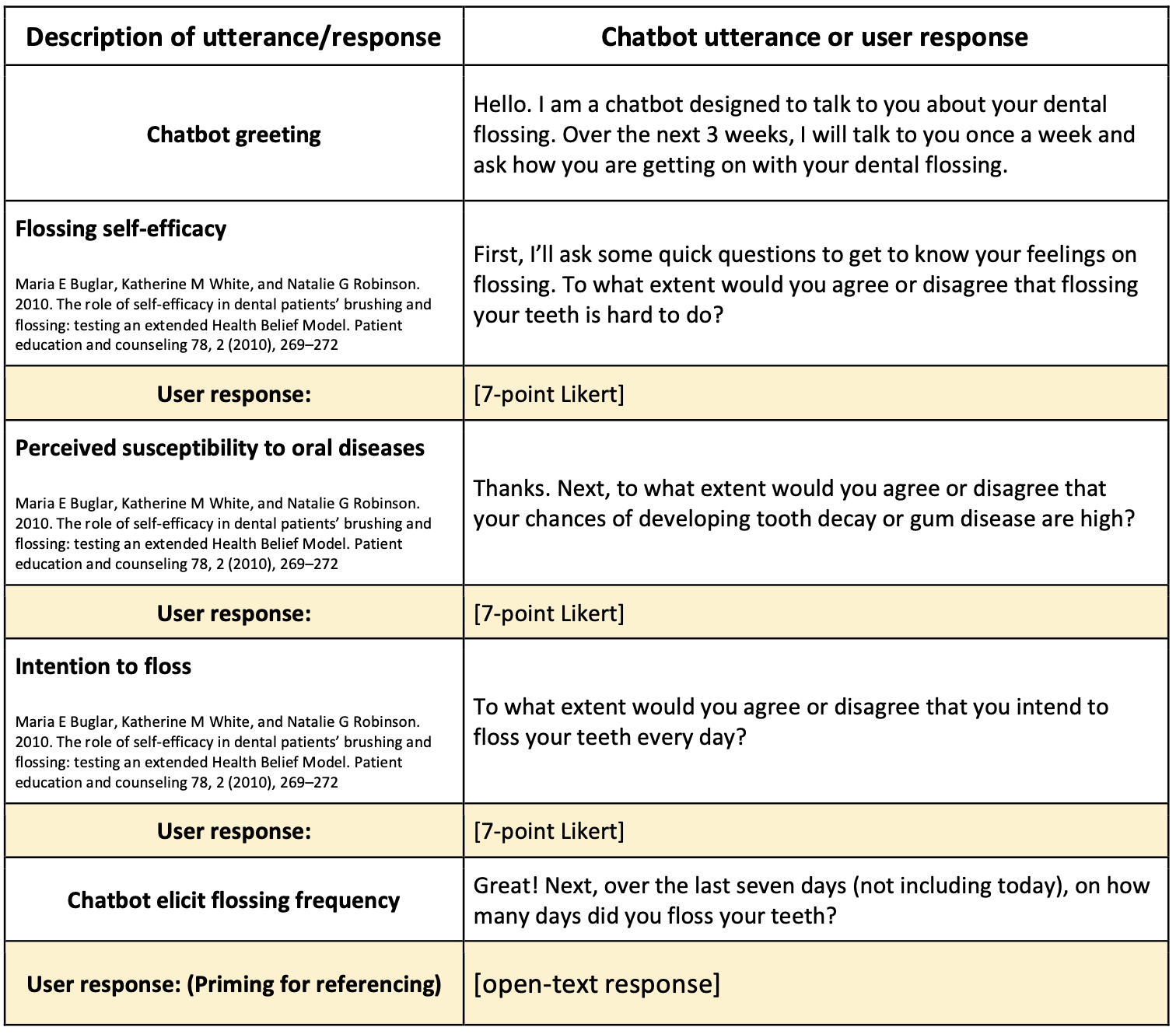}
    \label{fig:week1a}
\end{figure}
\begin{figure}[H]
    \centering
    \includegraphics[width=1\textwidth]{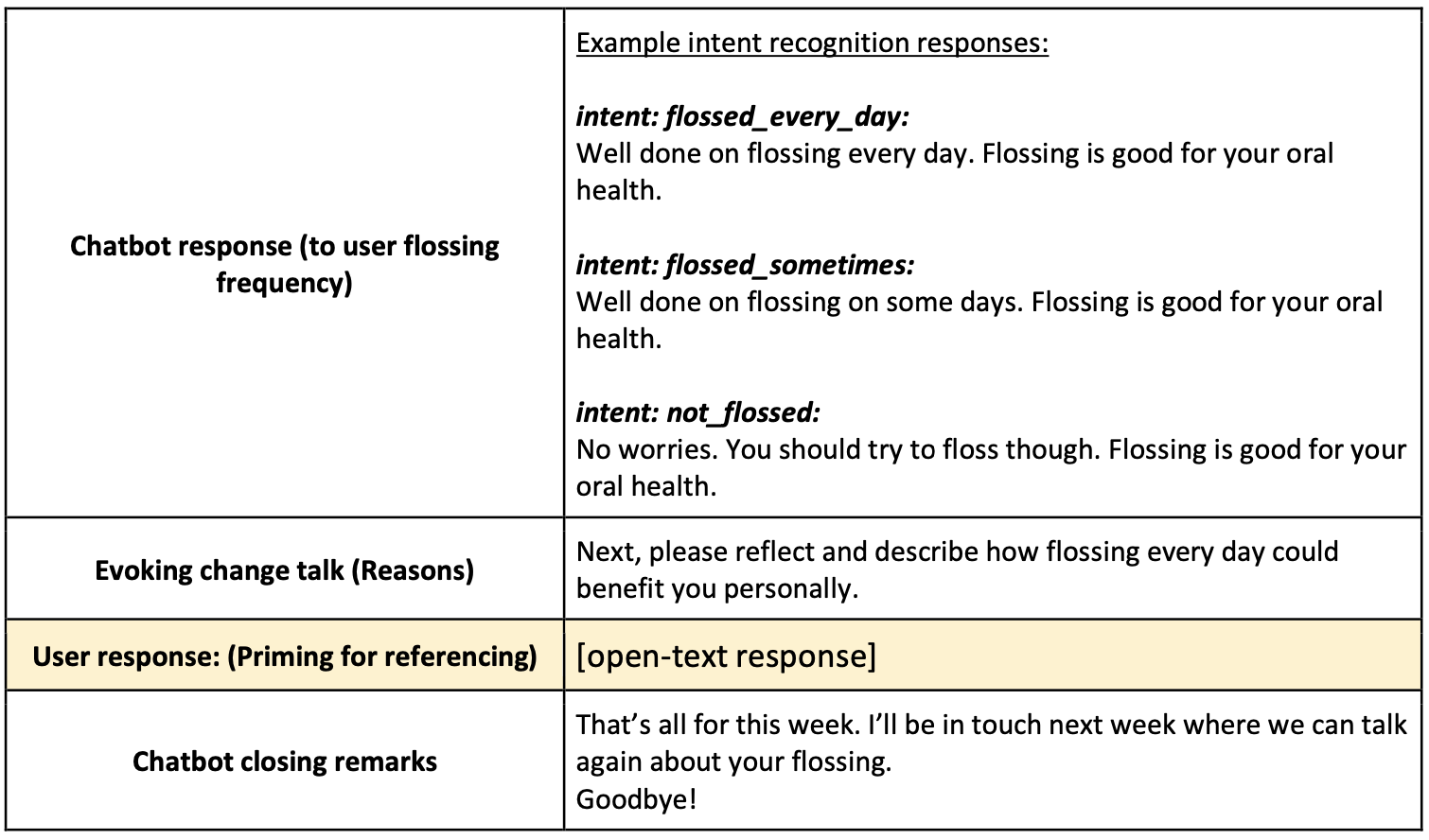}
    \label{fig:week1b}
\end{figure}

\subsection{Week 2 Script}

\begin{figure}[H]
    \centering
    \includegraphics[width=1\textwidth]{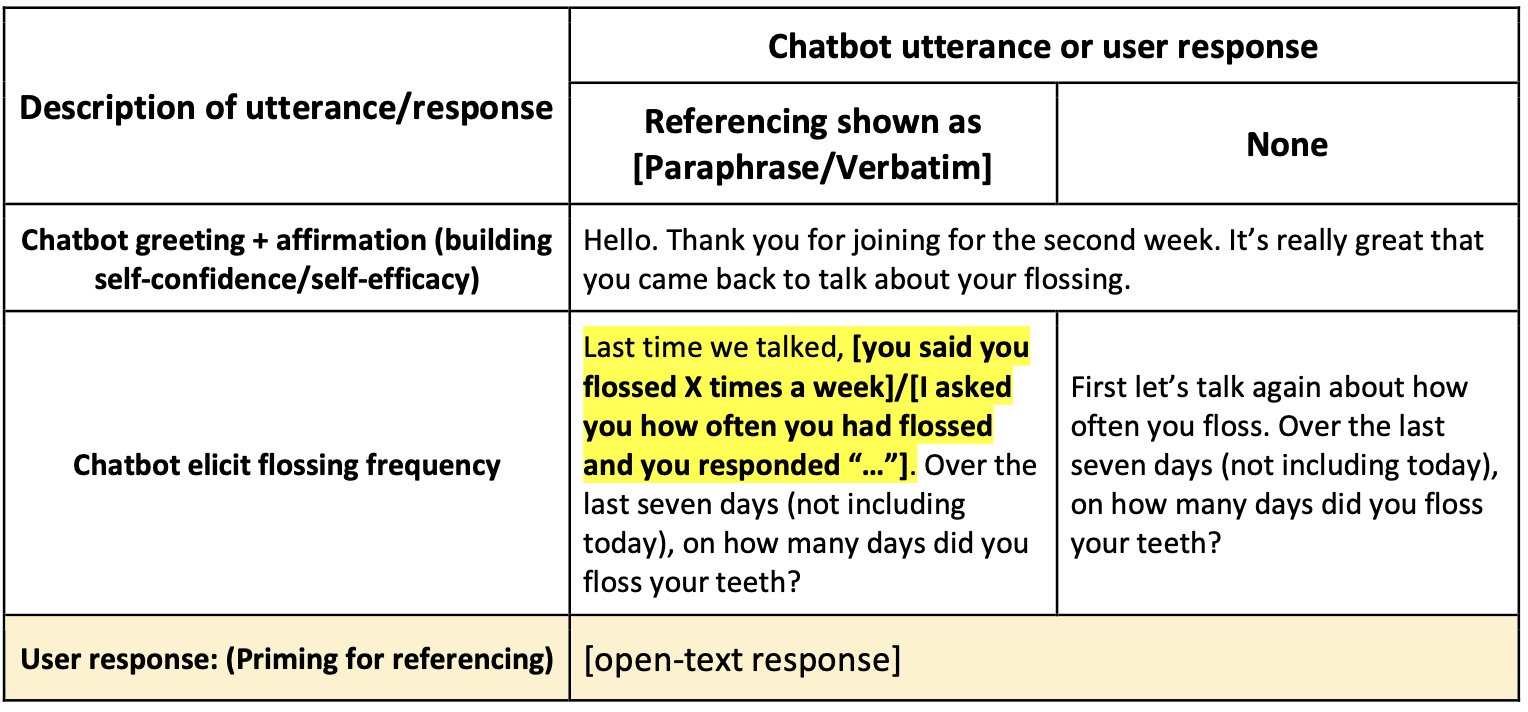}
    \label{fig:week2a}
\end{figure}
\begin{figure}[H]
    \centering
    \includegraphics[width=1\textwidth]{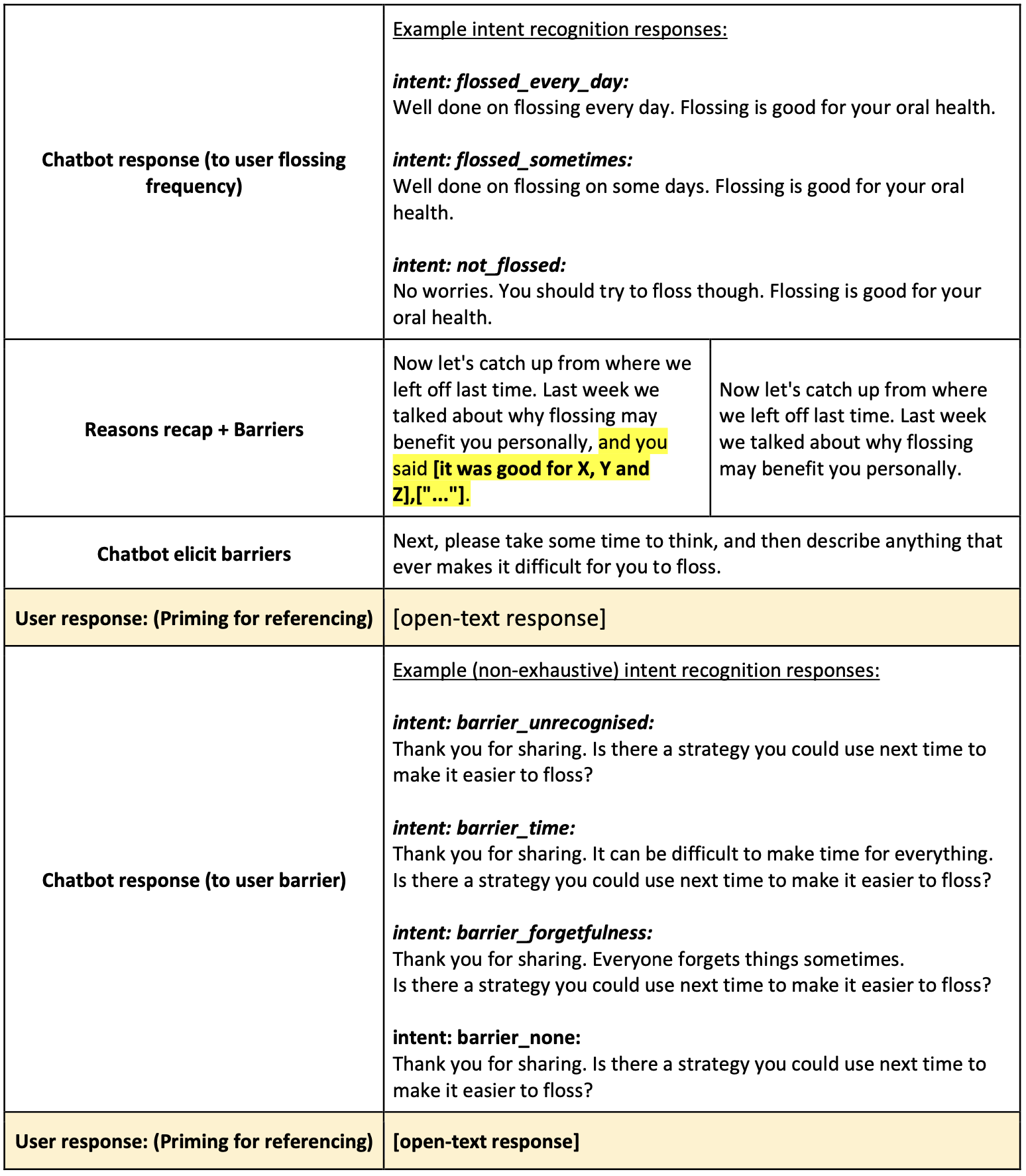}
    \label{fig:week2b}
\end{figure}
\begin{figure}[H]
    \centering
    \includegraphics[width=1\textwidth]{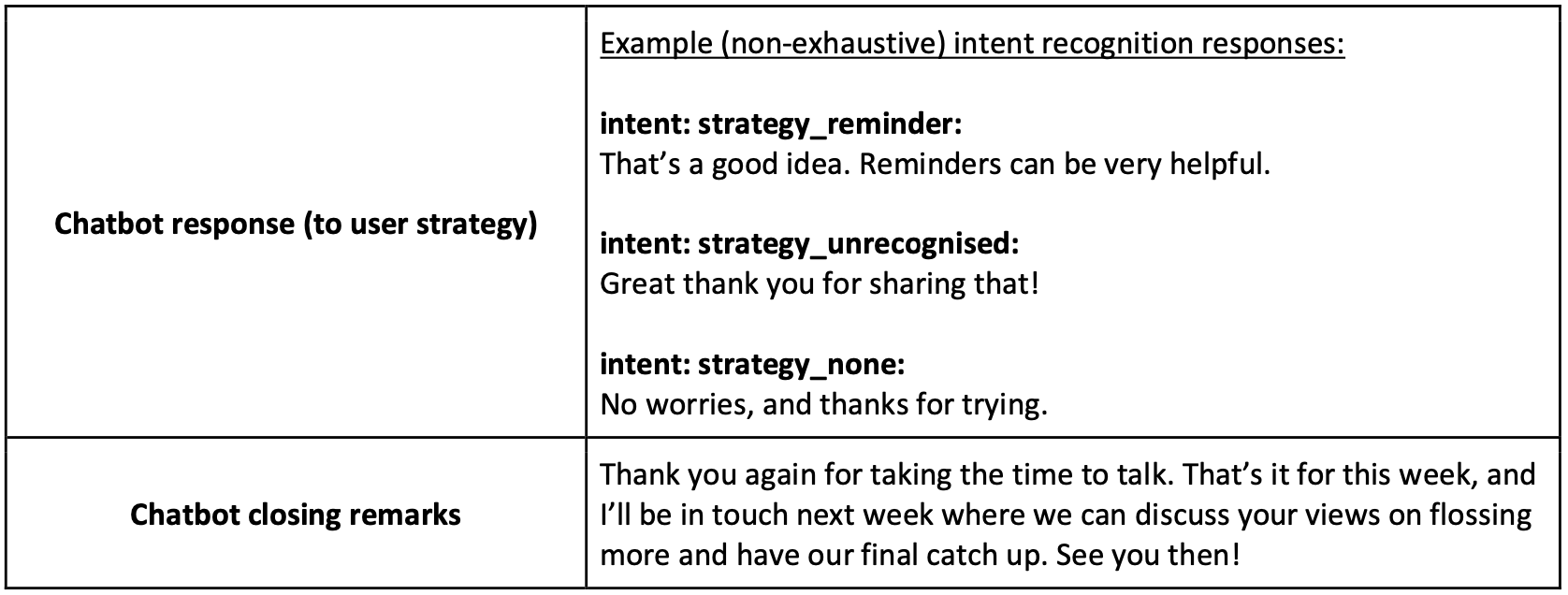}
    \label{fig:week2c}
\end{figure}

\subsection{Week 3 Script}

\begin{figure}[H]
    \centering
    \includegraphics[width=1\textwidth]{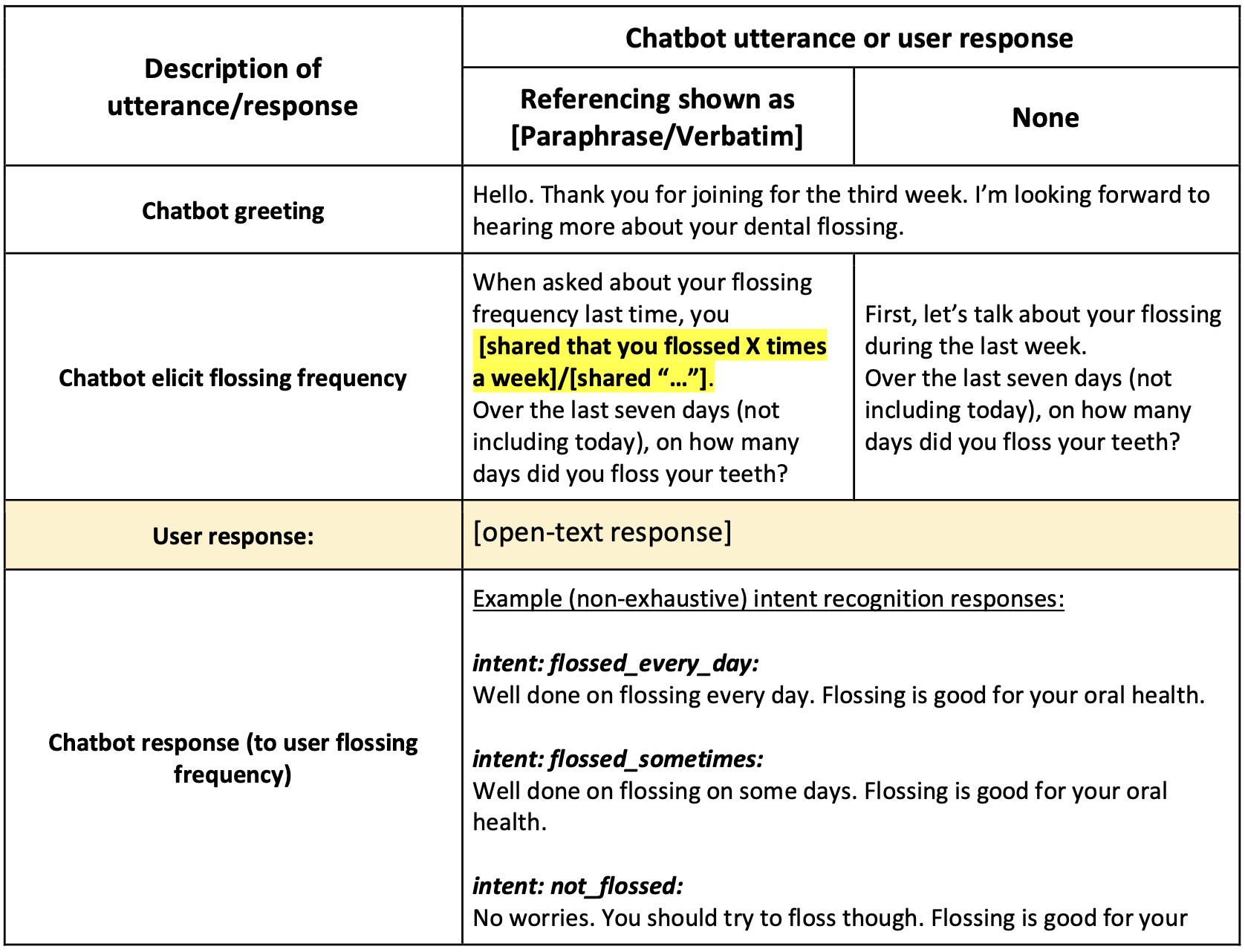}
    \label{fig:week3a}
\end{figure}
\begin{figure}[H]
    \centering
    \includegraphics[width=1\textwidth]{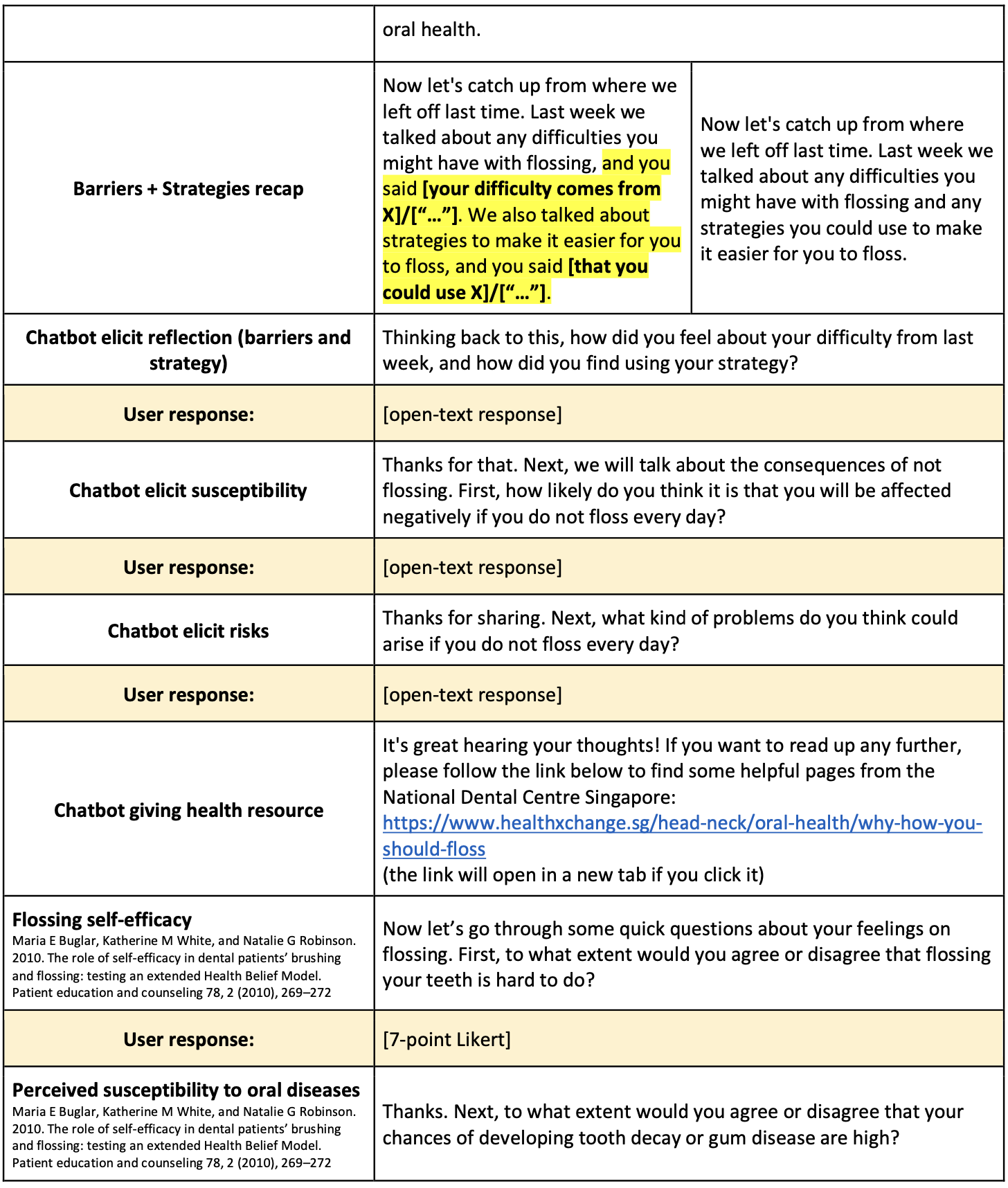}
    \label{fig:week3b}
\end{figure}
\begin{figure}[H]
    \centering
    \includegraphics[width=1\textwidth]{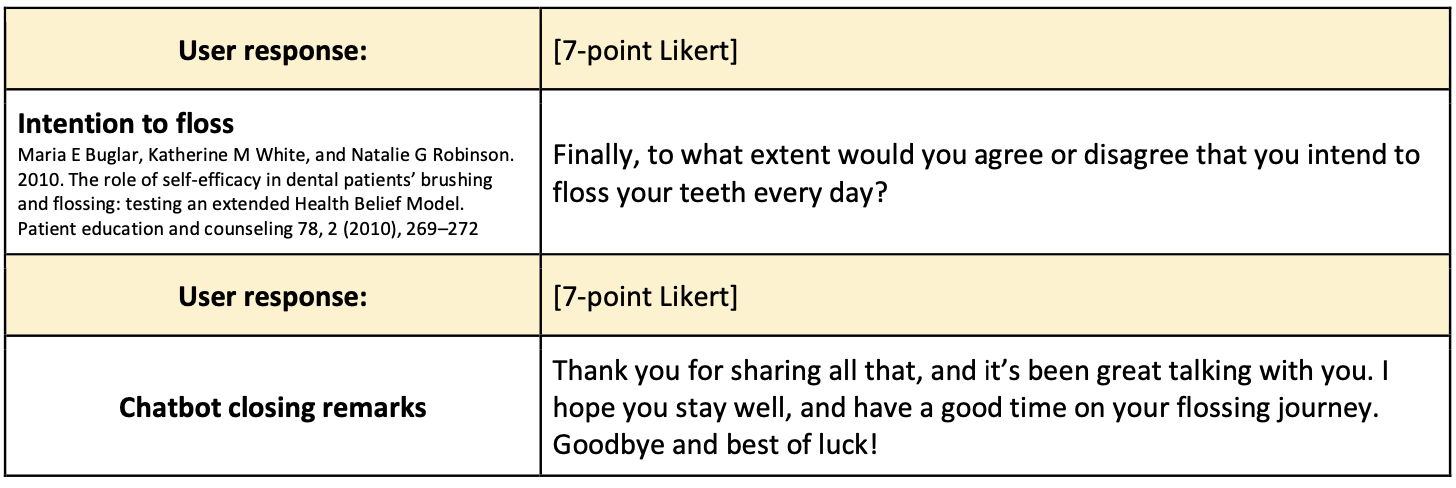}
    \label{fig:week3c}
\end{figure}

\section{Paraphrases Used}
\label{appendix:paraphrases_used}

Below are the paraphrases shown to Paraphrase condition participants in weeks 2 and 3.

\subsection{Paraphrases given in Week 2:}

\subsubsection{Flossing frequency paraphrases:}

\begin{itemize}
    \item you said you flossed 0 days a week
    \item you said you flossed 1 day a week
    \item you said you flossed 2 days a week
    \item you said you flossed 3 days a week
    \item you said you flossed 4 days a week
    \item you said you flossed 5 days a week
    \item you said you flossed 6 days a week
    \item you said you flossed 7 days a week
\end{itemize}

\subsubsection{Flossing benefits paraphrases:}

\begin{itemize}
    \item flossing helps you have cleaner teeth 
    \item flossing helps you feel good
    \item flossing helps you build habits
    \item flossing helps improve your appearance 
    \item flossing helps improve your breath 
    \item flossing helps improve gum health 
    \item flossing helps improve oral hygiene
    \item flossing has no benefits
    \item flossing helps prevent gum disease
    \item flossing helps prevent tooth ache
    \item flossing helps prevent tooth decay
    \item flossing helps remove bacteria from your mouth 
    \item flossing helps remove debris stuck in your teeth
    \item flossing helps remove plaque from your teeth
\end{itemize}

\subsection{Paraphrases given in Week 3:}

\subsubsection{Flossing frequency paraphrases:}

\begin{itemize}
    \item you shared that you flossed 0 times a week
    \item you shared that you flossed 1 time a week
    \item you shared that you flossed 2 times a week
    \item you shared that you flossed 3 times a week
    \item you shared that you flossed 4 times a week
    \item you shared that you flossed 5 times a week
    \item you shared that you flossed 6 times a week
    \item you shared that you flossed 7 times a week
\end{itemize}

\subsubsection{Flossing barrier paraphrases:}

\begin{itemize}
    \item you sometimes forget to floss
    \item you sometimes find flossing techniques difficult
    \item you sometimes don't have time to floss
    \item flossing is not one of your regular habits
    \item you don't have any floss on hand
    \item you sometimes don't feel motivated to floss
    \item you sometimes don't have the energy to floss
    \item you sometimes find it too much effort to floss
    \item you could not think of any difficulties
    \item you don't think about flossing
    \item you sometimes find it feels uncomfortable to floss
    \item you don't have any good floss on hand
    \item you sometimes find it inconvenient to have flossing tools on hand
    \item you sometimes don't have any floss on hand
\end{itemize}

\subsubsection{Flossing strategies paraphrases:}

\begin{itemize}
    \item you could leave floss in a noticeable place
    \item you could use more comfortable flossing tools
    \item you could floss at a certain time
    \item you could make flossing one of your regular habits
    \item you could use reminders
    \item you could buy more floss
    \item you could do you oral routine earlier at night
    \item you could use recommended flossing techniques
    \item you could research flossing techniques
    \item you could not think of anything yet
    \item you could use social support
    \item you could floss after brushing
    \item you could wake up earlier
    \item you could practice your flossing
    \item you could free up more time to floss
    \item you could remind yourself of why flossing is important
    \item you could keep floss in your bag
\end{itemize}

\newpage
\section{Semi-Structured Interview Guide}
\label{appendix:interview_guide}

\subsection*{Introduction}

Thank you for agreeing to participate in this interview. We are interviewing you to better understand what participants thought about their conversations with the chatbot, and how we can improve chatbot interactions. There are no right or wrong answers to any of our questions, and we are interested in your own experiences.
Participation in this study is voluntary and your decision to participate, or not participate, will not affect reimbursement you receive for talking to the chatbot. The interview should take up to 30 minutes depending on how much information you would like to share. With your permission, I would like to audio record the interview because I don’t want to miss any of your comments. All responses will be kept confidential. This means that your de-identified interview responses will only be shared with research team members and we will ensure that any information we include in our report does not identify you as the respondent. You may decline to answer any question or stop the interview at any time and for any reason. Are there any questions about what I have just explained?

\noindent May I turn on the digital recorder?

\vspace{10pt}
\hrule
\vspace{10pt}

\noindent\textit{\textbf{Note for interviewer:} Please note that this guide only represents the main themes to be discussed with the participants and as such does not include the various prompts that may also be used (examples given for each question). Non-leading and general prompts will also be used, such as “Can you please tell me a little bit more about that?” and “What does that look like for you”.}

\subsection*{[Effect on dental flossing]:}

• Can you tell me how concerned you were about dental flossing before talking to the chatbot? (and why?). Did this change as you talked to the chatbot?

\noindent• Can you tell me about your flossing behaviour before, during and after your conversations with the chatbot?

\subsection*{[Perception of privacy violations]:}

• Did the chatbot say anything to make you feel uncomfortable?

\noindent• Did you feel comfortable talking about yourself to the chatbot?

\noindent• Did you ever feel that you did not want to tell the chatbot anything? If so, what made you hesitant about talking to the chatbot?

\noindent• At any point did you feel that the chatbot was violating your privacy?

\noindent• Can you tell me overall how comfortable you felt talking to the chatbot?

\subsection*{[Thoughts on level of individuation]:}

\noindent• How did you find the way the chatbot referenced the content of your previous week’s discussion?

\noindent• If you had to ideally change how the chatbot referenced your previous messages, how might you change it?
 
\subsection*{[Perception of chatbot intelligence]:}

\noindent• Can you tell me how intelligent you felt the chatbot was?

\subsection*{[Engagement]:}

\noindent• Can you describe how engaged the felt the chatbot was in your conversations?

\noindent• When you were talking to the chatbot, did you feel that it was getting to know you well?

\subsection*{[Perception of chatbot warmth (friendliness)]:}

\noindent• Can you tell me how friendly you felt the chatbot was?

\subsection*{[General Questions]:}

Would you recommend this chatbot to someone who wants to floss their teeth more? 

\noindent• Can you explain why you would or would not recommend this chatbot? Is there anything else that you would like to comment on about the chatbot that we haven’t discussed today?

\subsection*{[Comparing all 3 conditions]:}

[\textit{Share screen to show and explain the 3 different conditions (from example screenshots), and ask for user preference and comments}]

\noindent For example:

\noindent• Would you see any pros and cons of different methods and which one would you prefer most do you think?

\noindent• How would you personally want the chatbot to reference your previous messages and why?

\noindent• Please can you walk through the 3 different conditions and explain your preferences?

\vspace{10pt}

\noindent Thank you very much for your time and the information you shared today

\chapter{Chapter \ref{ch:conclusion} Appendices}
\label{ch:appendix-conclusion}

Below are the appendices for \textit{Chapter \ref{ch:conclusion}: Conclusion and Future Work}.

\section{Generating exercise plans (using instructions from Agapie et al. \cite{agapie2016plansourcing})}
\label{ch:appendix-agapie}

\subsection{Prompt used to generate an exercise plan using GPT-4. Prompt uses instructions taken from Agapie et al. \cite{agapie2016plansourcing}.}
\label{sec:agapie-prompt}

\noindent Task: 
\\1. Create a 1-week exercise plan (Monday-Sunday) that the person described below can follow next week. They are a beginner at exercising. 
\\•	Your plan should have the appropriate level of detail. Include physical activities and any details the person might need to follow the plan. 
\\•	The plan should be something the person can achieve in the next week, and that they will want to follow. 
\\•	Format your plan the way that you would want the recipient to see it. 
\\•	All that you include will be shared with the person you are creating the plan for 
\\
\\
Person description:
\\Male, 28 years, 135 Ib
\\Graduate Student
\\
\\Goals:
\\Next Week's Goal:
\\At least 3 instances of at least 30 minutes of continuous exercise.
\\
\\Long Term Goal:
\\I'd like to create a habit out of exercising, something I've never been able to do in the past.
\\
\\Constraints:
\\
\\Time constraints mostly, and just general laziness.
\\I am a University of Washington student, so I have access to the gym. I also personally have a small set of hand weights.
\\
\\Activities:
\\
\\Likes:
\\Running, aerobics, dancing, strength training - I like activities with constant motion, and I'm most interested in cardio exercise. Something I can listen to podcasts while doing is also appealing
\\
\\Interested in, but have not tried:
\\More structured strength training activities - I rarely go to a gym with access to strength training equipment. I also don't know how to use most of that equipment
\\
\\Tried, but did not enjoy:
\\Yoga - While I see the benefits of it, it did not feel like I was improving my body
while doing it
\\
\\Other preferences:
\\I'd prefer to exercise in the morning.

\newpage
\subsection{GPT-4 exercise plan output:}

\begin{figure*}[h]
    \centering
    \includegraphics[width=0.6\textwidth]{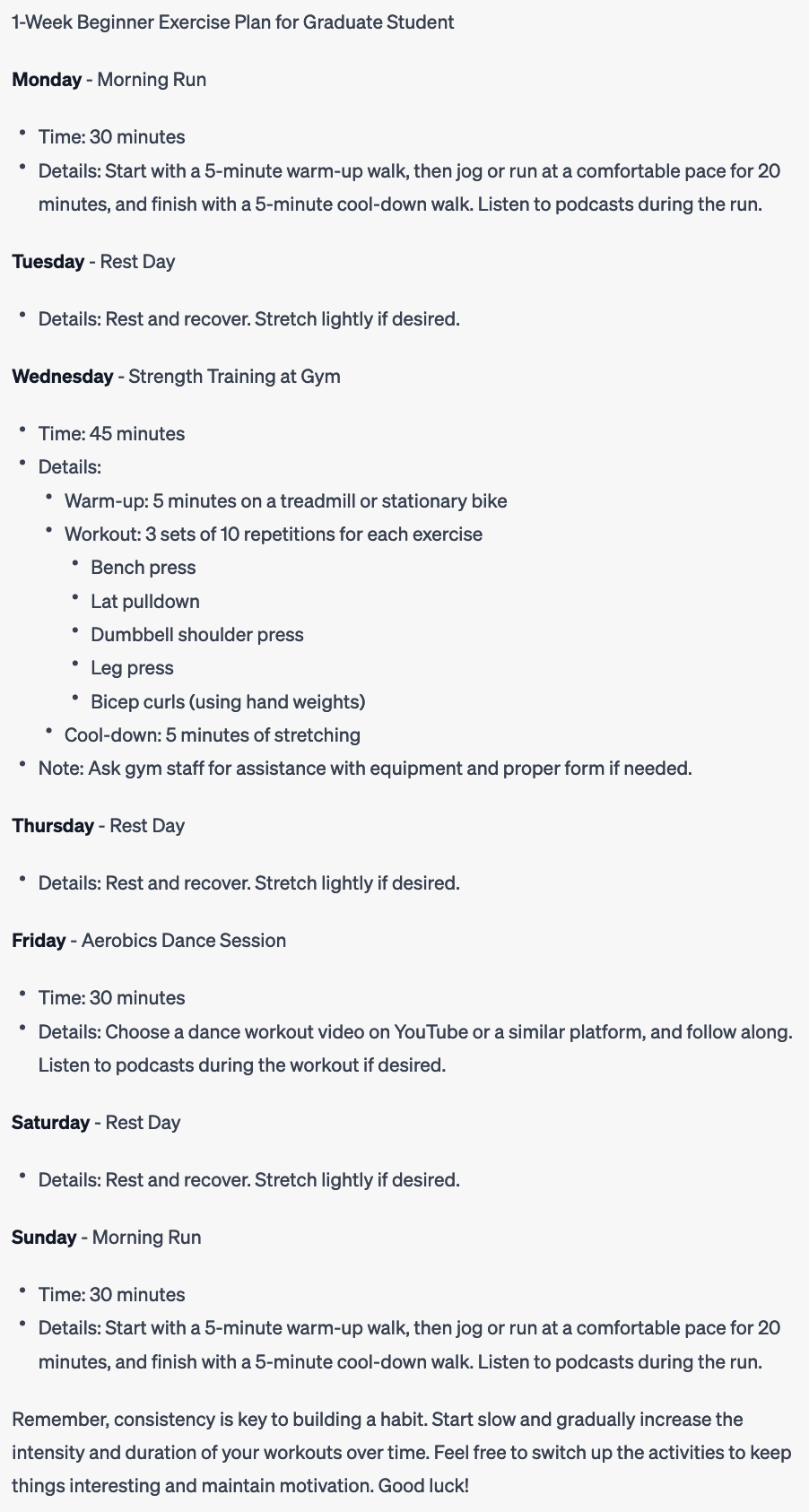}
    \caption{Exercise plan generated by GPT-4 in response to prompt from Section \ref{sec:agapie-prompt}.}
    \label{fig:agapie-GPT-output}
\end{figure*}

\end{document}